\documentclass[fleqn, 12pt]{article}

\usepackage[running]{lineno}
\usepackage{setspace}

\usepackage[top=30mm, bottom=30mm, left=30mm, right=30mm]{geometry}
\usepackage[]{amsmath}
\usepackage{amssymb, mathrsfs, bm}
\usepackage{amsthm}

\usepackage[]{graphicx}
\usepackage{natbib}
\bibpunct{(}{)}{;}{a}{,}{;}
\usepackage{authblk}
\usepackage{url}
\usepackage{titlesec}
\usepackage{cases}
\usepackage{empheq}
\usepackage{enumitem}
\usepackage{subcaption}
\usepackage{float}
\usepackage{tabularx}
\usepackage{booktabs}
\usepackage{multirow}
\usepackage{threeparttable}
\usepackage{caption}
\usepackage[hypertexnames=false, hidelinks]{hyperref}

\newif\ifarxiv
\arxivtrue

\title{Confidence Intervals for Extinction Risk: Validating Population Viability Analysis with Limited Data}

\author[1]{Hiroshi Hakoyama}
\affil[1]{Institute of Freshwater Biology, Nagano University,\par 1088 Komaki, Ueda, Nagano 386-0031, Japan \par \texttt{hiroshi-hakoyama@nagano.ac.jp}}
\date{}

\begin{document}
\maketitle

\begin{abstract}\noindent
\begin{enumerate}
\item The assessment of extinction risk remains a key component of IUCN and CITES evaluations.
However, it has been argued that, under realistic data limitations, confidence intervals (CIs) for extinction probability often span the entire 0--1 range, rendering such assessments meaningless.
I revisit the issue by analytically deriving accurate CIs under the drift--Wiener process, a canonical model of extinction dynamics.

\item In this model, extinction probability $G$ depends on the time horizon, growth rate, environmental variance, and initial population size relative to the threshold.
As functions of the original parameters, I define two transformed parameters, $w$ and $z$, whose maximum likelihood estimators follow noncentral $t$ distributions.
When the initial population is far above the threshold relative to expected drift and environmental variability ($z \gg 0$), $G$ asymptotically reduces to a function of $w$ alone, enabling exact CIs.
In contrast, for $z \ll 0$, the estimates of $w$ and $z$ are strongly negatively correlated, so I construct a CI for $G$ by evaluating $G(w,z)$ at opposite corners of the confidence rectangle for $(w,z)$.
This yields CIs with accurate nominal coverage.

\item Using this approach, I examine how the observation span of available time series affects the width of the CI for extinction probability.
A key finding is that the CI width depends on both the data and the parameters, particularly on effect size---the distance between the true $G$ and the maximally uncertain midpoint ($G\simeq 0.5$).
Even with limited time-series data, extinction probabilities that are sufficiently low or high can be reliably estimated, addressing a long-standing concern that population viability analysis (PVA) becomes unreliable under data scarcity.
I also propose an observation-error-and-autocovariance-robust (OEAR) estimator for $G(w,z)$ under additive observation error and short-run dependence.

\item Applying this method to two 64-year national harvest indices for Japanese eel (\textit{Anguilla japonica}), I find that Criterion~E extinction probabilities are far below the thresholds for the IUCN threatened categories, with narrow CIs.
Although the species is currently listed as Endangered under Criterion A (population decline), this discrepancy is consistent with theory showing that decline-based assessments can substantially overestimate extinction risk in large populations.
\end{enumerate}
\end{abstract}
 
 
\section{Introduction}
The reliability of extinction risk assessments has been debated for several decades.
In his paper \emph{Is it meaningful to estimate a probability of extinction?}, \citet{Ludwig:1999aa} pointed out that when models are poorly fitted to short time series, the CI for the extinction probability becomes wide, and warned that ignoring CIs in extinction risk assessments may lead to over-optimism.
Building on this concern, \citet{Fieberg:2000aa} investigated whether the extinction probability could be estimated with meaningful precision despite data limitations.
They concluded that accurate long-term predictions of extinction risk would require unrealistically long monitoring periods.
Using analyses based on diffusion approximations, they found that reliable estimates are possible only over short-term time horizons, typically 10--20\,\% of the monitoring period.
If this assertion were correct, then reliably estimating extinction risk over a century, as required by IUCN Criterion E for Vulnerable species, for example, would require centuries of monitoring, which is impractical.

While these theoretical concerns emphasize limitations in estimating extinction risk, empirical evidence offers a more optimistic view.
\citet{Brook:2000aa} tested PVA retrospectively on 21 long-monitored populations and found that model projections closely matched subsequent observations.
These results underscore the practical value of PVA for conservation.
\citet{ellner2002precision}, however, argued that, although PVA predictions appeared essentially unbiased when averaged across the ensemble of 21 species, the CIs for individual species remained so wide, especially when data were limited, that practical precision was lacking.
\citet{Brook:2002aa} replied that, despite this limitation, PVA remains the most transparent quantitative tool available.

A different perspective was introduced by \citet{Halley:2003aa}, who proposed that long-term trends or drift in model parameters, which are characteristic of reddened environmental variability, could actually improve the precision of extinction risk quantification.
This suggestion runs counter to earlier concerns about the high sensitivity of extinction risk to small differences in growth rate.
Halley argued that such drift weakens the dependence of extinction risk on growth rate and instead shifts it to other parameters, such as environmental variance, for which estimates may be less sensitive.
In contrast, \citet{Ellner:2003aa} critiqued this view by pointing out that although drift may mitigate sensitivity to growth rate, it also increases uncertainty in the parameter estimates themselves, particularly when the available data are limited.

Simulation studies have provided further support for the feasibility of estimating extinction risk with limited data.
\citet{holmes2007statistical} examined CIs derived from parametric bootstrap simulations of a simple extinction model, using 63 time series, each at least 30~years long.
Their simulations showed that 20--30-year time series can yield estimates with 95\,\% CIs that are substantially narrower than the full 0--1 range, particularly for declining populations and those exhibiting rapid fluctuations.
The authors further argued that forecasting a population's extinction risk does not necessarily require knowledge of population structure, as the stochastic nature of population growth means that population trajectories across various processes can often be described by a common stochastic model.
Previous work likewise demonstrates that structured models can often be approximated by simple diffusions; age-structured populations \citep{lande1988extinction} and metapopulations \citep{hakoyama2005extinction} are notable examples.
\citet{ellner2008commentary} reconciled the relatively narrow 95\,\% CIs for many vertebrate populations reported by \citet{holmes2007statistical} with earlier theoretical predictions of much wider intervals \citep{Fieberg:2000aa}.
By analytically mapping combinations of the time horizon and the initial distance to the extinction threshold, they identified when high precision can be achieved, albeit under the strong assumption that the growth rate is the sole estimated parameter and the environmental variance is fixed.
These analytical results provide initial insight into when high precision is attainable, although they do not explore how CI width varies with the risk itself.

Despite these advances, the statistical methods for estimating CIs for extinction probability are often inadequate in terms of coverage.
Approaches for constructing CIs, including the delta/logit method \citep[e.g.,][]{DENNIS:1991aa}, the percentile form of the parametric bootstrap \citep[e.g.,][]{efron1993}, and methods that ignore variance uncertainty \citep{ellner2008commentary}, often deliver poor coverage.
\citet{holmes2007statistical} also used a parametric bootstrap approach, although the details of their CI construction are not fully described, making it difficult to evaluate the coverage properties of their approach.
From a statistical perspective, a CI is the inverse of a hypothesis test, and its width is governed by both sample size and effect size, that is, a function of the parameters that quantifies the magnitude of a phenomenon \citep[e.g.,][]{cohen1988statistical}.
The key question is therefore not simply how many data points are needed, but which model quantities serve as effect sizes and, for each such effect size, how much data are required to attain a desired level of confidence.
In extinction models, I examine the possibility that a function of the extinction probability serves as an effect size: when risks are very high or very low, the parameter deviates strongly from its null value, permitting narrow CIs, as illustrated in \citet[][Figure~8e]{holmes2007statistical}.
Here, I derive an analytical CI with near-nominal coverage for the extinction probability under the Wiener process with drift, explicitly incorporating the effect size perspective to clarify when reliable estimation is feasible with limited data.
I then extend the framework to settings with additive observation error and short-run dependence by introducing an observation-error-and-autocovariance-robust (OEAR) effective-diffusion estimator based on the long-run variance.
I apply these methods to assess the extinction risk of the Japanese eel (\textit{Anguilla japonica}).
 
\section{Materials and Methods}
\label{sec:methods}

\begin{figure}[tb]
\centering
\includegraphics[width=1\linewidth]{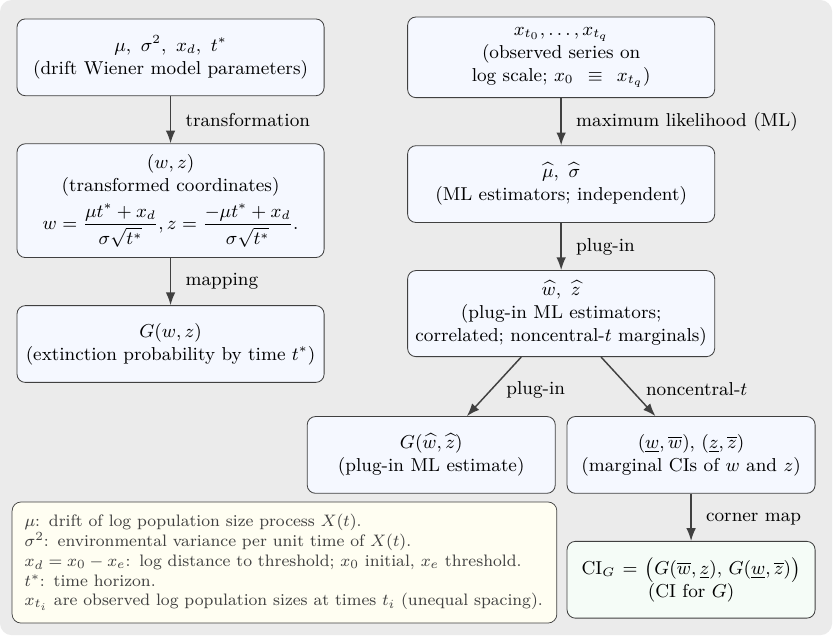}
\caption{The $w$--$z$ method for CIs of the extinction probability $G(w,z)$. Schematic overview of the $w$--$z$ transformation and the mapping to $G(w,z)$ (left), and estimation from an observed time series (right). Hats denote maximum-likelihood estimators, and underlines/overlines denote confidence limits. For OEAR, replace $\widehat{\sigma}^2$ by the long-run-variance-based estimator $\widetilde{\sigma}^2$ and use the corresponding $\widetilde{G}$.}
\label{fig:schematic_wz_ci}
\end{figure}

I model log population size as a Wiener process with drift and summarize the parameter transformation to $(w,z)$ and the associated extinction probability map (Section~\ref{sec:model}; Figure~\ref{fig:schematic_wz_ci}, left).
CIs for the extinction probability are constructed using the $w$--$z$ method developed here (Section~\ref{sec:CI}; Figure~\ref{fig:schematic_wz_ci}, right).
For settings with additive observation error and short-run dependence, I also consider an OEAR extension based on the long-run variance (Section~\ref{sec:model_relaxations}; Appendix~\ref{appendix:obs_error}).
The methods developed here are implemented in the accompanying R package \texttt{extr}, available from \href{https://doi.org/10.32614/CRAN.package.extr}{CRAN}.
Numerical calculations were performed in R (version~4.5.1); Mathematica was used to solve the implicit equations defining the required observation span.
I used ChatGPT (OpenAI, GPT-5 Thinking; accessed Aug--Sep 2025) to assist with code refactoring and phrasing; I verified all outputs and take full responsibility, and no copyrighted third-party text was incorporated.

\section{Model and Analytical Properties}\label{sec:model}
\subsection{Stochastic differential equation}

Population dynamics are modeled by a stochastic differential equation that describes the change in population size $N \in (0, \infty)$ over continuous time $t \in [0, \infty)$.
The equation is given by
\begin{equation}\label{eq:abs_SDE}dN = r N\, dt + \sigma N\, dW,
\end{equation}
where $r \in \mathbb{R}$ is the instantaneous per-capita growth rate, and $\sigma \in (0, \infty)$ denotes the intensity of environmental stochasticity.
The environmental variance is given by $\sigma^2$.
The term $W(t)$ denotes a Wiener process modeling continuous-time random fluctuations due to environmental stochasticity.
The stochastic term $\sigma N\, dW$ is interpreted in the It\^{o} sense, which results in different long-term behavior from the Stratonovich interpretation \citep[see e.g.,][]{Hakoyama:2000aa}.
This model does not account for density dependence, and is therefore more appropriate for non-stationary populations (those expected to grow or decline) than for populations near equilibrium, for which density-dependent models are preferable \citep[e.g.,][]{Hakoyama:2000aa}.

To facilitate analysis on a log scale, the natural logarithm of the population size is taken, $X = \log N$.
Applying It\^{o}'s lemma to Equation~(\ref{eq:abs_SDE}) yields the following simplified equation, which describes a Wiener process with drift:
\begin{equation}\label{eq:log_SDE}dX = \mu\, dt + \sigma\, dW,
\end{equation}
where $\mu = r - \tfrac{1}{2} \sigma^2$ is the drift parameter.
Equation~(\ref{eq:log_SDE}) has been extensively used as a simple extinction model \citep[e.g.,][]{lande1988extinction, DENNIS:1991aa}.

Equations (\ref{eq:abs_SDE}) and (\ref{eq:log_SDE}) represent the same stochastic process, expressed on the absolute scale (a geometric Brownian motion) and on the log scale (a Wiener process with drift), respectively.
The expectations of these processes evolve as follows:
\begin{equation*}
\mathbb{E}[N(t)] = n_0\,e^{rt}, \qquad
\mathbb{E}[X(t)] = x_0 + \mu t,
\end{equation*}
where $n_0$ and $x_0=\log n_0$ are the initial population sizes on the absolute and log scales, respectively.
Thus, $r$ governs the expected rate of change on the absolute scale (actual population numbers), whereas $\mu$ is the corresponding drift on the log scale.

In applications, the growth rate $r$ estimated from time-series data reflects both intrinsic population potential and external influences such as exploitation and environmental change.
When time-varying covariates are available, $r$ can be modeled as covariate-dependent to improve prediction \citep[e.g.,][]{dennis2000joint}.

\subsection{Extinction probability}

In practice, extinction is typically defined not as the population reaching exactly zero, but as falling below a small, biologically meaningful threshold (e.g., $n_e = 1$, representing a single individual).
This is because, under Equation~(\ref{eq:abs_SDE}), true extinction ($N = 0$) is theoretically unreachable: although $N(t)$ may asymptotically approach zero due to stochastic fluctuations, the noise term $\sigma N\, dW$ diminishes proportionally with population size, preventing extinction from occurring in finite time.
Under this definition, extinction is said to occur when the population size $N(t)$ falls at or below a threshold $n_e$, corresponding to $X(t) \leq \log(n_e)$ on the log scale.

Let $T$ denote the first-passage time at which the population size, started from $n_0$, first hits the lower threshold $n_e$, with $T=\infty$ if the threshold is never reached.
The probability that extinction occurs by time $t > 0$, given the growth rate $\mu \in \mathbb{R}$, environmental variance $\sigma^2 > 0$, and initial log-distance $x_d = \log(n_0/n_e) > 0$, is given by
\begin{equation}\label{eq:G_finite}\Pr[T \leq t] = G(t \mid x_d, \mu, \sigma^2) = \Phi\left( \frac{-\mu t - x_d}{\sigma \sqrt{t}} \right) + \exp \left( \frac{-2 \mu x_d}{\sigma^2} \right) \Phi\left( \frac{\mu t - x_d}{\sigma \sqrt{t}} \right),
\end{equation}
where $\Phi(\cdot)$ denotes the cumulative distribution function (CDF) of the standard normal distribution.
This expression originates from classical results on first-passage times of diffusion processes \citep[e.g.,][]{cox1965theory} and has been applied to extinction problems in ecology \citep[e.g.,][]{lande1988extinction}.

As shown by \citet{lande1988extinction}, the ultimate extinction probability is given by
\begin{subequations} \label{eq:G_infinite}\begin{empheq}[left = {G(\infty \mid x_d, \mu, \sigma^2) = \empheqlbrace \,}]{alignat = 2}
    & 1 &\qquad& \text{for} \qquad \mu \leq 0, \label{eq:G_infinite_a} \\
    & \exp\left( -\dfrac{2 \mu x_d}{\sigma^2} \right) & & \text{for} \qquad \mu > 0. \label{eq:G_infinite_b}
  \end{empheq}
\end{subequations}

That is, if the growth rate is non-positive, extinction is certain in the long run; if it is positive, the population may persist indefinitely.
Accordingly, the extinction probability $G(t \mid x_d, \mu, \sigma^2)$ in Equation~(\ref{eq:G_finite}) is a proper CDF for $\mu \leq 0$, but an improper one for $\mu > 0$.

The proper conditional distribution of extinction times, given that extinction eventually occurs, is obtained by dividing the unconditional extinction probability in Equation~(\ref{eq:G_finite}) by the ultimate extinction probability in Equation~(\ref{eq:G_infinite}): $\Pr[T \leq t \mid T < \infty]$.
This conditional distribution is the cumulative distribution function of the inverse Gaussian distribution.
However, when $\mu > 0$ and the ultimate extinction probability is sufficiently small, the unconditional extinction probability $\Pr[T \leq t]$ accurately reflects the true extinction risk over time, whereas the conditional distribution $\Pr[T \leq t \mid T < \infty]$ may significantly overestimate the extinction risk.
For example, with $\mu = 0.1$, $\sigma^2 = 0.1$, $n_0 = 100$, and $n_e = 1$, the unconditional extinction probability by $t = 50$ is extremely low, $G(50) = 6.6 \times 10^{-5}$, while the conditional probability among populations that eventually go extinct is much higher, $G(50)/G(\infty) = 0.66$.
In other words, while nearly all populations persist, the few that happen to be destined for extinction tend to disappear quickly.
For this reason, extinction risk is evaluated using the unconditional probability given by Equation~(\ref{eq:G_finite}), which, as in \citet{lande1988extinction, Fieberg:2000aa, ellner2008commentary}, accurately reflects the practical extinction risk relevant to finite-time assessments, such as those used by the IUCN.

\subsection{Extinction--risk landscape in transformed parameter space}

As in Equation~\eqref{eq:G_finite}, the extinction probability $G$ depends nonlinearly on all three biological parameters (the growth rate $\mu$, the environmental variance $\sigma^{2}$, and the initial log-distance $x_{d}$) and the time horizon $t$.
Here, $\mu$ and $\sigma^2$ have units of time$^{-1}$, while $x_d$ is dimensionless.
For analytical convenience and CI construction, I adopt the $(w,z)$ transformation rather than the earlier $(U,V)$ rescaling with $U=-\mu\sqrt{t}/\sigma$ and $V=x_d/(\sigma\sqrt{t})$ \citep{Fieberg:2000aa, ellner2008commentary}.
For a finite horizon $t>0$, define
\begin{equation}\label{eq:wz}w = \frac{\mu t + x_d}{\sigma\sqrt{t}}, \qquad
z = \frac{-\,\mu t + x_d}{\sigma\sqrt{t}}.
\end{equation}
Each variable is a signal-to-noise ratio expressed in dimensionless form: the numerator represents net deterministic displacement toward or away from the threshold, and the denominator, $\sigma \sqrt{t}$, characterizes the scale of stochastic fluctuation.
The transformation from $(\mu t, x_d)$ to $(w, z)$ can be interpreted geometrically as a $45^\circ$ clockwise rotation, followed by isotropic scaling by $\sqrt{2}/(\sigma\sqrt{t})$:
\begin{equation*}
\begin{pmatrix}
w \\
z
\end{pmatrix}
=
\frac{1}{\sigma \sqrt{t}}
\begin{pmatrix}
1 & 1 \\
-1 & 1
\end{pmatrix}
\begin{pmatrix}
\mu t \\
x_d
\end{pmatrix}.
\end{equation*}
Applying the change of variables in Equation~\eqref{eq:wz} to Equation~\eqref{eq:G_finite} yields a compact expression for the extinction probability:
\begin{equation}\label{eq:G_wz}G(w, z) = \Phi(-w) + \exp\left(\frac{z^2 - w^2}{2}\right)\Phi(-z)
  \quad \text{for} \quad w + z > 0, \quad w, z \in \mathbb{R},
\end{equation}
which depends solely on the transformed coordinates $(w,z)$ and expresses extinction probability as a function of signal-to-noise ratios.
The condition $w + z > 0$ follows from $w + z = 2x_d / (\sigma\sqrt{t})$, and reflects the assumption that the initial population size $n_0$ exceeds the extinction threshold $n_e$.
The introduced parameters $w$ and $z$ can be estimated via maximum likelihood, and their estimators have tractable sampling distributions, specifically noncentral $t$, as derived in Section~\ref{sec:MLE_wz}.
This tractability underlies the CI construction in Section~\ref{sec:CI}; stable evaluation of $G(w,z)$ is discussed in Appendix~\ref{appendix:numerical_stability_G}.

\begin{figure}[H]
\centering
\includegraphics[width=0.7 \linewidth]{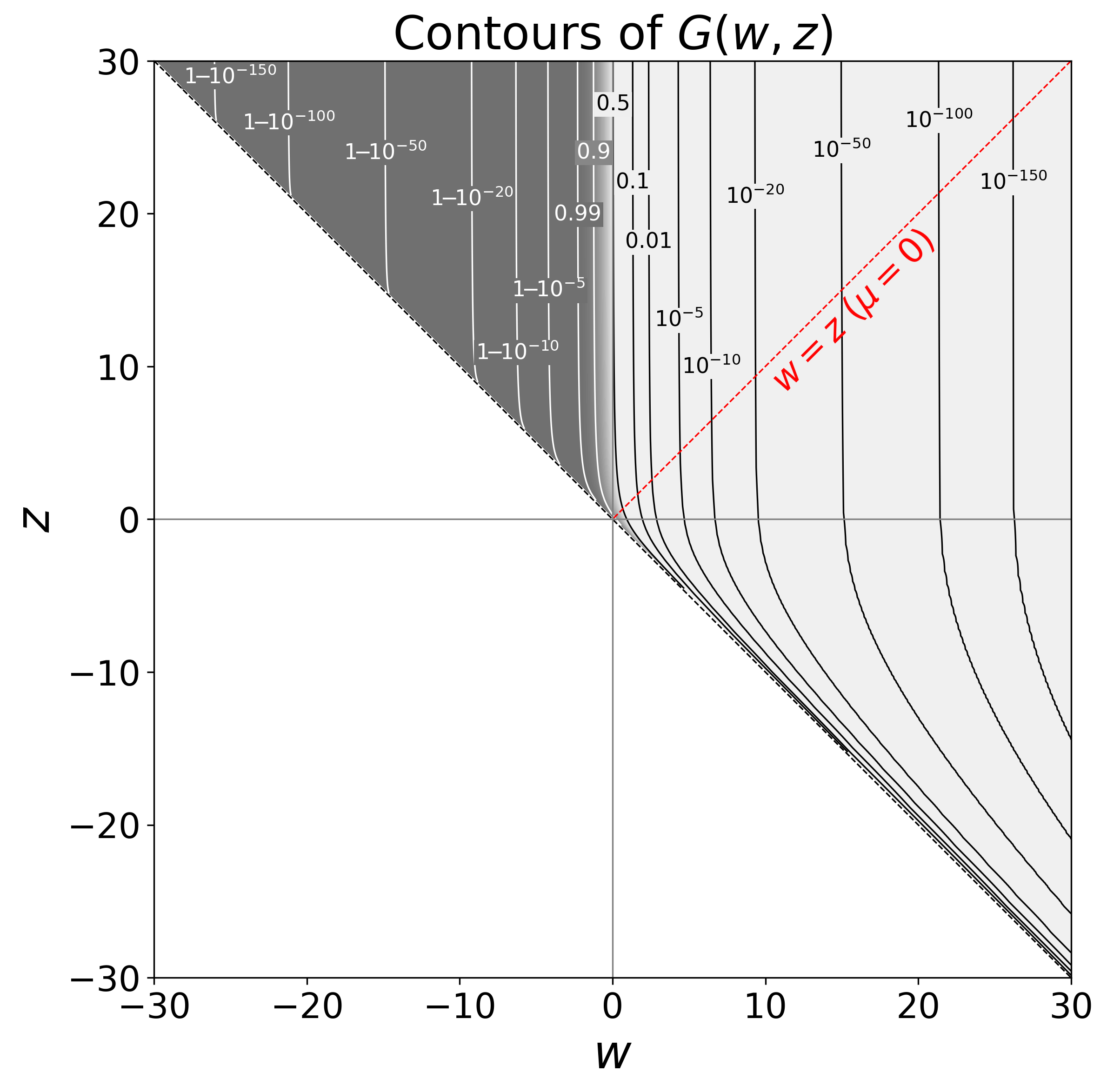}
\caption{
Contour plot of extinction probability $G(w, z)$ in the $(w, z)$ plane, restricted to $w + z > 0$.
Shading indicates extinction risk (light: low, dark: high), with contours labeled by $G(w, z)$.
The red dotted line ($w = z$) separates regions of positive drift ($\mu > 0$, below) and negative drift ($\mu < 0$, above).
}
\label{fig:G_ws}\end{figure}

While the function $G(w, z)$ is strictly decreasing in both arguments throughout the domain $w + z > 0$ (Appendix~\ref{appendix:G_monotonicity}), the nature of this decline is not uniform across the domain.
In the two extreme regimes within the region $w + z > 0$, one can apply the leading-order Mills ratio approximation $\Phi(-x) \sim \phi(x)/x$ as $x \to +\infty$ (with $\phi(x) = \tfrac{1}{\sqrt{2\pi}} \mathrm{e}^{-x^2/2}$) to obtain the following asymptotic form:
\begin{empheq}[left = {G(w,z) = \empheqlbrace \,}]{alignat*=2}
  & \Phi(-w) &\qquad& \text{for } \qquad z \gg 0, \\
  & \exp\left( \dfrac{z^2 - w^2}{2} \right) & & \text{for } \qquad z \ll 0.
\end{empheq}
When $z \gg 0$, the second term in Equation~\eqref{eq:G_wz} becomes negligible, and extinction risk is determined primarily by $w$.
Conversely, when $z \ll 0$, the second term $\exp((z^{2}-w^{2})/2)[1+o(1)]$ dominates; hence $w^{2}-z^{2}=\text{const}$ along contours of constant $G(w,z)$.
Along such contours in $z \ll 0$, the derivative satisfies ${dw}/{dz} = z/w \to -1$, indicating that extinction risk is most sensitive to changes along the diagonal direction $z \approx -w$.
This asymmetry reflects the distinct ways in which drift and stochasticity interact to shape extinction risk.
See Appendix~\ref{appendix:asymptotics_G} for a derivation of this asymptotic behavior.
Notably, \(\exp((z^2-w^2)/2)=\exp(-2\mu x_d/\sigma^2)\), so in the $z\ll0$ regime the asymptotic form coincides with the ultimate extinction probability in Equation~\eqref{eq:G_infinite_b}, as shown by \citet{lande1988extinction}.

Figure~\ref{fig:G_ws} shows that the equal-risk contours are nearly vertical over a broad range of positive $z$ values ($z > 0$; equivalently, $x_d > \mu t$), indicating that the extinction probability is primarily governed by horizontal variation in $w$.
This confirms that the approximation $G(w, z) \simeq \Phi(-w)$ holds well in this region.
In contrast, for $z<0$ the contours lie close to the diagonal direction and are steeper than $w=-z$, with slopes approaching $-1$ as $z\to-\infty$, consistent with Appendices~\ref{appendix:asymptotics_G} and \ref{appendix:slope_G}.
This pattern aligns with Appendix~\ref{appendix:asymptotics_G}: $G(w,z)\simeq\Phi(-w)$ for $z\gg0$ with relative error $O(1/z)$, whereas $G(w,z)\simeq \exp((z^2-w^2)/2)$ is exponentially accurate for $z\ll0$.
These asymptotic accuracies support their use in the construction and interpretation of CIs.
As evident in Figure~\ref{fig:G_ws}, the equal-risk contours are steeper than the diagonal $w=-z$ throughout $w+z>0$, and their slopes approach $-1$ from below as $z\to-\infty$, consistent with the analytical result $\,\mathrm{d}z/\mathrm{d}w<-1$ on $w+z>0$ proved in Appendix~\ref{appendix:slope_G}.
 
\section{Estimation and Confidence Intervals}\label{sec:CI}
To quantify extinction risk from time-series data, I derive estimators for the parameters of the diffusion model and construct CIs for the finite-horizon extinction probability.
The analysis is based on maximum likelihood (ML) estimation and rescaled coordinates that lead to a tractable sampling distribution.
This section develops the estimation procedure, derives the associated sampling laws, and describes a method for constructing CIs without linear approximation.

\subsection{Maximum likelihood estimation of \texorpdfstring{$\mu$}{mu} and \texorpdfstring{$\sigma^{2}$}{sigma-squared}}\label{sec:MLE_mu_sigma}

Let $\mathbf{n}_{\mathrm{obs}}=\{n_{t_0}, n_{t_1}, \dots, n_{t_q}\}$ be a univariate time series observed at unequally spaced times $t_0 = 0 < t_1 < \dots < t_q$, arising from the stochastic process in Equation~(\ref{eq:abs_SDE}).
Define the log-transformed values $x_i = \log n_{t_i}$ and denote the sampling intervals by $\tau_i = t_i - t_{i-1} > 0$.
Let $X_i = X(t_i)$ denote the corresponding random variables.
Under model~\eqref{eq:log_SDE}, the increments $X_i-X_{i-1}$ are independent and satisfy
\begin{equation*}
  X_i - X_{i-1} \sim \mathcal{N}(\mu\,\tau_i,\, \sigma^{2}\tau_i)
  \quad \text{for } i = 1, \dots, q.
\end{equation*}
Here $q$ denotes the number of increments (so the sample size of the series is $q+1$), and $t_q$ is the length of the observation span.

The Gaussian likelihood based on these increments leads to the ML estimators \citep[][Equations~24--26]{DENNIS:1991aa}
\begin{equation}\label{eq:mle_orig}\widehat{\mu}
  = \frac{\sum_{i=1}^{q} (x_i - x_{i-1})}{\sum_{i=1}^{q} \tau_i}
  = \frac{x_q - x_0}{t_q},
  \qquad
  \widehat{\sigma}^{2}
  = \frac{1}{q} \sum_{i=1}^{q} \frac{(x_i - x_{i-1} - \widehat{\mu}\,\tau_i)^2}{\tau_i}.
\end{equation}

The sampling distributions of the estimators follow from standard arguments:
\begin{equation}\label{eq:sampling_muhat_sigmahat}\widehat{\mu} \sim \mathcal{N}\left(\mu,\, \frac{\sigma^{2}}{t_q} \right),
  \qquad
  \frac{q\,\widehat{\sigma}^{2}}{\sigma^{2}} \sim \chi^{2}_{q - 1}, \quad q>1,
\end{equation}
with $\widehat{\mu}$ and $\widehat{\sigma}^{2}$ mutually independent, a property unique to the normal distribution \citep[][]{geary1936distribution}.

\subsection{Sampling distributions of the rescaled statistics \texorpdfstring{$\widehat{w}$}{w-hat} and \texorpdfstring{$\widehat{z}$}{z-hat}}\label{sec:MLE_wz}

Applying the transformation in Equation~\eqref{eq:wz} with $t = t^{\ast}$, the finite time horizon for extinction risk prediction, to the ML estimators in Equation~\eqref{eq:mle_orig} yields the following rescaled statistics.
In the error-free formulation considered here, prediction starts at the terminal observation, so I set $x_d:=x_{t_q}-x_e$ and treat it as known when constructing $(w,z)$ and $G(w,z)$.
\begin{equation}\label{eq:wz_hat}\widehat{w} =
    \frac{\widehat{\mu}\,t^{\ast} + x_d}{\widehat{\sigma}\sqrt{t^{\ast}}},
  \qquad
  \widehat{z} =
    \frac{-\,\widehat{\mu}\,t^{\ast} + x_d}{\widehat{\sigma}\sqrt{t^{\ast}}}, \qquad q>1.
\end{equation}

Since $\widehat{\mu}$ is normally distributed, $\widehat{\sigma}^{2}$ is scaled-$\chi^{2}$, and the two are independent, each follows a noncentral $t$ distribution with $q-1$ degrees of freedom ($q>1$):
\begin{equation}\label{eq:wz_tdist}
  \widehat{w}\,\sqrt{\frac{q-1}{q}}\,\sqrt{\frac{t_q}{t^{\ast}}}
  \sim t(\delta_w, q-1),
  \qquad
  \widehat{z}\,\sqrt{\frac{q-1}{q}}\,\sqrt{\frac{t_q}{t^{\ast}}}
  \sim t(\delta_z, q-1),
\end{equation}
where the noncentrality parameters are
\begin{equation}\label{eq:wz_delta}\delta_w = w \,\sqrt{\frac{t_q}{t^{\ast}}},
  \qquad
  \delta_z = z \,\sqrt{\frac{t_q}{t^{\ast}}}.
\end{equation}

\subsection{Correlation between \texorpdfstring{$\widehat{w}$}{w-hat} and \texorpdfstring{$\widehat{z}$}{z-hat}}\label{sec:wz_corr}
To construct reliable CIs for $G(w,z)$ it is necessary to understand how the estimators $\widehat{w}$ and $\widehat{z}$ vary together.
This joint behavior is described by their covariance matrix, but it can be summarized more compactly through a closed-form expression of the correlation coefficient after some algebra (see Appendix~\ref{appendix:wz_corr_details}):
\begin{equation}\label{eq:corr-k}\rho(w,z,k) \coloneqq \operatorname{Corr}(\widehat{w},\widehat{z})
  =\frac{-k+w z}{\sqrt{(k+w^{2})(k+z^{2})}}, \qquad w+z>0,\; k>0.
\end{equation}
where
\begin{equation}\label{eq:k-def}k = \frac{t^{\ast}/t_q}{\,1-\tfrac12\,(q-3)\left\{\Gamma\!\left[(q-2)/2\right]/\Gamma\!\left[(q-1)/2\right]\right\}^{2}\,}
    \;\approx\; \frac{2q t^{\ast}}{t_q},
    \quad (q \ \text{moderately large}).
\end{equation}
Here $k$ depends only on $(q,t_q,t^{\ast})$ and is strictly positive for $q>3$.

\begin{figure}[H]
\centering
\includegraphics[width=0.84\linewidth]{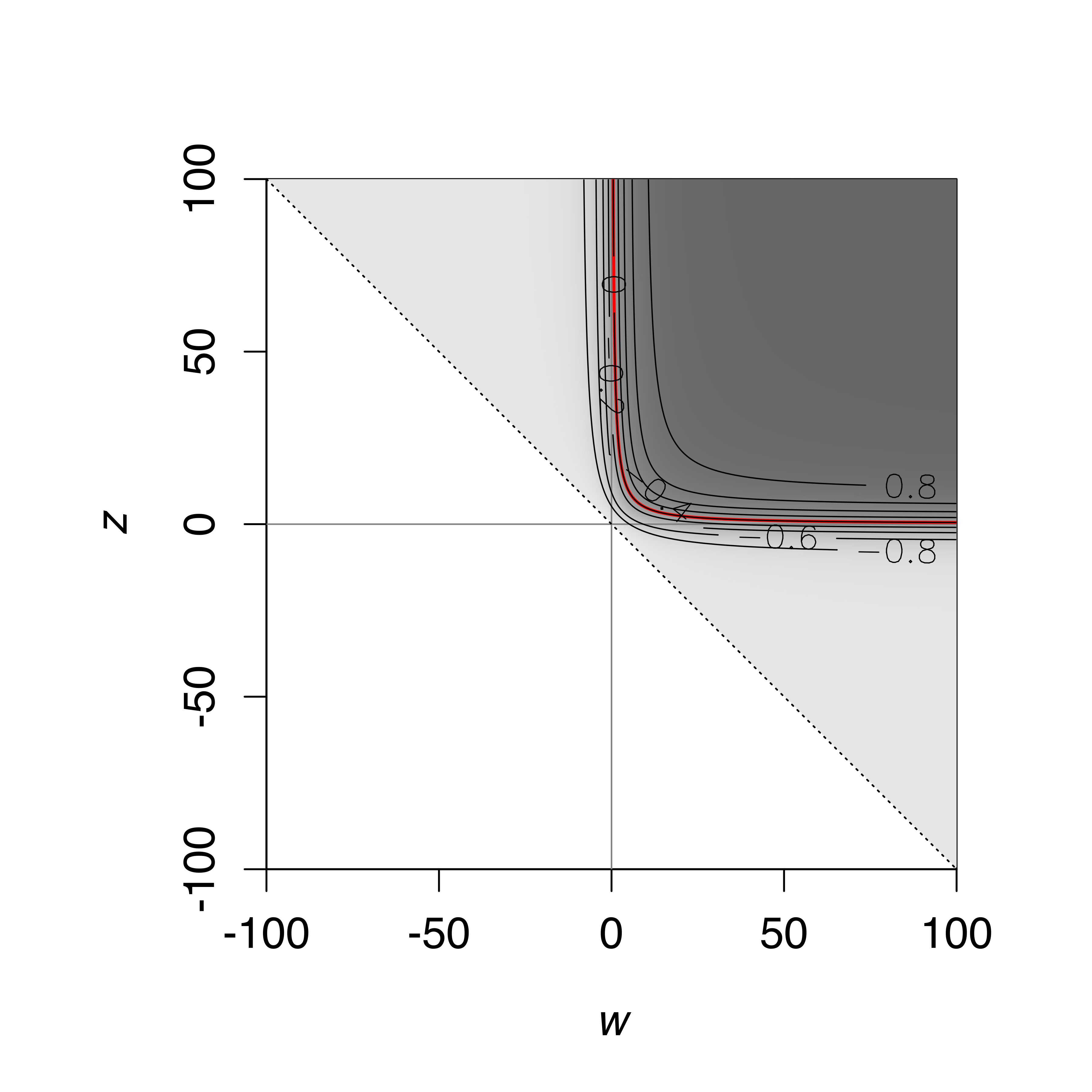}
\caption{Correlation landscape of ML estimators $\widehat{w}$ and $\widehat{z}$ in the $(w, z)$ plane, evaluated for $q = t_q = 63$ and time horizon $t^\ast = 25$. Shading indicates $\operatorname{Corr}(\widehat{w}, \widehat{z})$, from negative (light gray) to positive (dark gray). The red curve shows the zero-correlation hyperbola $w z = k$, where $k \approx 47.82$ (Equation~\ref{eq:k-def}). Contours follow Equation~\eqref{eq:corr-k}. }
\label{fig:ws_corr}
\end{figure}

Equation~\eqref{eq:corr-k} reveals a simple geometry within the admissible region $w+z>0$.
In the first quadrant ($w, z>0$), the sign changes along the hyperbola $w z=k$: inside this curve ($w z<k$) the correlation is negative, while outside ($w z>k$) it becomes positive.
In the second and fourth quadrants ($wz \le 0$) the correlation is negative throughout.
As $|w|$ and $|z|$ grow large within $w+z>0$, the magnitude of the correlation approaches one, with the sign determined by $wz$.
In particular, for $z<0$ the correlation is strongly negative; this feature is exploited later when constructing CIs for $G(w,z)$.
Figure~\ref{fig:ws_corr} illustrates the analytic structure implied by Equation~\eqref{eq:corr-k}.

The constant $k$ increases with the number of increments $q$ and time horizon $t^\ast$, and decreases with observation span $t_q$.
Because the sign change occurs on the curve $wz=k$, larger values of $k$ move the zero-correlation hyperbola farther from the origin, shrinking the region of positive correlation in the first quadrant and strengthening the negative correlation elsewhere.
These effects are quantified analytically in Appendix~\ref{appendix:wz_corr_details}.

\subsection{Confidence intervals for \texorpdfstring{$w$}{w} and \texorpdfstring{$z$}{z}}
Let $\mathbf{N} = \{N_{t_0}, \dots, N_{t_q}\}$ be the random vector of observations from Equation~(\ref{eq:abs_SDE}), and $\mathbf{n}_{\mathrm{obs}}$ its realization.
An equal-tailed $(1-\alpha)\times 100\%$ CI $(\underline{w}, \overline{w})$ for $w$ is defined by
\begin{equation*}
  \Pr\bigl( w < \underline{w}(\mathbf{N}) \bigr)
  = \Pr\bigl( w > \overline{w}(\mathbf{N}) \bigr) = \alpha/2.
\end{equation*}
Equivalently, in terms of the observed estimator $\widehat{w}(\mathbf{n}_{\mathrm{obs}})$,
\begin{equation*}
\begin{aligned}
  \Pr\!\left( \widehat{w}(\mathbf{N}) > \widehat{w}(\mathbf{n}_{\mathrm{obs}})
  \,\mid\, w = \underline{w}(\mathbf{n}_{\mathrm{obs}}) \right) &= \alpha/2, \\
  \Pr\!\left( \widehat{w}(\mathbf{N}) < \widehat{w}(\mathbf{n}_{\mathrm{obs}})
  \,\mid\, w = \overline{w}(\mathbf{n}_{\mathrm{obs}}) \right) &= \alpha/2,
\end{aligned}
\end{equation*}
because $\widehat{w}(\mathbf{N})$ is monotone increasing in $w$.

Using the noncentral-$t$ distribution (Eqs.~\eqref{eq:wz_hat}--\eqref{eq:wz_delta}), let
\begin{equation*}
  s_{\mathrm{obs}}
  = \widehat{w}(\mathbf{n}_{\mathrm{obs}})\,\sqrt{\tfrac{q-1}{q}}\,\sqrt{\tfrac{t_q}{t^{\ast}}},
\end{equation*}
and denote the noncentrality parameter for a given $w$ by $\delta$.
The bounds $\underline{\delta}_{\mathrm{obs}},\overline{\delta}_{\mathrm{obs}}$ solve
\begin{equation*}
  \Pr(S > s_{\mathrm{obs}} \,\mid\, \delta = \underline{\delta}_{\mathrm{obs}}) = \alpha/2,
  \qquad
  \Pr(S < s_{\mathrm{obs}} \,\mid\, \delta = \overline{\delta}_{\mathrm{obs}}) = \alpha/2,
\end{equation*}
with $S \sim t(\delta,q-1)$.
Since the noncentral-$t$ distribution is stochastically increasing in $\delta$, the solutions are unique, and the bounds for $w$ follow as
\begin{equation*}
  \underline{w}=\underline{\delta}_{\mathrm{obs}}\sqrt{\tfrac{t^{\ast}}{t_q}},\qquad
  \overline{w}=\overline{\delta}_{\mathrm{obs}}\sqrt{\tfrac{t^{\ast}}{t_q}}.
\end{equation*}

The CI for $z$ is constructed analogously.

\subsection{Confidence intervals for the extinction probability \texorpdfstring{$G$}{G}}\label{subsec:w-z_method}

A $(1-\alpha)\times 100\%$ confidence interval for $G=G(w,z)$ is defined as
\begin{equation*}
  \mathrm{CI}_{G}
  =
  \Bigl(
    G(\overline{w},\,\underline{z}),
    \;
    G(\underline{w},\,\overline{z})
  \Bigr),
\end{equation*}
hereafter referred to as the $w$--$z$ method.

\paragraph{Simplification for large $z$.}
When $z$ is sufficiently large, the exponential term in Equation~\eqref{eq:G_wz} becomes negligible, and $G(w,z)$ is well approximated by $\Phi(-w)$.
The CI then simplifies to
\begin{equation*}
  \mathrm{CI}_{G} = \bigl( \Phi(-\overline{w}),\; \Phi(-\underline{w}) \bigr),
\end{equation*}
which avoids the need for evaluating exponentials and is asymptotically exact as $z\to+\infty$.

\begin{figure}[H]
\centering
\begin{subfigure}[t]{0.48\linewidth}
  \centering
  \includegraphics[height=0.24\textheight,keepaspectratio]{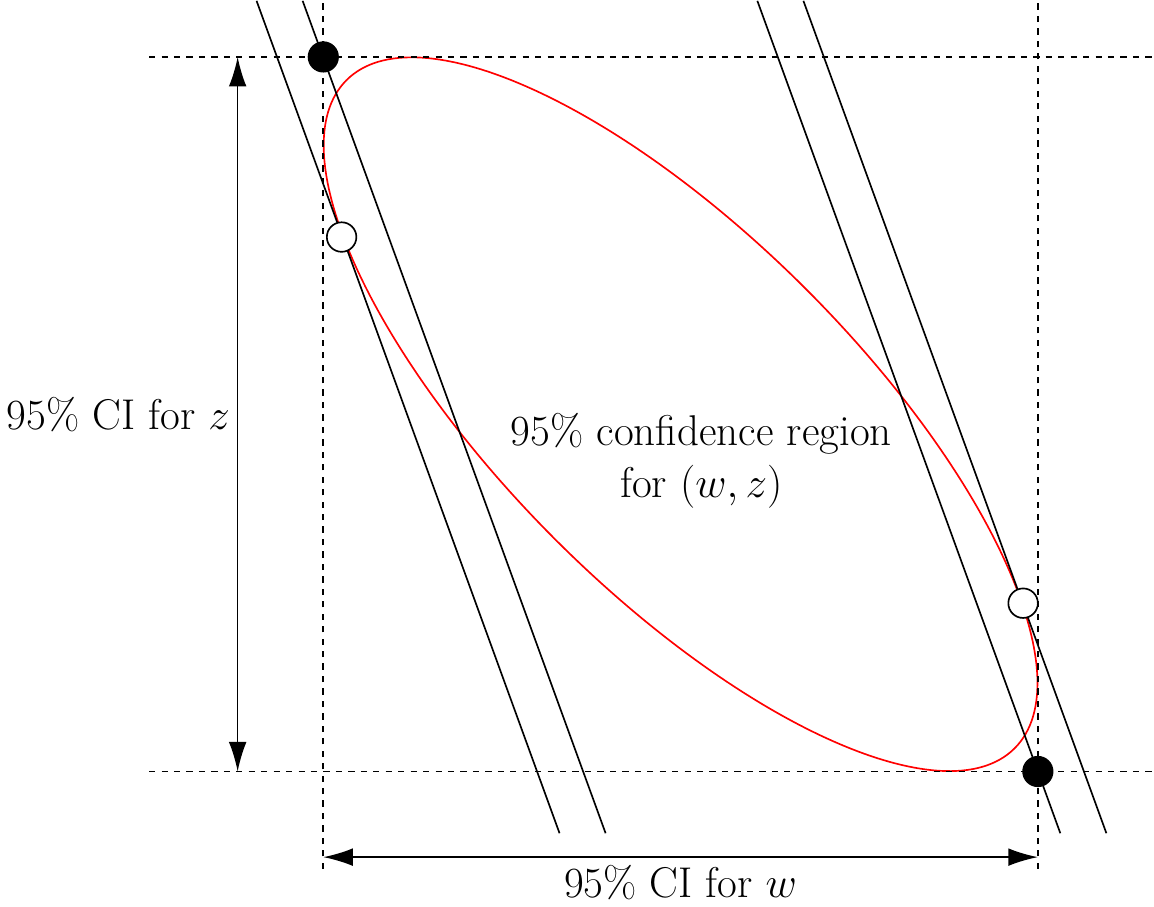}
  \caption{}
  \label{fig:wz_ci_region_concept}
\end{subfigure}\hfill
\begin{subfigure}[t]{0.48\linewidth}
  \centering
  \includegraphics[height=0.24\textheight,keepaspectratio]{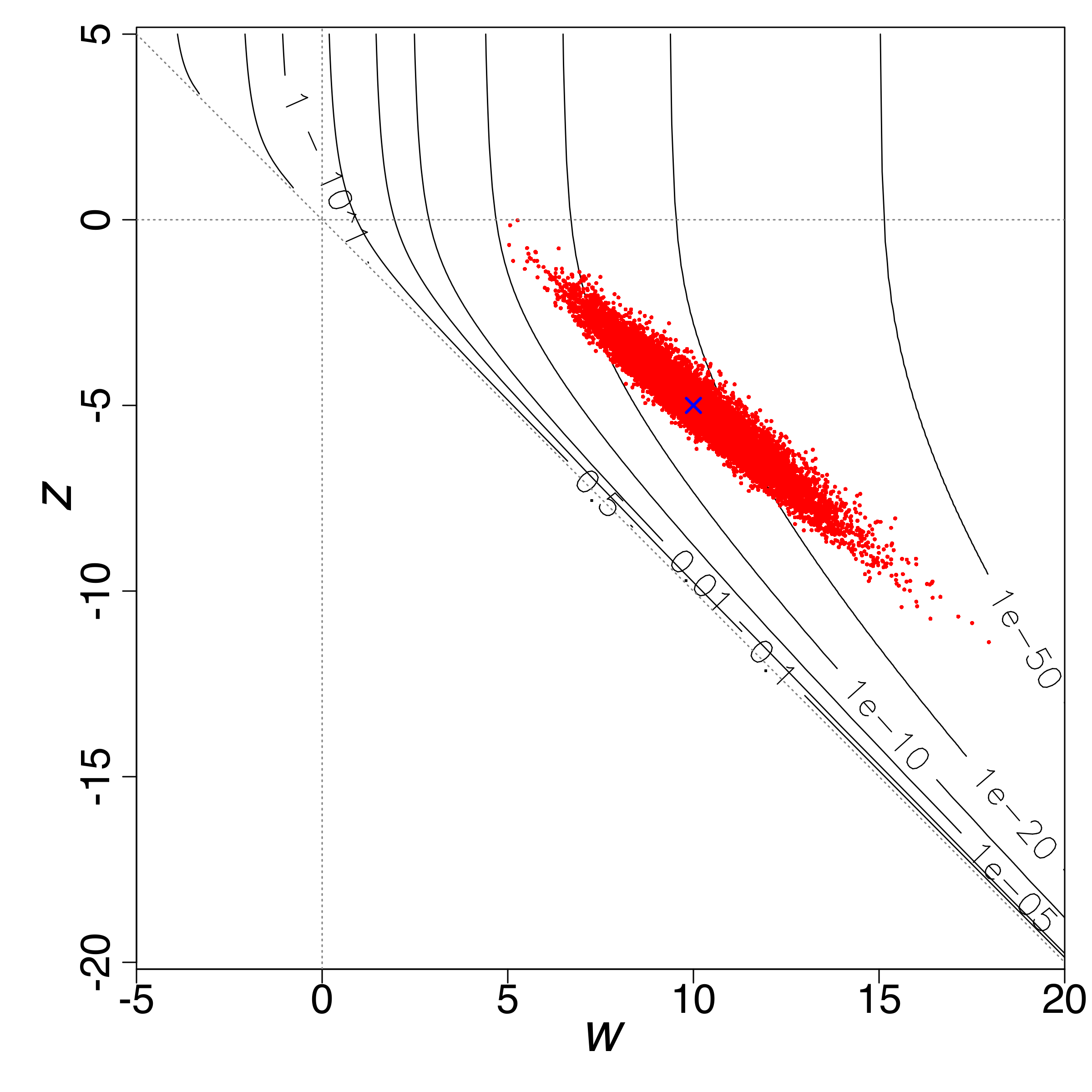}
  \caption{}
  \label{fig:wz_ci_region_sample}
\end{subfigure}
\caption{
Geometry underlying the $w$--$z$ method when $z < 0$.
(a) The red ellipse conceptually illustrates the joint 95\% confidence region for the parameters $(w, z)$ when $z \ll 0$, typically elongated along slope $-1$ due to negative correlation between the ML estimators $\widehat{w}$ and $\widehat{z}$.
The extinction probability $G(w,z)$ is shown as contours of constant $G$.
As shown analytically in Appendix~\ref{appendix:slope_G}, these contours have slope strictly less than $-1$, implying they are steeper than the ellipse orientation.
Open white circles mark the two points where the ellipse is tangent to a $G$ contour, determining the exact CI for $G$.
Filled black circles mark the diagonally opposite corners of the dashed rectangle formed by the 95\% CIs for $w$ and $z$, which approximate the CI for $G$ (the $w$--$z$ method).
(b) Monte Carlo sample cloud of $(\widehat{w},\widehat{z})$ for $(w,z) = (10, -5)$, overlaid on $G$ contours.
$q = t_q = 63$ and time horizon $t^\ast = 100$.
Although not a true confidence region, this bootstrap-type sample cloud approximates the joint confidence region for $(w, z)$.
}
\label{fig:wz_ci_region}
\end{figure}

\paragraph{Justification for small $z$.}
When $z \ll 0$, the second term in Equation~\eqref{eq:G_wz} dominates, and the equal-risk contours of $G$ have slope strictly less than $-1$ throughout the region $w + z > 0$ (Appendix~\ref{appendix:slope_G}).
The joint confidence region for $(\widehat{w}, \widehat{z})$ is elongated along a direction close to slope $-1$, due to strong negative correlation (see Section~\ref{sec:wz_corr}).
In this case, evaluating $G$ at the diagonally opposite corners $(\overline{w}, \underline{z})$ and $(\underline{w}, \overline{z})$ captures the range of variation in $G$ over the joint confidence region with high accuracy (see Figure~\ref{fig:wz_ci_region}).

\subsection{Monte Carlo validation of confidence interval accuracy}\label{sec:MC}

\begin{figure}[H]
\centering
\includegraphics[width=0.7\linewidth]{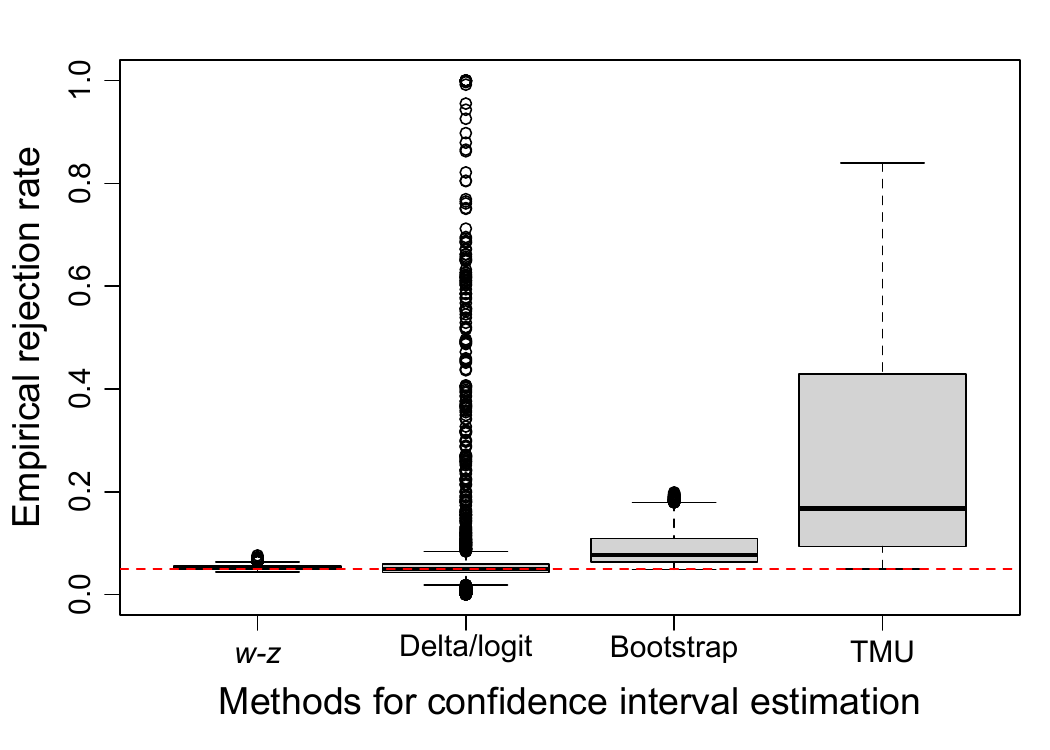}
\caption{Empirical rejection rates of CI methods, based on Monte Carlo simulations over 2{,}688 parameter combinations.
Each point shows the proportion of 10{,}000 replicates in which the true extinction probability $G(w,z)$ fell outside the estimated interval.
Parameters were varied over $\mu \in \{-0.5,-0.3,-0.1,0,0.1,0.3,0.5\}$, $\sigma^2 \in \{0.001,0.01,0.1,1\}$, $x_d \in \{3,5,7,9,11,13\}$, $t^{\ast} \in \{10,20,50,100\}$, and number of increments $q \in \{10,20,50,100\}$.
The dashed red line marks the nominal significance level $\alpha=0.05$.
Among the methods compared, the $w$--$z$ interval maintains rejection rates consistently close to the nominal level; the delta/logit method deviates in both directions depending on the parameter regime; and the percentile bootstrap and TMU (theoretical minimum uncertainty) methods are predominantly anti-conservative.}
\label{fig:ci_comparison}
\end{figure}

I conducted extensive Monte Carlo simulations to evaluate the accuracy of the proposed CI method for the extinction probability $G(w, z)$.
For each of the 2,688 parameter combinations, 10{,}000 replicate pairs $(\widehat{\mu},\widehat{\sigma}^2)$ were generated from their exact sampling distributions, and the empirical rejection rate (i.e., one minus the coverage probability) was computed as the proportion of intervals failing to cover the true value of $G$.
The $w$--$z$ method consistently achieved rejection rates very close to the nominal level $\alpha=0.05$, with mean $0.053$ (s.d.\ $0.0045$, range $0.044$--$0.076$), demonstrating accurate coverage across the full parameter range (Figure~\ref{fig:ci_comparison}).
In contrast, the delta/logit method was unstable, with rejection rates deviating widely across parameter regimes, particularly in the tails where $G$ is near $0$ or $1$ (mean $0.085$, s.d.\ $0.14$, range $0.0002$--$1.00$).
The percentile bootstrap was systematically anti-conservative (mean $0.090$, s.d.\ $0.036$, range $0.049$--$0.20$), and the theoretical minimum uncertainty (TMU) CI \citep{Fieberg:2000aa, ellner2008commentary} was strongly anti-conservative (mean $0.27$, s.d.\ $0.23$, range $0.050$--$0.84$).
Further details of the simulation design are provided in Appendix~\ref{appendix:MC_details}.

\subsection{Precision of extinction risk estimation: dependence on model parameters and data}\label{sec:precision_dependence}

The CI width of the extinction probability $G$ depends systematically on both the model parameters and the data characteristics.
As a function of $G$, the CI width necessarily vanishes at the boundaries $G=0$ and $G=1$, because $\nabla G(w,z)\to 0$ in these limits and the variance at the boundaries is zero.
Since the CI width varies continuously with $G$, there must be at least one interior maximum (or a short plateau) between these two boundaries.

\subsubsection{Dependence on model parameters}
\begin{figure}[H]
  \centering
  \begin{subfigure}[t]{0.48\textwidth}
    \centering
    \includegraphics[width=\linewidth]{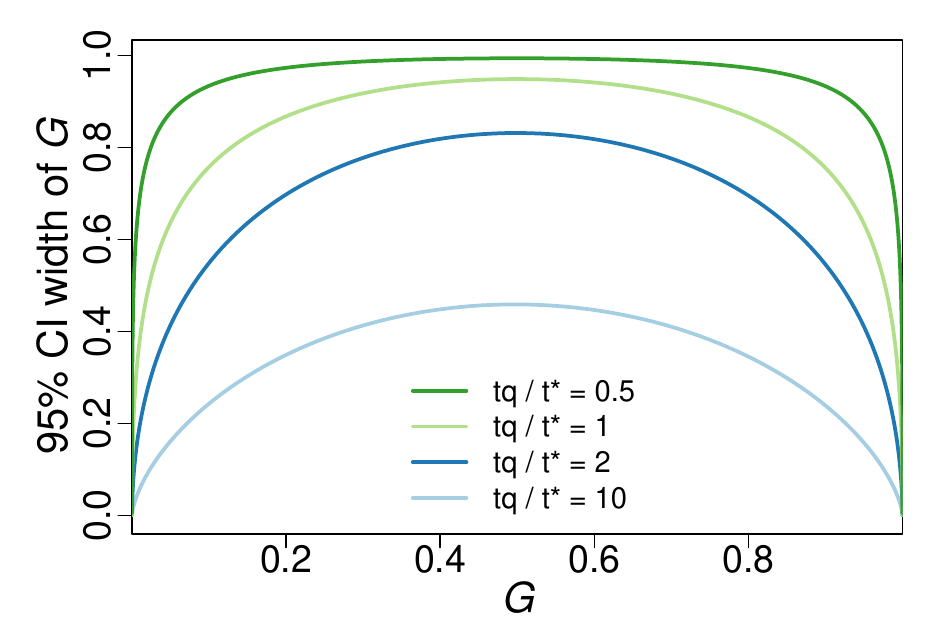}
    \caption{$z=20$.}
    \label{fig:CI_vs_G:a}
  \end{subfigure}\hfill
  \begin{subfigure}[t]{0.48\textwidth}
    \centering
    \includegraphics[width=\linewidth]{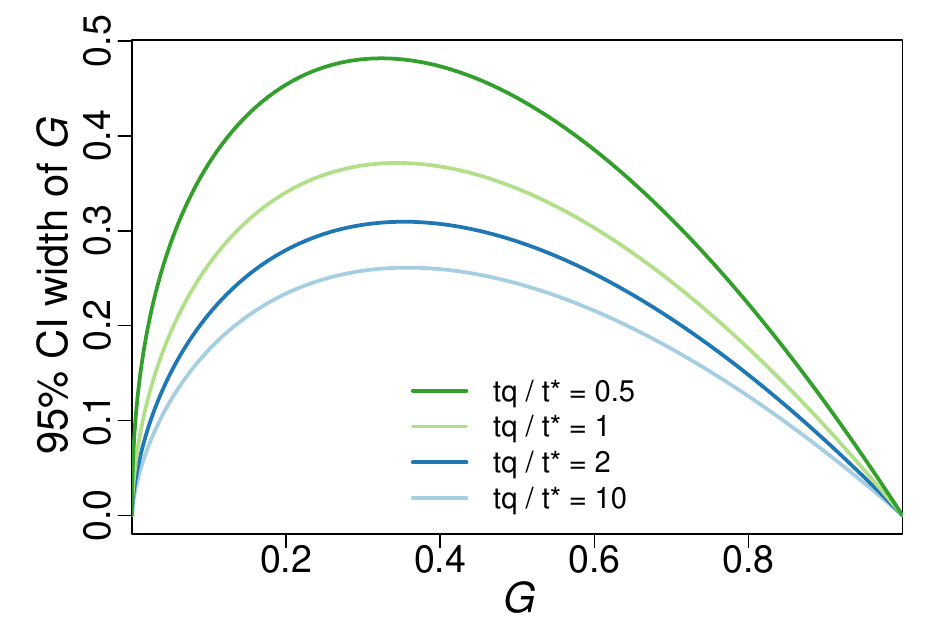}
    \caption{$z=-5$.}
    \label{fig:CI_vs_G:b}
  \end{subfigure}
\caption{95\% CI width of extinction probability $G$ as a function of $G$, shown for four values of the relative observation span $t_q/t^\ast$ (where $t^\ast$ is the time horizon), with $q = 63$ increments.
Panels show representative cases for (a) $z = 20$ (positive-$z$ regime) and (b) $z = -5$ (negative-$z$ regime), computed by the $w$--$z$ method.
}
  \label{fig:CI_vs_G}
\end{figure}

Consistent with these constraints, Figure~\ref{fig:CI_vs_G} shows a marked contrast between the positive-$z$ case ($z=20$) and the negative-$z$ case ($z=-5$).
For $z=20$, the CI width is maximized near $G\simeq 0.5$ in the typical regime $k\ge 1$; as $t_q/t^\ast$ decreases the peak broadens and the width becomes nearly flat across intermediate $G$, rendering the interval informative mainly near the boundaries.
For $z=-5$, the CI width is clearly unimodal with the peak around $G\simeq 0.3$--$0.4$, and even at the peak it remains comparatively small (about $0.3$--$0.4$).
A minor exception occurs when $0<k<1$ for moderate $z>0$, where the interior maximizer splits into two close peaks (cf.\ Figure~\ref{fig:varG_vs_G}\,c); no such split is seen for $z<0$ (Figure~\ref{fig:varG_vs_G}\,d).
Numerical calculations using the exact mixture representation confirm these patterns (Appendix~\ref{appendix:ci_width}).
In practice, however, $k$ is usually $\ge 1$, so the two-peak case is rarely encountered.

\begin{figure}[H]
  \centering
  \includegraphics[width=0.7\textwidth]{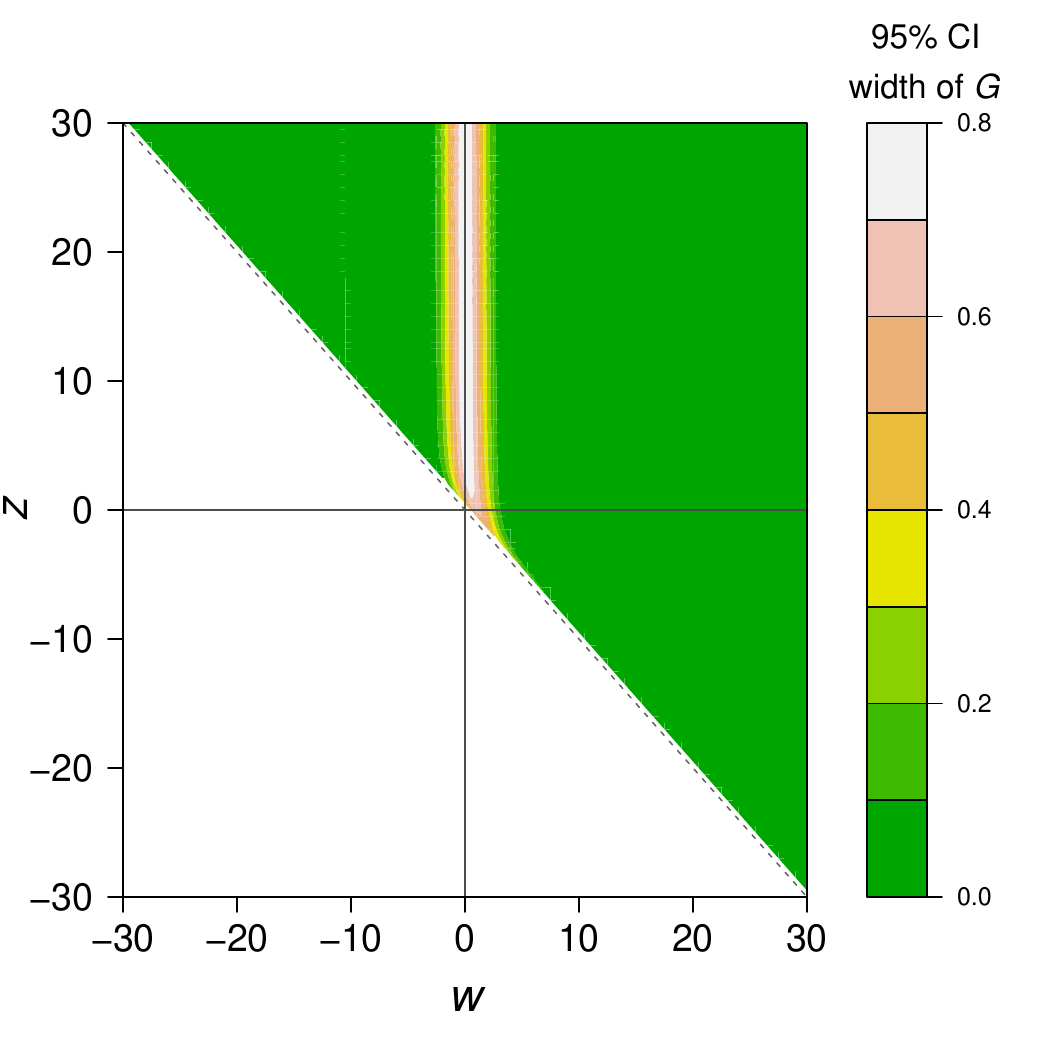}
  \caption{95\% CI width of extinction probability $G$ over the $(w, z)$ plane. The contour plot shows how the CI width of $G$ varies as a function of $w$ and $z$, highlighting regions where the estimation uncertainty is highest (near $w=0$, i.e., $G \approx 0.5$) and lowest (near $G = 0$ and $G = 1$). $q=t_q=63$ and $t^\ast=25$. The CI width of $G$ was calculated using the $w$--$z$ method.}
  \label{fig:ci_width_wz_plane}
\end{figure}

When interpreting CI width, it is important to recognize that the $(w,z)$ parameter space does not generate extinction probabilities uniformly across the full range of $G$.
In fact, the vast majority of the $(w,z)$ plane corresponds to regions where extinction is either extremely unlikely ($G\approx 0$) or almost certain ($G\approx 1$), while the zone of high uncertainty is confined to a narrow band near $w=0$ when $z>0$ (Figure~\ref{fig:ci_width_wz_plane}).

Accordingly, although CI-width or variance curves plotted against $G$ may suggest poor estimability across much of the range in the $z>0$ regime, this impression is misleading.
As noted above, the intervals remain extremely tight near the boundaries ($G=0,1$), ensuring informative assessments in these regions.
Reliable estimation therefore depends on where the system actually resides in $(w,z)$ space: in most realistic cases the estimates of $(w,z)$ are unlikely to fall exactly in the maximally uncertain region around $w=0$, and more often lie in regions where extinction risk can be judged with confidence.

\subsubsection{Effect-size perspective}
It is useful to consider the reliability of extinction-risk estimation from the viewpoint of effect size. 
Following the effect-size literature \citep[e.g.,][]{cohen1988statistical}, an effect size indexes the magnitude of the phenomenon as determined by the model parameters; all else equal, larger effects are easier to detect and hence yield tighter CIs. 
Although one might be tempted to take the extinction probability itself as an effect size, the CI width is typically unimodal in $G$ and is small near $0$ and $1$; therefore $G$ alone is not a monotone index of inferential difficulty.
Outside the exceptional two-peak case ($0<k<1$ for moderate $z>0$), let $G_{\max}=G_{\max}(z,q,t_q,t^\ast)$ denote the value of $G$ at which the CI width is maximized for the given configuration, and define
\begin{equation*}
\Delta(G):=|G-G_{\max}|.
\end{equation*}
This distance increases as extinction becomes nearly certain or nearly impossible, and decreases in the ambiguous middle.
In the central $z>0$ regime, numerical computations indicate $G_{\max}\simeq 0.5$ (attained at $w=0$).
With this choice, estimation precision depends not only on the data $(q,t_q)$ but also on the effect size $\Delta(G)$ and the underlying parameters $(w,z,t^\ast)$: for fixed data, larger $\Delta(G)$ yields narrower CIs, while the width is largest near $G\simeq 0.5$.
Equal values of $G$ (or $\Delta(G)$) can nevertheless arise from different $(w,z)$ configurations, which differ in sensitivity and in the sampling correlation $\rho(w,z,k)$, so that the effect-size index $\Delta(G)$ captures the main pattern but does not fully determine CI width.

\subsubsection{Dependence on data}
Dependence on the data enters in complementary ways.
With $t_q$ fixed, increasing the number of increments $q$ improves precision primarily by stabilizing the variance estimator $\widehat{\sigma}^2$ (Equation~\ref{eq:sampling_muhat_sigmahat}), yielding an initial $\sim q^{-1/2}$ reduction in CI width; however, the gain quickly saturates because $\operatorname{Var}(\widehat{\mu})$ is governed by $t_q$ (not by $q$) and the noncentral-$t$ law approaches its normal limit as the degrees of freedom grow.
Beyond this point, additional increments no longer shrink the interval, which levels off at a value set by the ratio $t^\ast/t_q$ (i.e., of order $\sqrt{t^\ast/t_q}$).

\begin{figure}[H]
\centering
\includegraphics[width=0.98\linewidth]{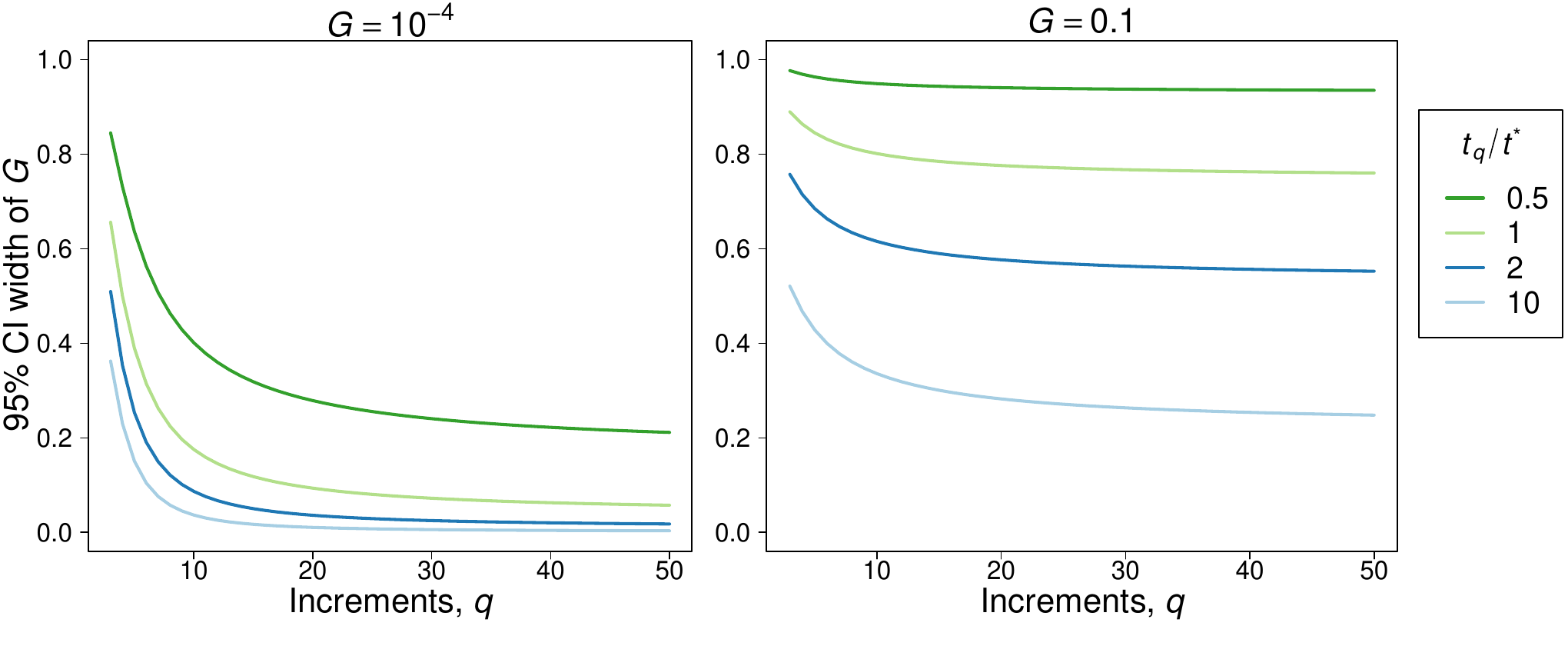}
\caption{
95\% CI width of extinction probability $G(w,z)$ versus the number of increments $q$, computed by the $w$--$z$ method, shown for four values of the relative observation span $t_q/t^\ast$ (where $t^\ast$ is the time horizon).
Panels use operating points with $G=10^{-4}$ (left) and $G=0.1$ (right), matched at $z=20$.
The qualitative dependence on $q$ and $t_q$ is similar across $z$ (Appendix~\ref{appendix:ci_width}).
}
\label{fig:design_chart}
\end{figure}

By contrast, lengthening the observation span $t_q$ directly reduces $\operatorname{Var}(\widehat{\mu})$ (proportional to $1/t_q$) and increases $t_q/t^\ast$, both of which narrow the CI and yield a smaller asymptotic plateau.
Consequently, for fixed $t_q$ further increases in $q$ cannot push the width below this plateau, whereas increasing $t_q$ continues to deliver substantial gains.
This practical tradeoff is visualized in Figure~\ref{fig:design_chart}.

Further details on the data dependence of CI width are given in Appendix~\ref{appendix:ci_width}.

\subsubsection{Dependence on the original parameters \texorpdfstring{$\mu$}{mu} and \texorpdfstring{$\sigma^2$}{sigma\textasciicircum2}}

On the $(w,z)$ plane (Figure~\ref{fig:ci_width_wz_plane}), the sign of $\mu$ is split by $w=z$.
For $\mu>0$, CIs are small almost everywhere; the uncertainty ridge lies near $w\approx 0$ (for fixed $\mu>0$ and $x_d>0$ the line $w=0$ itself is not reached).
For $\mu<0$, the CI is maximized at $w=0$ (that is, $-\mu t^\ast=x_d$) and is small elsewhere.
With $(\mu,x_d,t^\ast)$ fixed, increasing $\sigma$ moves $(w,z)$ toward the origin along a fixed ray, widening the CI up to a data-limited plateau of order $\sqrt{t^\ast/t_q}$.
Decreasing $\sigma$ pushes $(w,z)$ outward and the CI collapses rapidly at rate $\exp(-\text{const}/\sigma^{2})$, except along the $w=0$ ray (reachable only if $\mu=-x_d/t^\ast$).
See Appendix~\ref{appendix:ci_width}.

\subsubsection{Dependence on time horizon}
\begin{figure}[H]
  \centering
  \includegraphics[width=0.7\textwidth]{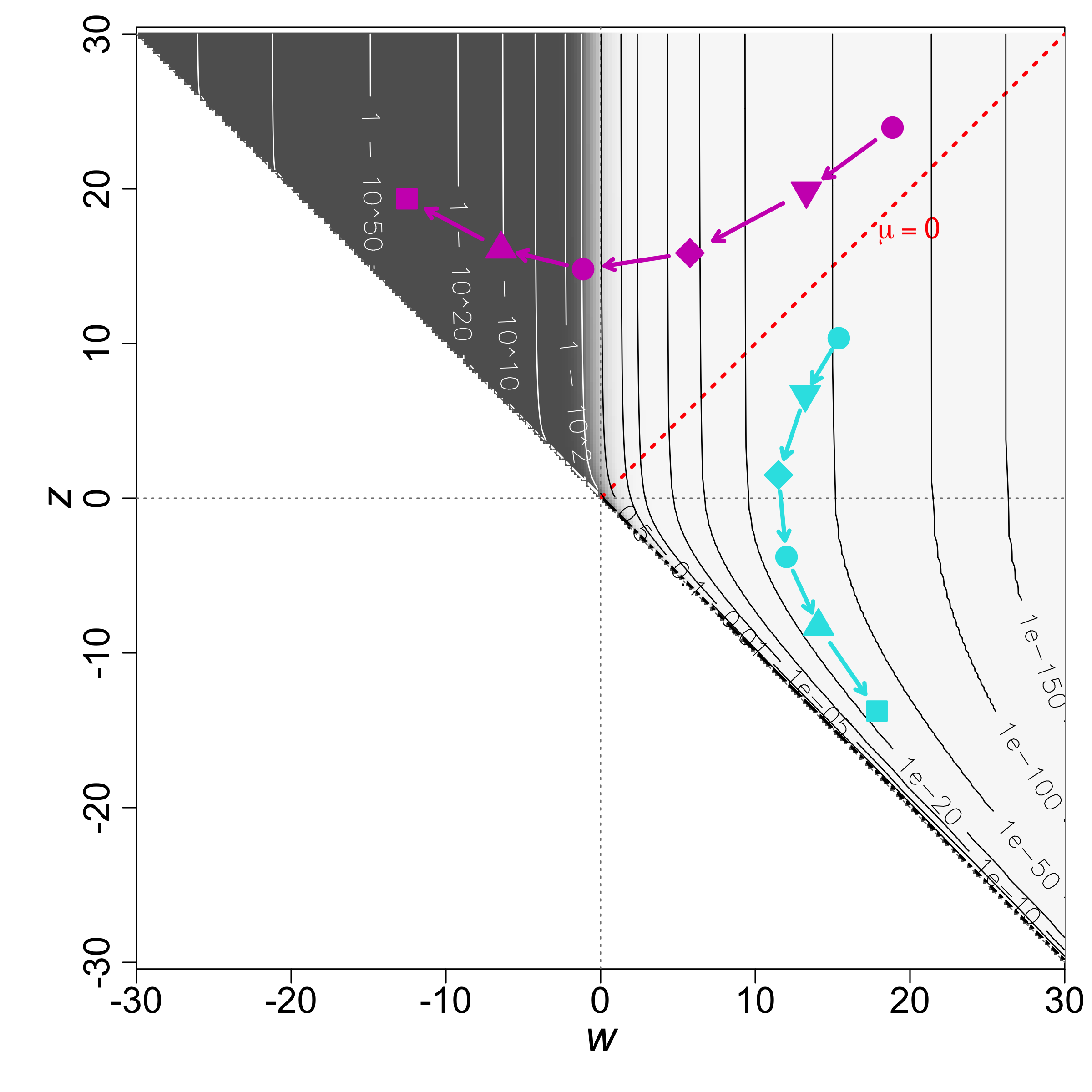}
\caption{
  Time-horizon dependence of extinction risk $G(w,z)$. 
  Varying the horizon $t^\ast$ (with $\mu,\sigma,x_d$ fixed) shifts the system's location in the $(w,z)$ plane and thereby alters $G(w,z)$. 
  The curves trace $(w,z)$ as $t^\ast$ increases for two parameter sets: 
  (a, magenta) $\mu<0$ case (crosses $w=0$; $\mu=-0.0589$, $\sigma=0.1169$, $x_d=12.64$); and 
  (b, cyan) $\mu>0$ case (does not cross $w=0$; $\mu=0.10$, $\sigma=0.20$, $x_d=13$). 
  Symbols mark $t^\ast \in \{25.5, 42.5, 100, 250, 500, 1000\}$; arrows indicate increasing $t^\ast$.
}
  \label{fig:G_contours_tstar_paths}
\end{figure}

As the time horizon $t^\ast$ increases, the system moves along the $(w,z)$-plane trajectories in Figure~\ref{fig:G_contours_tstar_paths}.
When the growth rate is negative ($\mu<0$), extinction risk shows a characteristic pattern.
Even if near-term extinction is highly unlikely, extending the horizon eventually brings the trajectory to the line $w=0$, which corresponds to the zone of maximal uncertainty, at which point the CI widens substantially.
Beyond this point, uncertainty decreases again as extinction becomes virtually certain at longer horizons.
By contrast, when the growth rate is positive ($\mu>0$), the trajectory does not cross $w=0$, so any increase in uncertainty is small; in the representative example shown here, extinction risk remains small across horizons and the CIs remain uniformly tight.
Numerical values are provided in Table~\ref{tab:ciwidth_vs_tstar_SI}.

\subsection{Observation span required for precise estimation of extinction risk}\label{sec:obs_span}

The required data length depends strongly on the true extinction probability $G(w,z)$.
When $G$ lies well below $G_{\mathrm{target}}$, shorter series can suffice, whereas the required span increases as $G$ approaches the threshold from below.
To quantify this, I computed the minimal observation span $t_q$ (unit-interval sampling, so $t_q=q$) needed for the upper bound of the 95\% CI to lie below a management threshold $G_{\mathrm{target}}=0.1$.
For each calculation $z$ was fixed, $w$ was adjusted to achieve the target $G$, and $t_q$ was then solved numerically (details in Appendix~\ref{appendix:req_span}).

Figure~\ref{fig:ci_width_vs_G} shows the required span relative to the horizon ($t^\ast=100$~years) over the range $G<G_{\mathrm{target}}$.
The two curves ($z=20$ and $z=-5$) are representative of the $z>0$ and $z<0$ regimes, and the results are broadly robust to the choice of $t^\ast$ when expressed as $t_q/t^\ast$ (Appendix~\ref{appendix:req_span}).
In both cases, only 10--20\% of the horizon suffices when $G$ is very small, while the requirement increases as $G$ approaches the threshold from below.
By definition, no finite required span exists for $G\ge G_{\mathrm{target}}$.

\begin{figure}[H]
\centering
\includegraphics[width=0.7\linewidth]{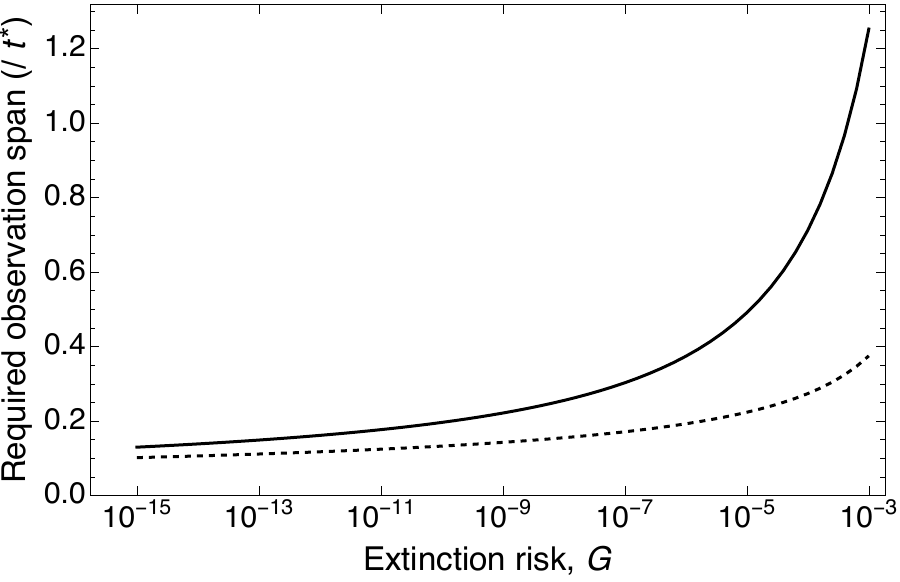}
\caption{Required observation span $t_q$ for confirming that extinction risk lies below the management threshold $G_{\mathrm{target}}=0.1$.
Shown is the minimal $t_q/t^\ast$ (with $t^\ast=100$~years and $t_q=q$) needed for the upper bound of the 95\% CI for $G(w,z)$ to lie below $G_{\mathrm{target}}$, over the range $G<G_{\mathrm{target}}$.
The solid line corresponds to $z=20$, and the dashed line to $z=-5$.
Even relatively short spans can be sufficient when the true risk is very small.}
\label{fig:ci_width_vs_G}
\end{figure}
 
\section{The observation-error-and-autocovariance-robust (OEAR) estimator and validity under model relaxations}
\label{sec:model_relaxations}
The CI construction in Section~\ref{sec:CI} is derived for a drifted Wiener process observed without error.
Rather than redefining the CI under alternative data-generating mechanisms, I keep it fixed and assess its operating characteristics under plausible relaxations of the baseline assumptions using geometric analysis and Monte Carlo simulation.
The main comparison is between the naive drift--Wiener fit and the effective-diffusion OEAR fit in the $(w,z)$ plane; under additive observation error, I also include state-space fitting as a benchmark.
Full designs and results are reported in Appendices~\ref{appendix:obs_error} and \ref{appendix:env_noise_density}.

\medskip
The OEAR strategy is motivated by an effective-diffusion view in which the drift--Wiener model is interpreted as a large-scale approximation for weakly dependent growth increments, with diffusion coefficient given by the long-run variance (LRV) \citep{lande1988extinction}.
For a broad class of dependence mechanisms, additive observation error inflates short-run variance and induces negative short-run autocovariances but cancels in the LRV, so extinction-risk calculations based on an LRV-calibrated effective diffusion can remain robust to additive observation error \citep{mcnamara2004measurement}.
Accordingly, I propose an OEAR estimator that plugs a heteroskedasticity-and-autocorrelation-consistent (HAC) estimate of the LRV \citep{newey1987simple} into the $w$--$z$ map and hence into $G(w,z)$:
\begin{equation*}
\mathcal C := \sum_{j=-\infty}^{\infty} C_j,
\qquad
C_j := \mathbb E\!\left[U_i\,U_{i-j}\right],
\qquad
\widetilde{\sigma}^{2} := \widetilde{\mathcal C}_{\mathrm{NW}}(J),
\end{equation*}
where $U_i$ denotes the drift-corrected per-unit-time increments.
Here $\widetilde{\mathcal C}_{\mathrm{NW}}(J)$ is the Bartlett (Newey--West) HAC estimate of the LRV with lag truncation $J$ on the original scale (Appendix~\ref{appendix:obs_error}).
I then obtain the OEAR operating point by combining the naive drift estimate with the LRV-based diffusion-scale estimate in the $(w,z)$ map; thus OEAR differs from the naive fit only in the diffusion scale.

I show analytically that additive observation error shifts naive plug-in operating points toward the origin along nearly fixed rays in the $(w,z)$ plane (radial shrinkage).
Monte Carlo experiments in Appendix~\ref{appendix:obs_error} confirm this prediction (for observation error comparable to environmental variance): naive points exhibit radial shrinkage, whereas OEAR and the state-space fit largely correct the median by accounting for inflated variance.
CIs widen and coverage declines as error increases (most under naive fitting at the highest error level), yet even naive fitting leaves the estimated risks far from Criterion~E thresholds, so classification never changes.
A power-law CPUE check mainly rescales the same shrinkage geometry without altering the qualitative conclusions.

Additional Monte Carlo experiments under time-correlated (colored) environmental variation and weak density feedback show that OEAR remains comparatively stable as an effective-diffusion plug-in, whereas naive drift--Wiener fitting is more sensitive to departures from independent increments (Appendix~\ref{appendix:env_noise_density}).
Under OU forcing, OEAR more consistently tracks the reference $G_{\mathrm{OU}}(t^\ast)$ as forcing strength and persistence increase, while naive plug-in summaries can deviate substantially.
Under weak density feedback, both procedures remain well behaved, but OEAR does not improve interval calibration relative to the naive CI in the weak-feedback range examined.
In all cases considered here, the reference probabilities remain far below the decision threshold, so the threshold-based interpretation is unchanged.
Which plug-in strategy performs better depends on the relaxation and sample size.

\section{Empirical application: Extinction risk assessment of the Japanese eel}\label{sec:eel_estimation}
The Japanese eel (\textit{Anguilla japonica}) was listed as Endangered on the IUCN Red List of Threatened Species in 2014.
Abundance indices suggest substantial declines, and eel aquaculture continues to rely heavily on wild-caught juvenile glass eels.
This reliance has raised concerns about recruitment and long-term productivity.
The IUCN currently classifies the Japanese eel as Endangered under Criterion~A, based on population reduction inferred from abundance indices and exploitation.
While such indicators highlight serious population loss, they do not directly estimate the probability of extinction.
A more informative reassessment requires the application of Criterion~E, which uses quantitative analysis to estimate future extinction risk.
Using carefully compiled time-series data and a probabilistic framework with CIs, I estimate finite-horizon extinction probabilities under Criterion~E to assess threat severity and inform conservation strategies.

\begin{figure}[H]
\centering
\includegraphics[width=0.7\linewidth]{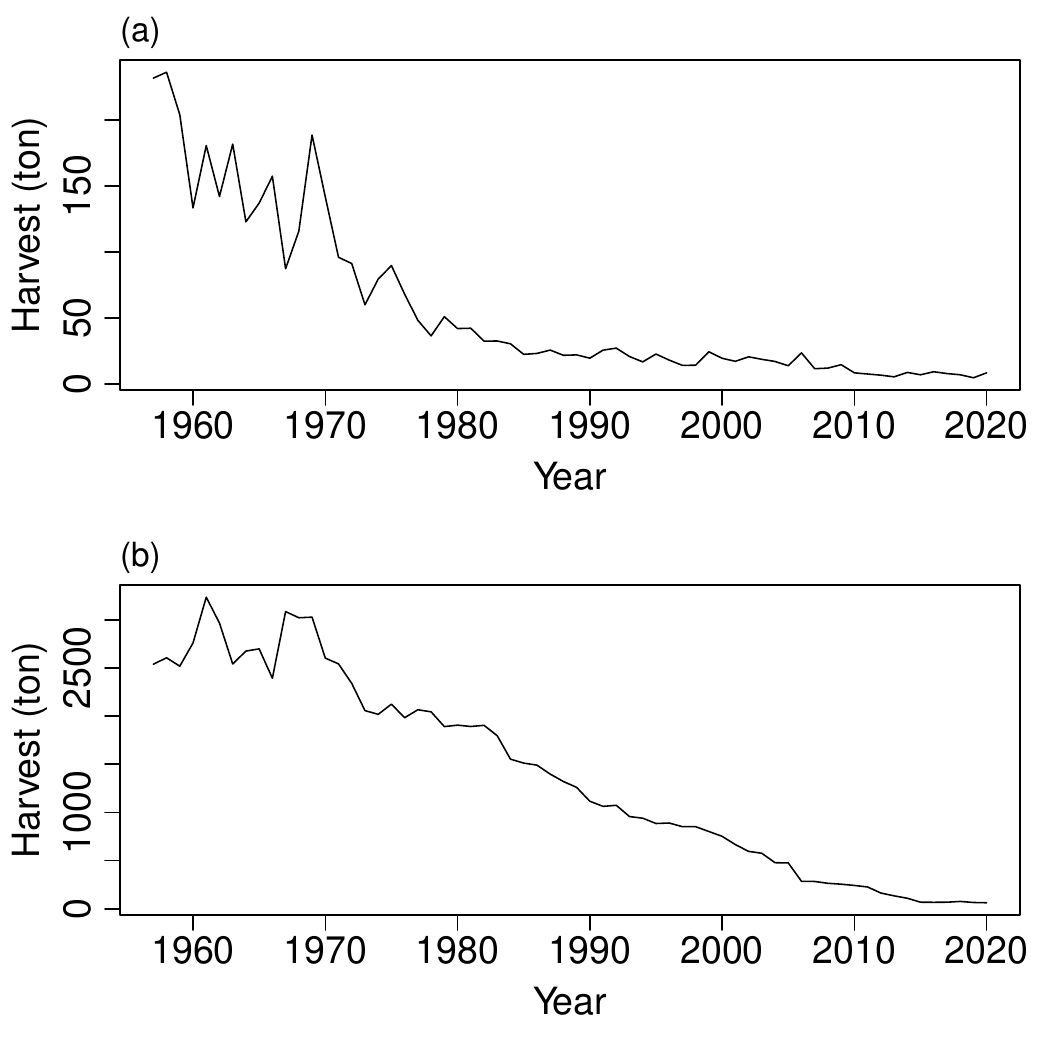}
\caption{National harvest time series used in the PVA (tonnes), 1957--2020.
(a) Common glass-eel seed (coastal $+$ inland).
(b) Inland eel harvest (yellow and silver eels).
The common-seed series combines official inland totals with a coastal component derived from prefectural data and rescaled to match official coastal totals (Appendix~\ref{appendix:eel_harvest}).}
\label{fig:eel_harvest}
\end{figure}

To evaluate extinction risk under Criterion~E, I analyzed two national time-series indices representing different life stages: a common glass-eel seed index (coastal $+$ inland) and an inland-eel (yellow and silver) index (Figure~\ref{fig:eel_harvest}; Appendix~\ref{appendix:eel_harvest}).

The Japanese eel population is biologically structured during the continental growth phase, with age and regional structure, despite broad-scale panmixia at the spawning population level.
For the present Criterion~E assessment, I therefore analyze aggregated harvest indices using a one-dimensional drift--Wiener approximation.
This choice follows the model-aggregation viewpoint: for extinction-risk approximation, a lower-dimensional stochastic summary can remain informative even when the underlying system is not exactly reducible to one dimension \citep{lande1988extinction,hakoyama2005extinction}.
Because the stock is broadly panmictic, recruitment variation is expressed broadly across the distribution range; consistent with this, glass-eel catches show uniformly positive interannual correlations across East Asia (Appendix~\ref{sec:S12_5_recruitment_synchrony}).
Accordingly, I analyze each index separately but use their agreement or discrepancy as a cross-stage robustness check rather than treating them as independent replicates for inference.
This assessment adopts a precautionary approach: I evaluate extinction risk on the local catch scale, which implies a much smaller $x_d$ than for the full panmictic stock and therefore errs on the side of higher risk.
If extinction risk is low even on the local catch scale, the overall risk for the species should be even lower.
Data sources and methodological details are provided in Appendix~\ref{appendix:eel_harvest}, the full PVA results and sensitivity analyses are reported in Appendix~\ref{appendix:eel_pva_sensitivity}, and the generation-time calculation with Criterion~E thresholds is given in Appendix~\ref{appendix:generation_time}.

The Criterion~E PVA for the Japanese eel, based on the two adopted national harvest indices and the drift--Wiener framework developed in this study, yields extinction probabilities that are far below the IUCN threatened-category thresholds at the CR/EN/VU horizons (Table~\ref{tab:estimates_oear}).
The OEAR-based summaries in the main text show that these probabilities remain very small, with CIs that stay well below the corresponding thresholds.
Although the main text emphasizes OEAR, the naive drift--Wiener likelihood analysis leads to the same qualitative conclusion; moreover, the effort-trend and abundance-scale calibration sensitivity analyses do not alter the threatened-category interpretation (Appendix~\ref{appendix:eel_pva_sensitivity}), and Table~\ref{tab:precautionary_summary} summarizes the main precautionary elements and robustness checks.

\begin{table}[H]
\centering
\caption{Estimated extinction probabilities (point estimates and 95\% CIs) for the Japanese eel (OEAR method).}
\label{tab:estimates_oear}

\resizebox{\textwidth}{!}{
\begingroup
\renewcommand{\arraystretch}{3}
\begin{tabular}{ccccccc}
\hline\hline
Time series & $\widehat{\mu}$ & $\widetilde{\sigma}^{2}$ & $\check{x}_d$ & CR: $\widetilde{G}(t^\ast=25.5)$ & EN: $\widetilde{G}(t^\ast=42.5)$ & VU: $\widetilde{G}(t^\ast=100)$\\
\hline
\shortstack{Glass eel} & \shortstack{-0.052 \\ (-0.124, 0.020)} & \shortstack{0.016 \\ (0.012, 0.024)} & 17.6 & \shortstack{$4 \times 10^{-140}$ \\ ($9 \times 10^{-191}$, $9 \times 10^{-93}$)} & \shortstack{$5 \times 10^{-76}$ \\ ($5 \times 10^{-106}$, $6 \times 10^{-49}$)} & \shortstack{$3 \times 10^{-22}$ \\ ($3 \times 10^{-36}$, $3 \times 10^{-11}$)}\\
\shortstack{Yellow and\\silver eel} & \shortstack{-0.059 \\ (-0.089, -0.029)} & \shortstack{0.016 \\ (0.011, 0.023)} & 12.6 & \shortstack{$4 \times 10^{-70}$ \\ ($10^{-96}$, $6 \times 10^{-46}$)} & \shortstack{$9 \times 10^{-36}$ \\ ($3 \times 10^{-51}$, $4 \times 10^{-22}$)} & \shortstack{$4 \times 10^{-8}$ \\ ($10^{-15}$, $4 \times 10^{-3}$)}\\
\hline
\end{tabular}
\endgroup
}

\vspace{4pt}
\caption*{\footnotesize\textit{Note.} Point estimates are shown with 95\% CIs in parentheses. Here $\widehat{\mu}$ is the ML estimate of the drift; for OEAR (Appendix~\ref{appendix:obs_error}), $\widetilde{\sigma}^{2}$ and $\widetilde{G}$ use tildes, with $\widetilde{\mu}=\widehat{\mu}$. Units: $\widehat{\mu}$ and $\widetilde{\sigma}^{2}$ in year$^{-1}$. Let $y_t:=\log n_t$ denote the log index constructed from harvest (Appendix~\ref{appendix:eel_harvest}). The threshold distance used in the calculation is $\check{x}_d:=y_q-\log n_e$ with $y_q:=\log n_{t_q}$, equivalently $\check{x}_d=\log(n_{t_q}/n_e)$, and is treated as known because $n_0=n_{t_q}$. Extinction probability is $\widetilde{G}(t^\ast)=\Pr[T\le t^\ast]$ with $t^\ast\in\{25.5,\,42.5,\,100\}$ years for CR, EN, and VU. Extinction probabilities are reported in scientific notation. }
\end{table}
 
\section{Discussion}

A central implication of these results is that reliable extinction-risk inference is often achievable even from relatively short time series.
Although uncertainty can be high at intermediate extinction risks, such cases are confined to limited regions of parameter space.
Outside those regions, observation spans as short as 10--20\% of the forecasting horizon can suffice to yield informative intervals and to confirm $G(t^\ast)\le G_{\mathrm{target}}$ when true risk is small (Section~\ref{sec:obs_span}; Figure~\ref{fig:ci_width_vs_G}).
The present analyses clarify the limits of earlier CI methods and the parameter regimes in which inference is most difficult, while also reconciling earlier debates by showing why empirical studies, simulations, and analytical work could report encouraging performance from relatively short time series \citep{Brook:2000aa, holmes2007statistical, ellner2008commentary}, despite concerns about long-term precision \citep{Fieberg:2000aa}.

These findings also bear on a broader methodological objection to PVA itself.
Critics often emphasize imprecision under limited data and recommend proxy-based alternatives, yet such proxies provide neither explicit extinction probabilities nor a principled measure of reliability.
As \citet{Brook:2002aa} argued, there is little reason to abandon PVA in favor of less transparent alternatives.
The robustness analyses show that extinction-risk inference based on the $w$--$z$ framework is reasonably robust to several plausible departures from the baseline error-free drift--Wiener model.
Under additive observation error, the HAC-based OEAR fit and the state-space fit largely corrected the radial-shrink bias of the naive baseline ML fit; under colored environmental variation, OEAR tracked the reference ordering more closely than this naive fit; and under the weak density-feedback range examined here, both plug-in strategies remained well behaved.
Taken together, these results support the effective-diffusion viewpoint as a useful and broadly robust, though not uniformly superior, extension of the drift--Wiener framework beyond the ideal error-free setting, especially for dependence-driven departures from the baseline model.
Residual limitations remain for departures such as additional stochasticity not represented in the baseline model, including demographic stochasticity at low abundance and rare catastrophic events \citep{lande2003stochastic}, as well as heavy-tailed noise, pronounced state dependence, structural breaks, or severe reporting artifacts, which would require separate robustness checks.

The Japanese eel is currently listed as Endangered under Criterion~A, but under Criterion~E the estimated extinction probabilities are far below the threatened-category thresholds (Table~\ref{tab:estimates_oear}), and this conclusion is unchanged under both the OEAR and naive fits, under the precautionary calibrations, and across the sensitivity analyses reported in Appendix~\ref{appendix:eel_pva_sensitivity}.
More broadly, the Japanese eel results exemplify a theoretical pattern established by the drift--Wiener comparison in Appendix~\ref{appendix:A_over_E_drift_wiener}: for sufficiently large populations, decline-based Criterion~A systematically overstates finite-horizon extinction risk relative to Criterion~E, and the accompanying numerical illustration shows that the population-size threshold for this divergence is surprisingly small.
This marks a departure from the original small-population context in which decline rates were intended as practical proxies for extinction risk \citep{mace1991assessing}.
Under the disjunctive listing rule (Appendix~\ref{appendix:A_over_E_drift_wiener}), this creates a risk of false-positive threat classification, which can in turn misdirect extinction-focused conservation effort away from taxa facing demonstrably higher extinction risk.
This implication is not specific to the Japanese eel: large marine populations assessed primarily from proportional declines may be especially prone to this problem.

\section{Conclusion}
The $w$--$z$ framework shows that finite-horizon extinction-risk inference can remain reliable even with limited data, because uncertainty is concentrated in restricted regions of parameter space rather than being uniformly large.
The $w$--$z$ representation turns inferential difficulty into a geometric question, clarifying how precision depends jointly on effect size, horizon, and data length.

For practical PVA, both the naive error-free ML and OEAR estimators are useful: the naive fit provides a simple baseline, whereas OEAR offers a more robust HAC-based effective-diffusion extension under observation error and dependence-driven departures from the baseline model.

In the Japanese eel application, the estimated extinction probabilities under Criterion~E remain far below the threatened-category thresholds under both the OEAR and naive fits, and this conclusion is unchanged across the precautionary calibrations and sensitivity analyses.

\section*{Author contributions}
Hiroshi Hakoyama conceived and designed the study, conducted the analyses, and wrote the manuscript.

\section*{Acknowledgements}
This research was funded by a grant from the Japan Fisheries Agency.

\section*{Conflict of interest statement}
The author declares no conflicts of interest.

\section*{Data availability statement}
Data and code available via \url{https://doi.org/10.5281/zenodo.18982178} (Hakoyama, 2026).
The baseline R package \texttt{extr} is available via \url{https://doi.org/10.32614/CRAN.package.extr} (Hakoyama, 2025).

\section*{Data sources}
The data sources used in this study are listed below.

Hakoyama, H., Fujimori, H., Okamoto, C. \& Kodama, S. (2016) Compilation of Japanese fisheries statistics for the Japanese eel, \textit{Anguilla japonica}, since 1894: a historical dataset for stock assessment. \textit{Ecological Research}, 31, 153. \url{https://doi.org/10.1007/s11284-015-1332-9}.

Ministry of Agriculture, Forestry and Fisheries, Japan. \textit{Annual Reports of Catch Statistics on Fishery and Aquaculture in Japan}. Relevant years cited in the text.

Shizuoka Prefectural Fisheries Experiment Station. \textit{Annual Reports of the Shizuoka Prefectural Fisheries Experiment Station}. 1979--2004. ISSN 2759-2340.

\textit{Hamana}. Public bulletin of the Shizuoka Prefectural Fisheries Experiment Station. 2005--2020. \url{https://fish-exp.pref.shizuoka.jp/hamanako/6_pro/hamana_backnumber.html}.

Eel Culture Research Council. Annual reports, Volumes 7--27 (1977--1997). Library holdings listed in CiNii (NCID: AN10187528).

Joint Statement of the Informal Consultation on International Cooperation for Conservation and Management of Japanese Eel Stock and Other Relevant Eel Species. Regional glass-eel catch data for China, Japan, Korea, and Chinese Taipei. \url{https://www.mofa.go.jp/files/100866466.pdf}.

\appendix
\titleformat{\section}{\normalfont\Large\bfseries}
  {Appendix \thesection.}{0.3em}{}
\renewcommand{\thesection}{S\arabic{section}} \renewcommand{\theequation}{\thesection.\arabic{equation}}
\renewcommand{\thefigure}{\thesection.\arabic{figure}}
\renewcommand{\thetable}{\thesection.\arabic{table}}
\renewcommand{\theproposition}{\thesection.\arabic{proposition}}
\renewcommand{\thelemma}{\thesection.\arabic{proposition}}
\renewcommand{\thecorollary}{\thesection.\arabic{proposition}}

\section{Numerical Stability of \texorpdfstring{$G(w, z)$}{G(w, z)}}\label{appendix:numerical_stability_G}
\setcounter{equation}{0}
\setcounter{figure}{0}

This appendix studies numerically stable evaluation of
\begin{subequations}\label{Seq:GQ-def}\begin{align}
  G(w, z) &= \Phi(-w) + \exp\!\left(\frac{z^2 - w^2}{2}\right)\Phi(-z) \quad \text{for} \quad w + z > 0, \quad w, z \in \mathbb{R}, \label{Seq:G-def}\\
  Q(w, z) &= 1 - G(w,z), \label{Seq:Q-def}
\end{align}
\end{subequations}
for real $w$ and $z$ satisfying $w + z > 0$.
Here, $\Phi(x)$ denotes the standard normal cumulative distribution function.
The two functions are complementary by definition.

The key numerical challenges in evaluating \eqref{Seq:GQ-def} in double-precision arithmetic are:
\begin{enumerate}[label=(\roman*)]
  \item Overflow: for large $z$, the exponential $\exp\!\left((z^{2}-w^{2})/2\right)$ can overflow.
Overflow occurs whenever $(z^{2}-w^{2})/2 > \log(\text{\texttt{DBL\_MAX}})$; in the worst case $w=0$ this corresponds to $z \gtrsim 37.7$.
  \item Loss of Information (LOI): The summands of \eqref{Seq:GQ-def} can be highly unbalanced, causing smaller terms to be rounded to zero.
This is particularly problematic in the region $w\ll0$, where the extinction probability $G$ is very close to 1 but is rounded to unity in double precision.
\item Probabilities beyond double-precision limits: when $G$ or $Q$ falls below the minimum positive number representable in double precision (\texttt{DBL\_TRUE\_MIN} $\approx 4.94\times10^{-324}$) the linear-scale arithmetic path underflows to~0.
\end{enumerate}

Accurate evaluation in cases (ii) and (iii) is particularly important because extinction probabilities inferred from data can be extremely small or extremely large, so values of $G$ near $0$ or $1$ naturally arise.
Although probabilities such as $10^{-9}$ and $10^{-20}$ may both appear negligible, they lead to markedly different confidence interval widths and therefore must be resolved with numerical accuracy.

I adopt three complementary strategies to address these issues.  
First, the use of a Mills ratio representation for large $z$ (next subsection) replaces the numerically unstable product $\exp((z^{2}-w^{2})/2)\,\Phi(-z)$ with a product of moderate-magnitude factors.  
Second, for large negative $w$, the complementary probability $Q(w,z) = 1 - G(w,z)$ is evaluated instead of $G(w,z)$ itself, thereby avoiding loss of information near unity.  
Third, I evaluate sums and differences on the log scale using stable identities (log-sum-exp and log-difference), which preserve information even when $G$ lies outside the representable range $4.94\times 10^{-324} < G < 1 - 4.94\times 10^{-324}$.

\subsection{Linear-scale formulation, Mills switch, and validation}

The asymptotic expansion of the Mills ratio and the identity used below are standard facts and are also employed in the appendix of \citet{DENNIS:1991aa}.

Let $\phi(x)=\Phi'(x)=\tfrac{1}{\sqrt{2\pi}}\exp(-x^{2}/2)$ denote the standard normal density.
The Mills ratio, defined for positive arguments $z>0$, is
\begin{equation*}
 R(z)=\frac{\Phi(-z)}{\phi(z)}.
\end{equation*}
As $z\to\infty$, it admits the asymptotic expansion
\begin{equation}\label{Seq:Mills-series}
  R(z) \sim \sum_{j=0}^{\infty}\frac{(-1)^{j}(2j-1)!!}{z^{2j+1}}.
\end{equation}
The eight-term truncation of \eqref{Seq:Mills-series} (i.e., $j = 0, \dots, 7$)
\begin{equation*}
  R_7(z)\coloneqq\frac{1}{z}-\frac{1}{z^{3}}+\frac{3}{z^{5}}
  -\frac{15}{z^{7}}+\frac{105}{z^{9}}
  -\frac{945}{z^{11}}+\frac{10395}{z^{13}}
  -\frac{135135}{z^{15}}
\end{equation*}
is used in the numerical implementation whenever $z\ge z_{\mathrm{thr}}$.

The identity relating the two terms is
\begin{equation*}\exp\!\left(\tfrac{z^2-w^2}{2}\right)\Phi(-z) = \phi(w) R(z).
\end{equation*}
This motivates the switched linear-scale definitions
\begin{equation}\label{Seq:GQ-linear-switch}\begin{aligned}
  G_{\mathrm{lin}}(w,z;z_{\mathrm{thr}}) &=
  \begin{cases}
    \Phi(-w) + \exp((z^2-w^2)/2)\,\Phi(-z), & z< z_{\mathrm{thr}},\\[3pt]
    \Phi(-w) + \phi(w)\,R_7(z), & z\ge z_{\mathrm{thr}},
  \end{cases} \\[6pt]
  Q_{\mathrm{lin}}(w,z;z_{\mathrm{thr}}) &=
  \begin{cases}
    \Phi(w) - \exp((z^2-w^2)/2)\,\Phi(-z), & z< z_{\mathrm{thr}},\\[3pt]
    \Phi(w) - \phi(w)\,R_7(z), & z\ge z_{\mathrm{thr}}.
  \end{cases}
\end{aligned}
\end{equation}
$G_{\mathrm{lin}}$ and $Q_{\mathrm{lin}}$ denote the double-precision linear-scale implementations, which switch to the Mills ratio expansion beyond the threshold $z_{\mathrm{thr}}$.
The Mills ratio representation never overflows and, for large $z$, provides higher accuracy by avoiding the cancellation between the exponential and $\Phi(-z)$ in \eqref{Seq:G-def}.

\begin{figure}[H]
  \centering
  \begin{subfigure}[t]{0.48\textwidth}
    \centering
    \includegraphics[width=\linewidth]{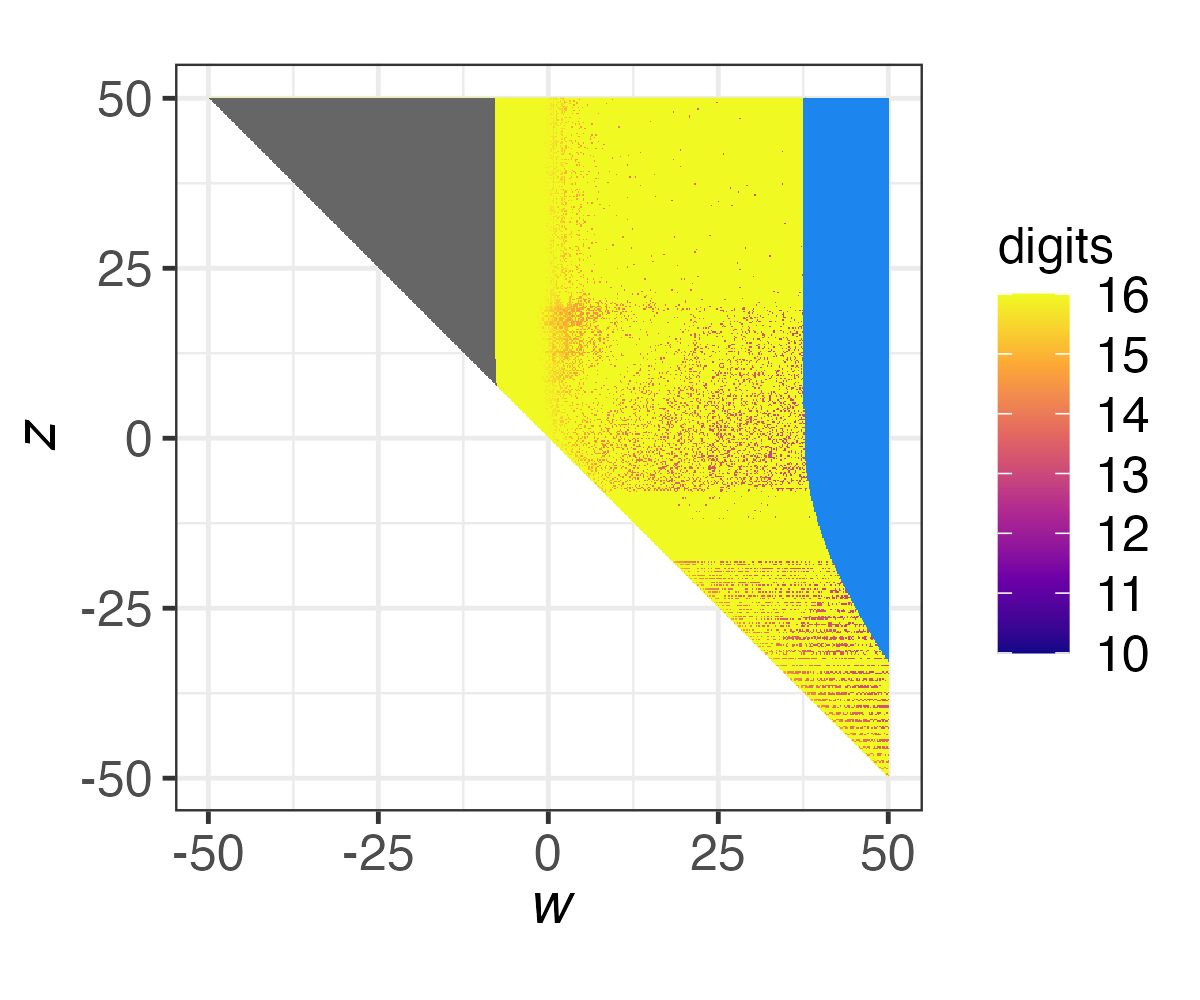}
    \caption{Accuracy in digits for $G_{\mathrm{lin}}$.}
    \label{fig:G-lin-heat}
  \end{subfigure}\hfill
  \begin{subfigure}[t]{0.48\textwidth}
    \centering
    \includegraphics[width=\linewidth]{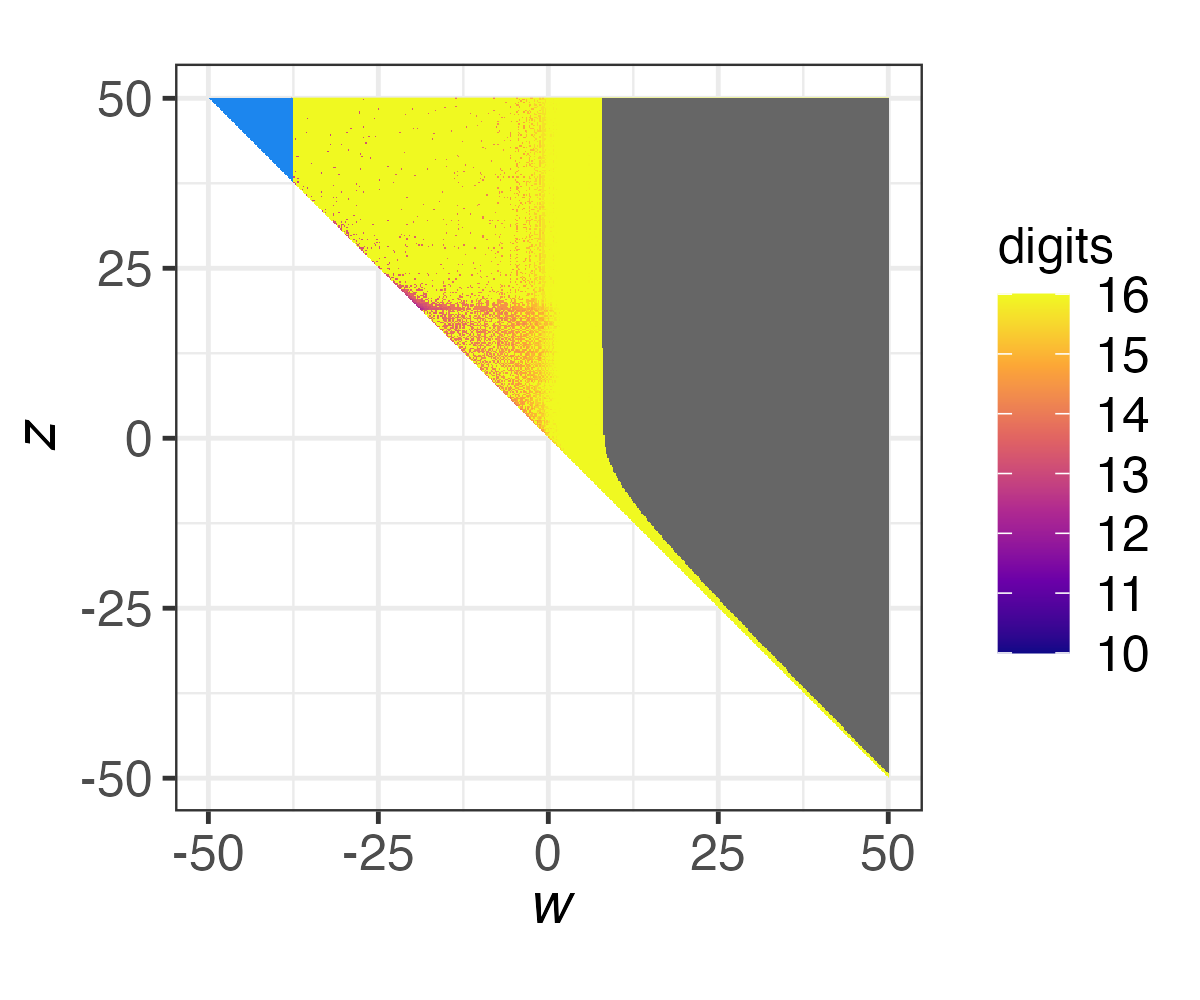}
    \caption{Accuracy in digits for $Q_{\mathrm{lin}}$.}
    \label{fig:Q-lin-heat}
  \end{subfigure}
\caption{
Digits of accuracy for $G_{\mathrm{lin}}$ and $Q_{\mathrm{lin}}$ on a regular grid in the $(w,z)$ plane (restricted to $w+z>0$).
Heatmap colors show digits of accuracy (capped at 16) computed from relative error against MPFR references, using $z_{\mathrm{thr}}=19$.
Loss-of-information regions are masked (gray: $\mathrm{LOI}_1$; light blue: $\mathrm{LOI}_0$).}
  \label{fig:lin-heatmaps}
\end{figure}

\paragraph{Numerical accuracy validation on the linear scale.}
Numerical accuracy of the double-precision implementations is assessed by comparison with high-precision references computed with MPFR.
In the double-precision path, $\Phi$ and $\phi$ are evaluated via R's \texttt{pnorm()} and \texttt{dnorm()}.
High-precision references are computed with \texttt{Rmpfr} by calling the same functions on \texttt{mpfr} arguments, with the precision explicitly set to \texttt{precBits = 128}.
The references are always based on the exact analytic forms \eqref{Seq:G-def}--\eqref{Seq:Q-def}, whereas the double-precision path may employ approximations such as the Mills expansion \eqref{Seq:GQ-linear-switch}; discrepancies therefore reflect both numerical precision and the chosen approximation.
To reduce cancellation, the expression $(z^2 - w^2)/2$ is evaluated in factored form $(z+w)(z-w)/2$ in numerical implementations.
All arithmetic operations for the references, including the summation of terms such as $\Phi(-w)+\phi(w)R(z)$ and the evaluation of logarithms, are carried out entirely in MPFR so that no rounding to double occurs during intermediate steps.
The result is converted back to double precision only after these operations are completed, for comparison with the double-precision approximations.
Validation then compares the high-precision MPFR references with double-precision results on a regular grid, defined by $w \in [w_{\min}, w_{\max}]$ and $z \in [z_{\min}, z_{\max}]$, subject to the strict mask $w+z>0$ (Fig.~\ref{fig:lin-heatmaps}).

Errors are evaluated with a log-scale metric based on $\log G_{\mathrm{lin}}$ and the MPFR reference $\log G_{\mathrm{ref}}$, applied consistently to both the linear-scale implementations and the log-scale formulation described in the next subsection.

For each grid point the relative error is recorded as
\begin{equation*}\varepsilon_{\mathrm{lin}} \;=\; \left|\,\exp\!\left(\log G_{\mathrm{lin}}-\log G_{\mathrm{ref}}\right) - 1\,\right|, 
  \qquad
  d_{\mathrm{lin}} \;=\; -\log_{10}\!\left(\max\!\left(\varepsilon_{\mathrm{lin}},10^{-16}\right)\right).
\end{equation*}
Here $\varepsilon_{\mathrm{lin}}$ coincides with the conventional relative error
$\lvert G_{\mathrm{lin}}-G_{\mathrm{ref}}\rvert/G_{\mathrm{ref}}$ whenever both values are positive and well resolved.
The same definitions are applied to $Q$ in place of $G$.
Digits of accuracy $d_{\mathrm{lin}}$ are capped at $16$ to avoid divergence in the case of exact matches.
In the implementation, the function $\exp(x)-1$ is evaluated using the numerically stable routine \texttt{expm1} in R.

Because relative error becomes uninformative near unity and in the extreme lower tail, two practical loss-of-information (LOI) flags are introduced based on the reference logs, where ``ref'' denotes either $G_{\mathrm{ref}}$ or $Q_{\mathrm{ref}}$:
\begin{equation*}
  \mathrm{LOI}_1:\ \log(1-\mathrm{ref}) < \log \vartheta_1,
  \qquad
  \mathrm{LOI}_0:\ \log(\mathrm{ref}) < \log \vartheta_0,
\end{equation*}
with thresholds $\vartheta_1=10^{-15}$ and the conservative lower-tail cutoff $\vartheta_0=10^{-308}$.
Points flagged as LOI are excluded from accuracy evaluation.

\begin{figure}[H]
  \centering
  \includegraphics[width=0.6\linewidth]{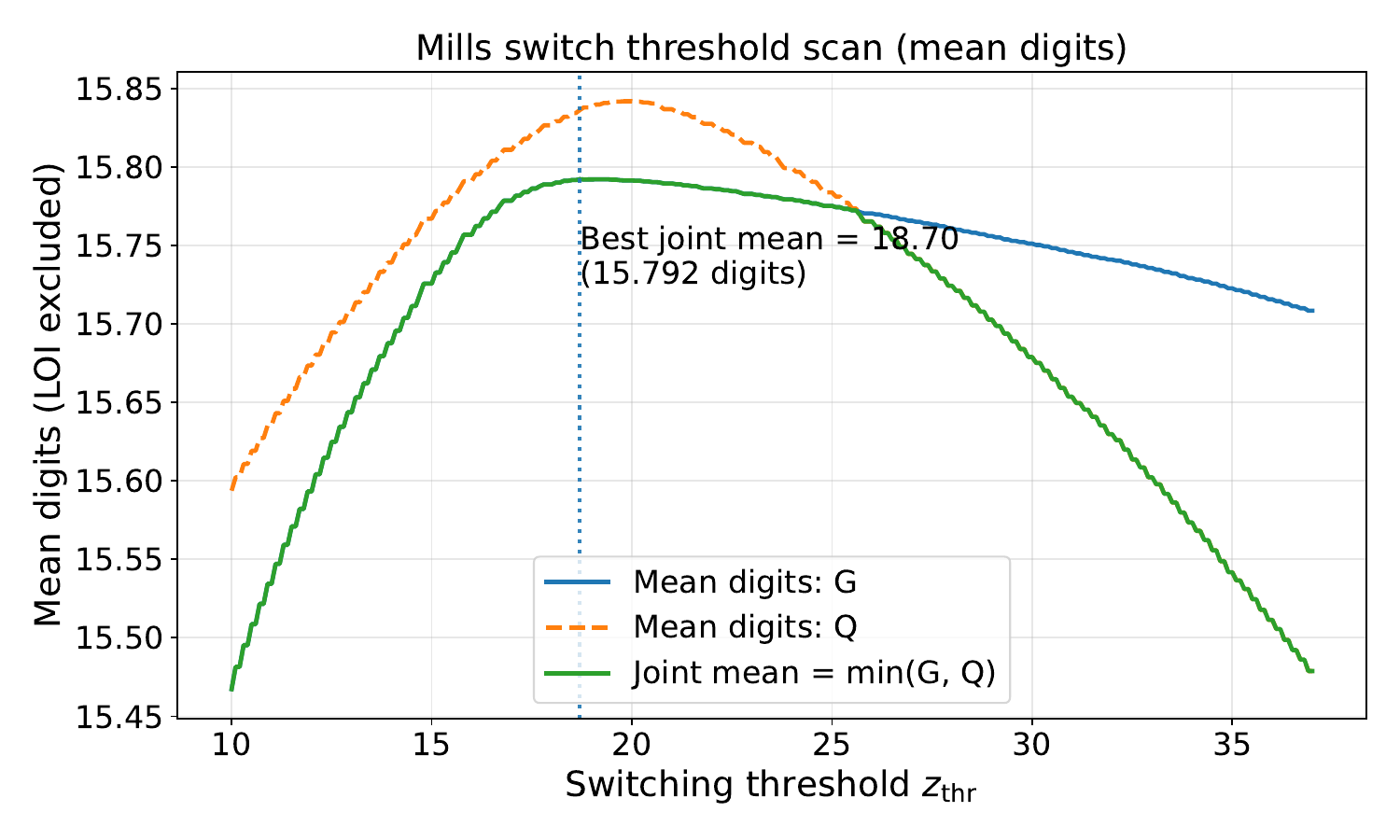}
\caption{
Mean digits of accuracy for $G_{\mathrm{lin}}$ and $Q_{\mathrm{lin}}$ (log-transformed relative error, capped at 16) versus the $z$-direction Mills switch threshold $z_{\mathrm{thr}}$.
Digits are computed against MPFR reference values, and LOI regions are excluded.
The vertical dashed line indicates the threshold that maximizes the joint criterion $\min\{\mathrm{mean}_G,\mathrm{mean}_Q\}$, which occurs near $z_{\mathrm{thr}} \approx 18.7$ for the grid and precision used here.}
  \label{fig:zthr-scan}
\end{figure}

A scan over $z_{\mathrm{thr}}$ (Fig.~\ref{fig:zthr-scan}) maximizes the joint mean digits $\min\{\text{mean}_G,\text{mean}_Q\}$ and exhibits a broad plateau with values above 13 digits, with an optimum near $z_{\mathrm{thr}}\approx 18.7$.

As shown in Fig.~\ref{fig:lin-heatmaps}\subref{fig:G-lin-heat}--\subref{fig:Q-lin-heat}, in the region $w\ll0$, $G_{\mathrm{lin}} \approx 1$ and its linear evaluation is ill-conditioned.
This is because the second term in $G_{\mathrm{lin}}$ is extremely small, while the first term is nearly one.
On the other hand, $Q_{\mathrm{lin}}$ remains representable because both terms are small but their difference avoids the same loss of precision.
Therefore, in the $w\ll0$ region I evaluate $Q_{\mathrm{lin}}$ and return the extinction probability $G$ in the form $1-Q_{\mathrm{lin}}$.
This notation indicates that $Q_{\mathrm{lin}}$ is computed directly and then used symbolically in $1-Q_{\mathrm{lin}}$, rather than numerically recombining two floating-point numbers close to $1$.
Symmetrically, for $w\gg0$ I evaluate $G_{\mathrm{lin}}$ and (if necessary) return $Q$ as $1-G_{\mathrm{lin}}$.

For clarity, define the linear-scale hybrid return value
\begin{equation*}
G_{\mathrm{lin,hyb}}(w,z) \coloneqq
\begin{cases}
G_{\mathrm{lin}}(w,z;z_{\mathrm{thr}}), & w \ge 0,\\
1 - Q_{\mathrm{lin}}(w,z;z_{\mathrm{thr}}), & w < 0.
\end{cases}
\end{equation*}

This complementary $w$ switching is independent of and applied simultaneously with the $z$ Mills switch, thereby extending the reliable evaluation region to most of the $(w,z)$ domain.
The only exception is the $\mathrm{LOI}_0$ region, where values of $G$ fall outside the representable range of double precision.
Since the smallest positive double is $2^{-1074}\approx 4.94\times 10^{-324}$, the admissible domain is bounded by
\begin{equation*}
4.94\times 10^{-324} < G_{\mathrm{lin,hyb}} < 1 - 4.94\times 10^{-324}.
\end{equation*}

The remaining challenge is estimation in the $\mathrm{LOI}_0$ region, which will be addressed in the next section by logarithmic evaluation.

\paragraph{Range extension on the log scale.}
Working on the log scale removes the hard double-precision bound $G>4.94\times10^{-324}$ imposed on the linear representation.
Because the most negative finite double that can be represented is $-\texttt{DBL\_MAX}\approx -1.80\times10^{308}$, the admissible window expands from this linear-scale interval to
\begin{equation*}\exp\!\left(-1.80\times10^{308}\right)
  \;<\;
  G(w,z)
  \;<\;
  1-\exp\!\left(-1.80\times10^{308}\right),
\end{equation*}
where the endpoint $\exp(-\texttt{DBL\_MAX})$ is kept \emph{symbolically}, because its direct evaluation in double precision would underflow to~0.
This widens the range by more than $7.8\times10^{307}$ orders of magnitude (base-10), i.e., $\Delta\log_{10}\approx 7.8\times10^{307}$.

All subsequent operations are carried out in the log scale, so values as small as $\log G\approx-1.80\times10^{308}$ remain representable.
Although converting such extremes back to linear scale inevitably rounds to 0, the probabilities themselves can still be \emph{reported} in scientific form $a\times10^{b}$ by formatting $\log_{10}G$ instead of evaluating $G$.  Consequently, every $(w,z)$ pair encountered in population-viability applications is evaluable without underflow, and the $\mathrm{LOI}_{\mathrm{LOG0}}$ region lies far outside the biologically relevant domain.

\subsection{Log-scale formulation and stable transforms}

The log scale is introduced to address challenge (iii), namely probabilities beyond the limits of double precision.
When $G$ or $Q$ falls below the minimum positive number representable in double precision (\texttt{DBL\_TRUE\_MIN} $\approx 4.94\times10^{-324}$), linear-scale arithmetic underflows to zero, and no departure from zero can be resolved.
If numbers smaller than this bound were representable, then in combination with the complementary evaluation described in the previous subsection, it would be possible to evaluate extinction probabilities closer to one than $1-4.94\times10^{-324}$.
To preserve information in such tails, I evaluate $G$ (and similarly $Q$) on the log scale, where sums and differences of exponentials are computed using stable identities such as the log-sum-exp and log-difference forms.
These transforms have been analyzed for rounding error, for example by \citet{blanchard2021accurately}, and are stable under floating-point arithmetic.

Let $a=\log \Phi(-w)$ and $b=(z^{2}-w^{2})/2+\log \Phi(-z)$, which are strictly negative throughout the admissible domain $w+z>0$.
Then
\begin{equation}\label{Seq:logG_ab}\log G(w,z)
  = \log\!\left(\Phi(-w) + \exp\!\left(\frac{z^{2}-w^{2}}{2}\right)\Phi(-z)\right)
  = \log\!\left(e^{a}+e^{b}\right),
\end{equation}
which is evaluated via the log-sum-exp identity
\begin{equation}\label{Seq:logsumexp}\log\!\left(e^{a}+e^{b}\right)
  = \max(a,b) + \log\!\left(1+\exp\!\left(-|a-b|\right)\right).
\end{equation}
By definition $Q(w,z)=1-G(w,z)$.
With $c=\log \Phi(w)$, and since $Q(w,z)=\Phi(w)-e^{b}\ge 0$ for $w+z>0$, it follows that $e^{b}\le \Phi(w)$ and thus $c>b$.
Accordingly, the log-difference identity can be applied directly, without case distinctions:
\begin{equation}\label{Seq:logdiffexp}\log Q(w,z)=\log\!\left(e^{c}-e^{b}\right)
= c + \log\!\left(1-\exp(b-c)\right).
\end{equation}
In practice, \texttt{log1p} is used in \eqref{Seq:logsumexp}, and \texttt{log1p}\slash\texttt{expm1} are used in \eqref{Seq:logdiffexp} when $b-c$ is near zero to avoid loss of significance.

Overflow cannot occur in the log scale.
Nevertheless, for $z \gg 0$ the Mills ratio approximation $R_7(z)$ is accurate, and it is numerically convenient to retain in the log scale the same large-$z$ branch as in the linear domain (Eq.~\ref{Seq:GQ-linear-switch}).
Reusing $a$ and $c$ from above, and introducing
\begin{equation*}
b_7 = \log \phi(w) + \log R_7(z),
\end{equation*}
one has $c>b_7$, so the log scale switching forms are
\begin{equation}\label{Seq:GQ-log-switch}\begin{aligned}
  \log G_{\mathrm{log}}(w,z;z_{\mathrm{thr}}) &=
  \begin{cases}
    \max(a,b) + \log\!\left(1+\exp\!\left(-|a-b|\right)\right), & z< z_{\mathrm{thr}},\\[3pt]
    \max(a,b_7) + \log\!\left(1+\exp\!\left(-|a-b_7|\right)\right), & z\ge z_{\mathrm{thr}},
  \end{cases} \\[6pt]
  \log Q_{\mathrm{log}}(w,z;z_{\mathrm{thr}}) &=
  \begin{cases}
    c + \log\!\left(1-\exp(b-c)\right), & z< z_{\mathrm{thr}},\\[3pt]
    c + \log\!\left(1-\exp(b_7 - c)\right), & z\ge z_{\mathrm{thr}}.
  \end{cases}
\end{aligned}
\end{equation}

\begin{figure}[H]
  \centering
  \begin{subfigure}[t]{0.48\textwidth}
    \centering
    \includegraphics[width=\linewidth]{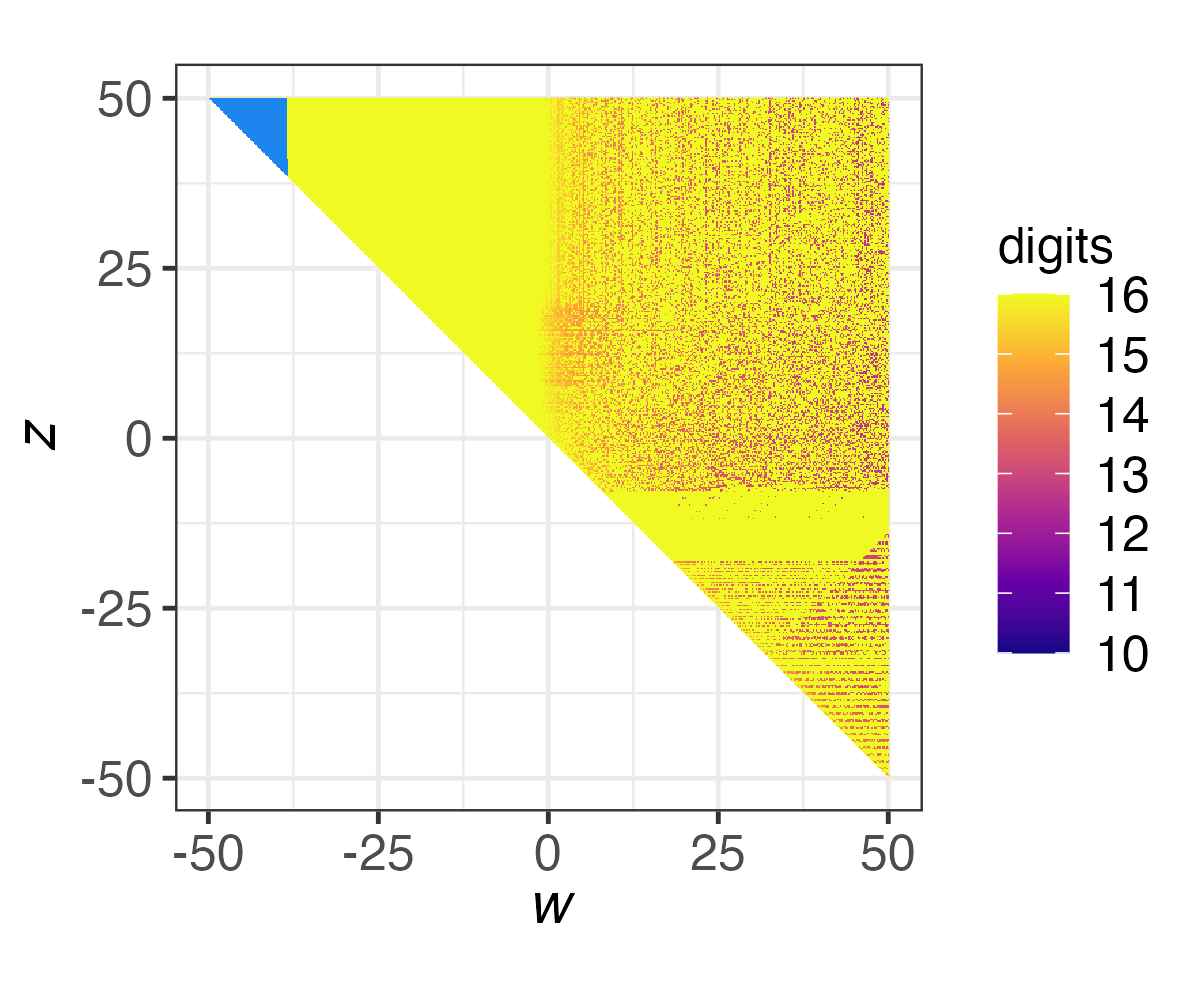}
    \caption{Accuracy in digits for $\log G_{\mathrm{log}}$.}
    \label{fig:G-log-heat}
  \end{subfigure}\hfill
  \begin{subfigure}[t]{0.48\textwidth}
    \centering
    \includegraphics[width=\linewidth]{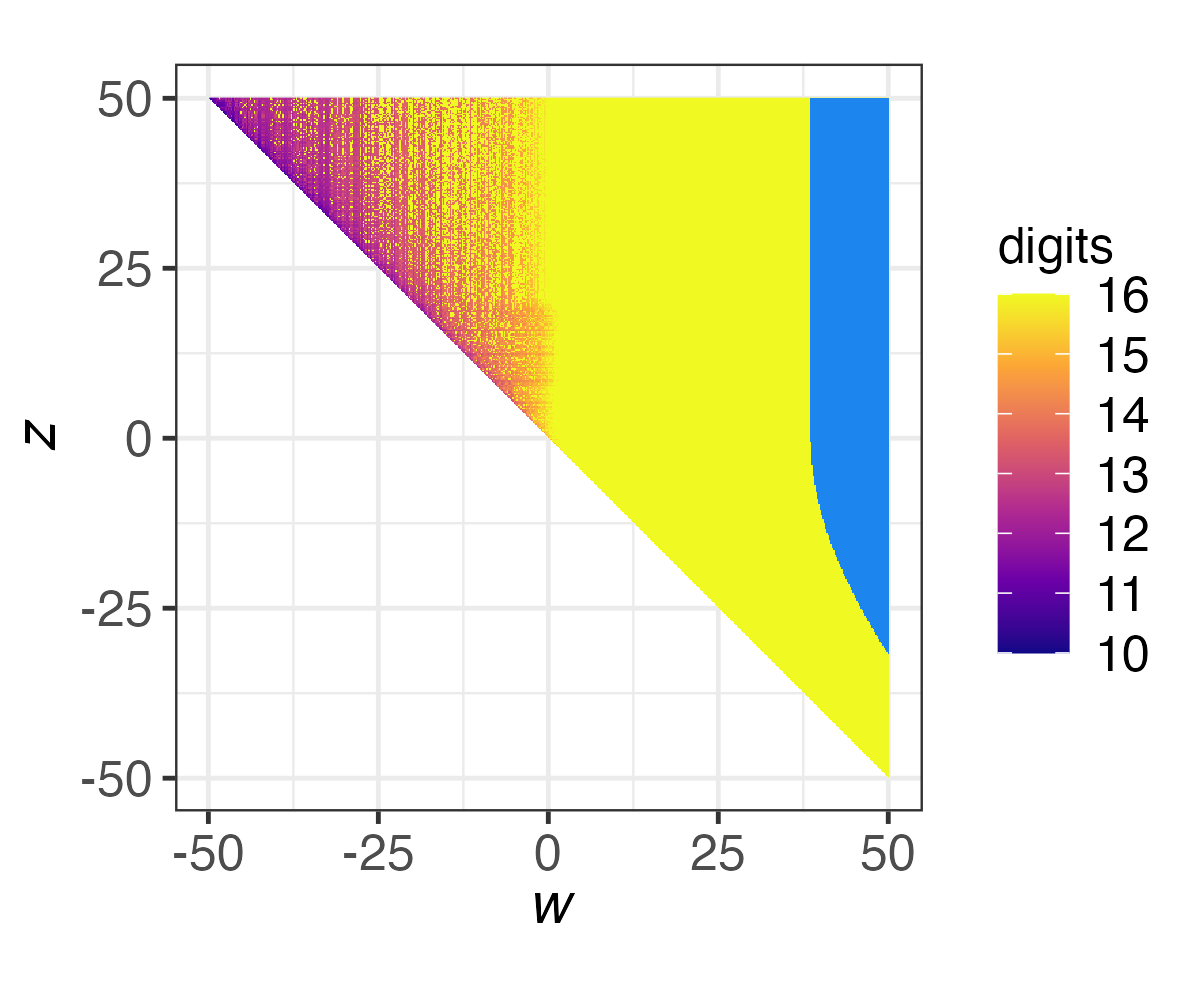}
    \caption{Accuracy in digits for $\log Q_{\mathrm{log}}$.}
    \label{fig:Q-log-heat}
  \end{subfigure}
\caption{
Digits of accuracy for $\log G_{\mathrm{log}}$ and $\log Q_{\mathrm{log}}$ on a regular grid in the $(w,z)$ plane (restricted to $w+z>0$).
Heatmap colors show digits of accuracy (capped at 16) computed from relative error against MPFR references, using $z_{\mathrm{thr}}=19$.
Loss-of-information regions on the log scale are masked (light blue: $\mathrm{LOI}_\mathrm{LOG0}$, gray: $\mathrm{LOI}_\mathrm{LOG\infty}$).}
  \label{fig:log-heatmaps}
\end{figure}

\paragraph{Numerical accuracy validation on the log scale.}
As in the linear scale, accuracy on the log scale is measured by the relative error computed from the log scale outputs,
\begin{equation*}\varepsilon_{\mathrm{log}} \;=\; \left|\,\exp\!\left(\log G_{\mathrm{log}}-\log G_{\mathrm{ref}}\right) - 1\,\right|, 
  \qquad
  d_{\mathrm{log}} \;=\; -\log_{10}\!\left(\max\!\left(\varepsilon_{\mathrm{log}},10^{-16}\right)\right).
\end{equation*}
Here $\log G_{\mathrm{log}}$ denotes the log scale evaluations defined in \eqref{Seq:GQ-log-switch}, while $\log G_{\mathrm{ref}}$ denotes high-precision MPFR reference values computed from the defining expression \eqref{Seq:logG_ab} without the Mills approximation. 
The computation of references, the precision setting (\texttt{precBits = 128}), the digits cap at 16, and the evaluation of $(z^2-w^2)/2$ in factored form $(z+w)(z-w)/2$ are identical to the linear-scale validation.
On the log scale, certain outputs carry no tail information and are flagged as LOI: $\log G=0$ or $\log Q=0$; $\log G=-\infty$ or $\log Q=-\infty$; and \texttt{NaN}, which is internally mapped to $-\infty$ and treated as the boundary case $G=0$ or $Q=0$.
Validation then compares the high-precision MPFR references with double-precision log scale results on a regular grid, defined by $w \in [w_{\min}, w_{\max}]$ and $z \in [z_{\min}, z_{\max}]$, subject to the mask $w+z>0$ (Fig.~\ref{fig:log-heatmaps}).

\begin{figure}[H]
  \centering
  \includegraphics[width=0.6\linewidth]{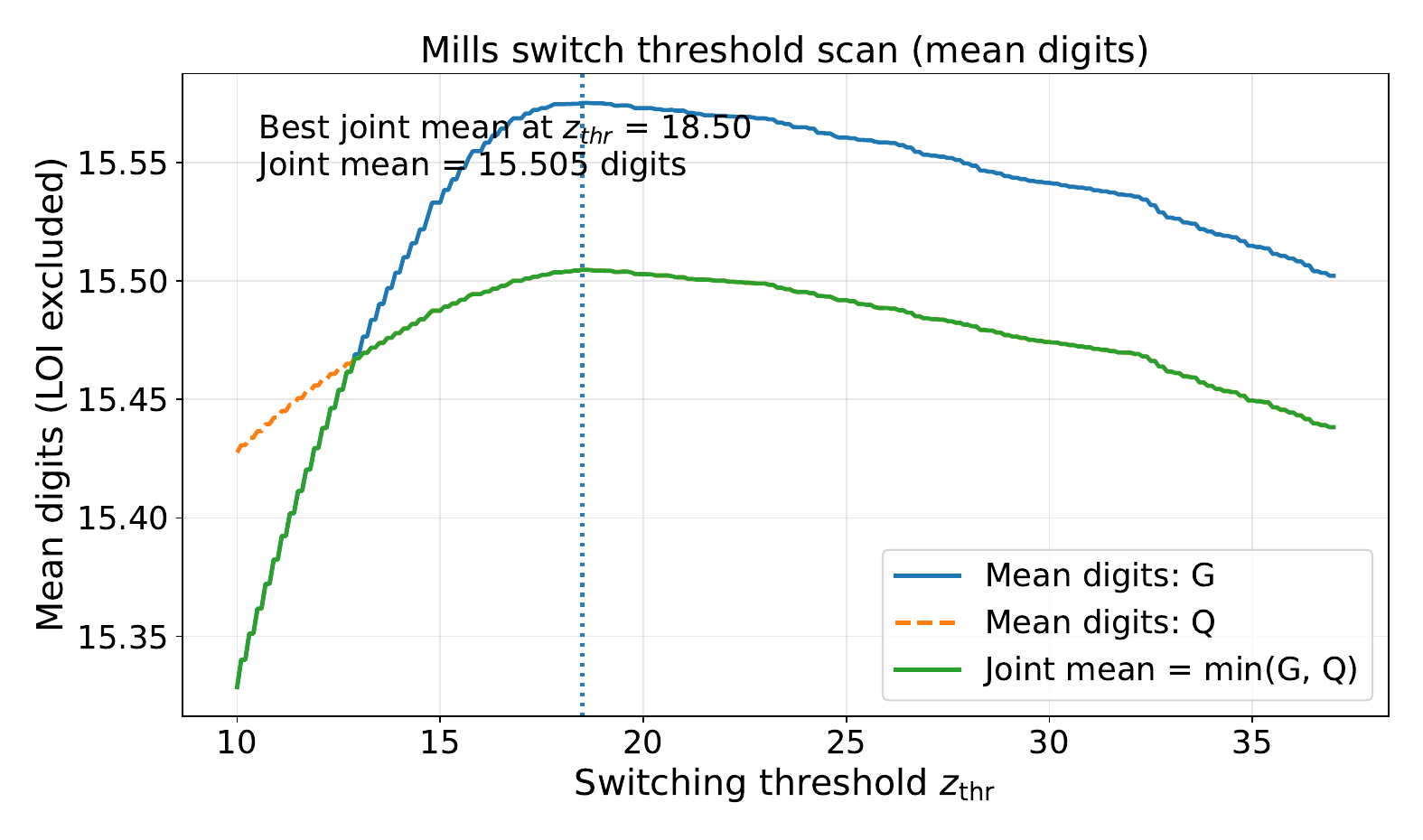}
\caption{
Mean digits of accuracy for $\log G_{\mathrm{log}}$ and $\log Q_{\mathrm{log}}$ (log transformed relative error, capped at 16; LOI excluded) versus the Mills switch threshold $z_{\mathrm{thr}}$ in the $z$ direction.
Digits are computed against MPFR reference values.
The vertical dashed line marks the maximizer of the joint criterion $\min\{\mathrm{mean}_G,\mathrm{mean}_Q\}$, which occurs at $z_{\mathrm{thr}}=18.50$ for the grid and precision used here.}
  \label{fig:zthr-scan-log}
\end{figure}

As shown in Figs~\ref{fig:log-heatmaps}a and b, the two complementary evaluations together allow extinction probabilities to be computed across the entire domain. 
On the log scale, regions of loss of information remain, but their locations mirror those observed on the linear scale.
In Fig.~\ref{fig:log-heatmaps}a, corresponding to $\log G_{\mathrm{log}}$, the upper-left corner ($w\ll0$) exhibits a $\mathrm{LOI}_\mathrm{LOG0}$ region, which corresponds to the $\mathrm{LOI}_0$ zone of $Q_{\mathrm{lin}}$ in Fig.~\ref{fig:Q-lin-heat}.
Conversely, in Fig.~\ref{fig:log-heatmaps}b, corresponding to $\log Q_{\mathrm{log}}$, the far right-hand side ($w\gg0$) contains a $\mathrm{LOI}_\mathrm{LOG0}$ region, aligned with the $\mathrm{LOI}_0$ region of $G_{\mathrm{lin}}$ in Fig.~\ref{fig:G-lin-heat}.
A grid scan in the log scale over $z_{\mathrm{thr}}$ shows that the optimum threshold for minimizing the mean relative error in the hybrid estimation with the Mills approximation is $z_{\mathrm{thr}} \approx 18.5$, which is very close to the linear-scale optimum (Fig.~\ref{fig:zthr-scan-log}).

\begin{figure}[H]
  \centering
  \includegraphics[width=0.7\linewidth]{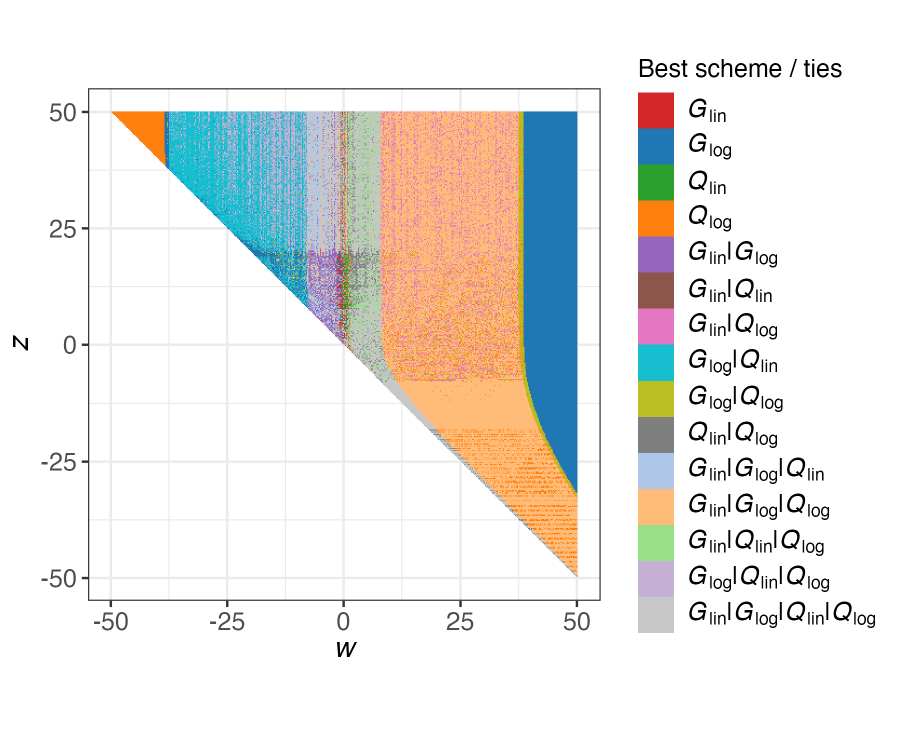}
\caption{
Optimal scheme map on the $(w,z)$ plane (restricted to $w+z>0$) comparing digits of accuracy among four evaluation schemes,
$G_{\mathrm{lin}}$, $G_{\mathrm{log}}$, $Q_{\mathrm{lin}}$, and $Q_{\mathrm{log}}$ (MPFR reference, digits capped at 16, $z_{\mathrm{thr}}=19$).
Each point is colored by the scheme attaining the highest digits of accuracy, with ties grouped into combined categories.
Loss-of-information regions are excluded.}
  \label{fig:optimal-scheme-map}
\end{figure}

\paragraph{Conclusions}
The linear- and log-scale schemes ($G_{\mathrm{lin}}$, $Q_{\mathrm{lin}}$, $\log G_{\mathrm{log}}$, $\log Q_{\mathrm{log}}$) exhibit different relative errors, but all are sufficiently accurate except in regions of loss of information.
Across the non-LOI region, the best-performing scheme typically attains about 15--16 digits of accuracy relative to MPFR references, i.e., relative error at the $\sim 10^{-16}$ level.
Thus extremely small probabilities are numerically well defined on double precision; for example, $G\approx 10^{-79}$ with relative error $\sim 10^{-16}$ corresponds to an absolute error of order $10^{-95}$ (approximately $10^{-79}(1\pm 10^{-16})$, i.e.\ $\bigl[10^{-79}-10^{-95},\,10^{-79}+10^{-95}\bigr]$).
The practical limitation is statistical uncertainty, not floating-point underflow.
Figure~\ref{fig:optimal-scheme-map} shows the regions of superiority in the $(w,z)$ domain, with many areas displaying ties.
In the left region, where $G$ is difficult to represent in double precision, $\log Q_{\mathrm{log}}$ provides stable values, whereas in the right region with analogous challenges, $\log G_{\mathrm{log}}$ achieves the highest accuracy.
Tie regions, such as $G_{\mathrm{log}}|Q_{\mathrm{lin}}$ or $G_{\mathrm{lin}}|G_{\mathrm{log}}|Q_{\mathrm{log}}$, occupy broad areas where all schemes perform adequately.
Overall, the log scale implementations effectively compensate for the information lost in the linear formulas, demonstrating that logarithmic formulations are robust across the domain.
Accordingly, $G$ is evaluated on the linear scale (with the $z$-direction Mills switch), while probabilities and confidence intervals are reported on the log scale via $\log G$ or $\log Q$ when $G$ is near $0$ or $1$ to avoid loss of information.

\subsection{Absolute error at the IUCN Criterion E thresholds}
While relative error is the natural metric for scanning accuracy over the whole $(w,z)$ domain, Red List category assignments hinge on whether the extinction probability crosses \emph{fixed cut-offs}.
In such decision regimes, even a small absolute bias near a threshold can invert the category; absolute error must therefore be evaluated locally around the IUCN Criterion~E thresholds $G_{\text{thr}}\in\{0.5,0.2,0.1\}$.

To define a symmetric and decision relevant neighborhood around each threshold on the probability scale,
I use logit bands
$\left|\operatorname{logit}(G_{\text{ref}})-\operatorname{logit}(G_{\text{thr}})\right|
\le \log(1.5)$.
Equivalently, the odds of $G_{\text{ref}}$ are within a factor $1.5$ of the odds at $G_{\text{thr}}$.
Within each band, selection is performed on a fixed regular grid in $(w,z)$ with
$w\in[-50,50]$, $z\in[-50,50]$, step $0.2$ on both axes (restricted to $w+z>0$).
Grid points whose MPFR reference satisfies $\mathrm{lo}\le G_{\mathrm{ref}}<\mathrm{hi}$ are retained, while loss-of-information cases (e.g., linear-scale underflow to $0$ or rounding to $1$, or non-finite/zero log outputs) are excluded.
For the scheme under test, the linear-scale absolute error $|\widehat{G}-G_{\mathrm{ref}}|$ is then evaluated at the retained points; for log scale implementations, errors are computed after exponentiating the log outputs.
Table~\ref{tab:abs-iucn-logit} reports, for each scheme, the sample size $N$, the maximum, the $95$th percentile ($p_{95}$), and the median absolute error.
Across all thresholds and schemes, the worst-case error is $\lesssim 5\times10^{-16}$, i.e., on the order of machine epsilon ($\varepsilon\approx 2.2\times10^{-16}$) and far below the $10^{-3}$--$10^{-2}$ tolerances commonly used in Criterion~E simulations.
Combined with the relative-error results, these bounds are more than sufficient for correct CR/EN/VU assignments in the threshold bands examined here.

\begin{table}[H]
  \centering
  \begin{threeparttable}
  \caption{Absolute-error statistics inside logit bands around the IUCN thresholds.
           The column ``Band [lo, hi]'' indicates the probability interval examined.}
  \label{tab:abs-iucn-logit}
  \begin{tabular}{@{}lllrccc@{}}
    \hline
    Threshold & Band [lo, hi]      & Scheme         & \multicolumn{1}{c}{$N$} & max                 & $p_{95}$             & median \\
    \hline
    $G_{\text{thr}}=0.5$ & $[0.400,\,0.600]$ & $G_{\text{lin}}$ & 699 & $3.33{\times}10^{-16}$ & $1.11{\times}10^{-16}$ & $0$ \\
                         &                         & $G_{\text{log}}$ & 699 & $5.00{\times}10^{-16}$ & $2.22{\times}10^{-16}$ & $5.55{\times}10^{-17}$ \\
                         &                         & $Q_{\text{lin}}$ & 699 & $3.33{\times}10^{-16}$ & $1.11{\times}10^{-16}$ & $5.55{\times}10^{-17}$ \\
                         &                         & $Q_{\text{log}}$ & 699 & $4.44{\times}10^{-16}$ & $2.22{\times}10^{-16}$ & $5.55{\times}10^{-17}$ \\
    \hline
    $G_{\text{thr}}=0.2$ & $[0.143,\,0.273]$ & $G_{\text{lin}}$ & 561 & $2.50{\times}10^{-16}$ & $1.39{\times}10^{-16}$ & $2.78{\times}10^{-17}$ \\
                         &                         & $G_{\text{log}}$ & 561 & $4.16{\times}10^{-16}$ & $1.67{\times}10^{-16}$ & $2.78{\times}10^{-17}$ \\
                         &                         & $Q_{\text{lin}}$ & 561 & $3.33{\times}10^{-16}$ & $1.11{\times}10^{-16}$ & $0$ \\
                         &                         & $Q_{\text{log}}$ & 561 & $4.44{\times}10^{-16}$ & $2.22{\times}10^{-16}$ & $0$ \\
    \hline
    $G_{\text{thr}}=0.1$ & $[0.069,\,0.143]$ & $G_{\text{lin}}$ & 557 & $1.94{\times}10^{-16}$ & $5.55{\times}10^{-17}$ & $1.39{\times}10^{-17}$ \\
                         &                         & $G_{\text{log}}$ & 557 & $2.22{\times}10^{-16}$ & $1.11{\times}10^{-16}$ & $2.78{\times}10^{-17}$ \\
                         &                         & $Q_{\text{lin}}$ & 557 & $2.22{\times}10^{-16}$ & $1.11{\times}10^{-16}$ & $0$ \\
                         &                         & $Q_{\text{log}}$ & 557 & $3.33{\times}10^{-16}$ & $1.11{\times}10^{-16}$ & $0$ \\
    \hline
  \end{tabular}
\begin{tablenotes}[flushleft]
\footnotesize
\item[] Bands: $\lvert \operatorname{logit}(G_{\mathrm{ref}})-\operatorname{logit}(G_{\mathrm{thr}})\rvert\le\log(1.5)$; LOI cases excluded.
\item[\dag] For log scale schemes ($G_{\text{log}}$ and $Q_{\text{log}}$), errors are assessed on the linear scale after exponentiation:
$G_{\text{log}}=\exp(\log G_{\text{log}})$ and $Q_{\text{log}}=\exp(\log Q_{\text{log}})$.
\item[\ddag] For $Q$ schemes, the absolute error is computed against $Q_{\text{ref}}=1-G_{\text{ref}}$ as
$|Q_{\text{scheme}}-Q_{\text{ref}}|$; $1-Q_{\text{scheme}}$ is not formed in double precision prior to comparison.
\end{tablenotes}
  \end{threeparttable}
\end{table}
 
\section{Monotonicity of \texorpdfstring{$G(w, z)$}{G(w, z)}}
\label{appendix:G_monotonicity}
\setcounter{equation}{0}

This appendix proves that the extinction probability function
\begin{equation}\label{Seq:G_recap}G(w, z) = \Phi(-w) + \exp\left( \frac{z^2 - w^2}{2} \right)\Phi(-z)
  \qquad \text{for} \quad w + z > 0, \quad w, z \in \mathbb{R}
\end{equation}
is strictly decreasing in both variables $w$ and $z$ throughout its domain.
Here, $\Phi(x)$ denotes the standard normal cumulative distribution function and $\phi(x) = \Phi'(x)$ its corresponding probability density function.

\subsection{Derivative with respect to \texorpdfstring{$w$}{w}}
Differentiating \eqref{Seq:G_recap} with respect to $w$ gives
\begin{align}\label{Seq:dGdw}\frac{\partial G}{\partial w}
  &= \frac{\partial}{\partial w} \Phi(-w) 
   + \frac{\partial}{\partial w} \left[ \exp\left( \frac{z^2 - w^2}{2} \right) \right] \Phi(-z), \notag \\
  &= -\phi(w) - w\, \exp\left( \frac{z^2 - w^2}{2} \right) \Phi(-z).
\end{align}

\paragraph{Case 1: $w > 0$.} Both terms on the right-hand side of \eqref{Seq:dGdw} are negative, so $\partial G/\partial w < 0$ directly.

\paragraph{Case 2: $w < 0$.}
Because $w+z>0$, it follows that $z>-w>0$.
Rewrite \eqref{Seq:dGdw} as
\begin{equation*}
\frac{\partial G}{\partial w}
  = -\phi(w) +(-w)\exp\left(\frac{z^{2}-w^{2}}{2}\right)\Phi(-z),\qquad (-w>0).
\end{equation*}
Apply the Mills ratio bound \citep{gordon1941values}
\begin{equation}\label{Seq:mills_bound}\Phi(-z) < \frac{\phi(z)}{z}, \qquad z>0,
\end{equation}
and use the identity
\begin{equation*}\exp\left(\frac{z^{2}-w^{2}}{2}\right)\phi(z) = \phi(w).
\end{equation*}
Then
\begin{equation*}
(-w)\exp\left(\frac{z^{2}-w^{2}}{2}\right)\Phi(-z) < \frac{-w}{z}\phi(w)<\phi(w),
\end{equation*}
since $-w < z$.
Substituting this strict inequality back into the derivative gives
\begin{equation*}
\frac{\partial G}{\partial w} < -\phi(w)+\phi(w) = 0.
\end{equation*}
Hence $\displaystyle \partial G/\partial w < 0$ for all $w<0$ with $w+z>0$.

\subsection{Derivative with respect to \texorpdfstring{$z$}{z}}

Differentiating \eqref{Seq:G_recap} with respect to $z$ yields
\begin{equation}\label{Seq:dGdz}\frac{\partial G}{\partial z}
  = \exp\left( \frac{z^2 - w^2}{2} \right)\left[ z\,\Phi(-z) - \phi(z) \right].
\end{equation}

\paragraph{Case 1: $z > 0$.} From inequality~\eqref{Seq:mills_bound}, one has
\begin{equation*}
z\,\Phi(-z) < \phi(z).
\end{equation*}
Therefore the bracket in \eqref{Seq:dGdz} is negative.

\paragraph{Case 2: $z < 0$.} In this case, $z\,\Phi(-z) < 0$ while $\phi(z) > 0$, so the expression in brackets is again negative.

Therefore, \eqref{Seq:dGdz} shows that $\partial G/\partial z < 0$ for all $w + z > 0$.

\paragraph*{Remarks at $w=0$ and $z=0$.}
At $w=0$ (hence $z>0$), one has $\partial G/\partial w=-\phi(0)<0$.
At $z=0$ (hence $w>0$), one has $\partial G/\partial z=\exp \bigl(-(w^2)/2\bigr)\,[0\cdot \Phi(0)-\phi(0)]<0$.
These confirm strict decrease also on these coordinate axes within $w+z>0$.

\paragraph*{Boundary continuity at $w+z=0$.}
On the boundary $z=-w$ one has
\begin{equation*}
G(w,-w)=\Phi(-w)+\exp(0)\,\Phi(w)=\Phi(-w)+\Phi(w)=1.
\end{equation*}
By continuity of $\Phi$ and the exponential, it follows that $G(w,z)\to 1$ as $w+z\downarrow 0$ with $(w,z)$ restricted to the admissible region $w+z>0$.

\subsection{Conclusion}

The function $G(w, z)$ is strictly decreasing in both $w$ and $z$ throughout its domain:
\begin{equation*}\frac{\partial G}{\partial w}(w, z) < 0
  \quad \text{and} \quad
  \frac{\partial G}{\partial z}(w, z) < 0
  \qquad \text{for all } w + z > 0.
\end{equation*}
 
\section{Asymptotic Behavior of \texorpdfstring{$G(w,z)$}{G(w,z)} as \texorpdfstring{$z \to \pm\infty$}{z to +/- infinity}}
\label{appendix:asymptotics_G}
\setcounter{equation}{0}

The extinction probability from the main text is restated below:
\begin{equation}\label{Seq:G2}G(w, z) = \Phi(-w) + \exp\left(\frac{z^2 - w^2}{2}\right)\Phi(-z)
  \quad \text{for} \quad w + z > 0, \quad w, z \in \mathbb{R}.
\end{equation}
This appendix establishes the two limiting forms
\begin{align}
  G(w,z) \simeq
  \begin{cases}\label{Seq:G_asymptotics}\Phi(-w) & \qquad \text{for} \qquad z \gg 0,\\
    \exp\left(\frac{z^2 - w^2}{2}\right) & \qquad \text{for} \qquad z \ll 0.
  \end{cases}
\end{align}
and shows that, in the lower-right quadrant ($z<0$, $w+z>0$), the equal-risk contours satisfy $w^{2}-z^{2}=\text{const}$, with slope $dz/dw=w/z\to-1$ as $z \to -\infty$.

Throughout, write $\phi(x)=\Phi'(x)=\tfrac{1}{\sqrt{2\pi}}\exp(-x^{2}/2)$ for the standard normal density and recall the Mills ratio bounds \citep{gordon1941values}
\begin{equation}\label{Seq:mills_bounds}\frac{\phi(x)}{x+1/x} < \Phi(-x) < \frac{\phi(x)}{x}, \qquad x > 0.
\end{equation}
All limits below are taken within the admissible region $w+z>0$.

\subsection{Asymptotics for \texorpdfstring{$z \gg 0$}{z >> 0}}

Consider the second term on the right-hand side of Eq.~\eqref{Seq:G2}:
\begin{equation*}
\exp\left( \frac{z^{2}-w^{2}}{2} \right)\Phi(-z).
\end{equation*}
For $z > 0$, the Mills ratio bound~\eqref{Seq:mills_bounds} gives
\begin{equation*}
\Phi(-z) \le \frac{\phi(z)}{z}.
\end{equation*}
Hence,
\begin{align*}
\exp\left( \frac{z^{2}-w^{2}}{2} \right)\Phi(-z)
&\le \frac{\phi(z)}{z}\exp\left( \frac{z^{2}-w^{2}}{2} \right) \nonumber\\
&= \frac{1}{z} \cdot \frac{1}{\sqrt{2\pi}} \exp\left( -\frac{w^2}{2} \right) \nonumber \\
&= \frac{\phi(w)}{z},
\end{align*}
which tends to $0$ as $z \to \infty$ for any fixed $w$.

To establish that the second term is negligible relative to the first term $\Phi(-w)$, consider the ratio
\begin{equation*}
\frac{\exp\left( (z^{2}-w^{2})/2 \right)\Phi(-z)}{\Phi(-w)}
      \le
      \frac{\phi(w)}{z \Phi(-w)}
      \longrightarrow 0, \qquad \text{as} \qquad z\to\infty,
\end{equation*}
because $\Phi(-w)$ is a $z$-independent positive constant, while the numerator decays like $1/z$.

Therefore, in the limit $z \gg 0$ with fixed $w$, the second term becomes asymptotically negligible, and
\begin{equation*}
G(w, z) \sim \Phi(-w).
\end{equation*}
This establishes the first line of Eq.~\eqref{Seq:G_asymptotics}.

\subsection{Asymptotics for \texorpdfstring{$z \ll 0$}{z << 0}}

Now take the limit $z \to -\infty$ within the admissible region $w+z>0$.
Since $z < 0$, it follows that $w > -z$, hence $w > |z|$ and therefore $w \to \infty$ as $z \to -\infty$.

In this limit, the first term $\Phi(-w)$ becomes negligible compared to the second.
The second term can be written as
\begin{equation*}
\exp\left(\frac{z^2 - w^2}{2}\right)\Phi(-z)
      =\exp\left(\frac{z^2 - w^2}{2}\right)\left[1+O\left(\frac{\phi(|z|)}{|z|}\right)\right],
\end{equation*}
using that $\Phi(-z)=1-\Phi(z)$ and $\Phi(z)=O\!\bigl(\phi(|z|)/|z|\bigr)$ as $z \to -\infty$.

For the first term, the Mills bound implies
\begin{equation*}
\Phi(-w)=O\left(\frac{\phi(w)}{w}\right)=O\left(\frac{e^{-w^{2}/2}}{w}\right),
\qquad w \to \infty.
\end{equation*}
Hence, the ratio of the first term to the second is
\begin{equation*}
\frac{\Phi(-w)}{\exp\left((z^2-w^2)/2\right)}
   =O\left(\frac{e^{-w^{2}/2}}{w}\right)e^{(w^2-z^2)/2}
   =O\left(\frac{e^{-z^{2}/2}}{w}\right)
   \longrightarrow 0,
\end{equation*}
as $z \to -\infty$.

It follows that the first term is negligible, and
\begin{equation*}
G(w,z) \sim \exp\left(\frac{z^2-w^2}{2}\right),
\qquad z \to -\infty.
\end{equation*}
This completes the proof of the second line of Eq.~\eqref{Seq:G_asymptotics}.

\subsection{Asymptotic form of equal-risk contours for \texorpdfstring{$z \ll 0$}{z << 0}}

Fix $G(w,z)=g_0 \in (0,1)$ in the region $z < 0$ and $w + z > 0$.
Applying the asymptotic approximation derived in the previous subsection for $z \ll 0$ yields
\begin{equation*}
g_0 = \exp\left(\frac{z^{2}-w^{2}}{2}\right)+o(1)
      \quad\Longrightarrow\quad
      w^{2}-z^{2}= -2\log g_0 + o(1).
\end{equation*}
Differentiating the level set $w^{2}-z^{2} = \text{const}$ with respect to $w$ gives
\begin{equation*}
2w\,dw-2z\,dz=0 \quad \Longrightarrow \quad \frac{dz}{dw}=\frac{w}{z}.
\end{equation*}

In the domain $z < 0$ and $w > |z|$, it follows that $w/z < 0$ and $|w/z| > 1$.
Hence,
\begin{equation*}
  \frac{dz}{dw}=\frac{w}{z}=-\left|\frac{w}{z}\right|<-1
  \quad\text{and}\quad
  \frac{dz}{dw}\longrightarrow-1 \quad\text{as}\quad z\to-\infty,
\end{equation*}
Therefore, the equal-risk contours are steeper than the diagonal line $w = -z$ and approach it asymptotically from below.

\paragraph{Remark.}
For $z\to+\infty$,
\begin{equation*}
  G(w,z)=\Phi(-w)+\frac{\phi(w)}{z}\Bigl(1-\frac{1}{z^{2}}+O(z^{-4})\Bigr).
\end{equation*}
Thus the one-term approximation $G\simeq\Phi(-w)$ has relative error $O(1/z)$, while including the next term $G\simeq \Phi(-w)+\phi(w)/z$ reduces the residual relative error to $O(1/z^{3})$.
For $z\to-\infty$ with $w+z>0$, since $\Phi(-z)=1-O(\phi(|z|)/|z|)$ and $\Phi(-w)$ is exponentially smaller than
$\exp((z^{2}-w^{2})/2)$, the approximation $G\simeq \exp((z^{2}-w^{2})/2)$ has relative error
$O(\phi(|z|)/|z|)=O(e^{-z^{2}/2}/|z|)$, i.e., exponentially small.
These accuracies ensure that the asymptotic forms are highly reliable even for moderate values of $|z|$, and hence suitable for use in the CI construction in Section~\ref{sec:CI}.
 
\section{Slope of Equal-Risk Contours for \texorpdfstring{$w + z > 0$}{w + z > 0}}
\label{appendix:slope_G}
\setcounter{equation}{0}

This appendix shows that the equal-risk contours $G(w,z)=g_0$ in the region $w + z > 0$ are steeper than the diagonal line $w = -z$.

The extinction-probability function under analysis is restated below:
\begin{equation*}G(w, z) = \Phi(-w) + \exp\left(\frac{z^{2}-w^{2}}{2}\right)\Phi(-z)
  \quad \text{for} \quad w + z > 0, \quad w, z \in \mathbb{R}.
\end{equation*}
Here, $\Phi(x)$ denotes the standard normal cumulative distribution function and $\phi(x)=\Phi'(x)$ is the corresponding probability density function.

\medskip
\noindent
Implicit differentiation of $G(w,z)=g_0$ gives the slope
\begin{equation*}\frac{\mathrm{d}z}{\mathrm{d}w}
  = -\frac{\partial G/\partial w}{\partial G/\partial z}
  = -\frac{G_w}{G_z}
  = \frac{\phi(z) + w\,\Phi(-z)}{z\,\Phi(-z) - \phi(z)}
  \eqqcolon \ell(w,z).
\end{equation*}

\subsection{Auxiliary function}

Define
\begin{equation*}h(z) \coloneqq \phi(z) - z\,\Phi(-z),
\end{equation*}
where $\phi$ and $\Phi$ denote the standard normal density and distribution function, respectively.

Since $\phi'(z) = -z\,\phi(z)$ and $\Phi'(-z) = -\phi(z)$, it follows that
\begin{equation*}
  h'(z) = -\Phi(-z) < 0 \quad \text{for all } z \in \mathbb{R},
\end{equation*}
so $h(z)$ is strictly decreasing.

To establish that $h(z) > 0$ for all $z$, consider first the case $z < 0$.
In this region, both $\phi(z)$ and $\Phi(-z)$ are positive, and the negative sign on $z$ implies that $-z\,\Phi(-z) > 0$, so $h(z) = \phi(z) + |z|\,\Phi(-z) > 0$.
On the other hand, for $z \ge 0$, the Mills upper bound $\Phi(-z) < \phi(z)/z$ \citep{gordon1941values} yields
\begin{equation*}
  z\,\Phi(-z) < \phi(z) \quad \Longrightarrow \quad h(z) = \phi(z) - z\,\Phi(-z) > 0.
\end{equation*}

Therefore $h(z)$ is strictly positive for all $z \in \mathbb{R}$, strictly decreasing in $z$, and satisfies $\lim_{z\to+\infty} h(z)=0$ and $\lim_{z\to-\infty} h(z)=+\infty$.

\subsection{Comparison with the diagonal}

To compare $\ell(w,z)$ with $-1$, multiply both sides of $\ell(w,z)<-1$ by the negative denominator $z\,\Phi(-z)-\phi(z)=-h(z)$:
\begin{align*}\ell(w,z) < -1
  &\;\Longleftrightarrow\;
    \bigl(\phi(z) + w\,\Phi(-z)\bigr)
    + \bigl(z\,\Phi(-z) - \phi(z)\bigr) > 0 \nonumber\\[2pt]
  &\;\Longleftrightarrow\;
    \Phi(-z)\bigl(w + z\bigr) > 0.
\end{align*}
Because $\Phi(-z)>0$ and $w+z>0$, the inequality holds, yielding
\begin{equation*}\frac{\mathrm{d}z}{\mathrm{d}w} = \ell(w,z) < -1
  \qquad \text{throughout the region } w + z > 0.
\end{equation*}
Therefore the equal-risk contours are strictly steeper than the diagonal $w = -z$ across the entire domain $w + z > 0$.
 
\section{Details of the Sampling Correlation of \texorpdfstring{$\widehat{w}$}{w-hat} and \texorpdfstring{$\widehat{z}$}{z-hat}}\label{appendix:wz_corr_details}
\setcounter{equation}{0}
This appendix outlines the derivation of the closed-form expression for the correlation coefficient between the estimators $\widehat{w}$ and $\widehat{z}$.

\subsection{Definition of \texorpdfstring{$w$}{w} and \texorpdfstring{$z$}{z} and their estimators}
Consider two mutually independent random variables
$\widehat{\mu}$ and $\widehat{\sigma}^{2}$, which satisfy
\begin{equation*}
  \widehat{\mu} \sim \mathcal{N}\!\left(\mu,\; \frac{\sigma^{2}}{t_q}\right), 
  \qquad
  \frac{q\,\widehat{\sigma}^{2}}{\sigma^{2}} \sim \chi^{2}_{\,q-1},
\end{equation*}
and serve as estimators of the drift $\mu \in \mathbb{R}$ and the variance $\sigma^{2}>0$ 
of the Wiener process with drift.
Suppose that observations are available at $q$ intervals over a total observation time span $t_q>0$.

For any $t>0$ and initial log-distance $x_d>0$, define
\begin{equation*}
  w=\frac{\mu t+x_d}{\sigma\sqrt{t}},
  \qquad
  z=\frac{-\mu t+x_d}{\sigma\sqrt{t}}.
\end{equation*}
For inference at a finite time horizon $t^\ast$, the plug-in estimators are
\begin{equation*}
  \widehat{w}=\frac{\widehat{\mu}\,t^\ast+x_d}{\widehat{\sigma}\sqrt{t^\ast}},
  \qquad
  \widehat{z}=\frac{-\widehat{\mu}\,t^\ast+x_d}{\widehat{\sigma}\sqrt{t^\ast}}.
\end{equation*}

\subsection{Reparameterization and auxiliary variables}
It is convenient to introduce the reparameterization
\begin{equation*}
  m \coloneqq t^{\ast}\widehat{\mu},\qquad \eta \coloneqq \widehat{\sigma}^{-1},
  \qquad
  \bar m \coloneqq \mathbb{E}[m]=t^{\ast}\mu,\qquad
  \sigma_m^{2} \coloneqq \operatorname{Var}(m)=\frac{t^{\ast 2}\sigma^{2}}{t_q}.
\end{equation*}
Then $m \sim \mathcal{N}(\bar m, \sigma_m^{2})$.
Since $\widehat{\mu}$ and $\widehat{\sigma}^{2}$ are independent, it follows that $m$ and $\eta$ are independent (as functions of independent variables), and
\begin{equation*}
  \widehat{w}=\frac{x_d+m}{\sqrt{t^{\ast}}}\,\eta,
  \qquad
  \widehat{z}=\frac{x_d-m}{\sqrt{t^{\ast}}}\,\eta.
\end{equation*}

\paragraph{Moments of $\eta=\widehat{\sigma}^{-1}$.}

Write $\Xi \coloneqq q\,\widehat{\sigma}^{2}/\sigma^{2}\sim\chi^{2}_{q-1}$.
Then
\begin{equation*}
  \eta=\widehat{\sigma}^{-1}=\frac{1}{\sigma}\sqrt{\frac{q}{\Xi}}.
\end{equation*}

For a chi-squared random variable $\Xi\sim\chi^{2}_{q-1}$ with $q-1$ degrees of freedom, 
the following identity for inverse moments holds whenever $a<(q-1)/2$:
\begin{equation*}
  \mathbb{E}[\Xi^{-a}]
  = 2^{-a}\,\frac{\Gamma\!\left(\tfrac{q-1}{2}-a\right)}{\Gamma\!\left(\tfrac{q-1}{2}\right)}.
\end{equation*}

Applying this formula yields, for $q>2$ (for $\mathbb{E}[\eta]$) and $q>3$ (for $\mathbb{E}[\eta^{2}]$),
\begin{equation*}
  \mathbb{E}[\eta]
  =\frac{\sqrt{q}}{\sqrt{2}\,\sigma}\,
    \frac{\Gamma\!\left((q-2)/2\right)}{\Gamma\!\left((q-1)/2\right)},
  \qquad
  \mathbb{E}[\eta^{2}]
  =\frac{q}{(q-3)\,\sigma^{2}}.
\end{equation*}
Hence
\begin{equation*}
  \sigma_{\eta}^{2}
  :=\operatorname{Var}(\eta)
  =\frac{q}{\sigma^{2}}
     \left[
       \frac{1}{q-3}
       -\frac12
        \left\{\frac{\Gamma\!\left((q-2)/2\right)}{\Gamma\!\left((q-1)/2\right)}\right\}^{2}
     \right],
  \qquad q>3.
\end{equation*}

The following formulas require $q>2$ (for $\mathbb{E}[\eta]$) and for $q>3$ (for $\mathbb{E}[\eta^{2}]$ and $\operatorname{Var}(\eta)$).

\subsection{Variances and covariance of \texorpdfstring{$(\widehat{w},\widehat{z})$}{(w-hat, z-hat)}}

Using independence of $m$ and $\eta$ and elementary identities for moments of products,
\begin{equation*}
  \operatorname{Var}(\widehat{w})
  =\frac{\,\mathbb{E}[\eta^{2}]\,\sigma_{m}^{2}
         +\sigma_{\eta}^{2}\,(x_d+\bar m)^{2}}{t^{\ast}},
  \qquad
  \operatorname{Var}(\widehat{z})
  =\frac{\,\mathbb{E}[\eta^{2}]\,\sigma_{m}^{2}
         +\sigma_{\eta}^{2}\,(x_d-\bar m)^{2}}{t^{\ast}},
\end{equation*}
\begin{equation*}
  \operatorname{Cov}(\widehat{w},\widehat{z})
  =\frac{-\,\mathbb{E}[\eta^{2}]\,\sigma_{m}^{2}
          +\sigma_{\eta}^{2}\,(x_d^{2}-\bar m^{2})}{t^{\ast}}.
\end{equation*}

\subsection{Closed-form correlation}

Using
\begin{equation*}
  x_d=\tfrac{\sigma\sqrt{t^{\ast}}}{2}\,(w+z),\qquad
  \bar m=\tfrac{\sigma\sqrt{t^{\ast}}}{2}\,(w-z),
\end{equation*}
one has
\begin{equation*}
  (x_d+\bar m)^{2}=\sigma^{2}t^{\ast}w^{2},\qquad
  (x_d-\bar m)^{2}=\sigma^{2}t^{\ast}z^{2},\qquad
  x_d^{2}-\bar m^{2}=\sigma^{2}t^{\ast}w z.
\end{equation*}
Introduce the design-dependent constants
\begin{equation*}
  A:=\mathbb{E}[\eta^{2}]\,\sigma_{m}^{2}
   =\frac{q\,t^{\ast 2}}{(q-3)\,t_q},
  \qquad
  D:=\sigma_{\eta}^{2}\,\sigma^{2}\,t^{\ast}
   =q\,t^{\ast}\!\left[
       \frac{1}{q-3}
       -\frac12
        \left\{\frac{\Gamma\!\left((q-2)/2\right)}{\Gamma\!\left((q-1)/2\right)}\right\}^{2}
     \right],
\end{equation*}
which depend only on $(q,t_q,t^{\ast})$ and not on $(\mu,\sigma,x_d)$.
Then
\begin{equation*}
  \operatorname{Var}(\widehat{w})=\frac{A+D\,w^{2}}{t^{\ast}},\qquad
  \operatorname{Var}(\widehat{z})=\frac{A+D\,z^{2}}{t^{\ast}},\qquad
  \operatorname{Cov}(\widehat{w},\widehat{z})=\frac{-A+D\,w z}{t^{\ast}},
\end{equation*}
and therefore, for $w+z>0$,
\begin{equation*}
  \operatorname{Corr}(\widehat{w},\widehat{z})
  =\frac{-A+D\,w z}{\sqrt{(A+D\,w^{2})(A+D\,z^{2})}}.
\end{equation*}
Defining $k:=A/D$ yields
\begin{equation}\label{Seq:corr-k}
  \operatorname{Corr}(\widehat{w},\widehat{z})
  =\rho(w,z;k)
  :=\frac{-k+w z}{\sqrt{(k+w^{2})(k+z^{2})}},
  \qquad w+z>0,\ k>0.
\end{equation}

\subsection{Positivity of \texorpdfstring{$D$}{D} and \texorpdfstring{$k$}{k}}

Let
\begin{equation*}
  \psi_q:=\frac{\Gamma\!\left((q-2)/2\right)}{\Gamma\!\left((q-1)/2\right)},\qquad q>3.
\end{equation*}
Gautschi's inequality states that for $u>0$ and $0<s<1$,
\begin{equation*}
  u^{\,1-s} < \frac{\Gamma(u+1)}{\Gamma(u+s)} < (u+1)^{\,1-s}.
\end{equation*}
Applying it with $s=\tfrac12$ and $u=(q-3)/2$ gives
\begin{equation*}
  \frac{1}{\sqrt{u+1}} < \frac{\Gamma(u+\tfrac12)}{\Gamma(u+1)} < \frac{1}{\sqrt{u}},
\end{equation*}
and therefore
\begin{equation*}
  \psi_q^{2} < \frac{2}{q-3}.
\end{equation*}
Hence
\begin{equation*}
  \frac{1}{q-3} - \frac12\,\psi_q^{2} > 0,
\end{equation*}
so $D>0$. Since $A>0$ trivially, $k=A/D>0$ whenever $q>3$, $t_q>0$, and $t^{\ast}>0$.

\subsection{Asymptotic behavior of \texorpdfstring{$k$}{k} and parameter dependence}

Using a Stirling-type expansion for the Gamma ratio, one has
\begin{equation*}
  \psi_q:=\frac{\Gamma\!\left((q-2)/2\right)}{\Gamma\!\left((q-1)/2\right)}
  =\sqrt{\frac{2}{q}}\left(1+\frac{5}{4q}+O\!\left(q^{-2}\right)\right),
\end{equation*}
and hence
\begin{equation*}
  \frac{1}{q-3}-\frac12\,\psi_q^{2}
  =\frac{1}{2q^{2}}\left(1+O\!\left(q^{-1}\right)\right).
\end{equation*}
Therefore,
\begin{equation*}
  A=\frac{q\,t^{\ast 2}}{(q-3)\,t_q}
    =\frac{t^{\ast 2}}{t_q}\left(1+O\!\left(q^{-1}\right)\right),
  \qquad
  D=q\,t^\ast\!\left[\frac{1}{q-3}-\frac12\,\psi_q^2\right]
    =\frac{t^{\ast}}{2q}\left(1+O\!\left(q^{-1}\right)\right).
\end{equation*}
Consequently,
\begin{equation*}
  k=\frac{A}{D}
    =\frac{2q\,t^{\ast}}{t_q}\left(1+O\!\left(q^{-1}\right)\right),
\end{equation*}
so $k$ grows linearly with $q$ up to $O(q^{-1})$ corrections, and is exactly proportional to $t^{\ast}$ and inversely proportional to $t_q$.

\subsection{Geometric implications}

On the admissible half-plane $w+z>0$, the correlation landscape in \eqref{Seq:corr-k} has the following features.
In the first quadrant ($w,z>0$), the sign changes on the hyperbola $w z=k$: outside this curve the correlation is positive, and inside it is negative.
Along the axes ($w=0$ or $z=0$), the correlation is strictly negative and tends to $0$ as the other coordinate diverges, reflecting the fact that the curve $wz=k$ is asymptotic to the axes.
For $w z < 0$ the correlation is negative throughout.
Moreover, if both $|w|$ and $|z|$ grow without bound (e.g., $w=ta$, $z=tb$ with $t\to\infty$ for fixed $a,b$), then
\begin{equation*}
  \operatorname{Corr}(\widehat{w},\widehat{z}) \longrightarrow
  \begin{cases}
    +1, & \text{when } w \text{ and } z \text{ have the same sign},\\[0.3ex]
    -1, & \text{when } w \text{ and } z \text{ have opposite signs}.
  \end{cases}
\end{equation*}
 
\section{Monte Carlo evaluation of confidence interval methods}
\label{appendix:MC_details}
\setcounter{equation}{0}
\setcounter{figure}{0}

\subsection{Monte Carlo procedure}
Recall that the extinction probability is defined by
\begin{align*}
\Pr[T \leq t \mid x_d, \mu, \sigma^2]
&= G(t \mid x_d, \mu, \sigma^2)\\
&= G(w,z)\\
&=\Phi(-w)+\exp\!\left(\tfrac{z^{2}-w^{2}}{2}\right)\Phi(-z) \quad \text{for} \quad w + z > 0, \quad w, z \in \mathbb{R},
\end{align*}
where $w={(\mu t +x_d)}/{\sigma\sqrt{t}}$ and $z={(-\mu t+x_d)}/{\sigma\sqrt{t}}$ are transformed parameters.
With $t = t^{\ast}$, the finite time horizon for extinction risk prediction, the ML estimators from Equation~\eqref{eq:mle_orig} yield transformed statistics $\widehat w, \widehat z$.
Independent random variables $\widehat\mu$ and $\widehat\sigma^2$ were generated according to their exact sampling distributions under observation span $t_q=q$:
\begin{equation*}
\widehat\mu \sim N\!\left(\mu,\;\sigma^2/t_q\right), 
\qquad
\widehat\sigma^2 \sim \frac{\sigma^2}{q}\,\chi^2_{\,q-1}.
\end{equation*}
Each replicate pair $(\widehat\mu,\ \widehat\sigma^2)$ was substituted into the definitions
\begin{equation*}
\widehat w=\frac{\widehat\mu t^{\ast}+x_d}{\widehat\sigma\sqrt{t^{\ast}}}, 
\qquad
\widehat z=\frac{-\widehat\mu t^{\ast}+x_d}{\widehat\sigma\sqrt{t^{\ast}}},
\end{equation*}
and then used to compute $G(\widehat w,\widehat z)$ and the confidence intervals for each method.

\textit{Numerical evaluation.}
The function $G$ was evaluated using a log-stable implementation described in Appendix~\ref{appendix:numerical_stability_G}.
Specifically, sums and differences on the probability scale were computed on the log scale via log-sum-exp and log-difference identities; the factor $(z^{2}-w^{2})/2$ was formed as $((z+w)(z-w))/2$ to reduce cancellation; and an eight-term Mills expansion in the $z$-direction was applied beyond a threshold $z_{\mathrm{thr}}=19$ to avoid overflow and loss of precision.
In regions where $G$ is numerically ill-conditioned (notably $w\ll 0$ with $G\to 1$), the complementary probability $Q=1-G$ was evaluated for the coverage decision and recorded alongside $G$ (``$G$-for-$Q$'' hybrid); the switching rule was fixed per parameter cell (based on the true parameters) and never per replicate.
All methods were applied to identical sets of replicate statistics to ensure comparability of results.

In this way, Monte Carlo replicates of the extinction probability and its confidence intervals were obtained without simulating full time series.

\subsection{Confidence interval methods}
Confidence intervals for $G$ were evaluated using four methods: the $w$--$z$ method, the theoretical minimum uncertainty (TMU) CI, the delta/logit method, and the percentile bootstrap.
The latter three methods are described below.

\subsubsection{Theoretical minimum uncertainty (TMU) CI}
The TMU confidence interval \citep{Fieberg:2000aa, ellner2008commentary} assumes that the environmental variance $\sigma^2$ is known and ignores uncertainty in its estimation.
Writing extinction risk as
\begin{equation*}
G(w,z)=\Phi(U-V)+\exp(2UV)\,\Phi\!\bigl(-(U+V)\bigr),
\end{equation*}
where $U=-\mu\sqrt{t^{\ast}}/\sigma$ and $V=x_d/(\sigma\sqrt{t^{\ast}})$,
the TMU interval for $q$ increments takes the form
\begin{equation*}
G(\widehat U - z_{1-\alpha/2}\sqrt{t^{\ast}/q},\,\widehat V) \;\leq\; G \;\leq\; 
G(\widehat U + z_{1-\alpha/2}\sqrt{t^{\ast}/q},\,\widehat V).
\end{equation*}
This construction perturbs only the $U$ component, corresponding to growth rate uncertainty, while $V$ is treated as fixed.
In the limit of large $V$, the expression reduces to a simple probit shift, 
$\Phi^{-1}(\widehat G)\pm z_{1-\alpha/2}\sqrt{t^{\ast}/q}$.
Because variance uncertainty is excluded, the TMU interval is generally anti-conservative, and Monte Carlo validation shows empirical rejection rates that often exceed the nominal level.

\subsubsection{Delta/logit method}
The delta/logit method \citep{DENNIS:1991aa} applies a first-order delta approximation to the logit of the extinction probability. 
Let
\begin{equation*}
H(t \mid x_d, \mu, \sigma^2)=\log\left(\frac{G(t \mid x_d, \mu, \sigma^2)}{1-G(t \mid x_d, \mu, \sigma^2)}\right).
\end{equation*}
The variance of $\widehat H = H(t \mid x_d, \widehat \mu, \widehat \sigma^2)$ is approximated as
\begin{align*}
\mathrm{Var}(\widehat H) &\approx
\left(\frac{\partial H}{\partial \mu}\right)^2 \mathrm{Var}(\widehat\mu)
+\left(\frac{\partial H}{\partial \sigma^2}\right)^2 \mathrm{Var}(\widehat\sigma^2),\\
& = 
\left(\frac{{\partial G}/{\partial \mu}}{G(1-G)}\right)^2 \mathrm{Var}(\widehat\mu)
+\left(\frac{{\partial G}/{\partial \sigma^2}}{G(1-G)}\right)^2 \mathrm{Var}(\widehat\sigma^2),
\end{align*}
with
\begin{equation*}
\mathrm{Var}(\widehat\mu)=\frac{\sigma^2}{t_q}=\frac{\sigma^2}{q},\qquad
\mathrm{Var}(\widehat\sigma^2)=\frac{2\sigma^4\,(q-1)}{q^2}.
\end{equation*}
In implementation, the variance terms and derivatives in this approximation were evaluated at the plug-in values $(\widehat\mu,\widehat\sigma^2)$.

The derivatives are
\begin{align*}
\frac{\partial G}{\partial \mu}
&= -\,\frac{2x_d}{\sigma^{2}}\; \exp\!\Bigl(\tfrac{z^{2}-w^{2}}{2}\Bigr)\,\Phi(-z),\\[8pt]
\frac{\partial G}{\partial \sigma^{2}}
&= \frac{x_d}{\sigma^{3}\sqrt{t^{\ast}}}\,\phi(w)
\;+\; \frac{2\mu x_d}{\sigma^{4}}\;\exp\!\Bigl(\tfrac{z^{2}-w^{2}}{2}\Bigr)\,\Phi(-z).
\end{align*}

The two-sided Wald interval on the logit scale,
$\widehat H \pm z_{1-\alpha/2}\sqrt{\mathrm{Var}(\widehat H)}$,
is mapped back to the probability scale by the inverse logit.

\textit{Implementation details.}
The evaluation of $G$ and its derivatives was carried out on the log scale using stable transforms, with the same Mills expansion as in Appendix~\ref{appendix:numerical_stability_G}, in order to avoid underflow or cancellation.
Despite these precautions, the delta/logit method remains sensitive in extreme regions ($w\ll 0$ or $w\gg 0$), where $G$ or $Q$ approaches the limits of double precision.
In Monte Carlo experiments this method often yields empirical rejection rates below the nominal level in some regions and above it in others, reflecting the limitations of a first-order linearization.

\subsubsection{Percentile parametric bootstrap}
The percentile parametric bootstrap treats $(\widehat\mu,\widehat\sigma^2)$ as plug-in truths and draws independent bootstrap replicates from their sampling laws with span $t_q=q$.
For $j=1,\dots,B$,
\begin{equation*}
\mu^{(j)} \sim N\!\left(\widehat\mu,\;\frac{\widehat\sigma^2}{t_q}\right),\qquad
\sigma^{2\,(j)} \sim \frac{\widehat\sigma^2}{q}\,\chi^2_{\,q-1},
\end{equation*}
and define $\sigma^{(j)}=\sqrt{\sigma^{2\,(j)}}$.
For each draw, the extinction probability is recomputed as
\begin{equation*}
G^{(j)}=G\!\left(\frac{\mu^{(j)}t^{\ast}+x_d}{\sigma^{(j)}\sqrt{t^{\ast}}},\;
\frac{-\mu^{(j)}t^{\ast}+x_d}{\sigma^{(j)}\sqrt{t^{\ast}}}\right).
\end{equation*}
After $B$ replicates, the percentile interval is taken as the empirical $\alpha/2$ and $1-\alpha/2$ quantiles of $\{G^{(j)}\}_{j=1}^B$.
In the implementation, these were computed using \texttt{quantile()} in R with its default rule.

Typical runs used $B=2{,}000$, which provides sufficient resolution for the percentile estimates while keeping computation time moderate. 
Additional checks with larger values ($B=5{,}000$--$10{,}000$) yielded essentially identical results, indicating that $B=2{,}000$ is adequate for stable estimation. 
All bootstrap evaluations of $G$ were carried out using the log-stable implementation described in Appendix~\ref{appendix:numerical_stability_G}, so that very small or very large probabilities were handled without numerical underflow or loss of precision.

\subsection{Evaluation metric}
For each parameter combination, $n_{\mathrm{MC}}=10{,}000$ replicates were generated. 
The empirical rejection rate (that is, one minus the coverage probability) was computed as the proportion of replicates in which the true $G$ was not contained in the estimated CI. 
The binomial standard error of this estimate is at most $0.005$ for the reported ranges.

\subsection{Implementation notes}
Evaluation of $G(w,z)$ relied on log-scale formulas and an 8-term Mills expansion for large $z$ values; details are provided in Appendix~\ref{appendix:numerical_stability_G}. 
Noncentral-$t$ inversions for the $w$--$z$ method used bracketing solvers with tolerance $10^{-10}$. 
All methods were applied to identical sets of replicate statistics to ensure comparability of results.

A bias-corrected and accelerated (BCa) bootstrap was also tested in preliminary runs \citep[see][]{efron1993}.
Although mean rejection rates were marginally lower than those of the percentile bootstrap, the BCa intervals exhibited greater variability with occasional extreme outliers, and remained anti-conservative overall.
For this reason, BCa results are not reported in the main text.
 
\section{Confidence-interval width as a function of data and model parameters}\label{appendix:ci_width}
\setcounter{equation}{0}
\setcounter{figure}{0}

This appendix analyzes how the confidence interval for the extinction probability depends on the data and on the model parameters.
Although $G(t\mid\cdot)$ is defined for arbitrary $t>0$, in what follows I fix $t=t^\ast$, a finite time horizon for extinction risk assessment; consequently, even though $G$ itself is a function of $(w,z)$ only, the distribution of its plug-in estimator depends on $t^\ast$ through the sampling distribution of $(\widehat w,\widehat z)$.
The aim is to disentangle these contributions and to describe the asymptotic behavior of the interval across different regions of the $(w,z)$ plane.
Two complementary approaches are used: the delta method, which provides an analytically transparent quadratic-form decomposition linking sampling design and model sensitivity under large-sample asymptotic normality, and a mixture representation, which yields more accurate results at finite $q$.
In particular, the mixture approach gives semi-exact formulas in the region $z>0$, while the delta method remains useful for structural interpretation across the full domain.

\subsection{Setup and definitions}
I first recall the relevant definitions.
For any $t>0$ and initial log-distance $x_d>0$, define
\begin{equation*}
  w=\frac{\mu t + x_d}{\sigma\sqrt{t}},\qquad
  z=\frac{-\,\mu t + x_d}{\sigma\sqrt{t}}.
\end{equation*}
The extinction probability in transformed coordinates was defined in Equation~\eqref{eq:G_wz} of the main text as
\begin{equation*}
  G(w,z)=\Phi(-w)+\exp\!\left(\tfrac{z^{2}-w^{2}}{2}\right)\Phi(-z), \quad w+z>0,
\end{equation*}
with gradient components (Appendix~\ref{appendix:G_monotonicity})
\begin{align*}
  \frac{\partial G}{\partial w}=G_w
  &= -\phi(w)-w\exp\!\left(\tfrac{z^{2}-w^{2}}{2}\right)\Phi(-z),\\
  \frac{\partial G}{\partial z}=G_z
  &= \exp\!\left(\tfrac{z^{2}-w^{2}}{2}\right)\bigl(z\Phi(-z)-\phi(z)\bigr).
\end{align*}
Throughout, $\Phi$ and $\phi$ denote the standard normal distribution and density.

For inference at a fixed finite time horizon $t^\ast$, the plug-in estimators are
\begin{equation*}
  \widehat w=\frac{\widehat\mu\,t^\ast + x_d}{\widehat\sigma\sqrt{t^\ast}},\qquad
  \widehat z=\frac{-\,\widehat\mu\,t^\ast + x_d}{\widehat\sigma\sqrt{t^\ast}}, \qquad q>1,
\end{equation*}
where $\widehat\mu$ and $\widehat\sigma^{2}$ are mutually independent, a property specific to the normal model (Subsection~\ref{sec:MLE_mu_sigma}).
As established in Subsection~\ref{sec:MLE_wz}, these follow noncentral-$t$
distributions with $q-1$ degrees of freedom:
\begin{equation*}
  \widehat{w}\,\sqrt{\tfrac{q-1}{q}}\,\sqrt{\tfrac{t_q}{t^{\ast}}}
    \;\sim\; t\!\left(w\sqrt{\tfrac{t_q}{t^{\ast}}},\,q-1\right),
  \qquad
  \widehat{z}\,\sqrt{\tfrac{q-1}{q}}\,\sqrt{\tfrac{t_q}{t^{\ast}}}
    \;\sim\; t\!\left(z\sqrt{\tfrac{t_q}{t^{\ast}}},\,q-1\right).
\end{equation*}

Their joint covariance structure, derived in Subsection~\ref{sec:MLE_wz} and Appendix~\ref{appendix:wz_corr_details}, is
\begin{equation}\label{Seq:Sigma-k}
  \Sigma=\upsilon(q)
  \begin{pmatrix}
    k+w^{2} & -k+w z\\[2pt]
    -k+w z & k+z^{2}
  \end{pmatrix},
\end{equation}
where $\upsilon(q)$ depends only on the sample size (minus one) $q$, while $k$ depends on $(q,t_q,t^{\ast})$:
\begin{align*}
  \upsilon(q)&=\frac{D}{t^{\ast}}=q\!\left[\frac{1}{q-3}
    -\tfrac12\left\{\tfrac{\Gamma((q-2)/2)}{\Gamma((q-1)/2)}\right\}^{2}\right] > 0,\\
  k(q,t_q,t^{\ast})&=\frac{t^{\ast}}{t_q}\,
    \left[1-\tfrac12(q-3)\left\{\tfrac{\Gamma((q-2)/2)}{\Gamma((q-1)/2)}\right\}^{2}\right]^{-1} > 0,\qquad q>3.
\end{align*}
From this structure, Appendix~\ref{appendix:wz_corr_details} also showed that
\begin{equation*}
  \mathrm{Corr}(\widehat w,\widehat z)=\rho(w,z;k)
  =\frac{-\,k + w z}{\sqrt{(k+w^{2})(k+z^{2})}}.
\end{equation*}

\paragraph{Remark on the roles of $q$ and $t_q$.}
From Equation~(\ref{eq:sampling_muhat_sigmahat}) in the main text, one has $\widehat{\mu} \sim \mathcal{N}\!\left(\mu,\,\sigma^2/t_q\right)$, so $\operatorname{Var}(\widehat{\mu})$ decreases in proportion to $1/t_q$ and is independent of $q$.
Thus increasing $q$ primarily stabilizes $\widehat{\sigma}^2$ and raises the degrees of freedom in the noncentral-$t$ laws for $(\widehat w,\widehat z)$, while increasing $t_q$ directly improves the signal for $\widehat{\mu}$.
Equivalently, in the notation of \eqref{Seq:Sigma-k}, $\upsilon(q)\sim 1/q$ but $\upsilon(q)\,k(q,t_q,t^\ast)=\tfrac{q}{q-3}\tfrac{t^\ast}{t_q}$, so with $t_q$ fixed the leading entries of $\Sigma$ converge to $t^\ast/t_q$ as $q\to\infty$, explaining the plateau in CI width; extending $t_q$ lowers this plateau.
This plateau behavior is quantitatively similar across $z$: while the projection of $(w,z)$ uncertainty onto $G(w,z)$ differs by regime, the qualitative effects of $q$ and $t_q$ on CI width remain essentially unchanged, because the asymptotic scale of $\Sigma$ is set by $t^\ast/t_q$ regardless of $z$.

\subsection{Delta-method expression}

Let $\widehat G = G(\widehat w,\widehat z)$ be the estimator of the extinction probability.
Under large-sample asymptotics in $q$ (with $t^\ast$ and $t_q$ fixed),
\begin{equation*}
(\widehat w-w,\;\widehat z-z)^{\top}\;\Rightarrow\;\mathcal N\left(0,\;\Sigma(q,t_q,t^\ast)\right).
\end{equation*}
By the delta method, which applies under this asymptotic normality,
\begin{equation*}\operatorname{Var}(\widehat G)\;\approx\;\nabla G(w,z)^{\top}\,\Sigma\,\nabla G(w,z),
\end{equation*}
where $\nabla G(w,z)=\left(G_w, G_z\right)^{\top}$ and $\Sigma$ is given in \eqref{Seq:Sigma-k}.
Hence the half-width of the two-sided $100(1-\alpha)\%$ Wald interval is
\begin{equation}\label{Seq:delta_halfwidth}h_G(w,z)=z_{1-\alpha/2}\,\|\nabla G(w,z)\|_{\Sigma},
\end{equation}
with $\|v\|_{\Sigma}=\sqrt{v^{\top}\Sigma v}$.
Explicitly,
\begin{equation*}
\|\nabla G(w,z)\|_{\Sigma}^2
= G_w^2\,\Sigma_{11} + 2 G_w G_z\,\Sigma_{12} + G_z^2\,\Sigma_{22}.
\end{equation*}
Thus both the data $(q,t_q)$ and the model parameters $(w,z,t^\ast)$ enter through $\Sigma$, while the sensitivity of $G$ itself contributes through the gradient components $G_w,G_z$.
As $G\to 0$ or $G\to 1$, both $G_w$ and $G_z$ vanish, so that $h_G\to 0$.

For moderate $z>0$, the approximation $G\simeq \Phi(-w)$ yields $\nabla G\simeq(-\phi(w),0)^\top$ and thus
\begin{equation*}
\|\nabla G(w,z)\|_\Sigma^2 \simeq \phi(w)^2\,\Sigma_{11}
= \upsilon(q)\,\phi(w)^2\,(k+w^2).
\end{equation*}
In this approximation the squared width satisfies
\begin{equation*}
h_G^2 \;\propto\; \phi(w)^2 (k+w^2)
= \tfrac{1}{2\pi}\,e^{-w^2}(k+w^2).
\end{equation*}
Maximizing $e^{-w^2}(k+w^2)$ with respect to $w$ therefore determines the point of maximal interval width.
The maximum occurs at $w=0$ when $k\ge 1$, whereas for $0<k<1$ it is attained at $w=\pm\sqrt{1-k}$, because $e^{-(1-k)} > k$ for $0<k<1$.
This two-hump structure for $k<1$ is clearly seen in finite-$q$ calculations; see Fig.~\ref{fig:varG_vs_G}c (note that for $z<0$ no such splitting occurs; cf.\ Fig.~\ref{fig:varG_vs_G}d).
In typical applications $k$ scales like $q\,(t^\ast/t_q)$ and is usually $\ge 1$, so the maximal width occurs at $w=0$, i.e., at intermediate extinction probability ($G\simeq 0.5$) in this regime.

In the opposite tail, $z\ll 0$, the extinction probability behaves as $G\left(w,z\right)\approx \exp\left(\left(z^{2}-w^{2}\right)/2\right)$, and its gradient as $\nabla G\left(w,z\right)\approx\left(-wG,\ zG\right)^{\top}$.
Substitution into \eqref{Seq:delta_halfwidth} then yields
\begin{equation*}
h_{G}\approx z_{1-\alpha/2}\,\sqrt{\upsilon(q)}\,G\,
\sqrt{\left(w^{2}-z^{2}\right)^{2}+k\left(q,t_{q},t^{\ast}\right)\left(w+z\right)^{2}}.
\end{equation*}
Introducing $\kappa=w+z>0$ as the distance from the boundary and writing $w=\kappa-z$ gives
\begin{equation*}
G=\exp\left(\kappa z-\frac{\kappa^{2}}{2}\right),
\end{equation*}
so for fixed $\kappa$ the width $h_{G}$ decays exponentially in $\left|z\right|$, with only polynomial modulation from the square-root factor involving $\upsilon(q)$ and $k\left(q,t_{q},t^{\ast}\right)$.
No interior maximum arises in this regime.

\paragraph{Upper bound and overestimation.}
Since $\widehat G \in [0,1]$, its variance is bounded by $1/4$ regardless of the parameter values.
In the $z>0$ regime where $G\simeq\Phi(-w)$, the delta-method variance attains
\begin{equation*}
\max_w \operatorname{Var}_\Delta(\widehat G)
=\begin{cases}
\dfrac{\upsilon(q)}{2\pi}\,k, & \text{attained at } w=0 \text{ when } k\ge 1,\\[6pt]
\dfrac{\upsilon(q)}{2\pi}\,e^{-(1-k)}, & \text{attained at } w=\pm\sqrt{1-k} \text{ when } 0<k<1.
\end{cases}
\end{equation*}
using the identity $\upsilon(q)k=(q/(q-3))(t^\ast/t_q)$.
Therefore, for $k\ge 1$, whenever
\begin{equation*}
\frac{q}{q-3}\,\frac{t^\ast}{t_q}>\frac{\pi}{2},
\end{equation*}
its maximal value necessarily exceeds the universal upper bound $1/4$, implying overestimation of the true variance and the corresponding CI width.
At the symmetry point $w=0$ and in the large-$q$ limit, the delta variance is $\operatorname{Var}_\Delta(\widehat G)\to \frac{1}{2\pi}\,\frac{t^\ast}{t_q}$, whereas the mixture (true) variance (see next subsection) is $\operatorname{Var}_{\mathrm{mix}}(\widehat G)\to \frac{1}{2\pi}\,\arcsin\!\Bigl(\frac{t^\ast/t_q}{1+t^\ast/t_q}\Bigr)$.
Since $\frac{t^\ast}{t_q}>\arcsin\!\bigl(\frac{t^\ast/t_q}{1+t^\ast/t_q}\bigr)$ for all $t^\ast/t_q>0$, the delta method overestimates the variance at $w=0$ for any $t^\ast/t_q>0$.

In the opposite tail ($z\ll 0$), $G(w,z)\approx \exp\!\bigl((z^2-w^2)/2\bigr)$ decays exponentially, and both the delta and true variances are proportional to $G^2$ up to polynomial factors in $(w,z)$; hence no systematic overestimation arises.

In contrast, as $G\to 0$ or $G\to 1$ both gradient components vanish, so that $\operatorname{Var}_\Delta(\widehat G)\to 0$ consistently with the true variance.
Thus the overestimation phenomenon is confined to the central $z>0$ region, where $G\simeq\Phi(-w)$ and the delta method can exceed the variance bound.

For completeness, in the $0<k<1$ branch the maximum is $\tfrac{\upsilon(q)}{2\pi}e^{-(1-k)}<\tfrac{\upsilon(q)}{2\pi}$; since $\tfrac{\upsilon(q)}{2\pi}<\tfrac14$ for all $q\ge4$, the upper-bound argument above does not apply (the delta method need not overestimate).

\subsection{Mixture representation}
Although the variance of $\widehat G=G(\widehat w,\widehat z)$ can in principle be expressed as a two-dimensional integral over the joint distribution of $(\widehat w,\widehat z)$, the density is intractable. A more convenient approach uses the standard mixture representation of noncentral $t$ statistics.

\begin{figure}[H]
  \centering
  \begin{subfigure}[t]{0.48\textwidth}
    \centering
    \includegraphics[width=\textwidth]{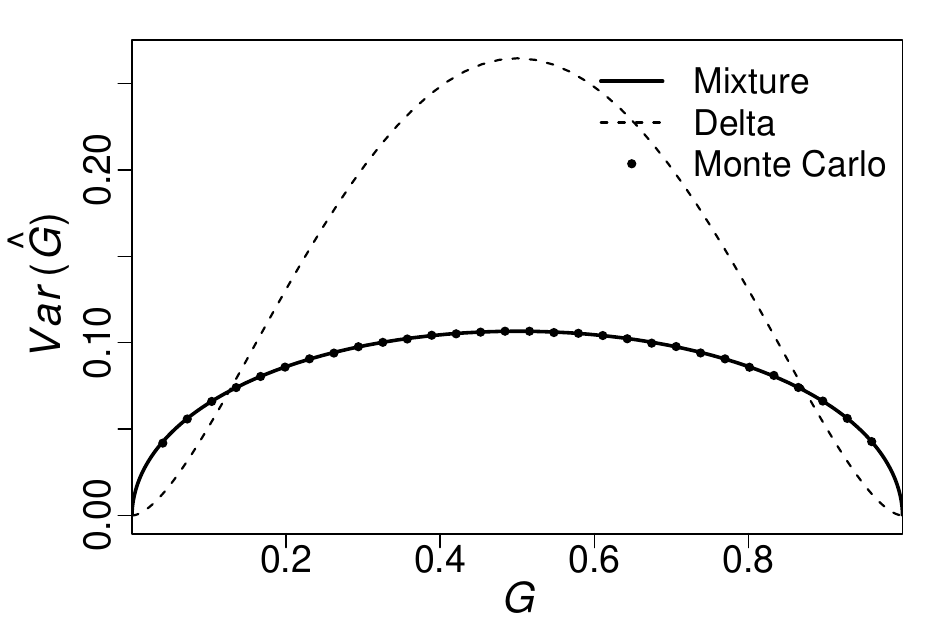}
    \caption{$z=20$, $t^\ast=100$ (central $z>0$ regime)}
    \label{fig:varG_vs_G_a}
  \end{subfigure}
  \hfill
  \begin{subfigure}[t]{0.48\textwidth}
    \centering
    \includegraphics[width=\textwidth]{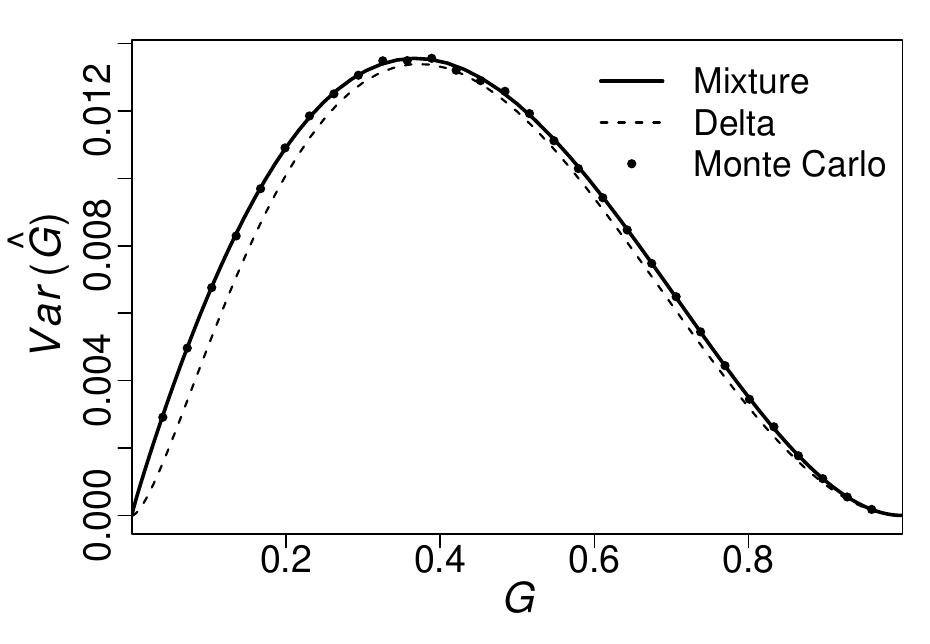}
    \caption{$z=-5$, $t^\ast=100$ (negative-tail regime)}
    \label{fig:varG_vs_G_b}
  \end{subfigure}
  \vspace{2mm}
  \begin{subfigure}[t]{0.48\textwidth}
    \centering
    \includegraphics[width=\textwidth]{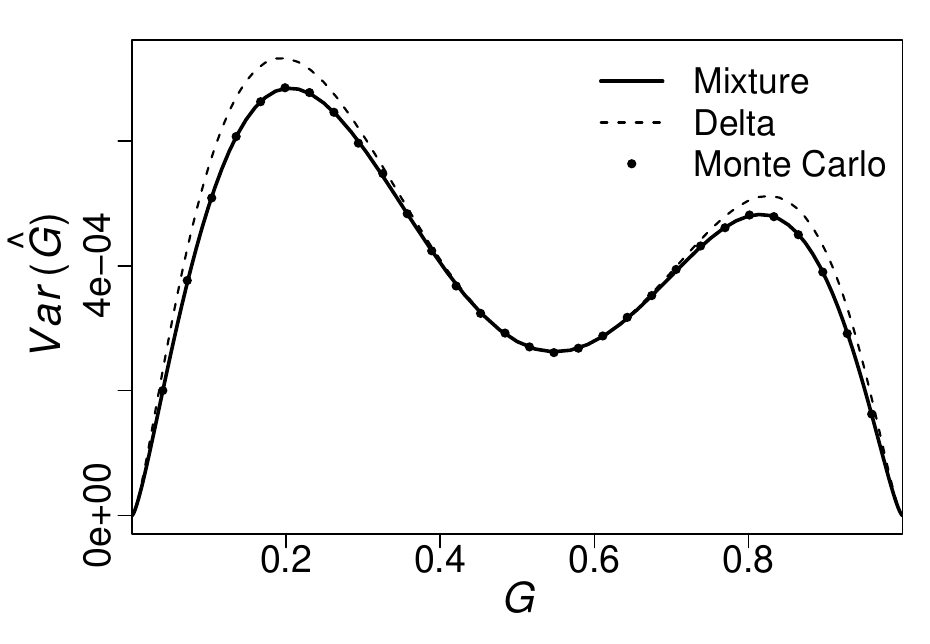}
    \caption{$z=20$, $t^\ast=0.1$ ($k<1$, two-hump structure)}
    \label{fig:varG_vs_G_c}
  \end{subfigure}
  \hfill
  \begin{subfigure}[t]{0.48\textwidth}
    \centering
    \includegraphics[width=\textwidth]{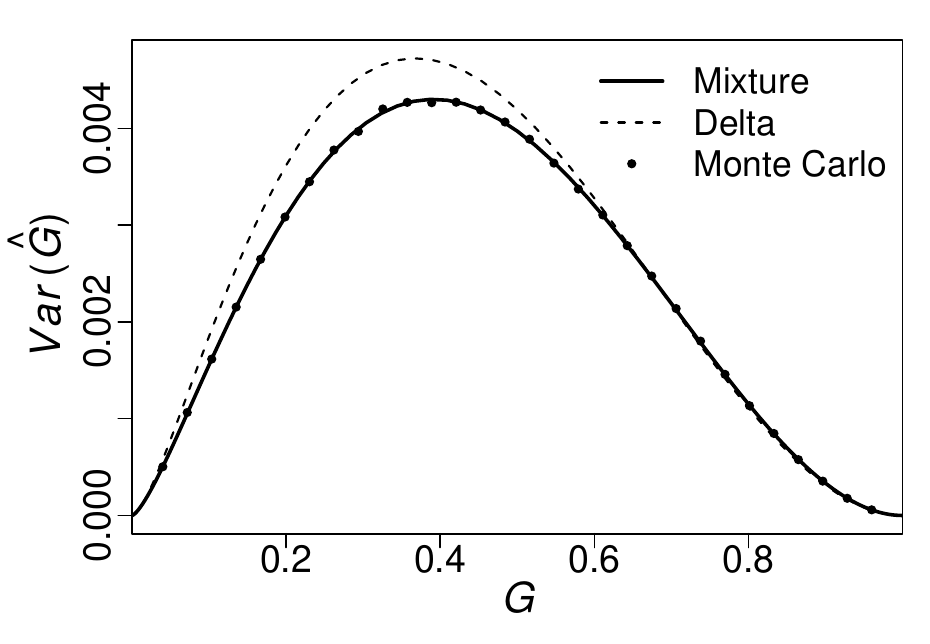}
    \caption{$z=-5$, $t^\ast=0.1$ ($k<1$, single-peak)}
    \label{fig:varG_vs_G_d}
  \end{subfigure}
\caption{
Variance of the extinction probability estimator $\operatorname{Var}(\widehat{G})$ as a function of the true extinction probability $G$.
In each panel, $z$ is fixed and $w$ is varied to sweep $G(w,z)$ over $(0,1)$ (here $q=63$).
Solid line: mixture evaluation (finite $q$ reference); dashed line: delta method; dots: Monte Carlo simulations ($n=120{,}000$).
Mixture and Monte Carlo curves nearly coincide, indicating the finite $q$ variance.
(a,b) $t^\ast=100$: (a) central regime ($z>0$), (b) tail regime ($z<0$).
(c,d) $t^\ast=0.1$ with $k<1$, where $k=k(q,t_q,t^\ast)$ is defined in Eq.~\eqref{Seq:Sigma-k}:
(c) $z>0$ shows a two hump structure, whereas (d) $z<0$ remains single peaked.
The delta method overestimates variance for intermediate $G$ when $z>0$, reflecting a finite $q$ effect.
}
  \label{fig:varG_vs_G}
\end{figure}

As introduced in Appendix~\ref{appendix:wz_corr_details}, the estimators can be written as
\begin{equation*}
\widehat w=\frac{x_d+m}{\sqrt{t^\ast}}\,\eta,
\qquad
\widehat z=\frac{x_d-m}{\sqrt{t^\ast}}\,\eta,
\end{equation*}
where $m=t^\ast\widehat\mu$ and $\eta=\widehat\sigma^{-1}$ are independent.
It is convenient to set
\begin{equation*}
m=\bar m+\sigma_m Z_0,\qquad Z_0\sim\mathcal N(0,1),
\quad \bar m=t^\ast\mu,\quad \sigma_m^2=t^{\ast2}\sigma^2/t_q,
\end{equation*}
and introduce
\begin{equation*}
s=\sqrt{t^\ast/t_q},\qquad
\Lambda=\sigma\eta=\sqrt{q/\chi^2_{q-1}},\qquad
Z_0\sim\mathcal N(0,1)\text{ and }Z_0\perp\!\!\!\perp\Lambda.
\end{equation*}
Then the estimators admit the mixture representation in terms of a standard normal $Z_0\sim\mathcal N(0,1)$ and an independent random scale $\Lambda$:
\begin{equation*}
(\widehat w,\widehat z)\ \stackrel{d}{=}\ \bigl(\Lambda(w+sZ_0),\ \Lambda(z-sZ_0)\bigr).
\end{equation*}
For any fixed $\Lambda=\lambda$, the conditional distribution is
\begin{equation*}
(\widehat w,\widehat z)\bigm|_{\Lambda=\lambda}
= \bigl(\lambda(w+sZ_0),\ \lambda(z-sZ_0)\bigr),
\qquad Z_0\sim\mathcal N(0,1).
\end{equation*}
Hence $(\widehat w,\widehat z)\bigm|_{\Lambda=\lambda}$ is a degenerate bivariate normal distribution with mean $(\lambda w,\lambda z)$, marginal variances $\lambda^2 s^2$, and perfect negative correlation $-1$.
Conditioning on $\Lambda$ therefore reduces the randomness to the single standard normal $Z_0$. For any measurable function $f:\mathbb R^2\to\mathbb R$,
\begin{equation*}
\mathbb E\!\left[f(\widehat w,\widehat z)\right]
= \mathbb E_{\Lambda}\,\mathbb E_{Z_0}\!\left[
  f\!\left(\Lambda(w+sZ_0),\,\Lambda(z-sZ_0)\right)
\right],
\qquad Z_0\sim\mathcal N(0,1)\text{ and } Z_0\perp\!\!\!\perp\Lambda.
\end{equation*}
In other words, integrals with respect to the joint distribution of $(\widehat w,\widehat z)$ reduce to a one-dimensional inner expectation over $Z_0$ and a one-dimensional outer expectation over $\Lambda$.
In particular, for subsequent calculations define, for $j=1,2$,
\begin{equation*}
\mathcal M_j(\lambda;w,z)
:=\mathbb E_{Z_0}\!\left[
G\!\bigl(\lambda(w+sZ_0),\,\lambda(z-sZ_0)\bigr)^{j}
\right],\qquad \lambda>0.
\end{equation*}
Then the unconditional mean and variance of $\widehat G:=G(\widehat w,\widehat z)$ are
\begin{equation*}
\mathbb E[\widehat G]=\mathbb E_{\Lambda}\!\left[\mathcal M_1(\Lambda;w,z)\right],
\qquad
\operatorname{Var}(\widehat G)
=\mathbb E_{\Lambda}\!\left[\mathcal M_2(\Lambda;w,z)\right]
-\left(\mathbb E_{\Lambda}\!\left[\mathcal M_1(\Lambda;w,z)\right]\right)^{2}.
\end{equation*}
These representations are numerically stable and agree well with Monte Carlo simulations across a broad range of $(w,z)$ values (Fig.~\ref{fig:varG_vs_G}).

\paragraph{The case $z>0$.}
When $z$ is moderately large and positive (formally $z\to\infty$), the second term in $G(w,z)$ is exponentially small, so $\widehat G\simeq \Phi(-\widehat w)$ and the inner expectation is one-dimensional in closed form:
\begin{equation*}
\mathcal M_1(\lambda;w,z)\simeq
\mathbb E_{Z_0}\!\left[\Phi\bigl(-\lambda(w+sZ_0)\bigr)\right]
=\Phi\!\left(\frac{-\lambda w}{\sqrt{1+\lambda^{2}s^{2}}}\right),
\end{equation*}
\begin{equation*}
\mathcal M_2(\lambda;w,z)\simeq
\mathbb E_{Z_0}\!\left[\Phi\bigl(-\lambda(w+sZ_0)\bigr)^{2}\right]
=\Phi_{2}\!\left(\varpi(\lambda),\,\varpi(\lambda);\ \varrho(\lambda)\right),
\end{equation*}
with
\begin{equation*}
\varpi(\lambda)=\frac{-\lambda w}{\sqrt{1+\lambda^{2}s^{2}}},\qquad
\varrho(\lambda)=\frac{\lambda^{2}s^{2}}{1+\lambda^{2}s^{2}},
\end{equation*}
and $\Phi_{2}(\cdot,\cdot;\varrho)$ the standard bivariate normal cdf with
correlation $\varrho$.
Therefore
\begin{equation*}
\operatorname{Var}_{z>0}\!\left(\widehat G\right)\;\simeq\;
\mathbb E_{\Lambda}\!\left[\Phi_{2}\!\left(\varpi(\Lambda),\,\varpi(\Lambda);\ \varrho(\Lambda)\right)\right]
-\Bigl(\mathbb E_{\Lambda}\!\left[\Phi\!\left(\varpi(\Lambda)\right)\right]\Bigr)^{2},\end{equation*}
and the corresponding Wald half-width is
\begin{equation*}
h_{G}^{\mathrm{mix}}(w;z>0)
= z_{1-\alpha/2}\,\sqrt{\operatorname{Var}_{z>0}\!\left(\widehat G\right)}.
\end{equation*}

As $q\to\infty$ with $t_q$ fixed, one has $\Lambda\to 1$ in probability while
$s=\sqrt{t^\ast/t_q}$ is fixed. Hence the finite-$q$ mixture converges to the
$\Lambda\equiv 1$ limit
\begin{equation*}
\operatorname{Var}_{z>0}(\widehat G)\ \longrightarrow\
\Phi_{2}\!\left(\varpi(1),\,\varpi(1);\ \varrho(1)\right)
-\Phi\!\left(\varpi(1)\right)^{2},
\qquad (q\to\infty),
\end{equation*}
where $\varpi(1)=-w/\sqrt{1+s^{2}}$ and $\varrho(1)=s^{2}/(1+s^{2})$.
Consequently, with $t_q$ fixed, increasing $q$ alone cannot shrink the width to $0$; rather,
\begin{equation*}
h_{G}^{\mathrm{mix}}(w;z>0,\,q\to\infty)
=
z_{1-\alpha/2}\,
\sqrt{
\Phi_{2}\!\left(\varpi(1),\,\varpi(1);\ \varrho(1)\right)
-\Phi\!\left(\varpi(1)\right)^{2}
}.
\end{equation*}
Moreover, for small $s$ (equivalently $t^\ast/t_q\downarrow 0$), a Taylor expansion yields
\begin{equation*}
\operatorname{Var}_{z>0}(\widehat G)
= \phi(w)^{2}\,s^{2}+O(s^{4}),
\end{equation*}
so that
\begin{equation*}
h_{G}^{\mathrm{mix}}(w;z>0,\,q\to\infty)
=
z_{1-\alpha/2}\,\phi(w)\,s
\;+\;O\!\left(s^{3}\right).
\end{equation*}

$\operatorname{Var}_{z>0}(\widehat G)$ is an even function of $w$ (by the symmetry $Z_0\mapsto -Z_0$) and decays to zero as $|w|\to\infty$; accordingly, its maximum occurs near $w=0$, corresponding to $G\simeq 0.5$.
At the symmetry point $w=0$ one has
\begin{equation*}
\operatorname{Var}_{z>0}\!\left(\widehat G\right)\Big|_{w=0}
=\frac{1}{2\pi}\,\mathbb E_{\Lambda}\!\left[\arcsin\!\left(\frac{\Lambda^{2}s^2}{1+\Lambda^{2}s^2}\right)\right]
\ \longrightarrow\ \frac{1}{2\pi}\,\arcsin\!\left(\frac{s^2}{1+s^2}\right)
\qquad (q\to\infty).
\end{equation*}
In particular, if $s^2=1$ (i.e., $t^\ast=t_q$), the limit equals $1/12$.

\subsection{Dependence on growth rate and environmental variance}
\label{sec:dep_mu_sigma}
For fixed $(\mu,x_d,t^\ast)$,
\begin{equation*}
w^{2}+z^{2}
=\frac{2\bigl((\mu t^\ast)^{2}+x_d^{2}\bigr)}{t^\ast}\,\frac{1}{\sigma^{2}}
\;\propto\;\frac{1}{\sigma^{2}}.
\end{equation*}
Hence iso-$\sigma^{2}$ contours are concentric circles centered at the origin in the $(w,z)$ plane. Varying $\sigma$ moves $(w,z)$ radially with constant polar angle $\arctan(z_{0}/w_{0})$, where $w_{0}=(\mu t^\ast+x_d)/\sqrt{t^\ast}$ and $z_{0}=(-\,\mu t^\ast+x_d)/\sqrt{t^\ast}$. As $\sigma$ increases, $(w,z)$ approaches the origin and the CI width typically increases, but the growth saturates at a plateau governed by $(q,t_q)$ through $s=\sqrt{t^\ast/t_q}$. As $\sigma$ decreases, $(w,z)$ moves outward and $G$ approaches $0$ or $1$, which drives the CI to shrink.

The qualitative split by the sign of $\mu$ follows from the boundary $w=z$. For $\mu<0$, the locus $w=0$ (equivalently $-\mu t^\ast=x_d$) marks a band of maximal uncertainty. For $\mu>0$, $w_0=(\mu t^\ast+x_d)/\sqrt{t^\ast}>0$, so $w=w_0/\sigma$ cannot approach $0$ as $\sigma\downarrow0$; hence the $w=0$ maximal-uncertainty ray is not attainable in the small-$\sigma$ limit. As $\sigma\downarrow 0$ with $(\mu,x_d,t^\ast)$ fixed, write $w=w_{0}/\sigma$ and $z=z_{0}/\sigma$. If $z>0$ (that is, $z_{0}>0$) then
\begin{equation*}
G(w,z)\simeq \Phi(-w)\sim \frac{\phi(w)}{w}
=O\!\left(\sigma\,e^{-\,w_{0}^{2}/(2\sigma^{2})}\right).
\end{equation*}
If $z<0$ (with $w+z>0$) then
\begin{equation*}
G(w,z)\simeq \exp\!\left(\frac{z^{2}-w^{2}}{2}\right)
=\exp\!\left(-\,\frac{w_{0}^{2}-z_{0}^{2}}{2\sigma^{2}}\right).
\end{equation*}
Thus, away from the $w=0$ ray, the CI width shrinks at rate $\exp(-\text{const}/\sigma^{2})$ as $\sigma\downarrow 0$.
The data-limited plateau of order $\sqrt{t^\ast/t_q}$ instead arises in the high-variance regime where $(w,z)$ approaches the origin.
On the $w=0$ ray (reachable only if $\mu=-\,x_d/t^\ast$), $G\to 1/2$ as $\sigma\downarrow 0$ and the CI remains maximal.

\subsection{Summary}
The delta method provides a general analytic framework, cleanly separating data and model contributions and describing asymptotic dependence across regimes.
The mixture representation, however, yields sharper results for finite $q$ and is particularly powerful in the region $z>0$, where the extinction probability reduces to a one-dimensional form and the distribution of $\widehat G$ is available in semi-exact mixture form.
In the central region, mixture calculations reveal that the delta method exaggerates the variance, while in the high-drift tail the delta method captures the exponential decay of the confidence-interval width.
Together these approaches offer both interpretability and accuracy across the $(w,z)$ plane.

From the standpoint of data dependence, the two methods highlight complementary roles of sample size (minus one) $q$ and observation span $t_q$.
Although $\upsilon(q)\sim 1/q$, the identity $\upsilon(q)\,k(q,t_q,t^\ast)=\tfrac{q}{q-3}\,\tfrac{t^\ast}{t_q}$ implies that, with $t_q$ fixed, the leading entries of $\Sigma$ converge to $s^2=t^\ast/t_q$.
Hence the width does not keep shrinking with $q$; it plateaus at order $\sqrt{t^\ast/t_q}$.
In the mixture representation, the random scale $\Lambda$ captures the extra finite-$q$ variability and converges to 1 as $q\to\infty$, while the rescaling factor $s=\sqrt{t^\ast/t_q}$ makes explicit how extending $t_q$ reduces uncertainty.
Thus $q$ controls replication-driven precision, and $t_q$ controls the per-observation signal-to-noise ratio; together they explain both the plateau (for fixed $t_q$ as $q\to\infty$) and the substantial gains from increasing $t_q$.
 
\section{Table: Extinction probability and CI width versus time horizon \texorpdfstring{$t^\ast$}{t*} for \texorpdfstring{$\mu<0$}{mu<0} and \texorpdfstring{$\mu>0$}{mu>0} examples}
\label{appendix:ciwidth_vs_tstar_SI}
\setcounter{table}{0}
\begin{table}[H]
\centering
\caption{
Representative examples of extinction probability and 95\% CI width versus the time horizon $t^\ast$.
The $\mu<0$ case ($\mu=-0.058925$, $\sigma=0.116939$, $x_d=12.64433$; path crosses $w=0$) 
contrasts with the $\mu>0$ case ($\mu=0.10$, $\sigma=0.20$, $x_d=13$; path does not cross $w=0$).
Here $q=t_q=63$. $G$ denotes the point estimate and ``CI width'' the two--sided interval width (the $w$--$z$ method).
}
\label{tab:ciwidth_vs_tstar_SI}
\begin{tabular}{@{}rcc@{\qquad}cc@{}}
\toprule
& \multicolumn{2}{c}{$\mu<0$ (crosses $w=0$)} & \multicolumn{2}{c}{$\mu>0$ (does not cross $w=0$)} \\
\cmidrule(lr){2-3}\cmidrule(l){4-5}
$t^\ast$ & $G$ & CI width & $G$ & CI width \\
\midrule
25.5  & $1.87\times10^{-79}$ & $3.16\times10^{-52}$ & $2.13\times10^{-53}$ & $8.62\times10^{-35}$ \\
42.5  & $1.91\times10^{-40}$ & $2.87\times10^{-25}$ & $8.64\times10^{-40}$ & $8.15\times10^{-25}$ \\
100   & $5.32\times10^{-9}$  & $1.38\times10^{-3}$  & $4.60\times10^{-30}$ & $3.59\times10^{-16}$ \\
250   & $0.88$              & $9.97\times10^{-1}$  & $5.90\times10^{-29}$ & $4.16\times10^{-13}$ \\
500   & $1 - 3.74\times10^{-11}$               & $2.04\times10^{-1}$  & $5.90\times10^{-29}$ & $6.64\times10^{-13}$ \\
1000  & $1 - 1.08\times10^{-36}$               & $5.91\times10^{-6}$  & $5.90\times10^{-29}$ & $6.21\times10^{-13}$ \\
\bottomrule
\end{tabular}
\end{table}
 
\section{Determination of the required observation span}
\label{appendix:req_span}
\setcounter{equation}{0}
\setcounter{figure}{0}

The numerical analysis in Subsection~\ref{sec:obs_span} rests on an explicit definition of the \textit{required observation span}.
Here the time horizon $t^\ast$ is fixed in advance, while the observation span $t_q$ is determined by the sampling design.
A time series of population size with $q+1$ observations at times $t_i = i$ ($i = 0,\dots,q$) is considered under unit-interval, equally spaced sampling; thus an actual design has $t_q=q$ with $q\in\mathbb{N}$.
For the purpose of tracing smooth required-span curves, however, I treat $t_q$ as a positive real variable and recover the corresponding integer design requirement by rounding up.
Given a true extinction probability $P = G(w,z)$ and a prescribed confidence level $1-\alpha$, the required span $t_q$ is defined as the smallest observation span such that the upper confidence bound of $G(w,z)$ falls below a management threshold $G_{\mathrm{target}}$:
\begin{equation}
  \label{Seq:required_span_def}
  t_q^\ast
  \;=\;
  \inf\Bigl\{\,t_q>1:\ G_{\text{upper}}(t_q; w, z, \alpha)\;\le\; G_{\mathrm{target}} \Bigr\}.
\end{equation}
For an actual unit-interval design, the required number of increments is then $q^\ast=\lceil t_q^\ast\rceil$.

To specify $G_{\text{upper}}$, note that the rescaled estimators $\widehat{w},\widehat{z}$ follow noncentral-$t$ distributions (see Subsection~\ref{sec:MLE_wz}):
\begin{equation*}\widehat{w}\,\sqrt{\frac{t_q-1}{t^\ast}}
  \;\sim\; t(\delta_w, t_q-1),
  \qquad
  \widehat{z}\,\sqrt{\frac{t_q-1}{t^\ast}}
  \;\sim\; t(\delta_z, t_q-1),
\end{equation*}
with noncentrality parameters
\begin{equation}\label{Seq:wz_delta_app}
  \delta_w = w\sqrt{\tfrac{t_q}{t^\ast}},
  \qquad
  \delta_z = z\sqrt{\tfrac{t_q}{t^\ast}}.
\end{equation}

Let $F_{t(\delta,\,t_q-1)}(\cdot)$ denote the CDF of the noncentral $t$ distribution with degrees of freedom $t_q-1$ and noncentrality parameter $\delta$.
Define $(\delta_{\text{low}}, \delta_{\text{high}})$ implicitly by
\begin{align*}
  F_{t(\delta_{\text{low}},\,t_q-1)}\!\left(\sqrt{\tfrac{t_q-1}{t^\ast}}\,w\right) &= 1 - \tfrac{\alpha}{2}, \\
  F_{t(\delta_{\text{high}},\,t_q-1)}\!\left(\sqrt{\tfrac{t_q-1}{t^\ast}}\,z\right) &= \tfrac{\alpha}{2}.
\end{align*}
Let $(w_L, z_U)$ be the parameter values induced by $(\delta_{\text{low}}, \delta_{\text{high}})$ via \eqref{Seq:wz_delta_app}.
Because $G$ decreases strictly in both arguments when $w+z>0$ (Appendix~\ref{appendix:G_monotonicity}), the upper bound is
\begin{equation*}
  G_{\text{upper}}(t_q; w, z, \alpha) \;=\; G(w_L, z_U).
\end{equation*}

For the numerical results in Subsection~\ref{sec:obs_span}, the horizon was set to $t^{\ast}=100$, the confidence level to $1-\alpha=0.95$, and the management threshold to $G_{\mathrm{target}}=0.1$.
The continuous relaxation \eqref{Seq:required_span_def} was then solved across a range of true extinction probabilities $P=G(w,z)$, with $z$ fixed and $w$ chosen to match the desired $P$.

\paragraph{Representativeness of the two $z$-regime examples.}
The two examples used in Fig.~\ref{fig:ci_width_vs_G} ($z=20$ and $z=-5$) are intended to represent the qualitatively distinct $z>0$ and $z<0$ regimes.
For sufficiently large $z>0$, the asymptotic form of $G(w,z)$ implies
$G(w,z)\simeq \Phi(-w)$ up to a small correction (Appendix~\ref{appendix:asymptotics_G}),
so iso-$G$ contours (and CI width at fixed $G$) depend primarily on $w$ and only weakly on $z$ in the $z>0$ regime.
In the opposite tail $z\ll 0$, letting $\kappa=w+z>0$ yields $G=\exp\!\left(\kappa z-\frac{\kappa^2}{2}\right)$, so equal-$G$ contours are asymptotically close to lines $w=-z+\mathrm{const}$ (slope $\approx -1$), and the CI width decreases as $z$ becomes more negative at fixed $G$; hence the $z=-5$ curve is conservative within the $z<0$ regime.
The horizon $t^\ast$ enters the required-span calculation primarily through the dimensionless ratio $t_q/t^\ast$.
Indeed, the noncentralities satisfy $\delta_w=w\sqrt{t_q/t^\ast}$ and $\delta_z=z\sqrt{t_q/t^\ast}$ \eqref{Seq:wz_delta_app}, while the confidence-limit construction depends on the rescaled observations $\sqrt{(t_q-1)/t^\ast}\,w$ and $\sqrt{(t_q-1)/t^\ast}\,z$.
Accordingly, expressing the required span as $t_q/t^\ast$ largely reduces sensitivity to $t^\ast$, unless changing $t^\ast$ moves the evaluation point between the $z>0$ and $z<0$ regimes.

\section{Observation error and robust inference: the OEAR estimator and implications for \texorpdfstring{$G(w,z)$}{G(w,z)}}
\label{appendix:obs_error}
\setcounter{equation}{0}
\setcounter{figure}{0}
\setcounter{table}{0}

The abundance time series used for PVA are often population-size indices rather than direct measurements, and can be affected by sampling and reporting errors, as well as by changes in detectability or catchability.
When such indices are observed with additive observation error on the log scale, estimated growth increments contain differenced observation error, inflating the marginal variance and inducing a negative lag-one covariance (while leaving the growth rate unbiased) \citep{staples2004estimating, mcnamara2004measurement}.
Naive plug-in of the inflated marginal variance into risk formulas can misestimate extinction risk.
However, early simulation work showed a relatively optimistic sensitivity pattern for such plug-in risk calculations: moderate random observation error often had little effect on estimated risk, except in regimes where extinction probability is intrinsically most uncertain, and systematic sampling bias rarely changed the results \citep{meir2000will}.
A general analytic explanation reconciling potentially large variance inflation with this empirical robustness remains limited, motivating a closer examination of how observation error propagates through risk calculations in the settings studied here.

One remedy is to model observation error explicitly and to estimate process and observation components jointly, rather than absorbing all variability into an effective process variance.
In diffusion-based settings, this can be done by formulating the latent dynamics and the observation equation as a linear-Gaussian state-space model (SSM) and evaluating the likelihood via Kalman-filter recursions, thereby targeting $(\mu,\sigma^2)$ under an explicit measurement-error specification \citep{lindley2003estimation}.
Related developments for abundance indices incorporate the implied covariance structure of observed increments, combine moment-type corrections with Gaussian likelihoods, and accommodate features such as unequal sampling intervals while keeping process and observation variances distinct \citep{staudenmayer2006measurement}.
Simulation studies further indicate that inference on $\mu$ and its CIs can be highly sensitive to misspecification of the error sources, motivating ML/REML implementations tailored to the combined observation-error model \citep{humbert2009better}.
Even when $(\mu,\sigma^2)$ can be estimated with good precision under an explicit observation-error model, their joint finite-sample distribution typically differs from the independent normal and scaled chi-square structure that underlies the $w$--$z$ method in the main text (i.e., the drift--Wiener likelihood fit without observation error).
Consequently, the induced $(w,z)$ need not inherit the noncentral-$t$ structure exploited by the $w$--$z$ method, so naive plug-in of state-space estimates into $G(w,z)$ may affect CI construction and finite-sample coverage.

More generally, state-space likelihood frameworks are now widely used in ecology to connect latent population dynamics and noisy observations under flexible model structures \citep[e.g.,][]{de2002fitting, Dennis:2006aa}.
However, short series and weak information about variance components can make separation of process noise and observation error difficult, so state-space methods are not universally stable or well identified in practice \citep{auger2016state}.
In such cases, replicated sampling can improve information about observation variance and reduce confounding \citep{dennis2010replicated}.

A complementary approach targets a different variance object when growth increments are serially dependent.
Following the diffusion-approximation perspective of \citet{lande1988extinction}, the drift--Wiener model can be viewed not as an exact description at the sampling interval, but as an effective large-scale approximation: under weak dependence, partial sums obey an invariance principle with diffusion coefficient equal to the long-run variance (LRV) of the growth-increment series, i.e., the spectral density at frequency zero.
Building on this scheme, \citet{mcnamara2004measurement} showed that, for a broad class of dependence mechanisms (including age structure and density dependence), additive observation error inflates short-run variance and induces negative short-run autocovariances but cancels in the LRV.
Accordingly, if the LRV is estimated by incorporating autocovariances, extinction-risk calculations based on an effective diffusion approximation can remain robust to additive observation error while accommodating a wide range of short-range dependence.
In specific AR(1) settings, \citet{staples2009effects} developed practical autocovariance-based estimators, although their finite-sample performance for estimating the diffusion scale was uneven in some regimes.

These considerations motivate a misspecification-robust effective diffusion approach based on direct estimation of the LRV.
As a simple new option in this context, I plug a heteroskedasticity-and-autocorrelation-consistent (HAC) estimator of the LRV (per unit time) \citep{newey1987simple, andrews1991heteroskedasticity, newey1994automatic} into $G(w,z)$.
Here I refer to such autocovariance-based LRV plug-in estimators, which are robust to additive observation error under the cancellation result of \citet{mcnamara2004measurement}, as \emph{observation-error-and-autocovariance-robust} (OEAR) estimators (Subsection~\ref{sec:oear_estimator}).
This OEAR strategy is best viewed as an effective diffusion approximation for dependent increments, rather than as an exact finite-sample likelihood procedure.

Thus, this appendix evaluates three strategies side by side: a naive drift--Wiener likelihood fit that ignores observation error, a state-space likelihood fit that models observation error explicitly, and an OEAR effective-diffusion fit (implemented here via HAC).
For each strategy, numerical experiments summarize the true $G(w,z)$ and the median plug-in estimate $G(\cdot,\cdot)$, along with CI width and coverage, under the additive observation-error drift--Wiener setting.
I also report a practical robustness criterion following \citet{meir2000will}: for each replicate time series and a given fitting approach, the replicate-level plug-in estimate from the error-laden series is flagged as ``degraded'' if
\begin{equation}
  \bigl|G(w_\bullet,z_\bullet)-G(w,z)\bigr|
  >
  \bigl|G(\widehat w,\widehat z)-G(w,z)\bigr|+0.1,
  \label{eq:degraded_def}
\end{equation}
where $G(w,z)$ is the model-implied extinction probability for the latent process, $(\widehat w,\widehat z)$ denotes the ML benchmark based on the paired error-free replicate, and $(w_\bullet,z_\bullet)$ denotes the fitted coordinates for the approach under consideration, using the notation defined below.
The slack value $0.1$ is taken directly from \citet{meir2000will}; it also aligns with the natural decision scale under IUCN Criterion~E, where category thresholds occur at extinction probabilities $0.1$, $0.2$, and $0.5$, so an additional absolute error of $0.1$ can be large enough to change a threshold-based interpretation.
More complex dependence structures (e.g., colored noise and density feedback) are examined separately in Appendix~\ref{appendix:env_noise_density}.

Throughout this appendix, estimators are distinguished by the data series used and the fitting approach.
Hats without subscripts, such as $(\widehat{\mu}, \widehat{\sigma}^2)$ and the corresponding $(\widehat{w}, \widehat{z})$, denote ML estimates obtained when the latent (error-free) series is available (used as a benchmark in simulations).
When the observed series contains additive observation error but is treated as error-free, the subscript $\mathrm{naive}$ is used: $(\widehat{\mu}_{\mathrm{naive}}, \widehat{\sigma}^2_{\mathrm{naive}})$ and $(\widehat{w}_{\mathrm{naive}}, \widehat{z}_{\mathrm{naive}})$.
State-space ML estimates that explicitly account for observation error are denoted by the subscript $\mathrm{SSM}$, as in $(\widehat{\mu}_{\mathrm{SSM}}, \widehat{\sigma}^2_{\mathrm{SSM}})$ and $(\widehat{w}_{\mathrm{SSM}}, \widehat{z}_{\mathrm{SSM}})$.
LRV-based OEAR quantities are denoted with tildes; in the baseline additive-observation-error setting, $\widetilde{\mu}:=\widehat{\mu}_{\mathrm{naive}}$, so the OEAR adjustment enters through $\widetilde{\sigma}^2$ and the resulting $(\widetilde{w},\widetilde{z},\widetilde{G})$.

A final subsection considers a power-law observation model for CPUE indices, which preserves linear-Gaussian structure on the log scale.

\subsection{Drift--Wiener model with observation error: setup and the \texorpdfstring{$(w,z)$}{w-z} representation}
\label{appendix:dwobs_setup_wz}
Let $0=t_0<t_1<\cdots<t_q$ be observation times and $\tau_i=t_i-t_{i-1}>0$ for $i=1,\dots,q$.
The latent log-abundance process $\{X(t)\}$ follows a drift Wiener process
\begin{equation}
  dX(t)=\mu\,dt+\sigma\,dW(t),\qquad t\ge 0,
  \label{Seq:dwobs_sde}
\end{equation}
so that $X(t_i)\mid X(t_{i-1})\sim\mathcal N\!\bigl(X(t_{i-1})+\mu\tau_i,\ \sigma^2\tau_i\bigr)$.
Observations on the log scale are subject to additive measurement error,
\begin{equation}
  Y_i = X(t_i) + E_i,\qquad
  E_i \overset{\mathrm{i.i.d.}}{\sim}\mathcal N(0,\omega^2),
  \qquad i=0,\dots,q,
  \label{Seq:dwobs_obs}
\end{equation}
with $\{E_i\}$ independent of $\{W(t)\}$, and realized values denoted by $y_i$.

Fix a biologically meaningful quasi-extinction threshold $n_e>0$ (e.g., $n_e=1$), and write $x_e = \log n_e$.
For any chosen initial time $t_q$ (typically the final observation time), define the initial log-distance to the threshold by
\begin{equation*}
  x_d := X(t_q)-x_e \;>\; 0.
  \label{Seq:dwobs_xd}
\end{equation*}
Under \eqref{Seq:dwobs_obs}, $X(t_q)$ is unobserved, so $x_d$ must be initialized from the data.
Throughout this appendix, I use the plug-in distance $\check{x}_d:=y_q-x_e$ to keep comparisons across fitting strategies aligned.
For state-space fitting, I compute an alternative plug-in distance as a sensitivity check, defined by $\check{x}_d^{(m)} := m_{q\mid q} - x_e$ with $m_{q\mid q} := \mathbb{E}[X(t_q)\mid Y_0, \ldots, Y_q]$ under the chosen initialization.
Implementation details (including initialization and any equivalent reparameterizations used for numerical convenience) are given in Section~\ref{sec:mc_ssm_inference}.
A fully integrated initialization that averages over the filtering distribution of $X(t_q)$ is possible but is not pursued here.

Under the drift Wiener model \eqref{Seq:dwobs_sde}, let $T$ be the (possibly infinite) first-passage time
\begin{equation*}
  T=\inf\{t\ge 0: X(t_q+t)\le x_e\}.
  \label{Seq:dwobs_T}
\end{equation*}

The finite-horizon extinction probability by time $t^\ast$ conditional on $(x_d,\mu,\sigma^2)$ is
\begin{equation}
  G(t^\ast\mid x_d,\mu,\sigma^2)
  =
  \Phi\!\left(\frac{-\mu t^\ast-x_d}{\sigma\sqrt{t^\ast}}\right)
  +
  \exp\!\left(\frac{-2\mu x_d}{\sigma^2}\right)
  \Phi\!\left(\frac{\mu t^\ast-x_d}{\sigma\sqrt{t^\ast}}\right),
  \label{Seq:dwobs_G}
\end{equation}
where $\Phi(\cdot)$ is the standard normal CDF.

Define the dimensionless coordinates
\begin{equation}
  w = \frac{\mu t^\ast + x_d}{\sigma\sqrt{t^\ast}},
  \qquad
  z = \frac{-\mu t^\ast + x_d}{\sigma\sqrt{t^\ast}},
  \qquad
  w + z = \frac{2x_d}{\sigma\sqrt{t^\ast}} > 0.
  \label{Seq:dwobs_wz}
\end{equation}
Under this transformation, the finite-horizon extinction probability can be written as a closed-form function of $(w,z)$ alone:
\begin{equation*}
  G(w,z)
  =
  \Phi(-w)
  +
  \exp\!\left(\frac{z^2 - w^2}{2}\right)\Phi(-z),
  \label{Seq:dwobs_G_wz}
\end{equation*}
which is algebraically equivalent to \eqref{Seq:dwobs_G}.
In the two extreme $z$-regimes within $w+z>0$,
\begin{empheq}[left = {G(w,z)\sim \empheqlbrace\,}]{alignat=2}
  & \Phi(-w), && \qquad z\gg 0, \nonumber\\
  & \exp\!\left(\frac{z^2-w^2}{2}\right), && \qquad z\ll 0,
  \label{Seq:dwobs_G_asym}
\end{empheq}
which will be convenient below.

\subsection{Increment moment identities under additive observation error}
Under \eqref{Seq:dwobs_obs}, the observed increments $\Delta Y_i := Y_i-Y_{i-1}$ $(i=1,\dots,q)$ satisfy
\begin{align}
  \mathbb E[\Delta Y_i] &= \mu\,\tau_i,
  \label{Seq:dwobs_E_dY}
  \\
  \mathrm{Var}[\Delta Y_i] &= \sigma^2\,\tau_i + 2\omega^2,
  \label{Seq:dwobs_Var_dY}
  \\
  \mathrm{Cov}[\Delta Y_i,\Delta Y_{i+1}] &= -\,\omega^2,
  \qquad
  \mathrm{Cov}[\Delta Y_i,\Delta Y_{i+k}] = 0 \quad (k\ge 2).
  \label{Seq:dwobs_Cov_dY}
\end{align}
Equation \eqref{Seq:dwobs_Cov_dY} shows that observation error induces a distinctive negative lag-one dependence in the increment series, even though the latent increments of $X(t)$ are independent.
These identities are standard in measurement-error analyses of random-walk-type models \citep[e.g.,][]{staudenmayer2006measurement}.

\subsection{Naive fitting inflates the environmental variance}
Here I analyze the naive ML fit, namely the drift--Wiener likelihood fit from the main text applied directly to the observed series $\{y_i\}$ generated under \eqref{Seq:dwobs_sde} and \eqref{Seq:dwobs_obs}.
Specifically,
\begin{equation}
  \widehat{\mu}_{\mathrm{naive}}
  = \frac{\sum_{i=1}^{q} (y_i-y_{i-1})}{\sum_{i=1}^{q}\tau_i}
  = \frac{y_q-y_0}{t_q},
  \qquad
  \widehat{\sigma}^{2}_{\mathrm{naive}}
  = \frac{1}{q}\sum_{i=1}^{q}\frac{(y_i-y_{i-1}-\widehat{\mu}_{\mathrm{naive}}\tau_i)^2}{\tau_i}.
  \label{Seq:dwobs_naive_def}
\end{equation}
Throughout, assume no systematic trend in observation bias, so that $\mathbb{E}[E_i]=0$.

The drift estimator $\widehat{\mu}_{\mathrm{naive}}$ remains unbiased:
\begin{equation*}
  \mathbb{E}[\widehat{\mu}_{\mathrm{naive}}]=\mu.
  \label{Seq:dwobs_naive_mu_unbiased}
\end{equation*}

To obtain a closed-form expression for $\mathbb{E}[\widehat{\sigma}^{2}_{\mathrm{naive}}]$, I assume equal spacing $\tau_i \equiv \tau$ (so $t_q = q\tau$).
Under this equal-spacing design, the naive estimator of the environmental variance is inflated by observation error:
\begin{equation}
  \mathbb{E}[\widehat{\sigma}^{2}_{\mathrm{naive}}]
  =
  \frac{q-1}{q}\,\sigma^{2}
  +
  \frac{2\omega^{2}}{\tau}\left(1-\frac{1}{q^{2}}\right),
  \label{Seq:dwobs_naive_sigma_exact}
\end{equation}
which already accounts for the finite-sample correction from estimating $\mu$.
Therefore, for moderately large $q$,
\begin{equation}
  \mathbb{E}[\widehat{\sigma}^{2}_{\mathrm{naive}}]
  \approx
  \sigma^{2}+\frac{2\omega^{2}}{\tau},
  \label{Seq:dwobs_naive_sigma_inflation}
\end{equation}
where $\approx$ indicates omission of $O(1/q)$ terms in \eqref{Seq:dwobs_naive_sigma_exact}.
Equation~\eqref{Seq:dwobs_naive_sigma_inflation} is the simplest equal-spacing special case of the more general bias expressions in \citet{staudenmayer2006measurement}.

This variance inflation alters plug-in estimates of $G(t^\ast\mid x_d,\mu,\sigma^2)$ through $\widehat{\sigma}^{2}_{\mathrm{naive}}$, but the direction of the resulting change in $\widehat{G}$ is not uniform over $(\mu,x_d,t^\ast)$, as clarified in the next subsection.

\subsection{A geometric understanding of naive variance inflation on the \texorpdfstring{$w$--$z$}{w-z} plane}
\label{sec:obs_wz_plane_view}

The $(w,z)$ representation clarifies how variance inflation under naive fitting perturbs plug-in evaluations of $G$ as well as CI width and coverage.
For fixed $(x_d,\mu,t^\ast)$, define the $\sigma$-free quantities $w_0=(\mu t^\ast+x_d)/\sqrt{t^\ast}$ and $z_0=(-\mu t^\ast+x_d)/\sqrt{t^\ast}$, so that
\begin{equation*}
  w=\frac{w_0}{\sigma},
  \qquad
  z=\frac{z_0}{\sigma}.
  \label{Seq:obs_wz_sigma_scaling}
\end{equation*}
Accordingly, any plug-in evaluation that replaces $\sigma$ by an inflated \emph{effective} diffusion scale
$\sigma_{\mathrm{eff}}>\sigma$ while holding $(x_d,\mu,t^\ast)$ fixed produces the contracted operating point
\begin{equation*}
  (w_{\mathrm{eff}},z_{\mathrm{eff}})
  :=
  \left(\frac{w_0}{\sigma_{\mathrm{eff}}},\,\frac{z_0}{\sigma_{\mathrm{eff}}}\right)
  =
  \varsigma\,(w,z),
  \qquad
  \varsigma:=\frac{\sigma}{\sigma_{\mathrm{eff}}}\in(0,1).
  \label{Seq:obs_wz_radial_shrink}
\end{equation*}
Thus scale inflation shrinks $(w,z)$ radially toward (but not to) the origin for fixed $x_d>0$; the ratio $z/w$ is unchanged, and the radius is multiplied by the shrinkage factor $\varsigma$ (see Subsection~\ref{sec:dep_mu_sigma}).

Under equal spacing $\tau_i\equiv\tau$ and moderately large $q$, naive fitting typically inflates the diffusion rate by approximately Eq.~\eqref{Seq:dwobs_naive_sigma_inflation}, $\sigma_{\mathrm{eff}}^2 \approx \sigma^2+2\omega^2/\tau$, so that
\begin{equation}
  \varsigma \approx
  \left(1+\frac{2\omega^2}{\sigma^2\,\tau}\right)^{-1/2}.
  \label{eq:mc_varsigma_eq}
\end{equation}
This shrinkage decreases monotonically but at a diminishing rate as $\omega^2/(\sigma^2\tau)$ increases, so larger observation error is required to achieve the same additional contraction in the $(w,z)$ plane.

On the $z>0$ side, when $z$ is sufficiently large that $G(w,z)\approx \Phi(-w)$ holds (main text; see also \eqref{Seq:dwobs_G_asym}), the operating-point effect of variance inflation is well captured by the induced change in $w$:
\begin{equation*}
  \Delta G_{\mathrm{op}}
  :=
  G(w_{\mathrm{eff}},z_{\mathrm{eff}})-G(w,z)
  \approx
  \Phi(-w_{\mathrm{eff}})-\Phi(-w)
  =
  \Phi(-\varsigma w)-\Phi(-w).
  \label{Seq:obs_DGop_def}
\end{equation*}
For $\varsigma$ near $1$, a first-order expansion gives
\begin{equation}
  \Delta G_{\mathrm{op}}
  \approx
  \phi(w)\,w\,(1-\varsigma),
  \label{Seq:obs_DGop_firstorder}
\end{equation}
making the direction and magnitude transparent.
First, shrinkage pulls $G$ toward $1/2$: for $w>0$ (small-risk side) one has $\Delta G_{\mathrm{op}}>0$, whereas for $w<0$ one has $\Delta G_{\mathrm{op}}<0$.
In other words, when \emph{observation} error inflates the variance under naive fitting, it \emph{overestimates} extinction risk for $w>0$ (true risk $<1/2$) and \emph{underestimates} risk for $w<0$ (true risk $>1/2$).
This contrasts with the usual monotone effect of increasing \emph{process} variance, which tends to increase extinction risk.
Second, the size of the effect is controlled by the product $\phi(w)\,|w|$.
Along a given ray $z/w$, the radius determines $|w|$ as its projection onto the $w$-axis, while the local sensitivity of $\Phi(-w)$ is $\phi(w)$.
Because $\phi(w)$ is maximized near $w\approx 0$, \eqref{Seq:obs_DGop_firstorder} shows that the operating-point shift is most pronounced when shrinkage moves the point toward the $w\approx 0$ band.

The same approximation clarifies how CI width can change geometrically.
When $G(w,z)\approx \Phi(-w)$, any equal-tailed CI for $w$ of the form $(\underline{w},\,\overline{w})$
maps approximately to
\begin{equation*}
  \mathrm{CI}_G \approx \bigl(\Phi(-\overline{w}),\,\Phi(-\underline{w})\bigr),
  \qquad
  \mathrm{width}(\mathrm{CI}_G)\approx \Phi(-\underline{w})-\Phi(-\overline{w}).
  \label{Seq:obs_CI_map_largez}
\end{equation*}
If $(\underline{w},\,\overline{w})$ is narrow, then
$\mathrm{width}(\mathrm{CI}_G)\approx \phi(w)\,(\overline{w}-\underline{w})$.
Radial shrinkage decreases $|w|$ and thus increases $\phi(w)$, but this increase is limited unless the operating point is shifted close to the $w\approx 0$ band where $\phi(w)$ is maximized.
This is consistent with the CI-width map on the $(w,z)$ plane (Fig.~\ref{fig:ci_width_wz_plane}), in which large widths are concentrated near $w\approx 0$.
Therefore, naive variance inflation need not dramatically widen the plug-in CI for $G$ unless the operating point is shifted into (or across) this high-uncertainty band.

These geometric considerations provide a constructive diagnostic for the practical robustness of naive plug-in inference.
Variance inflation moves the operating point through $\varsigma$, and thereby shifts $\widehat G$ and the mapped CI width.
In particular, when $\varsigma$ is close to $1$ and the operating point is not pushed into the high-uncertainty band around $w\approx 0$, these changes are expected to be limited, consistent with Fig.~\ref{fig:ci_width_wz_plane}.
The numerical experiments below then quantify how bias, CI width, and coverage change as $\varsigma$ decreases and as the operating point approaches $w\approx 0$.

\subsubsection{Monotonicity for \texorpdfstring{$w\ge 0$}{w>=0} and possible non-monotonicity for \texorpdfstring{$w<0$}{w<0}}

\begin{figure}[H]
\centering
\includegraphics[width=0.7\linewidth]{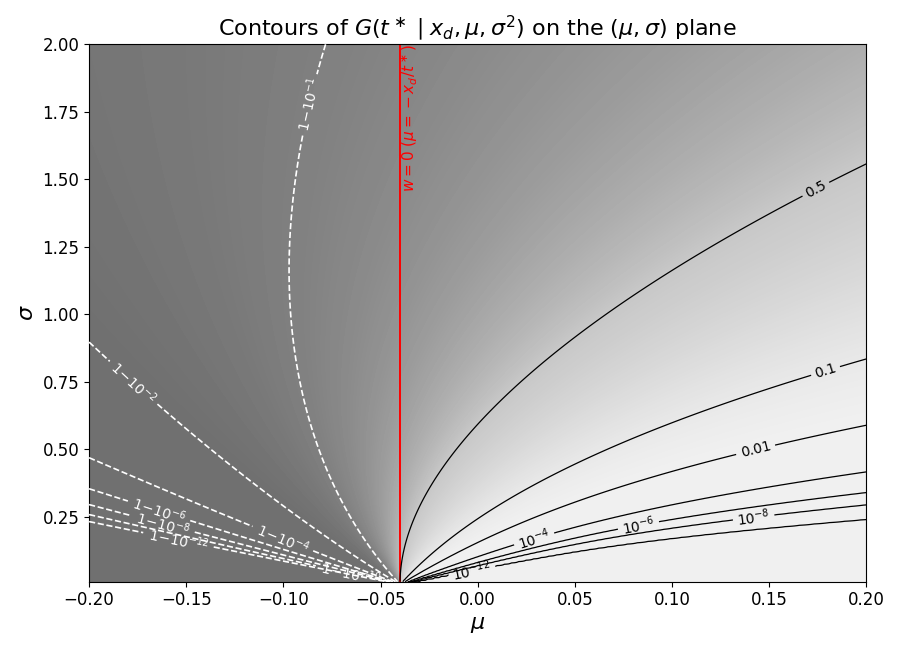}
\caption{
Contours of the finite-horizon extinction probability $G(t^\ast\mid x_d,\mu,\sigma^2)$ on the $(\mu,\sigma)$ plane for fixed $(t^\ast=100,x_d=4)$.
A red vertical line marks $w=0$, equivalently $\mu=\mu_c:=-x_d/t^\ast$.
For $\mu\ge\mu_c$, $G$ increases monotonically with $\sigma$, whereas for $\mu<\mu_c$ the dependence on $\sigma$ can be non-monotone over finite horizons.
}
\label{fig:obs_G_mu_sigma_contours}
\end{figure}

A useful global fact is that, on the $w\ge 0$ side (equivalently $\mu\ge -x_d/t^\ast$), $G(t^\ast\mid x_d,\mu,\sigma^2)$ is strictly increasing in $\sigma$ for all $\sigma>0$.
Consequently, in this regime, variance inflation under naive fitting shifts the plug-in estimate of $G$ upward and is therefore conservative for risk.
On the $w<0$ side, in contrast, $G$ need not be monotone decreasing in $\sigma$ over finite horizons; non-monotonic behavior occurs only near $(w,z)=(0,0)$, although the origin itself lies on the boundary $w+z=0$ and is excluded (Fig.~\ref{fig:obs_G_mu_sigma_contours}).

\medskip
\noindent
To see the monotonicity for $w\ge 0$, note that holding $(x_d,\mu,t^\ast)$ fixed implies $w,z\propto \sigma^{-1}$, hence
\begin{equation*}
  \frac{d}{d\sigma}G\!\left(t^\ast\mid x_d,\mu,\sigma^2\right)
  =
  -\frac{1}{\sigma}\Bigl(w\,G_w(w,z)+z\,G_z(w,z)\Bigr),
\end{equation*}
where $G_w<0$ and $G_z<0$ throughout $w+z>0$ (Appendix~\ref{appendix:G_monotonicity}).
If $z\ge 0$ and $w\ge 0$, then $wG_w+zG_z<0$ is immediate.
If $z<0$, write $u:=-z\in(0,w)$ (since $w+z>0$); substituting the derivative expressions from Appendix~\ref{appendix:G_monotonicity} shows that $wG_w+zG_z<0$ also holds.
Therefore $dG/d\sigma>0$ for all $\sigma>0$ when $w\ge 0$.

\subsubsection{A simple diagnostic for CI coverage distortion under naive fitting}

The geometry above addresses operating-point displacement and its implications for $\widehat G$ and the mapped CI width.
CI coverage can also be distorted because observation error induces serial dependence in the observed increments (Eq.~\eqref{Seq:dwobs_Cov_dY}), so the finite-sample calibration used under error-free increments is no longer exact.
A compact diagnostic for systematic displacement is the standardized bias of the naive variance component relative to its error-free sampling variability.
Under equal spacing $\tau_i\equiv\tau$ and moderately large $q$, $\mathrm{Bias}(\widehat{\sigma}^2_{\mathrm{naive}})\approx 2\omega^2/\tau$ \eqref{Seq:dwobs_naive_sigma_inflation}, while $\mathrm{se}(\widehat{\sigma}^2)\approx \sigma^2\sqrt{2/q}$ under the error-free model, giving
\begin{equation}
  \frac{\mathrm{Bias}(\widehat\sigma^2_{\mathrm{naive}})}{\mathrm{se}(\widehat\sigma^2)}
  \approx
  \sqrt{2q}\,\frac{\omega^2}{\sigma^2\,\tau}.
  \label{Seq:dwobs_sigma_SBias_scaling}
\end{equation}
When \eqref{Seq:dwobs_sigma_SBias_scaling} is small, variance inflation is typically dominated by sampling variability and naive CIs can remain practically close to nominal; values of order one (or larger) indicate that systematic displacement can be large enough to distort coverage.
For $w\ge 0$, monotonicity in $\sigma$ implies that variance inflation shifts $G$ upward, so any resulting coverage loss should arise chiefly on the lower tail unless CI widening offsets it.

\subsection{Naive fitting: Monte Carlo evaluation}
\label{sec:mc_naive}

I use Monte Carlo experiments to characterize the operating properties of the naive drift--Wiener plug-in approach under additive observation error, treating the observed series as error-free.
The experiments consider selected operating points spanning three quadrants that collectively cover the $(w,z)$ plane, together with several observation-error levels, to provide an overview of estimation and interval properties.

\subsubsection{Simulation design: operating points and observation-error levels}
\label{sec:mc_naive_design}

The extinction threshold and evaluation horizon are fixed at $x_d = 12$ and $t^\ast = 100$.
Sampling is equally spaced with $\tau = 1$, so the observation period is $t_q = q$, with $q \in \{30, 60\}$.
Simulation cases are defined by dimensionless coordinates $(w, z)$ (Table~\ref{tab:obs_sim_points}), selected to span three quadrants while keeping the radius approximately constant, $\sqrt{w^2 + z^2} \approx 25$.
Each $(w, z)$ is mapped to drift--Wiener parameters via
\begin{equation}
  x_d = \frac{w+z}{2}\,\sigma\sqrt{t^\ast},\qquad
  \mu t^\ast = \frac{w-z}{2}\,\sigma\sqrt{t^\ast},
  \label{eq:mc_wz_backmap_defs}
\end{equation}
which implies
\begin{equation}
  \sigma = \frac{2x_d}{(w+z)\sqrt{t^\ast}},\qquad
  \mu = \frac{w-z}{2}\,\frac{\sigma}{\sqrt{t^\ast}}.
  \label{eq:mc_wz_backmap_mu_sigma}
\end{equation}
Observation error is introduced as independent additive noise with variance $\omega^2$ at each time step.
I consider ratios $\omega^2/(\sigma^2\tau) = \omega^2/\sigma^2\in\{0,\,0.2,\,0.5,\,1\}$, with corresponding shrinkage factors $\varsigma=\sigma/\sigma_{\mathrm{eff}}$ taking approximate values $\{1,\,0.845,\,0.707,\,0.577\}$.
Because both $(w, z)$ and $\omega^2 / (\sigma^2 \tau)$ are dimensionless, operating characteristics of the naive plug-in procedure depend only on $(w, z)$, $q$, and the error ratio (equivalently, $\varsigma$), and are invariant to joint rescaling of $(x_d, \mu, \sigma, \omega)$.

\begin{table}[H]
\centering
\caption{Simulation points used to assess the impact of observation
error under naive fitting.}
\label{tab:obs_sim_points}
\setlength{\tabcolsep}{3pt}
\renewcommand{\arraystretch}{1.05}
\resizebox{\linewidth}{!}{\begin{tabular}{c c c c c c}
\hline
Case & $(w,z)$ & $G(w,z)$ & Angle $\theta$ & $(\mu,\sigma)$ &
Regime note \\
\hline
I-1 & $(21.65,\ 12.50)$ & $8.28\times 10^{-104}$ & $30.0$ & $(0.0322,\ 0.0703)$ & $w>0,\ z>0$ ($\mu>0$) \\
I-2 & $(12.50,\ 21.65)$ & $ 5.9\times 10^{-36}$ & $60.0$ & $(-0.0322,\ 0.0703)$ & $w>0,\ z>0$ ($\mu<0$) \\
II & $(-12.50,\ 21.65)$ & $1-1.57\times 10^{-36}$ & $120.0$ & $(-0.4479,\ 0.2623)$ & $w<0,\ z>0$ ($\mu<0$) \\
IV & $(21.65,\ -12.50)$ & $ 1.4\times 10^{-68}$ & $-30.0$ & $(0.4479,\ 0.2623)$ & $w>0,\ z<0$ ($\mu>0$) \\
\hline
\end{tabular}}

\smallskip
\footnotesize
\parbox{0.9\linewidth}{\emph{Note:} All operating points lie on the
same circle centered at the origin, $\sqrt{w^2+z^2}\approx 25$.
The displayed $(\mu,\sigma)$ values correspond to the
mapping in Eqs.~\eqref{eq:mc_wz_backmap_defs}--
\eqref{eq:mc_wz_backmap_mu_sigma} with $x_d=12$ and $t^\ast=100$.}
\end{table}

To generate one replicate, I simulate increments $\Delta X_i \sim \mathcal N(\mu\tau, \sigma^2 \tau)$ for $i = 1, \dots, q$ with $\tau = 1$, and form a latent path $X^{\mathrm{raw}}_0 = 0$, $X^{\mathrm{raw}}_i = \sum_{j=1}^i \Delta X_j$.
I then shift the path so that the true terminal state equals $x_d$:
\begin{equation*}
X_i = X^{\mathrm{raw}}_i - X^{\mathrm{raw}}_q + x_d,\qquad i = 0, \dots, q.
\end{equation*}
Observation error is added afterward: $Y_i = X_i + E_i$ with $E_i \stackrel{\mathrm{i.i.d.}}{\sim} \mathcal N(0, \omega^2)$.
Inference uses the plug-in distance $\check{x}_d:=y_q-x_e$ (here $x_e=0$, so $\check{x}_d=y_q$).
For each combination of operating point, error level, and $q$, I simulate 5{,}000 replicate series of length $q+1$.
Summaries report the Monte Carlo median of $G(\widehat w, \widehat z)$, the median 95\% CI width, empirical coverage, and the degraded rate.
The degraded rate is the Monte Carlo fraction of replicates flagged as degraded under \eqref{eq:degraded_def}.

\begin{figure}[H]
  \centering
  \includegraphics[width=0.82\linewidth]{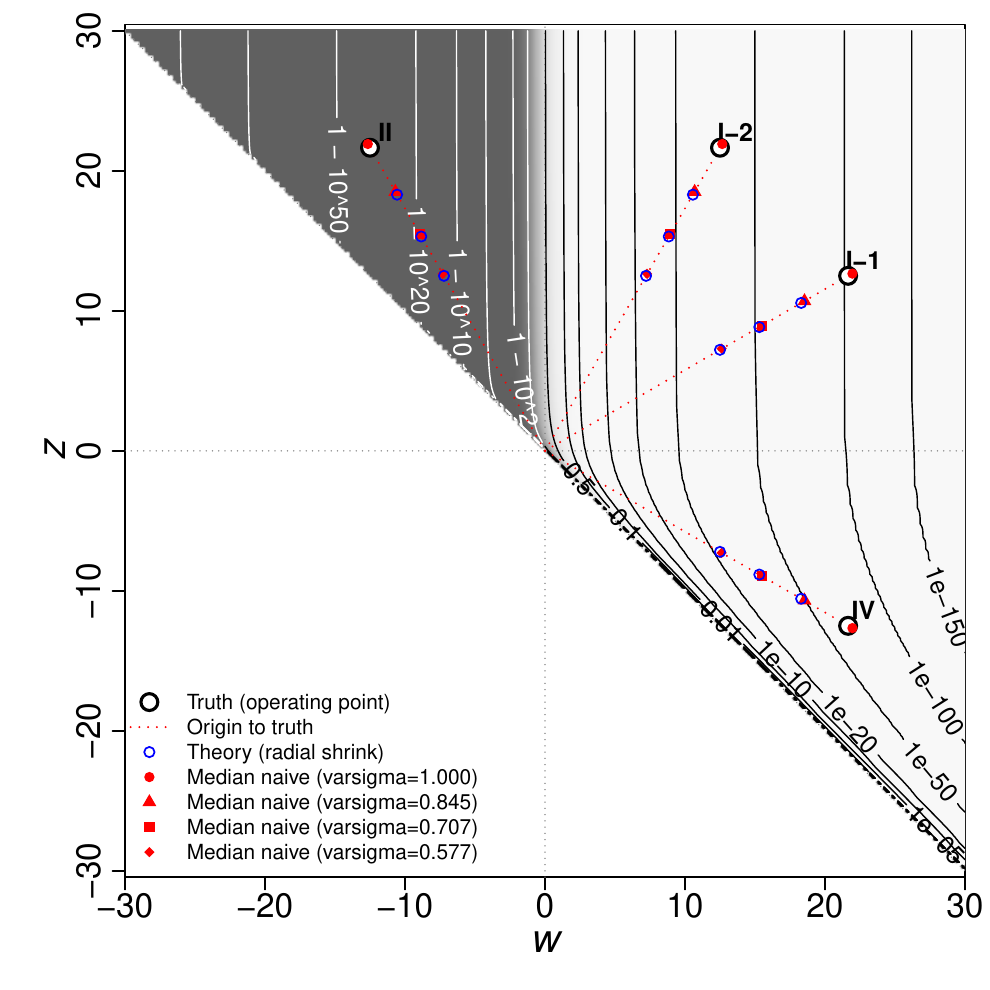}
  \caption{Operating-point shifts under naive fitting with additive observation error, shown on the $(w,z)$ plane for $q=60$. Black/white contours and grayscale shading show the geometry of $G(w,z)$, as in the background map. Open circles mark the true operating points; dashed red rays connect the origin to each truth point. Red symbols show the Monte Carlo median naive operating points for each observation-error level (indexed by $\varsigma$), and open blue circles show the corresponding radial-shrink theory points $(\varsigma w,\varsigma z)$.}
  \label{fig:wz_mc_overlay_q60}
\end{figure}

\begin{table}[H]
\caption{Naive fitting: Monte Carlo summary under additive observation error.}
\label{tab:mc_naive_summary}
\centering
\footnotesize
\setlength{\tabcolsep}{3.5pt}
\renewcommand{\arraystretch}{1.05}
\resizebox{\linewidth}{!}{\begin{tabular}{lrr r r r r r r}
\hline
Case & $q$ & $\omega^2/\sigma^2$ & Theory & MC estimate & MC SD &
CI width & Coverage & Degraded \\
\hline
I-1 & $30$ & $0.0$ & $8\times 10^{-104}$ & $7\times 10^{-109}$ & $1\times 10^{-42}$ & $1\times 10^{-51}$ & $0.946$ & $0.000$ \\
I-1 & $30$ & $0.2$ & $1\times 10^{-74}$ & $4\times 10^{-78}$ & $4\times 10^{-31}$ & $2\times 10^{-35}$ & $0.789$ & $0.000$ \\
I-1 & $30$ & $0.5$ & $9\times 10^{-53}$ & $3\times 10^{-55}$ & $4\times 10^{-22}$ & $1\times 10^{-23}$ & $0.374$ & $0.000$ \\
I-1 & $30$ & $1.0$ & $1\times 10^{-35}$ & $4\times 10^{-37}$ & $4\times 10^{-15}$ & $1\times 10^{-14}$ & $0.068$ & $0.000$ \\
\hline
I-1 & $60$ & $0.0$ & $8\times 10^{-104}$ & $1\times 10^{-106}$ & $4\times 10^{-50}$ & $2\times 10^{-65}$ & $0.952$ & $0.000$ \\
I-1 & $60$ & $0.2$ & $1\times 10^{-74}$ & $1\times 10^{-76}$ & $7\times 10^{-39}$ & $1\times 10^{-45}$ & $0.636$ & $0.000$ \\
I-1 & $60$ & $0.5$ & $9\times 10^{-53}$ & $6\times 10^{-54}$ & $8\times 10^{-28}$ & $6\times 10^{-31}$ & $0.114$ & $0.000$ \\
I-1 & $60$ & $1.0$ & $1\times 10^{-35}$ & $2\times 10^{-36}$ & $2\times 10^{-19}$ & $9\times 10^{-20}$ & $0.003$ & $0.000$ \\
\hline
I-2 & $30$ & $0.0$ & $6\times 10^{-36}$ & $9\times 10^{-38}$ & $3\times 10^{-12}$ & $7\times 10^{-15}$ & $0.949$ & $0.000$ \\
I-2 & $30$ & $0.2$ & $3\times 10^{-26}$ & $2\times 10^{-27}$ & $3\times 10^{-8}$ & $7\times 10^{-10}$ & $0.873$ & $0.000$ \\
I-2 & $30$ & $0.5$ & $8\times 10^{-19}$ & $1\times 10^{-19}$ & $1\times 10^{-6}$ & $3\times 10^{-6}$ & $0.602$ & $0.000$ \\
I-2 & $30$ & $1.0$ & $4\times 10^{-13}$ & $2\times 10^{-13}$ & $2\times 10^{-5}$ & $0.001$ & $0.240$ & $0.000$ \\
\hline
I-2 & $60$ & $0.0$ & $6\times 10^{-36}$ & $8\times 10^{-37}$ & $5\times 10^{-17}$ & $4\times 10^{-20}$ & $0.949$ & $0.000$ \\
I-2 & $60$ & $0.2$ & $3\times 10^{-26}$ & $9\times 10^{-27}$ & $1\times 10^{-12}$ & $9\times 10^{-14}$ & $0.773$ & $0.000$ \\
I-2 & $60$ & $0.5$ & $8\times 10^{-19}$ & $3\times 10^{-19}$ & $8\times 10^{-10}$ & $3\times 10^{-9}$ & $0.316$ & $0.000$ \\
I-2 & $60$ & $1.0$ & $4\times 10^{-13}$ & $2\times 10^{-13}$ & $2\times 10^{-7}$ & $8\times 10^{-6}$ & $0.031$ & $0.000$ \\
\hline
II & $30$ & $0.0$ & $1-2\times 10^{-36}$ & $1-2\times 10^{-38}$ & $1\times 10^{-10}$ & $2\times 10^{-15}$ & $0.951$ & $0.000$ \\
II & $30$ & $0.2$ & $1-9\times 10^{-27}$ & $1-6\times 10^{-28}$ & $4\times 10^{-10}$ & $3\times 10^{-10}$ & $0.876$ & $0.000$ \\
II & $30$ & $0.5$ & $1-2\times 10^{-19}$ & $1-4\times 10^{-20}$ & $3\times 10^{-8}$ & $1\times 10^{-6}$ & $0.608$ & $0.000$ \\
II & $30$ & $1.0$ & $1-1\times 10^{-13}$ & $1-4\times 10^{-14}$ & $2\times 10^{-6}$ & $5\times 10^{-4}$ & $0.243$ & $0.000$ \\
\hline
II & $60$ & $0.0$ & $1-2\times 10^{-36}$ & $1-2\times 10^{-37}$ & $6\times 10^{-18}$ & $2\times 10^{-20}$ & $0.947$ & $0.000$ \\
II & $60$ & $0.2$ & $1-9\times 10^{-27}$ & $1-2\times 10^{-27}$ & $3\times 10^{-13}$ & $3\times 10^{-14}$ & $0.785$ & $0.000$ \\
II & $60$ & $0.5$ & $1-2\times 10^{-19}$ & $1-9\times 10^{-20}$ & $2\times 10^{-10}$ & $1\times 10^{-9}$ & $0.326$ & $0.000$ \\
II & $60$ & $1.0$ & $1-1\times 10^{-13}$ & $1-6\times 10^{-14}$ & $8\times 10^{-8}$ & $3\times 10^{-6}$ & $0.033$ & $0.000$ \\
\hline
IV & $30$ & $0.0$ & $1\times 10^{-68}$ & $8\times 10^{-72}$ & $8\times 10^{-32}$ & $2\times 10^{-37}$ & $0.934$ & $0.000$ \\
IV & $30$ & $0.2$ & $3\times 10^{-49}$ & $1\times 10^{-51}$ & $2\times 10^{-23}$ & $2\times 10^{-26}$ & $0.702$ & $0.000$ \\
IV & $30$ & $0.5$ & $1\times 10^{-34}$ & $3\times 10^{-36}$ & $1\times 10^{-15}$ & $7\times 10^{-18}$ & $0.249$ & $0.000$ \\
IV & $30$ & $1.0$ & $2\times 10^{-23}$ & $3\times 10^{-24}$ & $7\times 10^{-11}$ & $3\times 10^{-11}$ & $0.031$ & $0.000$ \\
\hline
IV & $60$ & $0.0$ & $1\times 10^{-68}$ & $1\times 10^{-70}$ & $1\times 10^{-35}$ & $6\times 10^{-46}$ & $0.944$ & $0.000$ \\
IV & $60$ & $0.2$ & $3\times 10^{-49}$ & $2\times 10^{-50}$ & $8\times 10^{-28}$ & $2\times 10^{-32}$ & $0.531$ & $0.000$ \\
IV & $60$ & $0.5$ & $1\times 10^{-34}$ & $2\times 10^{-35}$ & $6\times 10^{-20}$ & $2\times 10^{-22}$ & $0.056$ & $0.000$ \\
IV & $60$ & $1.0$ & $2\times 10^{-23}$ & $8\times 10^{-24}$ & $5\times 10^{-14}$ & $1\times 10^{-14}$ & $0.001$ & $0.000$ \\
\hline
\end{tabular}
}
\smallskip
\footnotesize
\parbox{0.95\linewidth}{\emph{Note:} $G(w,z)$ denotes the model-implied extinction probability for the latent drift--Wiener process at the operating point $(w,z)$ in Table~\ref{tab:obs_sim_points}. The observation-error magnitude is reported as $\omega^2/\sigma^2$ (here $\tau=1$). ``Theory'' is $G(\varsigma w,\varsigma z)$, where, under equal spacing, the shrinkage factor is approximated by $\varsigma\approx(1+2\omega^2/\sigma^2)^{-1/2}$. ``MC estimate'' is the median plug-in estimate across Monte Carlo replicates, and ``MC SD'' is the corresponding standard deviation on the same scale. ``CI width'' is the median width of the nominal 95\% CI computed by the $w$--$z$ method across replicates, and ``Coverage'' is the corresponding empirical coverage. ``Degraded'' is the replicate-wise degraded rate defined in the text.}
\end{table}

Figure~\ref{fig:wz_mc_overlay_q60} provides a geometric view of how observation error affects naive fitting in the $(w,z)$ plane.
As observation error increases, the naive median operating points contract toward the origin along approximately the same rays, consistent with the predicted shrinkage effect; the Monte Carlo medians closely track the radial-shrink approximation $(\varsigma w,\varsigma z)$.
None of these median shifts moves a case across the relevant IUCN Criterion~E category thresholds.

Table~\ref{tab:mc_naive_summary} shows that this geometric stability carries over to quantitative performance.
Median plug-in estimates closely track the radial-shrink approximation, 95\% CIs widen as error grows, and empirical coverage declines, as expected under misspecification.
Nevertheless, the \texttt{MC SD} remains small on the probability scale and the \texttt{Degraded} rate stays at (or very near) zero, indicating that the error-laden plug-in estimates do not incur practically meaningful additional error relative to the error-free benchmark under \eqref{eq:degraded_def}.
Across the simulations in Table~\ref{tab:mc_naive_summary}, the CI endpoints remain on the same side of all Criterion~E category thresholds ($0.1$, $0.2$, and $0.5$), so category assignments do not change.
Additional runs with much larger observation error (e.g., $\omega^2/\sigma^2=10$, results not shown) lead to the same qualitative conclusion: the \texttt{Degraded} rate is zero in all cases except I-2 with $q=30$, where it is $0.015$; in that exception, the median CI width reaches $0.9$ while the \texttt{MC SD} is $0.02$ around a median estimate of $0.004$.

The behaviour is clearest in the extreme small-risk and large-risk regimes.
For the small-risk cases (I-1, I-2, and IV), the median estimate increases with $\omega^2/\sigma^2$ but remains far below the VU boundary, and the full CI stays below $0.1$.
For the large-risk case II ($G\approx 1$), the median remains close to one and the CI never drops beneath the CR threshold ($0.5$).

In brief, observation error inflates variance and reduces empirical coverage.
Nevertheless, the radial-shrink geometry keeps both the median estimate and the full CI on the same side of the relevant IUCN Criterion~E category thresholds, so classification does not change.
The CI can appear wide even when the plug-in $G$ estimates are not highly dispersed, which is an artifact of the nonlinear transformation from $w$ to $G$ (approximately $G\approx \Phi(-w)$ in the $z\gg 0$ regime) and has no practical consequence here.

\subsection{State-space fitting: Monte Carlo evaluation}
\label{sec:mc_ssm}

This subsection evaluates full adjustment by state-space ML under additive observation error, using the same Monte Carlo design as in Section~\ref{sec:mc_naive_design} so that results can be compared directly to naive fitting.

\subsubsection{Simulation design}
\label{sec:mc_ssm_design}

The operating points, observation-error levels, and replicate generation are identical to Section~\ref{sec:mc_naive_design}.
In brief, I fix $x_d=12$ and $t^\ast=100$, use equally spaced sampling with $\tau=1$ and $q\in\{30,60\}$, and simulate 5{,}000 replicate series for each combination of operating point (Table~\ref{tab:obs_sim_points}) and error ratio $\omega^2/\sigma^2\in\{0,0.2,0.5,1\}$.
Latent paths are generated from the drift--Wiener model and then contaminated by additive observation error $Y_i=X(t_i)+E_i$ with $E_i\stackrel{\mathrm{i.i.d.}}{\sim}\mathcal N(0,\omega^2)$, exactly as in the naive experiments.

To keep comparisons across fitting strategies aligned, I use the same plug-in distance to the threshold, $\check{x}_d:=y_q-x_e$ (with $n_e=1$ individual, so $x_e=0$ and $\check{x}_d=y_q$).
For state-space fitting, I compute an alternative plug-in distance as a sensitivity check, $\check{x}_d^{(m)}:=m_{q\mid q}-x_e$, where $m_{q\mid q}:=\mathbb{E}[X(t_q)\mid Y_0,\ldots,Y_q]$ under the chosen initialization.
In practice, I compute $m_{q\mid q}$ by running the Kalman filter at $(\widehat\mu_{\mathrm{SSM}},\widehat\sigma^2_{\mathrm{SSM}},\widehat\omega^2_{\mathrm{SSM}})$.

\subsubsection{State-space likelihood and inference}
\label{sec:mc_ssm_inference}

For each replicate series, I estimate $(\mu,\sigma^2,\omega^2)$ by maximum likelihood under the linear-Gaussian SSM \eqref{Seq:dwobs_sde}--\eqref{Seq:dwobs_obs}.
Marginally, the observation vector $y:=(y_0,\ldots,y_q)^\top$ is multivariate normal with mean $\mathbb E[y]$ and covariance matrix $\mathbf V_y(\mu,\sigma^2,\omega^2)$ implied by the SSM, after integrating out the unknown initial state using a diffuse initialization (as in \citealp{lindley2003estimation}).
Thus the log likelihood can be written as
\begin{equation*}
\log L(\mu,\sigma^2,\omega^2)
=
-\frac{1}{2}
\left\{
(q+1)\log(2\pi)
+
\log\bigl|\mathbf V_y\bigr|
+
(y-\mathbb E[y])^\top
\mathbf V_y^{-1}
(y-\mathbb E[y])
\right\}.
\end{equation*}

The same Gaussian likelihood admits an $O(q)$ evaluation via the Kalman-filter innovations factorization.
Under equal spacing $\tau=1$, the standard prediction and update recursions for $i=1,\ldots,q$ are
\begin{equation*}
  m_{i\mid i-1}=m_{i-1\mid i-1}+\mu,
  \qquad
  P_{i\mid i-1}=P_{i-1\mid i-1}+\sigma^2,
\end{equation*}
\begin{equation*}
  v_i=y_i-m_{i\mid i-1},
  \qquad
  F_i=P_{i\mid i-1}+\omega^2,
  \qquad
  K_i=P_{i\mid i-1}/F_i,
\end{equation*}
\begin{equation*}
  m_{i\mid i}=m_{i\mid i-1}+K_i v_i,
  \qquad
  P_{i\mid i}=(1-K_i)P_{i\mid i-1},
\end{equation*}
and the corresponding innovations log likelihood is
\begin{equation*}
  \log L(\mu,\sigma^2,\omega^2)
  =
  -\frac{1}{2}\sum_{i=1}^{q}
  \left\{
    \log(2\pi)
    +
    \log F_i
    +
    \frac{v_i^2}{F_i}
  \right\},
\end{equation*}
where $v_i$ and $F_i$ are the one-step-ahead innovations and their variances (with the contribution of $y_0$ absorbed into the chosen initialization).
Diffuse initialization is handled by the exact diffuse innovations likelihood implemented in the R package \texttt{KFAS} \citep{helske2017kfas}.

For numerical convenience in the optimization, I evaluate the same likelihood after drift-centering the observations as $y_i^{\circ}(\mu):=y_i-\mu t_i$ (i.e., absorbing the drift into the mean); this reparameterization does not change the likelihood.

Given $(\widehat\mu_{\mathrm{SSM}},\widehat\sigma^2_{\mathrm{SSM}})$, I map to $(\widehat w_{\mathrm{SSM}},\widehat z_{\mathrm{SSM}})$ via \eqref{Seq:dwobs_wz} using either $\check{x}_d$ or $\check{x}_d^{(m)}$ (based on $m_{q\mid q}$), and compute the plug-in extinction probability $G(\widehat w_{\mathrm{SSM}},\widehat z_{\mathrm{SSM}})$.
Uncertainty is summarized by the same $w$--$z$ CI procedure as in the main text.

\subsubsection{Monte Carlo summaries}
\label{sec:mc_ssm_summaries}
Reported summaries mirror Table~\ref{tab:mc_naive_summary} to enable direct comparison with naive fitting.
For each configuration, I report the Monte Carlo median of $G(\widehat w_{\mathrm{SSM}},\widehat z_{\mathrm{SSM}})$, the corresponding standard deviation across Monte Carlo replicates (\texttt{MC SD}, on the same scale), the median nominal 95\% CI width, empirical coverage, and the degraded rate defined in \eqref{eq:degraded_def}.
``Degraded'' is computed replicate-wise using the same rule as in the naive experiments.

\begin{figure}[H]
  \centering
  \includegraphics[width=0.82\linewidth]{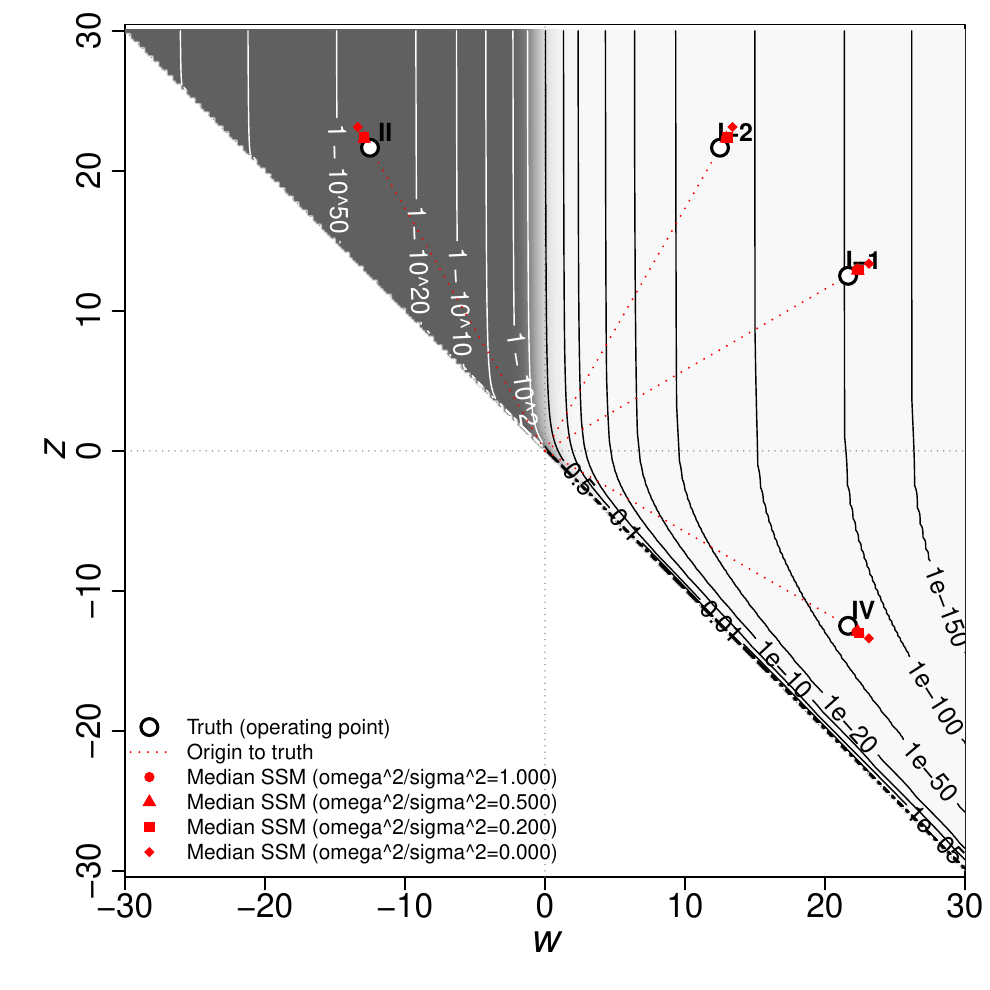}
  \caption{Operating-point recovery under state-space ML with additive observation error, shown on the $(w,z)$ plane for $q=60$. Black/white contours and grayscale shading show the geometry of $G(w,z)$, as in the background map. Open circles mark the true operating points. Red symbols show the Monte Carlo median state-space operating points obtained by maximizing the diffuse Gaussian innovations likelihood for $(\mu,\sigma^2,\omega^2)$ and mapping $(\widehat\mu,\widehat\sigma^2)$ to $(\widehat w,\widehat z)$ using the plug-in distance $\check{x}_d:=y_q-x_e$ (with $x_e=0$). Symbols are indexed by the observation-error ratio $\omega^2/\sigma^2$.}
  \label{fig:wz_mc_overlay_ssm_yq_q60}
\end{figure}

\begin{table}[H]
\caption{State-space fitting: Monte Carlo summary under additive observation error ($\check{x}_d:=y_q-x_e$).}
\label{tab:mc_ssm_summary_yq}
\centering
\footnotesize
\setlength{\tabcolsep}{3.5pt}
\renewcommand{\arraystretch}{1.05}
\resizebox{\linewidth}{!}{\begin{tabular}{lrr r r r r r r}
\hline
Case & $q$ & $\omega^2/\sigma^2$ & True $G$ & MC estimate & MC SD &
CI width & Coverage & Degraded \\
\hline
I-1 & $30$ & $0.0$ & $8\times 10^{-104}$ & $2\times 10^{-128}$ & $2\times 10^{-38}$ & $6\times 10^{-62}$ & $0.813$ & $0.000$ \\
I-1 & $30$ & $0.2$ & $8\times 10^{-104}$ & $5\times 10^{-120}$ & $1\times 10^{-31}$ & $1\times 10^{-57}$ & $0.726$ & $0.000$ \\
I-1 & $30$ & $0.5$ & $8\times 10^{-104}$ & $4\times 10^{-119}$ & $2\times 10^{-25}$ & $4\times 10^{-57}$ & $0.618$ & $0.000$ \\
I-1 & $30$ & $1.0$ & $8\times 10^{-104}$ & $2\times 10^{-119}$ & $7\times 10^{-18}$ & $3\times 10^{-57}$ & $0.489$ & $0.000$ \\
\hline
I-1 & $60$ & $0.0$ & $8\times 10^{-104}$ & $3\times 10^{-118}$ & $2\times 10^{-53}$ & $6\times 10^{-73}$ & $0.831$ & $0.000$ \\
I-1 & $60$ & $0.2$ & $8\times 10^{-104}$ & $2\times 10^{-111}$ & $2\times 10^{-43}$ & $2\times 10^{-68}$ & $0.737$ & $0.000$ \\
I-1 & $60$ & $0.5$ & $8\times 10^{-104}$ & $2\times 10^{-110}$ & $6\times 10^{-36}$ & $9\times 10^{-68}$ & $0.621$ & $0.000$ \\
I-1 & $60$ & $1.0$ & $8\times 10^{-104}$ & $1\times 10^{-109}$ & $1\times 10^{-28}$ & $3\times 10^{-67}$ & $0.496$ & $0.000$ \\
\hline
I-2 & $30$ & $0.0$ & $6\times 10^{-36}$ & $1\times 10^{-44}$ & $2\times 10^{-10}$ & $3\times 10^{-18}$ & $0.834$ & $0.000$ \\
I-2 & $30$ & $0.2$ & $6\times 10^{-36}$ & $1\times 10^{-41}$ & $7\times 10^{-7}$ & $8\times 10^{-17}$ & $0.772$ & $0.000$ \\
I-2 & $30$ & $0.5$ & $6\times 10^{-36}$ & $2\times 10^{-41}$ & $1\times 10^{-6}$ & $1\times 10^{-16}$ & $0.684$ & $0.000$ \\
I-2 & $30$ & $1.0$ & $6\times 10^{-36}$ & $6\times 10^{-42}$ & $2\times 10^{-6}$ & $6\times 10^{-17}$ & $0.573$ & $0.000$ \\
\hline
I-2 & $60$ & $0.0$ & $6\times 10^{-36}$ & $5\times 10^{-41}$ & $1\times 10^{-11}$ & $1\times 10^{-22}$ & $0.860$ & $0.000$ \\
I-2 & $60$ & $0.2$ & $6\times 10^{-36}$ & $7\times 10^{-39}$ & $6\times 10^{-13}$ & $2\times 10^{-21}$ & $0.792$ & $0.000$ \\
I-2 & $60$ & $0.5$ & $6\times 10^{-36}$ & $2\times 10^{-38}$ & $4\times 10^{-12}$ & $4\times 10^{-21}$ & $0.697$ & $0.000$ \\
I-2 & $60$ & $1.0$ & $6\times 10^{-36}$ & $3\times 10^{-38}$ & $5\times 10^{-10}$ & $5\times 10^{-21}$ & $0.586$ & $0.000$ \\
\hline
II & $30$ & $0.0$ & $1-2\times 10^{-36}$ & $1-4\times 10^{-45}$ & $7\times 10^{-8}$ & $1\times 10^{-18}$ & $0.836$ & $0.000$ \\
II & $30$ & $0.2$ & $1-2\times 10^{-36}$ & $1-6\times 10^{-42}$ & $1\times 10^{-9}$ & $4\times 10^{-17}$ & $0.773$ & $0.000$ \\
II & $30$ & $0.5$ & $1-2\times 10^{-36}$ & $1-9\times 10^{-42}$ & $4\times 10^{-8}$ & $5\times 10^{-17}$ & $0.680$ & $0.000$ \\
II & $30$ & $1.0$ & $1-2\times 10^{-36}$ & $1-8\times 10^{-42}$ & $6\times 10^{-7}$ & $5\times 10^{-17}$ & $0.569$ & $0.000$ \\
\hline
II & $60$ & $0.0$ & $1-2\times 10^{-36}$ & $1-2\times 10^{-41}$ & $5\times 10^{-11}$ & $4\times 10^{-23}$ & $0.855$ & $0.000$ \\
II & $60$ & $0.2$ & $1-2\times 10^{-36}$ & $1-6\times 10^{-39}$ & $1\times 10^{-13}$ & $2\times 10^{-21}$ & $0.790$ & $0.000$ \\
II & $60$ & $0.5$ & $1-2\times 10^{-36}$ & $1-9\times 10^{-39}$ & $3\times 10^{-13}$ & $2\times 10^{-21}$ & $0.698$ & $0.000$ \\
II & $60$ & $1.0$ & $1-2\times 10^{-36}$ & $1-1\times 10^{-38}$ & $4\times 10^{-11}$ & $2\times 10^{-21}$ & $0.582$ & $0.000$ \\
\hline
IV & $30$ & $0.0$ & $1\times 10^{-68}$ & $4\times 10^{-84}$ & $2\times 10^{-29}$ & $5\times 10^{-44}$ & $0.791$ & $0.000$ \\
IV & $30$ & $0.2$ & $1\times 10^{-68}$ & $6\times 10^{-79}$ & $2\times 10^{-24}$ & $3\times 10^{-41}$ & $0.687$ & $0.000$ \\
IV & $30$ & $0.5$ & $1\times 10^{-68}$ & $4\times 10^{-78}$ & $1\times 10^{-18}$ & $9\times 10^{-41}$ & $0.571$ & $0.000$ \\
IV & $30$ & $1.0$ & $1\times 10^{-68}$ & $2\times 10^{-78}$ & $4\times 10^{-13}$ & $7\times 10^{-41}$ & $0.437$ & $0.000$ \\
\hline
IV & $60$ & $0.0$ & $1\times 10^{-68}$ & $6\times 10^{-78}$ & $7\times 10^{-38}$ & $7\times 10^{-51}$ & $0.807$ & $0.000$ \\
IV & $60$ & $0.2$ & $1\times 10^{-68}$ & $3\times 10^{-73}$ & $2\times 10^{-30}$ & $9\times 10^{-48}$ & $0.688$ & $0.000$ \\
IV & $60$ & $0.5$ & $1\times 10^{-68}$ & $7\times 10^{-73}$ & $4\times 10^{-25}$ & $2\times 10^{-47}$ & $0.561$ & $0.000$ \\
IV & $60$ & $1.0$ & $1\times 10^{-68}$ & $1\times 10^{-72}$ & $5\times 10^{-21}$ & $3\times 10^{-47}$ & $0.446$ & $0.000$ \\
\hline
\end{tabular}
}
\smallskip
\footnotesize
\parbox{0.95\linewidth}{\emph{Note:} $G(w,z)$ denotes the model-implied extinction probability for the latent drift--Wiener process. For each replicate series with additive observation error, I fit the linear-Gaussian state-space model by ML and compute a plug-in extinction estimate $G(\widehat w,\widehat z)$ using $(\widehat\mu,\widehat\sigma^2)$ and the plug-in distance $\check{x}_d:=y_q-x_e$ (with $x_e=0$). The observation-error magnitude is reported as $\omega^2/\sigma^2$ (here $\tau=1$). ``True $G$'' is the true model-implied extinction probability at the operating point. ``MC estimate'' is the median extinction estimate across Monte Carlo replicates, and ``MC SD'' is the corresponding standard deviation on the same scale. ``CI width'' is the median width of the nominal 95\% CI computed by the $w$--$z$ method using the fitted parameters, and ``Coverage'' is the corresponding empirical coverage. ``Degraded'' is the replicate-wise degraded rate defined in the text.}
\end{table}

Figure~\ref{fig:wz_mc_overlay_ssm_yq_q60} summarizes the geometry under state-space fitting on the $(w,z)$ plane.
Unlike naive fitting, the Monte Carlo median operating points remain close to the true points across observation-error levels, indicating that joint ML estimation of $(\mu,\sigma^2,\omega^2)$ largely removes the shrinkage bias induced by treating noisy observations as error-free.
Using $\check{x}_d^{(m)}:=m_{q\mid q}-x_e$ instead of $\check{x}_d:=y_q-x_e$ produces virtually identical summaries (results not shown), so I report only the $y_q$ version for brevity.

Table~\ref{tab:mc_ssm_summary_yq} confirms the same pattern quantitatively.
Across all cases and both $q\in\{30,60\}$, the median plug-in estimates remain far from the Criterion~E decision thresholds on the probability scale, so the implied category is stable across observation-error levels.
As observation error increases, \texttt{MC SD} rises gradually, while the median CI width changes only modestly (and remains extremely small in absolute terms for the small-risk cases).
Empirical coverage declines with $\omega^2/\sigma^2$, but for $\omega^2/\sigma^2>0$ it remains substantially higher than under naive fitting.
Even when $\omega^2/\sigma^2=0$, coverage remains below the nominal level, indicating finite-sample undercoverage of this likelihood--plug-in CI pipeline in short series.
The degraded rate is zero in all configurations.

Overall, state-space fitting delivers the intended full adjustment to additive observation error in the present operating-point regime: it recovers operating points without the systematic radial contraction seen under naive fitting, yields stable extinction-risk estimates, and maintains CIs whose endpoints remain on the same side of the relevant Criterion~E thresholds (0.1, 0.2, and 0.5) as in the error-free benchmark.
Consequently, category assignments do not change in any case considered here.

\subsection{OEAR diffusion-scale estimator via HAC long-run variance}
\label{sec:oear_estimator}

This subsection describes an observation-error-and-autocovariance-robust (OEAR) diffusion-scale estimator constructed from a HAC estimate of the long-run variance (LRV) of drift-corrected increments.
Whereas SSM specifies an explicit observation model and estimates $(\mu,\sigma^2,\omega^2)$ by likelihood, the OEAR fit targets only the diffusion scale $\sigma^2$ (environmental variance per unit time) through the long-run variance (LRV) of drift-corrected increments.

The motivation has two parts. First, under additive observation error $E_i$, differencing introduces the error difference $E_i - E_{i-1}$, which inflates short-run variance and induces negative short-run autocovariances, yet these effects cancel in the long-run variance (LRV) under broad conditions (see \citealp{mcnamara2004measurement} for discussion of this phenomenon).
Second, even when the latent increment process has unknown serial dependence, the LRV is the quantity that governs the variance of long-run aggregated fluctuations, so it provides a natural effective diffusion scale without committing to a particular autocovariance model.
I operationalize these ideas by estimating the LRV using a HAC estimator, with an AR(1) pre-whitening step (and subsequent recoloring) and the Bartlett kernel.

Define the standardized centered increments (per unit time)
\begin{equation*}
U_i
:=
\frac{\Delta Y_i-\widehat{\mu}_{\mathrm{naive}}\,\tau_i}{\sqrt{\tau_i}},
\qquad i=1,\ldots,q,
\label{eq:oear_Ui_def}
\end{equation*}
with realized values denoted by $u_i$.
The standardization by $\sqrt{\tau_i}$ accommodates irregular observation
intervals $\tau_i>0$.

Under the additive observation-error model, differenced errors inflate $\mathrm{Var}(\Delta Y_i)$ but induce $\mathrm{Cov}(\Delta Y_i,\Delta Y_{i-1})=-\omega^2$ (and zero for longer lags), so the LRV equals the environmental variance $\sigma^2$ when centering by the true drift \citep{mcnamara2004measurement}.
For equally spaced sampling ($\tau_i=\tau$), letting $U_i^{(0)}:=(\Delta Y_i-\mu\tau)/\sqrt{\tau}$,
\begin{equation*}
\mathrm{LRV}\!\left(\{U_i^{(0)}\}\right)
=
\mathrm{Var}(U_i^{(0)})+2\,\mathrm{Cov}(U_i^{(0)},U_{i-1}^{(0)})
=
\left(\sigma^2+\frac{2\omega^2}{\tau}\right)+2\left(-\frac{\omega^2}{\tau}\right)
=
\sigma^2.
\end{equation*}

The target diffusion scale is the long-run variance (LRV) of $\{U_i\}$,
\begin{equation*}
\mathcal C := \sum_{j=-\infty}^{\infty} C_j,
\qquad
C_j := \mathbb E\!\left[U_i\,U_{i-j}\right],
\end{equation*}
which is interpreted as an environmental variance per unit time.
Then $\{U_i\}$ can be viewed as approximately standardized per-unit-time
increments, and the long-run variance $\mathcal C$ serves as an effective
diffusion scale under the observed sampling scheme.
When the $\tau_i$ are highly heterogeneous, however, $\mathcal C$ should be
interpreted as an approximation rather than an exact diffusion parameter.
In the settings considered here, the observation intervals are sufficiently
regular for this approximation to be adequate.

\medskip
\noindent
\textbf{AR(1) pre-whitening.}
Let $\bar u:=q^{-1}\sum_{i=1}^q u_i$ and define the centered series
$\tilde u_i:=u_i-\bar u$.
Estimate $\rho_{\mathrm{pw}}$ by OLS (equivalently, conditional Gaussian ML) in
\begin{equation*}
\tilde u_i=\rho_{\mathrm{pw}}\,\tilde u_{i-1}+\varepsilon_i,\qquad i=2,\ldots,q,
\label{eq:oear_ar1_prewhite}
\end{equation*}
yielding $\tilde\rho_{\mathrm{pw}}$ \citep{andrews1991heteroskedasticity}.
Define the pre-whitened residual sequence $\{\tilde \varepsilon_i\}_{i=1}^q$ by
\begin{equation*}
\tilde \varepsilon_1:=\tilde u_1,
\qquad
\tilde \varepsilon_i:=\tilde u_i-\tilde\rho_{\mathrm{pw}}\,\tilde u_{i-1}\quad (i=2,\ldots,q).
\end{equation*}
This pre-whitening step reduces short-run dependence in the series prior to HAC estimation, improving finite-sample stability; the subsequent recoloring step restores the implied long-run variance on the original scale.

\medskip
\noindent
\textbf{Bartlett HAC for the residual LRV.}
For $j=0,1,\ldots,q-1$, estimate the residual autocovariances by
\begin{equation*}
\widetilde C^{(\varepsilon)}_j
:=
\frac{1}{q}\sum_{i=j+1}^{q} \tilde \varepsilon_i\,\tilde \varepsilon_{i-j},
\label{eq:oear_chat_eps_def}
\end{equation*}
and set $\widetilde C^{(\varepsilon)}_{-j}:=\widetilde C^{(\varepsilon)}_j$ for $j\ge 1$.
Given a lag truncation $J\in\{0,1,\ldots,q-1\}$, the Bartlett (triangular) HAC estimator of the residual LRV is
\begin{equation*}
\widetilde{\mathcal C}^{(\varepsilon)}_{\mathrm{NW}}(J)
=
\widetilde C^{(\varepsilon)}_0
+
2\sum_{j=1}^{J}\left(1-\frac{j}{J+1}\right)\widetilde C^{(\varepsilon)}_j.
\label{eq:oear_NW_eps_def}
\end{equation*}
With Bartlett weights, $\widetilde{\mathcal C}^{(\varepsilon)}_{\mathrm{NW}}(J)$ is nonnegative in finite samples, which avoids negative variance estimates and stabilizes downstream variance-based calculations.
Equivalently, it can be viewed as a kernel (spectral-density) estimator at frequency zero, providing a standard nonparametric estimate of the residual LRV \citep{newey1987simple, andrews1991heteroskedasticity}.

\medskip
\noindent
\textbf{\citet{andrews1991heteroskedasticity} AR(1) plug-in bandwidth (Bartlett).}
I choose $J$ by the \citet{andrews1991heteroskedasticity} AR(1) plug-in form specialized to the Bartlett
window,
\begin{equation*}
J
:=
\min\!\left\{
q-1,\;
\left\lfloor
1.1447\left(
\frac{4\tilde\rho_{\mathrm{pw}}^{\,2}}{(1-\tilde\rho_{\mathrm{pw}}^{\,2})^{2}}
\,q
\right)^{1/3}
\right\rfloor
\right\}.
\label{eq:oear_A91_bandwidth}
\end{equation*}
This plug-in rule provides a simple data-driven choice of $J$ that adapts to the estimated short-run dependence while keeping the HAC estimator fully automatic. In the present Monte Carlo design, the selected truncation lags are small (range $1$--$3$ across all configurations), increasing mildly with $\omega^2/\sigma^2$ and with the number of increments $q$ in the time series.

\medskip
\noindent
\textbf{Recoloring and the OEAR diffusion-scale estimate.}
Under AR(1) pre-whitening, the LRV on the original centered scale is obtained by recoloring the residual LRV via $(1-\tilde\rho_{\mathrm{pw}})^{-2}$.
Accordingly,
\begin{equation*}
\widetilde{\sigma}^{2}
:= \widetilde{\mathcal C}_{\mathrm{NW}}(J)
:= \frac{\widetilde{\mathcal C}^{(\varepsilon)}_{\mathrm{NW}}(J)}{(1-\tilde\rho_{\mathrm{pw}})^2}.
\label{eq:oear_sigma2_tilde}
\end{equation*}
The corresponding OEAR plug-in estimator of extinction risk is obtained by replacing $(\mu,\sigma^2)$ in $(w,z)$ by $(\widehat\mu_{\mathrm{naive}},\widetilde{\sigma}^{2})$.

\medskip
\noindent
\textbf{Remark.}
In this construction, robustness to additive observation error comes from the LRV cancellation property noted by \citet{mcnamara2004measurement}, while robustness to unknown short-run autocovariance is handled by HAC estimation of the LRV.

\subsection{OEAR effective-diffusion fit: Monte Carlo evaluation}
\label{sec:mc_oear_hac}

\subsubsection{Simulation design}
\label{sec:mc_oear_design}

The operating points, observation-error levels, and replicate generation are identical to Section~\ref{sec:mc_naive_design}.

To keep comparisons across fitting strategies aligned, I use the same plug-in distance to the threshold, $\check{x}_d:=y_q-x_e$ (with $x_e=0$ so $\check{x}_d=y_q$).
For OEAR fitting, I estimate $\widehat\mu_{\mathrm{naive}}$ and the diffusion scale $\widetilde{\sigma}^2$ by the HAC LRV procedure in Subsection~\ref{sec:oear_estimator}, map $(\widehat\mu_{\mathrm{naive}},\widetilde{\sigma}^2)$ to $(w_\bullet,z_\bullet)$ via \eqref{Seq:dwobs_wz}, and compute the plug-in extinction probability $G(w_\bullet,z_\bullet)$.

\subsubsection{Monte Carlo summaries}
\label{sec:mc_oear_summaries}

Reported summaries mirror Tables~\ref{tab:mc_naive_summary} and \ref{tab:mc_ssm_summary_yq} to enable direct comparison with naive fitting and SSM. For each configuration, I report the Monte Carlo median of the OEAR plug-in estimate $G(w_\bullet,z_\bullet)$, where $(w_\bullet,z_\bullet)$ is obtained by mapping $(\widehat\mu_{\mathrm{naive}},\widetilde{\sigma}^2)$ into the $(w,z)$ coordinates, the corresponding standard deviation across Monte Carlo replicates (\texttt{MC SD}, on the same scale), the median nominal 95\% CI width, empirical coverage, and the degraded rate defined in \eqref{eq:degraded_def}. ``Degraded'' is computed replicate-wise using the same rule as in the naive and SSM experiments.

\begin{figure}[H]
\centering
\includegraphics[width=0.78\linewidth]{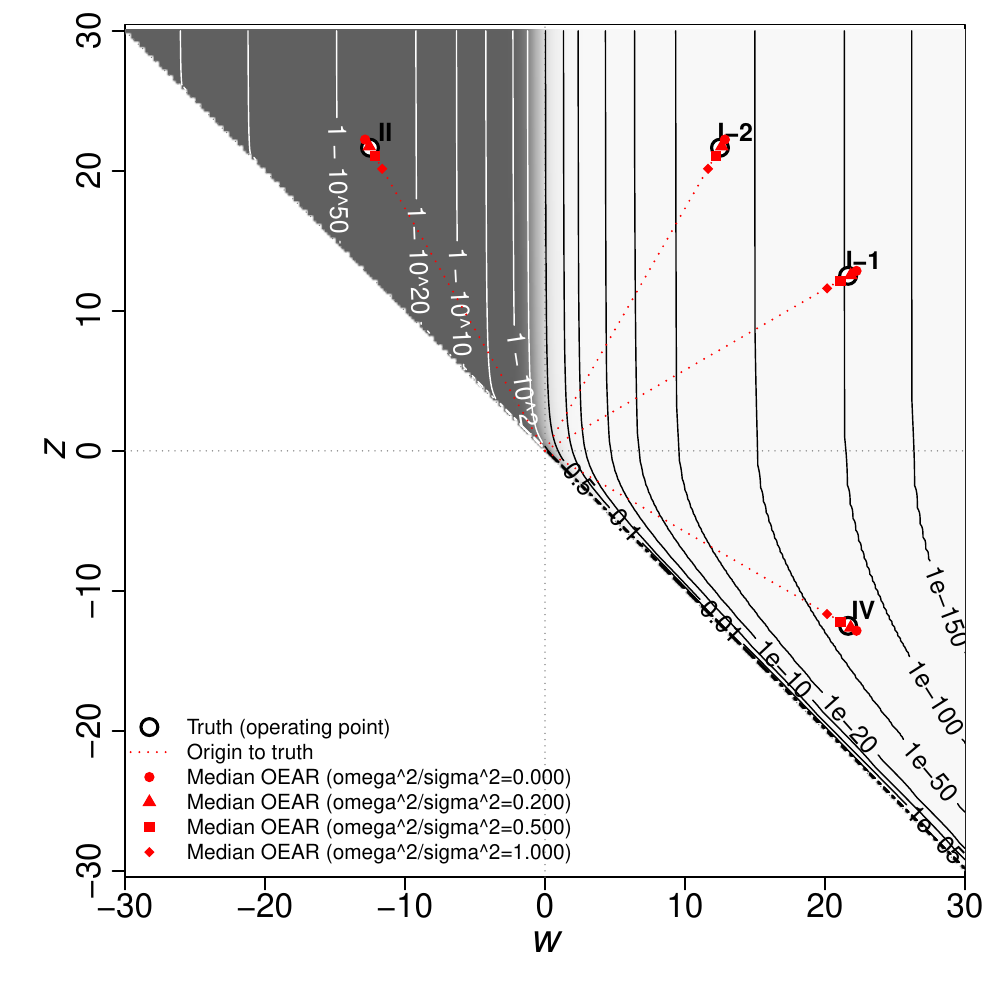}
\caption{Contours of the true extinction probability $G(w,z)$ (and its complement) in the $(w,z)$ plane, overlaid with Monte Carlo median operating points from the OEAR fit under additive observation error ($q=60$). White circles mark the true operating points for Cases I-1, I-2, II, and IV, and red symbols show the Monte Carlo medians of $(w_\bullet,z_\bullet)$ obtained by plugging $(\widehat\mu_{\mathrm{naive}},\widetilde{\sigma}^2)$ into the $w$--$z$ map; symbol shapes correspond to $\omega^2/\sigma^2\in\{0,0.2,0.5,1\}$. Dotted rays connect the origin to each true operating point.}
\label{fig:wz_oear_mc_overlay_q60}
\end{figure}

\begin{table}[H]
\caption{OEAR fitting: Monte Carlo summary under additive observation error.}
\label{tab:mc_oear_summary}
\centering
\footnotesize
\setlength{\tabcolsep}{3.5pt}
\renewcommand{\arraystretch}{1.05}
\resizebox{\linewidth}{!}{\begin{tabular}{lrr r r r r r r}
\hline
Case & $q$ & $\omega^2/\sigma^2$ & True $G$ & MC estimate & MC SD & CI width & Coverage & Degraded \\
\hline
I-1 & $30$ & $0.0$ & $8\times 10^{-104}$ & $1\times 10^{-116}$ & $1\times 10^{-14}$ & $8\times 10^{-56}$ & $0.787$ & $0.000$ \\
I-1 & $30$ & $0.2$ & $8\times 10^{-104}$ & $1\times 10^{-110}$ & $3\times 10^{-18}$ & $1\times 10^{-52}$ & $0.788$ & $0.000$ \\
I-1 & $30$ & $0.5$ & $8\times 10^{-104}$ & $2\times 10^{-101}$ & $1\times 10^{-18}$ & $1\times 10^{-47}$ & $0.762$ & $0.000$ \\
I-1 & $30$ & $1.0$ & $8\times 10^{-104}$ & $2\times 10^{-89}$ & $1\times 10^{-18}$ & $2\times 10^{-41}$ & $0.708$ & $0.000$ \\
\hline
I-1 & $60$ & $0.0$ & $8\times 10^{-104}$ & $2\times 10^{-109}$ & $6\times 10^{-40}$ & $3\times 10^{-67}$ & $0.794$ & $0.000$ \\
I-1 & $60$ & $0.2$ & $8\times 10^{-104}$ & $1\times 10^{-105}$ & $6\times 10^{-38}$ & $1\times 10^{-64}$ & $0.775$ & $0.000$ \\
I-1 & $60$ & $0.5$ & $8\times 10^{-104}$ & $9\times 10^{-99}$ & $2\times 10^{-32}$ & $4\times 10^{-60}$ & $0.744$ & $0.000$ \\
I-1 & $60$ & $1.0$ & $8\times 10^{-104}$ & $3\times 10^{-90}$ & $2\times 10^{-26}$ & $1\times 10^{-54}$ & $0.694$ & $0.000$ \\
\hline
I-2 & $30$ & $0.0$ & $6\times 10^{-36}$ & $2\times 10^{-40}$ & $4\times 10^{-5}$ & $3\times 10^{-16}$ & $0.835$ & $0.000$ \\
I-2 & $30$ & $0.2$ & $6\times 10^{-36}$ & $3\times 10^{-38}$ & $3\times 10^{-6}$ & $4\times 10^{-15}$ & $0.839$ & $0.000$ \\
I-2 & $30$ & $0.5$ & $6\times 10^{-36}$ & $3\times 10^{-35}$ & $3\times 10^{-6}$ & $1\times 10^{-13}$ & $0.830$ & $0.000$ \\
I-2 & $30$ & $1.0$ & $6\times 10^{-36}$ & $4\times 10^{-31}$ & $4\times 10^{-6}$ & $1\times 10^{-11}$ & $0.795$ & $0.000$ \\
\hline
I-2 & $60$ & $0.0$ & $6\times 10^{-36}$ & $7\times 10^{-38}$ & $3\times 10^{-13}$ & $1\times 10^{-20}$ & $0.839$ & $0.000$ \\
I-2 & $60$ & $0.2$ & $6\times 10^{-36}$ & $1\times 10^{-36}$ & $2\times 10^{-12}$ & $6\times 10^{-20}$ & $0.828$ & $0.000$ \\
I-2 & $60$ & $0.5$ & $6\times 10^{-36}$ & $2\times 10^{-34}$ & $6\times 10^{-11}$ & $2\times 10^{-18}$ & $0.814$ & $0.000$ \\
I-2 & $60$ & $1.0$ & $6\times 10^{-36}$ & $2\times 10^{-31}$ & $2\times 10^{-9}$ & $1\times 10^{-16}$ & $0.784$ & $0.000$ \\
\hline
II & $30$ & $0.0$ & $1-2\times 10^{-36}$ & $1-9\times 10^{-41}$ & $6\times 10^{-9}$ & $2\times 10^{-16}$ & $0.823$ & $0.000$ \\
II & $30$ & $0.2$ & $1-2\times 10^{-36}$ & $1-9\times 10^{-39}$ & $5\times 10^{-9}$ & $2\times 10^{-15}$ & $0.838$ & $0.000$ \\
II & $30$ & $0.5$ & $1-2\times 10^{-36}$ & $1-7\times 10^{-36}$ & $2\times 10^{-8}$ & $4\times 10^{-14}$ & $0.829$ & $0.000$ \\
II & $30$ & $1.0$ & $1-2\times 10^{-36}$ & $1-1\times 10^{-31}$ & $1\times 10^{-7}$ & $5\times 10^{-12}$ & $0.795$ & $0.000$ \\
\hline
II & $60$ & $0.0$ & $1-2\times 10^{-36}$ & $1-2\times 10^{-38}$ & $2\times 10^{-13}$ & $3\times 10^{-21}$ & $0.838$ & $0.000$ \\
II & $60$ & $0.2$ & $1-2\times 10^{-36}$ & $1-8\times 10^{-37}$ & $6\times 10^{-13}$ & $4\times 10^{-20}$ & $0.842$ & $0.000$ \\
II & $60$ & $0.5$ & $1-2\times 10^{-36}$ & $1-1\times 10^{-34}$ & $1\times 10^{-11}$ & $9\times 10^{-19}$ & $0.817$ & $0.000$ \\
II & $60$ & $1.0$ & $1-2\times 10^{-36}$ & $1-6\times 10^{-32}$ & $3\times 10^{-11}$ & $4\times 10^{-17}$ & $0.790$ & $0.000$ \\
\hline
IV & $30$ & $0.0$ & $1\times 10^{-68}$ & $3\times 10^{-77}$ & $8\times 10^{-12}$ & $2\times 10^{-40}$ & $0.736$ & $0.000$ \\
IV & $30$ & $0.2$ & $1\times 10^{-68}$ & $6\times 10^{-73}$ & $2\times 10^{-14}$ & $7\times 10^{-38}$ & $0.743$ & $0.000$ \\
IV & $30$ & $0.5$ & $1\times 10^{-68}$ & $5\times 10^{-67}$ & $1\times 10^{-14}$ & $9\times 10^{-35}$ & $0.704$ & $0.000$ \\
IV & $30$ & $1.0$ & $1\times 10^{-68}$ & $4\times 10^{-59}$ & $1\times 10^{-14}$ & $2\times 10^{-30}$ & $0.640$ & $0.000$ \\
\hline
IV & $60$ & $0.0$ & $1\times 10^{-68}$ & $2\times 10^{-72}$ & $4\times 10^{-27}$ & $4\times 10^{-47}$ & $0.747$ & $0.000$ \\
IV & $60$ & $0.2$ & $1\times 10^{-68}$ & $1\times 10^{-69}$ & $8\times 10^{-28}$ & $3\times 10^{-45}$ & $0.725$ & $0.000$ \\
IV & $60$ & $0.5$ & $1\times 10^{-68}$ & $3\times 10^{-65}$ & $8\times 10^{-24}$ & $2\times 10^{-42}$ & $0.681$ & $0.000$ \\
IV & $60$ & $1.0$ & $1\times 10^{-68}$ & $1\times 10^{-59}$ & $2\times 10^{-19}$ & $1\times 10^{-38}$ & $0.624$ & $0.000$ \\
\hline
\end{tabular}
}
\smallskip
\footnotesize
\parbox{0.95\linewidth}{\emph{Note:} The design matches Section~\ref{sec:mc_naive_design}. The observation-error magnitude is reported as $\omega^2/\sigma^2$ (with $\tau=1$). ``True $G$'' reports the corresponding true value $G(w,z)$ at the operating point in Table~\ref{tab:obs_sim_points}. ``MC estimate'' is the Monte Carlo median plug-in estimate using $(\widehat\mu_{\mathrm{naive}},\widetilde{\sigma}^2)$, and ``MC SD'' is its standard deviation on the same scale. ``CI width'' and ``Coverage'' summarize the nominal 95\% CI computed by the $w$--$z$ method, and ``Degraded'' is the replicate-wise degraded rate defined in the text.}
\end{table}

Figure~\ref{fig:wz_oear_mc_overlay_q60} provides a geometric view of how the OEAR fit behaves in the $(w,z)$ plane under additive observation error.
Across all cases, the Monte Carlo median operating points remain close to the true operating points, and their displacement changes smoothly as $\omega^2/\sigma^2$ increases.
Overall, the medians stay near the appropriate risk contours, consistent with the intended role of the LRV-based diffusion-scale estimate $\widetilde{\sigma}^2$ in correcting short-run variance inflation induced by differenced observation error.

Table~\ref{tab:mc_oear_summary} shows the same pattern on the probability scale.
Monte Carlo medians of $G(w_\bullet,z_\bullet)$ remain close to the true values $G(w,z)$, with a mild upward drift as $\omega^2/\sigma^2$ increases.
The Monte Carlo variability (\texttt{MC SD}) is comparable to, and in some settings slightly larger than, that under SSM, reflecting that OEAR targets the diffusion scale through the LRV rather than fitting an explicit observation model.
CI widths typically increase with $\omega^2/\sigma^2$ but remain small in absolute terms across all operating points considered here. 

Empirical coverage ranges from about $0.62$ to $0.84$ across configurations, with noticeable case dependence: coverage is relatively high for Case I-2 ($0.78$--$0.84$) and Case II ($0.79$--$0.84$), whereas it is lower for Case I-1 ($0.69$--$0.79$) and Case IV ($0.62$--$0.75$) and shows the clearest deterioration as $\omega^2/\sigma^2$ increases.
This pattern is consistent with a geometric explanation in the $(w,z)$ plane, where the operating-point direction (i.e., its angle from the origin) and local contour geometry affect how sampling variability in $(w_\bullet,z_\bullet)$ translates into probability-scale coverage through the nonlinear map $G(w,z)$.
Notably, coverage remains relatively high in Case I-2, which includes $\mu<0$ operating points that are still far from the quasi-extinction threshold (i.e., low extinction risk despite rapid decline).
Most importantly, the degraded rate is identically zero in every configuration, indicating that OEAR plug-in estimates never incur practically meaningful additional error relative to the error-free ML benchmark under \eqref{eq:degraded_def}.

Overall, OEAR fitting delivers a practically robust adjustment to additive observation error in the present operating-point regime: it replaces $\sigma^2$ by the LRV-based diffusion-scale estimate $\widetilde{\sigma}^2$, yields stable extinction-risk plug-in estimates, and maintains CIs whose endpoints remain on the same side of the relevant Criterion~E thresholds (0.1, 0.2, and 0.5) as in the error-free benchmark.
Consequently, category assignments do not change in any case considered here.

\subsection{CPUE nonlinearity as a power-law observation model}
\label{appendix:cpue_powerlaw_obs}

Up to this point, observation error enters additively as $Y_i=X(t_i)+E_i$ in \eqref{Seq:dwobs_obs}, and I compare fitting strategies under that common observation model.
CPUE indices, however, need not be proportional to abundance when catchability varies with stock size, yielding hyperstability or hyperdepletion.
Empirically, hyperstability (CPUE remaining high as abundance declines) appears to be common in many fisheries \citep{harley2001catch} and is illustrated by the northern cod case study of \citet{rose1999hyperaggregation}.
Here I therefore add a sensitivity check that introduces a power-law CPUE link in the data-generating process, while keeping the fitting procedures unchanged.

Specifically, the observed series on the log scale is generated as
\begin{equation}
  Y_i
  =
  \beta_{0,\mathrm{cpue}} + \beta_{1,\mathrm{cpue}}\,X(t_i) + E_i,
  \qquad
  E_i \overset{\mathrm{i.i.d.}}{\sim}\mathcal N(0,\omega^2),
  \qquad i=0,\dots,q.
  \label{Seq:dwobs_obs_cpue}
\end{equation}
The baseline case $\beta_{1,\mathrm{cpue}}=1$ reduces to \eqref{Seq:dwobs_obs}; $\beta_{1,\mathrm{cpue}}<1$ corresponds to hyperstability and $\beta_{1,\mathrm{cpue}}>1$ to hyperdepletion.

Under \eqref{Seq:dwobs_obs_cpue} and the drifted Wiener dynamics \eqref{Seq:dwobs_sde},
the increment identities analogous to \eqref{Seq:dwobs_E_dY}--\eqref{Seq:dwobs_Cov_dY}
become
\begin{align}
  \mathbb E[\Delta Y_i]
  &=
  \beta_{1,\mathrm{cpue}}\,\mu\,\tau_i,
  \label{Seq:dwobs_E_dY_cpue}
  \\
  \mathrm{Var}(\Delta Y_i)
  &=
  \beta_{1,\mathrm{cpue}}^{2}\sigma^{2}\tau_i + 2\omega^{2},
  \label{Seq:dwobs_Var_dY_cpue}
  \\
  \mathrm{Cov}(\Delta Y_i,\Delta Y_{i+1})
  &=
  -\,\omega^{2},
  \qquad
  \mathrm{Cov}(\Delta Y_i,\Delta Y_{i+k})=0 \ (k\ge 2),
  \label{Seq:dwobs_Cov_dY_cpue}
\end{align}
so the distinctive negative lag-one dependence induced by differenced observation error is unchanged by CPUE nonlinearity.

\subsubsection{Implications for naive fitting}
\label{appendix:cpue_powerlaw_obs_naive}

These identities imply that the naive estimators \eqref{Seq:dwobs_naive_def} are systematically rescaled. In particular,
\begin{equation}
  \mathbb E[\widehat{\mu}_{\mathrm{naive}}]=\beta_{1,\mathrm{cpue}}\,\mu.
  \label{Seq:dwobs_naive_mu_cpue}
\end{equation}
Under equal spacing $\tau_i\equiv\tau$, the same calculation leading to \eqref{Seq:dwobs_naive_sigma_exact} yields
\begin{equation}
  \mathbb{E}[\widehat{\sigma}^{2}_{\mathrm{naive}}]
  =
  \frac{q-1}{q}\,\beta_{1,\mathrm{cpue}}^{2}\sigma^{2}
  +
  \frac{2\omega^{2}}{\tau}\left(1-\frac{1}{q^{2}}\right),
  \label{Seq:dwobs_naive_sigma_exact_cpue}
\end{equation}
and hence, for moderately large $q$,
\begin{equation}
  \mathbb{E}[\widehat{\sigma}^{2}_{\mathrm{naive}}]
  \approx
  \beta_{1,\mathrm{cpue}}^{2}\sigma^{2}+\frac{2\omega^{2}}{\tau},
  \label{Seq:dwobs_naive_sigma_inflation_cpue}
\end{equation}
where $\approx$ omits $O(1/q)$ terms in \eqref{Seq:dwobs_naive_sigma_exact_cpue}.
Thus CPUE nonlinearity rescales the process component of the naive variance estimate by $\beta_{1,\mathrm{cpue}}^{2}$, while the observation-error inflation term is unchanged.
Notably, $\beta_{0,\mathrm{cpue}}$ drops out of all increment-based identities and hence does not affect \eqref{Seq:dwobs_E_dY_cpue}--\eqref{Seq:dwobs_naive_sigma_inflation_cpue}.

\medskip
\noindent
\textbf{Approximate radial-shrink geometry under CPUE slope.}
To clarify the dominant displacement pattern in the $(w,z)$ plane, I consider a first-order approximation that isolates the effect of CPUE slope $\beta_{1,\mathrm{cpue}}$ on the naive plug-in operating point.

Under the intercept convention
\begin{equation}
  \beta_{0,\mathrm{cpue}} = (1 - \beta_{1,\mathrm{cpue}})\,x_e,
  \label{Seq:dwobs_cpue_intercept_convention}
\end{equation}
the observation equation centers at the threshold as
\begin{equation}
  Y_i - x_e = \beta_{1,\mathrm{cpue}}\,\{X(t_i) - x_e\} + E_i.
  \label{Seq:dwobs_obs_cpue_centered}
\end{equation}
In particular, at the projection origin $t^\ast = t_q$, the naive plug-in distance satisfies
\begin{equation}
  \check{x}_d := Y_q - x_e = \beta_{1,\mathrm{cpue}}\,x_d + E_q,
  \qquad
  \mathbb{E}[\check{x}_d] = \beta_{1,\mathrm{cpue}}\,x_d.
  \label{Seq:dwobs_cpue_xd_plugin_scaling}
\end{equation}

Substituting the expectation-level rescaling results \eqref{Seq:dwobs_naive_mu_cpue} and \eqref{Seq:dwobs_naive_sigma_inflation_cpue}
into the plug-in map
\begin{equation*}
  w = \frac{\mu t^\ast + x_d}{\sigma\sqrt{t^\ast}},
  \qquad
  z = \frac{-\mu t^\ast + x_d}{\sigma\sqrt{t^\ast}},
\end{equation*}
suggests the following expectation-level approximation for the operating-point displacement:
\begin{align}
  \mathbb{E}[\widehat{w}_{\mathrm{naive}}]
  &\approx
  \frac{\beta_{1,\mathrm{cpue}}\,(\mu t^\ast + x_d)}{\sqrt{\beta_{1,\mathrm{cpue}}^{2}\sigma^{2} + 2\omega^{2}/\tau}\,\sqrt{t^\ast}}
  =
  \varsigma_{\mathrm{cpue}}\,w,
  \label{Seq:dwobs_cpue_radial_w}
  \\
  \mathbb{E}[\widehat{z}_{\mathrm{naive}}]
  &\approx
  \frac{\beta_{1,\mathrm{cpue}}\,(-\mu t^\ast + x_d)}{\sqrt{\beta_{1,\mathrm{cpue}}^{2}\sigma^{2} + 2\omega^{2}/\tau}\,\sqrt{t^\ast}}
  =
  \varsigma_{\mathrm{cpue}}\,z,
  \label{Seq:dwobs_cpue_radial_z}
\end{align}
where the shrink factor $\varsigma_{\mathrm{cpue}}$ takes the same form as \eqref{eq:mc_varsigma_eq}:
\begin{equation*}
  \varsigma_{\mathrm{cpue}}
  :=
  \frac{\beta_{1,\mathrm{cpue}}}{\sqrt{\beta_{1,\mathrm{cpue}}^{2}+2(\omega^{2}/\sigma^{2})/\tau}}
  =
  \left(
    1+\frac{2\omega^{2}}{\beta_{1,\mathrm{cpue}}^{2}\sigma^{2}\tau}
  \right)^{-1/2}.
  \label{Seq:dwobs_cpue_shrink_factor}
\end{equation*}
Equivalently, defining the CPUE-adjusted effective diffusion scale by
\begin{equation*}
  \sigma_{\mathrm{eff,cpue}}^{2}
  :=
  \sigma^{2}+\frac{2\omega^{2}}{\beta_{1,\mathrm{cpue}}^{2}\tau},
  \label{Seq:dwobs_cpue_sigma_eff}
\end{equation*}
one has $\varsigma_{\mathrm{cpue}}=\sigma/\sigma_{\mathrm{eff,cpue}}$.

Equations \eqref{Seq:dwobs_cpue_radial_w}--\eqref{Seq:dwobs_cpue_radial_z} show that, under the centered-intercept convention \eqref{Seq:dwobs_cpue_intercept_convention}, CPUE nonlinearity scales the naive $(w,z)$ operating point approximately along the ray from the origin to the true value. That is, it modifies the \emph{magnitude} of displacement while preserving the \emph{direction}.

Moreover, for fixed $\omega^{2}/\sigma^{2}$ the factor $\varsigma_{\mathrm{cpue}}$ is increasing in $\beta_{1,\mathrm{cpue}}$.
When $\omega^2=0$ it satisfies $\varsigma_{\mathrm{cpue}}=1$ for all $\beta_{1,\mathrm{cpue}}$.
When $\omega^2>0$ it satisfies $0<\varsigma_{\mathrm{cpue}}<1$, with $\varsigma_{\mathrm{cpue}}\to 1$ as $\beta_{1,\mathrm{cpue}}\to\infty$.
Relative to the baseline $\beta_{1,\mathrm{cpue}}=1$, whose shrink factor is
\begin{equation*}
\varsigma_{0}
:=
\frac{1}{\sqrt{1+2(\omega^{2}/\sigma^{2})/\tau}},
\end{equation*}
it follows that
\begin{equation*}
\beta_{1,\mathrm{cpue}}<1 \Rightarrow \varsigma_{\mathrm{cpue}}<\varsigma_{0},
\qquad
\beta_{1,\mathrm{cpue}}>1 \Rightarrow \varsigma_{\mathrm{cpue}}>\varsigma_{0},
\end{equation*}
so the contraction toward the origin is stronger under hyperstability and weaker under hyperdepletion.

\medskip
\noindent
\textbf{Remark (dependence on intercept convention).}
The centered form \eqref{Seq:dwobs_obs_cpue_centered} is equivalent to defining the observation-scale threshold
\begin{equation*}
y_e := \beta_{0,\mathrm{cpue}} + \beta_{1,\mathrm{cpue}}\,x_e
\end{equation*}
and working with $Y_i-y_e$.
If $\beta_{0,\mathrm{cpue}}$ is not chosen as in \eqref{Seq:dwobs_cpue_intercept_convention}, then centering at $x_e$ yields
\begin{equation*}
  Y_q-x_e
  =
  \beta_{1,\mathrm{cpue}}\,x_d
  +
  \Bigl\{\beta_{0,\mathrm{cpue}}-(1-\beta_{1,\mathrm{cpue}})\,x_e\Bigr\}
  +E_q,
  \label{Seq:dwobs_cpue_offset_xe}
\end{equation*}
so an additional constant shift appears and the exact common-factor structure in
\eqref{Seq:dwobs_cpue_radial_w}--\eqref{Seq:dwobs_cpue_radial_z} no longer holds under $x_e$-centering.
The radial-shrink approximation is recovered by centering at $y_e$ instead, in which case
$Y_q-y_e=\beta_{1,\mathrm{cpue}}\,x_d+E_q$.

\medskip
\noindent
\textbf{Monte Carlo settings and centering convention.}
For the numerical experiments below, I vary $\beta_{1,\mathrm{cpue}}$ over a small grid around $1$ (e.g., $\beta_{1,\mathrm{cpue}}\in\{0.8,1.0,1.25\}$).
For each value, I fix the intercept as $\beta_{0,\mathrm{cpue}}:=(1-\beta_{1,\mathrm{cpue}})\,x_e$ so that the expected index equals the threshold when $X(t_i)=x_e$; in particular, when $x_e=0$ this reduces to $\beta_{0,\mathrm{cpue}}=0$.

Each replicate dataset is generated under the nonlinear CPUE observation model \eqref{Seq:dwobs_obs_cpue}.
To isolate the effect of CPUE nonlinearity, I apply the same naive and OEAR fitting procedures as in Sections~\ref{sec:mc_naive} and \ref{sec:mc_oear_hac}, i.e., both methods fit the baseline observation model \eqref{Seq:dwobs_obs} and thus implicitly set $\beta_{1,\mathrm{cpue}}=1$ in fitting.

For each fitted replicate, I compute $(w_\bullet,z_\bullet)$ and the plug-in risk $G(w_\bullet,z_\bullet)$ as before, and summarize the impact of CPUE nonlinearity using the same performance metrics (CI width, coverage, and the degraded rate defined in \eqref{eq:degraded_def}).

\begin{figure}[H]
\centering
\includegraphics[width=1\linewidth]{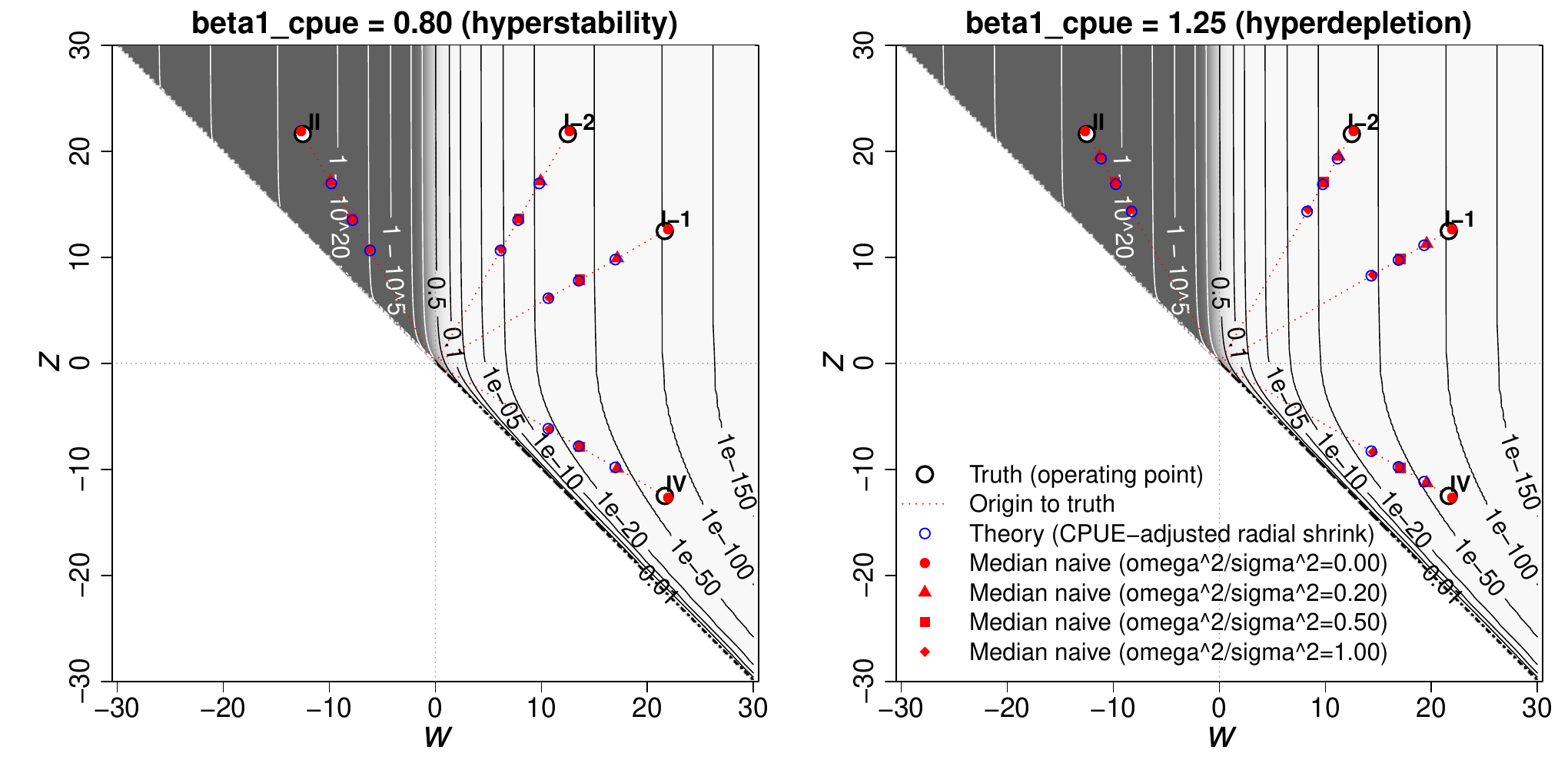}
\caption{CPUE nonlinearity under naive fitting ($q=60$): Monte Carlo medians in the $(w,z)$ plane for $\beta_{1,\mathrm{cpue}}=0.80$ (hyperstability) and $\beta_{1,\mathrm{cpue}}=1.25$ (hyperdepletion).
Contours show the true extinction probability $G(w,z)$.
For each operating point, the blue circles show the CPUE-adjusted radial-shrink prediction based on $\omega^2/\sigma^2$, and the red markers show the Monte Carlo median naive estimates at each tested observation-error level.
The baseline case $\beta_{1,\mathrm{cpue}}=1$ coincides with the additive-observation-error (naive) setting and is omitted as redundant.}
\label{fig:wz_cpue_mc_overlay_q60_beta08_125}
\end{figure}

\medskip
\noindent
\textbf{Monte Carlo results (naive fitting under CPUE nonlinearity).}
Figure~\ref{fig:wz_cpue_mc_overlay_q60_beta08_125} shows the Monte Carlo medians of the naive plug-in operating points in the $(w,z)$ plane for $q=60$ under CPUE nonlinearity, comparing hyperstability ($\beta_{1,\mathrm{cpue}}=0.80$) and hyperdepletion ($\beta_{1,\mathrm{cpue}}=1.25$).
Across operating points and observation-error levels (red markers), the median displacement follows the ray from the origin to the truth, indicating that CPUE nonlinearity primarily changes the \emph{magnitude} of the shift while leaving its \emph{direction} nearly unchanged.
The CPUE-adjusted radial-shrink prediction based on $\varsigma_{\mathrm{cpue}}$ (blue circles) closely matches these medians in both panels, confirming the first-order scaling approximation.
The corresponding $q=30$ results are qualitatively indistinguishable and are omitted for brevity.
The baseline case $\beta_{1,\mathrm{cpue}}=1$ coincides with the additive-observation-error (naive) setting and is omitted as redundant.

Tables~\ref{tab:mc_naive_cpue_q30_I12}--\ref{tab:mc_naive_cpue_q60_IIIV} report the same sensitivity check in Figure~\ref{fig:wz_cpue_mc_overlay_q60_beta08_125} in full numerical detail.
When $\omega^2/\sigma^2=0$, the results are essentially insensitive to $\beta_{1,\mathrm{cpue}}$: coverage stays near nominal and the fitted probabilities track the true extinction risk on the probability scale across the tested CPUE slopes.
Thus, under the centered-intercept convention and within these operating points, CPUE slope nonlinearity by itself does not compromise calibration; the deterioration observed in the tables is driven by observation error, not by the CPUE slope.

As $\omega^2/\sigma^2$ increases, the same qualitative effects as in the additive observation-error setting appear (CI widths expand and coverage declines), and CPUE nonlinearity mainly changes the \emph{rate} of this deterioration through the shrink factor $\varsigma_{\mathrm{cpue}}$:
hyperstability ($\beta_{1,\mathrm{cpue}}<1$) strengthens contraction toward the origin and tends to worsen coverage, whereas hyperdepletion ($\beta_{1,\mathrm{cpue}}>1$) weakens contraction and can partially mitigate it.
Despite these shifts, the \texttt{Degraded} rate remains essentially zero across all configurations, indicating no practically meaningful change in policy classification under \eqref{eq:degraded_def}.
Moreover, the fitted risks remain far from the policy threshold on the probability scale (close to $0$ in Cases~I-1/I-2/IV and close to $1$ in Case~II, reported as $1-Q$), and the associated CIs remain sufficiently narrow that they never straddle the decision boundary $G_{\mathrm{target}}=0.1$.

\begin{table}[H]
\centering
\footnotesize
\caption{Naive fitting: Monte Carlo summary under CPUE nonlinearity ($q=30$, Cases I-1 and I-2).}
\label{tab:mc_naive_cpue_q30_I12}
\setlength{\tabcolsep}{4pt}
\begin{tabular}{lrr rrr rrr}
\hline
Case & $\beta_{1,\mathrm{cpue}}$ & $\omega^2/\sigma^2$ & True $G$ & MC estimate & MC SD & CI width & Coverage & Degraded \\
\hline
\multicolumn{9}{l}{\textit{Case I-1}}\\
\multicolumn{9}{l}{$\beta_{1,\mathrm{cpue}}=0.80$ (hyperstability)}\\
 & 0.80 & 0.00 & $8\times 10^{-104}$ & $7\times 10^{-109}$ & $1\times 10^{-42}$ & $1\times 10^{-51}$ & 0.946 & 0.000 \\
 & 0.80 & 0.20 & $8\times 10^{-104}$ & $1\times 10^{-67}$ & $3\times 10^{-27}$ & $4\times 10^{-30}$ & 0.622 & 0.000 \\
 & 0.80 & 0.50 & $8\times 10^{-104}$ & $3\times 10^{-43}$ & $2\times 10^{-17}$ & $1\times 10^{-17}$ & 0.148 & 0.000 \\
 & 0.80 & 1.00 & $8\times 10^{-104}$ & $3\times 10^{-27}$ & $1\times 10^{-11}$ & $9\times 10^{-10}$ & 0.010 & 0.000 \\
\addlinespace[2pt]
\multicolumn{9}{l}{$\beta_{1,\mathrm{cpue}}=1.00$ (linear)}\\
 & 1.00 & 0.00 & $8\times 10^{-104}$ & $7\times 10^{-109}$ & $1\times 10^{-42}$ & $1\times 10^{-51}$ & 0.946 & 0.000 \\
 & 1.00 & 0.20 & $8\times 10^{-104}$ & $4\times 10^{-78}$ & $4\times 10^{-31}$ & $2\times 10^{-35}$ & 0.789 & 0.000 \\
 & 1.00 & 0.50 & $8\times 10^{-104}$ & $3\times 10^{-55}$ & $4\times 10^{-22}$ & $1\times 10^{-23}$ & 0.374 & 0.000 \\
 & 1.00 & 1.00 & $8\times 10^{-104}$ & $4\times 10^{-37}$ & $4\times 10^{-15}$ & $1\times 10^{-14}$ & 0.068 & 0.000 \\
\addlinespace[2pt]
\multicolumn{9}{l}{$\beta_{1,\mathrm{cpue}}=1.25$ (hyperdepletion)}\\
 & 1.25 & 0.00 & $8\times 10^{-104}$ & $7\times 10^{-109}$ & $1\times 10^{-42}$ & $1\times 10^{-51}$ & 0.946 & 0.000 \\
 & 1.25 & 0.20 & $8\times 10^{-104}$ & $4\times 10^{-87}$ & $4\times 10^{-34}$ & $3\times 10^{-40}$ & 0.872 & 0.000 \\
 & 1.25 & 0.50 & $8\times 10^{-104}$ & $5\times 10^{-67}$ & $4\times 10^{-27}$ & $9\times 10^{-30}$ & 0.611 & 0.000 \\
 & 1.25 & 1.00 & $8\times 10^{-104}$ & $2\times 10^{-48}$ & $2\times 10^{-19}$ & $3\times 10^{-20}$ & 0.239 & 0.000 \\
\addlinespace[2pt]
\hline
\multicolumn{9}{l}{\textit{Case I-2}}\\
\multicolumn{9}{l}{$\beta_{1,\mathrm{cpue}}=0.80$ (hyperstability)}\\
 & 0.80 & 0.00 & $6\times 10^{-36}$ & $9\times 10^{-38}$ & $3\times 10^{-12}$ & $7\times 10^{-15}$ & 0.949 & 0.000 \\
 & 0.80 & 0.20 & $6\times 10^{-36}$ & $9\times 10^{-24}$ & $2\times 10^{-7}$ & $4\times 10^{-8}$ & 0.786 & 0.000 \\
 & 0.80 & 0.50 & $6\times 10^{-36}$ & $1\times 10^{-15}$ & $9\times 10^{-6}$ & $1\times 10^{-4}$ & 0.367 & 0.000 \\
 & 0.80 & 1.00 & $6\times 10^{-36}$ & $3\times 10^{-10}$ & $1\times 10^{-4}$ & 0.02 & 0.069 & 0.000 \\
\addlinespace[2pt]
\multicolumn{9}{l}{$\beta_{1,\mathrm{cpue}}=1.00$ (linear)}\\
 & 1.00 & 0.00 & $6\times 10^{-36}$ & $9\times 10^{-38}$ & $3\times 10^{-12}$ & $7\times 10^{-15}$ & 0.949 & 0.000 \\
 & 1.00 & 0.20 & $6\times 10^{-36}$ & $2\times 10^{-27}$ & $3\times 10^{-8}$ & $7\times 10^{-10}$ & 0.873 & 0.000 \\
 & 1.00 & 0.50 & $6\times 10^{-36}$ & $1\times 10^{-19}$ & $1\times 10^{-6}$ & $3\times 10^{-6}$ & 0.602 & 0.000 \\
 & 1.00 & 1.00 & $6\times 10^{-36}$ & $2\times 10^{-13}$ & $2\times 10^{-5}$ & 0.001 & 0.240 & 0.000 \\
\addlinespace[2pt]
\multicolumn{9}{l}{$\beta_{1,\mathrm{cpue}}=1.25$ (hyperdepletion)}\\
 & 1.25 & 0.00 & $6\times 10^{-36}$ & $9\times 10^{-38}$ & $3\times 10^{-12}$ & $7\times 10^{-15}$ & 0.949 & 0.000 \\
 & 1.25 & 0.20 & $6\times 10^{-36}$ & $2\times 10^{-30}$ & $6\times 10^{-9}$ & $3\times 10^{-11}$ & 0.911 & 0.000 \\
 & 1.25 & 0.50 & $6\times 10^{-36}$ & $1\times 10^{-23}$ & $2\times 10^{-7}$ & $5\times 10^{-8}$ & 0.779 & 0.000 \\
 & 1.25 & 1.00 & $6\times 10^{-36}$ & $2\times 10^{-17}$ & $4\times 10^{-6}$ & $3\times 10^{-5}$ & 0.477 & 0.000 \\
\addlinespace[2pt]
\hline
\end{tabular}
\vspace{2pt}
\begin{minipage}{0.95\linewidth}
\footnotesize
\emph{Notes:} $\omega^2/\sigma^2$ is the observation-error to process-variance ratio (here $\tau=1$). True $G$ is the true value of the extinction probability $G$. ``MC estimate'' is the Monte Carlo median of the fitted extinction probability, and ``MC SD'' is the replicate SD. CI width and coverage are computed from the $w$--$z$ method. Degraded is defined in \eqref{eq:degraded_def}.
\end{minipage}
\end{table}

\begin{table}[H]
\centering
\footnotesize
\caption{Naive fitting: Monte Carlo summary under CPUE nonlinearity ($q=30$, Cases II and IV).}
\label{tab:mc_naive_cpue_q30_IIIV}
\setlength{\tabcolsep}{4pt}
\begin{tabular}{lrr rrr rrr}
\hline
Case & $\beta_{1,\mathrm{cpue}}$ & $\omega^2/\sigma^2$ & True $G$ & MC estimate & MC SD & CI width & Coverage & Degraded \\
\hline
\multicolumn{9}{l}{\textit{Case II}}\\
\multicolumn{9}{l}{$\beta_{1,\mathrm{cpue}}=0.80$ (hyperstability)}\\
 & 0.80 & 0.00 & $1-2\times 10^{-36}$ & $1-2\times 10^{-38}$ & $1\times 10^{-10}$ & $2\times 10^{-15}$ & 0.951 & 0.000 \\
 & 0.80 & 0.20 & $1-2\times 10^{-36}$ & $1-3\times 10^{-24}$ & $2\times 10^{-9}$ & $2\times 10^{-8}$ & 0.786 & 0.000 \\
 & 0.80 & 0.50 & $1-2\times 10^{-36}$ & $1-4\times 10^{-16}$ & $5\times 10^{-7}$ & $7\times 10^{-5}$ & 0.368 & 0.000 \\
 & 0.80 & 1.00 & $1-2\times 10^{-36}$ & $1-9\times 10^{-11}$ & $2\times 10^{-5}$ & 0.008 & 0.074 & 0.000 \\
\addlinespace[2pt]
\multicolumn{9}{l}{$\beta_{1,\mathrm{cpue}}=1.00$ (linear)}\\
 & 1.00 & 0.00 & $1-2\times 10^{-36}$ & $1-2\times 10^{-38}$ & $1\times 10^{-10}$ & $2\times 10^{-15}$ & 0.951 & 0.000 \\
 & 1.00 & 0.20 & $1-2\times 10^{-36}$ & $1-6\times 10^{-28}$ & $4\times 10^{-10}$ & $3\times 10^{-10}$ & 0.876 & 0.000 \\
 & 1.00 & 0.50 & $1-2\times 10^{-36}$ & $1-4\times 10^{-20}$ & $3\times 10^{-8}$ & $1\times 10^{-6}$ & 0.608 & 0.000 \\
 & 1.00 & 1.00 & $1-2\times 10^{-36}$ & $1-4\times 10^{-14}$ & $2\times 10^{-6}$ & $5\times 10^{-4}$ & 0.243 & 0.000 \\
\addlinespace[2pt]
\multicolumn{9}{l}{$\beta_{1,\mathrm{cpue}}=1.25$ (hyperdepletion)}\\
 & 1.25 & 0.00 & $1-2\times 10^{-36}$ & $1-2\times 10^{-38}$ & $1\times 10^{-10}$ & $2\times 10^{-15}$ & 0.951 & 0.000 \\
 & 1.25 & 0.20 & $1-2\times 10^{-36}$ & $1-6\times 10^{-31}$ & $1\times 10^{-10}$ & $1\times 10^{-11}$ & 0.921 & 0.000 \\
 & 1.25 & 0.50 & $1-2\times 10^{-36}$ & $1-5\times 10^{-24}$ & $2\times 10^{-9}$ & $2\times 10^{-8}$ & 0.780 & 0.000 \\
 & 1.25 & 1.00 & $1-2\times 10^{-36}$ & $1-7\times 10^{-18}$ & $1\times 10^{-7}$ & $1\times 10^{-5}$ & 0.485 & 0.000 \\
\addlinespace[2pt]
\hline
\multicolumn{9}{l}{\textit{Case IV}}\\
\multicolumn{9}{l}{$\beta_{1,\mathrm{cpue}}=0.80$ (hyperstability)}\\
 & 0.80 & 0.00 & $1\times 10^{-68}$ & $8\times 10^{-72}$ & $8\times 10^{-32}$ & $2\times 10^{-37}$ & 0.934 & 0.000 \\
 & 0.80 & 0.20 & $1\times 10^{-68}$ & $2\times 10^{-44}$ & $9\times 10^{-20}$ & $2\times 10^{-22}$ & 0.502 & 0.000 \\
 & 0.80 & 0.50 & $1\times 10^{-68}$ & $3\times 10^{-28}$ & $2\times 10^{-12}$ & $2\times 10^{-13}$ & 0.078 & 0.000 \\
 & 0.80 & 1.00 & $1\times 10^{-68}$ & $1\times 10^{-17}$ & $2\times 10^{-8}$ & $1\times 10^{-7}$ & 0.003 & 0.000 \\
\addlinespace[2pt]
\multicolumn{9}{l}{$\beta_{1,\mathrm{cpue}}=1.00$ (linear)}\\
 & 1.00 & 0.00 & $1\times 10^{-68}$ & $8\times 10^{-72}$ & $8\times 10^{-32}$ & $2\times 10^{-37}$ & 0.934 & 0.000 \\
 & 1.00 & 0.20 & $1\times 10^{-68}$ & $1\times 10^{-51}$ & $2\times 10^{-23}$ & $2\times 10^{-26}$ & 0.702 & 0.000 \\
 & 1.00 & 0.50 & $1\times 10^{-68}$ & $3\times 10^{-36}$ & $1\times 10^{-15}$ & $7\times 10^{-18}$ & 0.249 & 0.000 \\
 & 1.00 & 1.00 & $1\times 10^{-68}$ & $3\times 10^{-24}$ & $7\times 10^{-11}$ & $3\times 10^{-11}$ & 0.031 & 0.000 \\
\addlinespace[2pt]
\multicolumn{9}{l}{$\beta_{1,\mathrm{cpue}}=1.25$ (hyperdepletion)}\\
 & 1.25 & 0.00 & $1\times 10^{-68}$ & $8\times 10^{-72}$ & $8\times 10^{-32}$ & $2\times 10^{-37}$ & 0.934 & 0.000 \\
 & 1.25 & 0.20 & $1\times 10^{-68}$ & $3\times 10^{-57}$ & $8\times 10^{-26}$ & $2\times 10^{-29}$ & 0.820 & 0.000 \\
 & 1.25 & 0.50 & $1\times 10^{-68}$ & $5\times 10^{-44}$ & $2\times 10^{-19}$ & $4\times 10^{-22}$ & 0.489 & 0.000 \\
 & 1.25 & 1.00 & $1\times 10^{-68}$ & $1\times 10^{-31}$ & $8\times 10^{-14}$ & $2\times 10^{-15}$ & 0.137 & 0.000 \\
\addlinespace[2pt]
\hline
\end{tabular}
\vspace{2pt}
\begin{minipage}{0.95\linewidth}
\footnotesize
\emph{Notes:} See Table~\ref{tab:mc_naive_cpue_q30_I12}.
\end{minipage}
\end{table}

\begin{table}[H]
\centering
\footnotesize
\caption{Naive fitting: Monte Carlo summary under CPUE nonlinearity ($q=60$, Cases I-1 and I-2).}
\label{tab:mc_naive_cpue_q60_I12}
\setlength{\tabcolsep}{4pt}
\begin{tabular}{lrr rrr rrr}
\hline
Case & $\beta_{1,\mathrm{cpue}}$ & $\omega^2/\sigma^2$ & True $G$ & MC estimate & MC SD & CI width & Coverage & Degraded \\
\hline
\multicolumn{9}{l}{\textit{Case I-1}}\\
\multicolumn{9}{l}{$\beta_{1,\mathrm{cpue}}=0.80$ (hyperstability)}\\
 & 0.80 & 0.00 & $8\times 10^{-104}$ & $1\times 10^{-106}$ & $4\times 10^{-50}$ & $2\times 10^{-65}$ & 0.952 & 0.000 \\
 & 0.80 & 0.20 & $8\times 10^{-104}$ & $5\times 10^{-66}$ & $1\times 10^{-33}$ & $9\times 10^{-39}$ & 0.379 & 0.000 \\
 & 0.80 & 0.50 & $8\times 10^{-104}$ & $3\times 10^{-42}$ & $2\times 10^{-22}$ & $2\times 10^{-23}$ & 0.015 & 0.000 \\
 & 0.80 & 1.00 & $8\times 10^{-104}$ & $7\times 10^{-27}$ & $5\times 10^{-15}$ & $8\times 10^{-14}$ & 0.000 & 0.000 \\
\addlinespace[2pt]
\multicolumn{9}{l}{$\beta_{1,\mathrm{cpue}}=1.00$ (linear)}\\
 & 1.00 & 0.00 & $8\times 10^{-104}$ & $1\times 10^{-106}$ & $4\times 10^{-50}$ & $2\times 10^{-65}$ & 0.952 & 0.000 \\
 & 1.00 & 0.20 & $8\times 10^{-104}$ & $1\times 10^{-76}$ & $7\times 10^{-39}$ & $1\times 10^{-45}$ & 0.636 & 0.000 \\
 & 1.00 & 0.50 & $8\times 10^{-104}$ & $6\times 10^{-54}$ & $8\times 10^{-28}$ & $6\times 10^{-31}$ & 0.114 & 0.000 \\
 & 1.00 & 1.00 & $8\times 10^{-104}$ & $2\times 10^{-36}$ & $2\times 10^{-19}$ & $9\times 10^{-20}$ & 0.003 & 0.000 \\
\addlinespace[2pt]
\multicolumn{9}{l}{$\beta_{1,\mathrm{cpue}}=1.25$ (hyperdepletion)}\\
 & 1.25 & 0.00 & $8\times 10^{-104}$ & $1\times 10^{-106}$ & $4\times 10^{-50}$ & $2\times 10^{-65}$ & 0.952 & 0.000 \\
 & 1.25 & 0.20 & $8\times 10^{-104}$ & $4\times 10^{-85}$ & $3\times 10^{-43}$ & $3\times 10^{-51}$ & 0.799 & 0.000 \\
 & 1.25 & 0.50 & $8\times 10^{-104}$ & $2\times 10^{-65}$ & $2\times 10^{-33}$ & $2\times 10^{-38}$ & 0.367 & 0.000 \\
 & 1.25 & 1.00 & $8\times 10^{-104}$ & $2\times 10^{-47}$ & $9\times 10^{-25}$ & $9\times 10^{-27}$ & 0.040 & 0.000 \\
\addlinespace[2pt]
\hline
\multicolumn{9}{l}{\textit{Case I-2}}\\
\multicolumn{9}{l}{$\beta_{1,\mathrm{cpue}}=0.80$ (hyperstability)}\\
 & 0.80 & 0.00 & $6\times 10^{-36}$ & $8\times 10^{-37}$ & $5\times 10^{-17}$ & $4\times 10^{-20}$ & 0.949 & 0.000 \\
 & 0.80 & 0.20 & $6\times 10^{-36}$ & $3\times 10^{-23}$ & $2\times 10^{-11}$ & $1\times 10^{-11}$ & 0.599 & 0.000 \\
 & 0.80 & 0.50 & $6\times 10^{-36}$ & $2\times 10^{-15}$ & $3\times 10^{-8}$ & $6\times 10^{-7}$ & 0.086 & 0.000 \\
 & 0.80 & 1.00 & $6\times 10^{-36}$ & $4\times 10^{-10}$ & $4\times 10^{-6}$ & $5\times 10^{-4}$ & 0.002 & 0.000 \\
\addlinespace[2pt]
\multicolumn{9}{l}{$\beta_{1,\mathrm{cpue}}=1.00$ (linear)}\\
 & 1.00 & 0.00 & $6\times 10^{-36}$ & $8\times 10^{-37}$ & $5\times 10^{-17}$ & $4\times 10^{-20}$ & 0.949 & 0.000 \\
 & 1.00 & 0.20 & $6\times 10^{-36}$ & $9\times 10^{-27}$ & $1\times 10^{-12}$ & $9\times 10^{-14}$ & 0.773 & 0.000 \\
 & 1.00 & 0.50 & $6\times 10^{-36}$ & $3\times 10^{-19}$ & $8\times 10^{-10}$ & $3\times 10^{-9}$ & 0.316 & 0.000 \\
 & 1.00 & 1.00 & $6\times 10^{-36}$ & $2\times 10^{-13}$ & $2\times 10^{-7}$ & $8\times 10^{-6}$ & 0.031 & 0.000 \\
\addlinespace[2pt]
\multicolumn{9}{l}{$\beta_{1,\mathrm{cpue}}=1.25$ (hyperdepletion)}\\
 & 1.25 & 0.00 & $6\times 10^{-36}$ & $8\times 10^{-37}$ & $5\times 10^{-17}$ & $4\times 10^{-20}$ & 0.949 & 0.000 \\
 & 1.25 & 0.20 & $6\times 10^{-36}$ & $1\times 10^{-29}$ & $9\times 10^{-14}$ & $2\times 10^{-15}$ & 0.872 & 0.000 \\
 & 1.25 & 0.50 & $6\times 10^{-36}$ & $4\times 10^{-23}$ & $2\times 10^{-11}$ & $2\times 10^{-11}$ & 0.584 & 0.000 \\
 & 1.25 & 1.00 & $6\times 10^{-36}$ & $5\times 10^{-17}$ & $6\times 10^{-9}$ & $6\times 10^{-8}$ & 0.176 & 0.000 \\
\addlinespace[2pt]
\hline
\end{tabular}
\vspace{2pt}
\begin{minipage}{0.95\linewidth}
\footnotesize
\emph{Notes:} See Table~\ref{tab:mc_naive_cpue_q30_I12}.
\end{minipage}
\end{table}

\begin{table}[H]
\centering
\footnotesize
\caption{Naive fitting: Monte Carlo summary under CPUE nonlinearity ($q=60$, Cases II and IV).}
\label{tab:mc_naive_cpue_q60_IIIV}
\setlength{\tabcolsep}{4pt}
\begin{tabular}{lrr rrr rrr}
\hline
Case & $\beta_{1,\mathrm{cpue}}$ & $\omega^2/\sigma^2$ & True $G$ & MC estimate & MC SD & CI width & Coverage & Degraded \\
\hline
\multicolumn{9}{l}{\textit{Case II}}\\
\multicolumn{9}{l}{$\beta_{1,\mathrm{cpue}}=0.80$ (hyperstability)}\\
 & 0.80 & 0.00 & $1-2\times 10^{-36}$ & $1-2\times 10^{-37}$ & $6\times 10^{-18}$ & $2\times 10^{-20}$ & 0.947 & 0.000 \\
 & 0.80 & 0.20 & $1-2\times 10^{-36}$ & $1-8\times 10^{-24}$ & $5\times 10^{-12}$ & $5\times 10^{-12}$ & 0.607 & 0.000 \\
 & 0.80 & 0.50 & $1-2\times 10^{-36}$ & $1-5\times 10^{-16}$ & $1\times 10^{-8}$ & $2\times 10^{-7}$ & 0.093 & 0.000 \\
 & 0.80 & 1.00 & $1-2\times 10^{-36}$ & $1-9\times 10^{-11}$ & $2\times 10^{-6}$ & $2\times 10^{-4}$ & 0.002 & 0.000 \\
\addlinespace[2pt]
\multicolumn{9}{l}{$\beta_{1,\mathrm{cpue}}=1.00$ (linear)}\\
 & 1.00 & 0.00 & $1-2\times 10^{-36}$ & $1-2\times 10^{-37}$ & $6\times 10^{-18}$ & $2\times 10^{-20}$ & 0.947 & 0.000 \\
 & 1.00 & 0.20 & $1-2\times 10^{-36}$ & $1-2\times 10^{-27}$ & $3\times 10^{-13}$ & $3\times 10^{-14}$ & 0.785 & 0.000 \\
 & 1.00 & 0.50 & $1-2\times 10^{-36}$ & $1-9\times 10^{-20}$ & $2\times 10^{-10}$ & $1\times 10^{-9}$ & 0.326 & 0.000 \\
 & 1.00 & 1.00 & $1-2\times 10^{-36}$ & $1-6\times 10^{-14}$ & $8\times 10^{-8}$ & $3\times 10^{-6}$ & 0.033 & 0.000 \\
\addlinespace[2pt]
\multicolumn{9}{l}{$\beta_{1,\mathrm{cpue}}=1.25$ (hyperdepletion)}\\
 & 1.25 & 0.00 & $1-2\times 10^{-36}$ & $1-2\times 10^{-37}$ & $6\times 10^{-18}$ & $2\times 10^{-20}$ & 0.947 & 0.000 \\
 & 1.25 & 0.20 & $1-2\times 10^{-36}$ & $1-3\times 10^{-30}$ & $4\times 10^{-14}$ & $5\times 10^{-16}$ & 0.876 & 0.000 \\
 & 1.25 & 0.50 & $1-2\times 10^{-36}$ & $1-1\times 10^{-23}$ & $6\times 10^{-12}$ & $6\times 10^{-12}$ & 0.595 & 0.000 \\
 & 1.25 & 1.00 & $1-2\times 10^{-36}$ & $1-1\times 10^{-17}$ & $2\times 10^{-9}$ & $2\times 10^{-8}$ & 0.180 & 0.000 \\
\addlinespace[2pt]
\hline
\multicolumn{9}{l}{\textit{Case IV}}\\
\multicolumn{9}{l}{$\beta_{1,\mathrm{cpue}}=0.80$ (hyperstability)}\\
 & 0.80 & 0.00 & $1\times 10^{-68}$ & $1\times 10^{-70}$ & $1\times 10^{-35}$ & $6\times 10^{-46}$ & 0.944 & 0.000 \\
 & 0.80 & 0.20 & $1\times 10^{-68}$ & $2\times 10^{-43}$ & $4\times 10^{-24}$ & $9\times 10^{-28}$ & 0.262 & 0.000 \\
 & 0.80 & 0.50 & $1\times 10^{-68}$ & $9\times 10^{-28}$ & $5\times 10^{-16}$ & $3\times 10^{-17}$ & 0.005 & 0.000 \\
 & 0.80 & 1.00 & $1\times 10^{-68}$ & $1\times 10^{-17}$ & $8\times 10^{-11}$ & $2\times 10^{-10}$ & 0.000 & 0.000 \\
\addlinespace[2pt]
\multicolumn{9}{l}{$\beta_{1,\mathrm{cpue}}=1.00$ (linear)}\\
 & 1.00 & 0.00 & $1\times 10^{-68}$ & $1\times 10^{-70}$ & $1\times 10^{-35}$ & $6\times 10^{-46}$ & 0.944 & 0.000 \\
 & 1.00 & 0.20 & $1\times 10^{-68}$ & $2\times 10^{-50}$ & $8\times 10^{-28}$ & $2\times 10^{-32}$ & 0.531 & 0.000 \\
 & 1.00 & 0.50 & $1\times 10^{-68}$ & $2\times 10^{-35}$ & $6\times 10^{-20}$ & $2\times 10^{-22}$ & 0.056 & 0.000 \\
 & 1.00 & 1.00 & $1\times 10^{-68}$ & $8\times 10^{-24}$ & $5\times 10^{-14}$ & $1\times 10^{-14}$ & 0.001 & 0.000 \\
\addlinespace[2pt]
\multicolumn{9}{l}{$\beta_{1,\mathrm{cpue}}=1.25$ (hyperdepletion)}\\
 & 1.25 & 0.00 & $1\times 10^{-68}$ & $1\times 10^{-70}$ & $1\times 10^{-35}$ & $6\times 10^{-46}$ & 0.944 & 0.000 \\
 & 1.25 & 0.20 & $1\times 10^{-68}$ & $4\times 10^{-56}$ & $6\times 10^{-31}$ & $3\times 10^{-36}$ & 0.728 & 0.000 \\
 & 1.25 & 0.50 & $1\times 10^{-68}$ & $4\times 10^{-43}$ & $6\times 10^{-24}$ & $2\times 10^{-27}$ & 0.247 & 0.000 \\
 & 1.25 & 1.00 & $1\times 10^{-68}$ & $4\times 10^{-31}$ & $9\times 10^{-18}$ & $2\times 10^{-19}$ & 0.015 & 0.000 \\
\addlinespace[2pt]
\hline
\end{tabular}
\vspace{2pt}
\begin{minipage}{0.95\linewidth}
\footnotesize
\emph{Notes:} See Table~\ref{tab:mc_naive_cpue_q30_I12}.
\end{minipage}
\end{table}

\subsubsection{Implications for OEAR fitting}
\label{appendix:cpue_powerlaw_obs_oear}

Under the CPUE observation model \eqref{Seq:dwobs_obs_cpue} and the drifted Wiener dynamics \eqref{Seq:dwobs_sde},
the differenced series admits the decomposition
\begin{equation}
  \Delta Y_i
  =
  \beta_{1,\mathrm{cpue}}\,\mu\,\tau_i
  +
  \beta_{1,\mathrm{cpue}}\,\sigma\,\sqrt{\tau_i}\,\varepsilon_i
  +
  (E_i-E_{i-1}),
  \label{Seq:dwobs_cpue_dY_decomp}
\end{equation}
so CPUE nonlinearity rescales both the drift and diffusion components by $\beta_{1,\mathrm{cpue}}$, while the
differenced observation-error term is unchanged.

\medskip
\noindent
\textbf{Drift.}
For clarity in this CPUE-specific derivation, I temporarily write the OEAR drift and diffusion-scale quantities with an ``oear'' subscript; these correspond to the OEAR quantities denoted by tildes elsewhere in this appendix.
Because $\mathbb E[\Delta Y_i]=\beta_{1,\mathrm{cpue}}\,\mu\,\tau_i$ by \eqref{Seq:dwobs_E_dY_cpue}, the OEAR drift estimate (denoted here by $\widehat\mu_{\mathrm{oear}}$ for clarity) inherits the same rescaling:
\begin{equation}
  \mathbb E[\widehat\mu_{\mathrm{oear}}]
  =
  \beta_{1,\mathrm{cpue}}\,\mu,
  \label{Seq:dwobs_oear_mu_cpue}
\end{equation}
up to the same finite-sample effects as in the baseline OEAR setting.

\medskip
\noindent
\textbf{Effective diffusion scale via the LRV.}
For equal spacing $\tau_i\equiv\tau$, consider the OEAR standardized increments
\begin{equation}
  U_i
  :=
  \frac{\Delta Y_i-\widehat\mu_{\mathrm{oear}}\,\tau}{\sqrt{\tau}}.
  \label{Seq:dwobs_oear_Ui_cpue_def}
\end{equation}
At the expectation level, substituting $\widehat\mu_{\mathrm{oear}}\approx\beta_{1,\mathrm{cpue}}\mu$ into
\eqref{Seq:dwobs_cpue_dY_decomp} yields
\begin{equation}
  U_i
  \approx
  \beta_{1,\mathrm{cpue}}\,\sigma\,\varepsilon_i
  +
  \frac{E_i-E_{i-1}}{\sqrt{\tau}}.
  \label{Seq:dwobs_oear_Ui_cpue_decomp}
\end{equation}
Therefore,
\begin{equation}
  \mathrm{Var}(U_i)
  =
  \beta_{1,\mathrm{cpue}}^{2}\sigma^{2}
  +
  \frac{2\omega^{2}}{\tau},
  \qquad
  \mathrm{Cov}(U_i,U_{i+1})
  =
  -\,\frac{\omega^{2}}{\tau},
  \qquad
  \mathrm{Cov}(U_i,U_{i+k})=0\ (k\ge 2),
  \label{Seq:dwobs_oear_Ui_cpue_acov}
\end{equation}
so the long-run variance (LRV)
\begin{equation}
  \mathcal C_{\mathrm{cpue}}
  :=
  \sum_{j=-\infty}^{\infty}\mathrm{Cov}(U_i,U_{i-j})
  =
  \mathrm{Var}(U_i)+2\,\mathrm{Cov}(U_i,U_{i+1})
  =
  \beta_{1,\mathrm{cpue}}^{2}\sigma^{2}
  \label{Seq:dwobs_oear_LRV_cpue}
\end{equation}
is free of the observation-error inflation term.
Hence the OEAR diffusion-scale estimator targets the CPUE-rescaled process variance,
\begin{equation}
  \mathbb E[\widehat\sigma^{2}_{\mathrm{oear}}]
  \approx
  \beta_{1,\mathrm{cpue}}^{2}\sigma^{2},
  \label{Seq:dwobs_oear_sigma_cpue}
\end{equation}
with the same (HAC) estimation error as in the baseline OEAR setting.

\medskip
\noindent
\textbf{Implication for the $(w,z)$ operating point.}
Under the centered-intercept convention \eqref{Seq:dwobs_cpue_intercept_convention}, the plug-in distance satisfies
$\check{x}_d=Y_q-x_e=\beta_{1,\mathrm{cpue}}x_d+E_q$ as in \eqref{Seq:dwobs_cpue_xd_plugin_scaling}.
Using the OEAR plug-in map,
\begin{equation}
  \widehat w_{\mathrm{oear}}
  :=
  \frac{\widehat\mu_{\mathrm{oear}}\,t^\ast+\check{x}_d}{\widehat\sigma_{\mathrm{oear}}\sqrt{t^\ast}},
  \qquad
  \widehat z_{\mathrm{oear}}
  :=
  \frac{-\,\widehat\mu_{\mathrm{oear}}\,t^\ast+\check{x}_d}{\widehat\sigma_{\mathrm{oear}}\sqrt{t^\ast}},
  \label{Seq:dwobs_oear_wz_cpue_def}
\end{equation}
and substituting the corresponding expectation-level rescalings
$\widehat\mu_{\mathrm{oear}}\approx\beta_{1,\mathrm{cpue}}\mu$ and
$\widehat\sigma_{\mathrm{oear}}\approx\beta_{1,\mathrm{cpue}}\sigma$ from
\eqref{Seq:dwobs_oear_mu_cpue}--\eqref{Seq:dwobs_oear_sigma_cpue} suggests
\begin{equation*}
  \mathbb E[\widehat w_{\mathrm{oear}}]
  \approx
  \frac{\beta_{1,\mathrm{cpue}}(\mu t^\ast+x_d)}{\beta_{1,\mathrm{cpue}}\sigma\sqrt{t^\ast}}
  =
  w,
  \qquad
  \mathbb E[\widehat z_{\mathrm{oear}}]
  \approx
  \frac{\beta_{1,\mathrm{cpue}}(-\mu t^\ast+x_d)}{\beta_{1,\mathrm{cpue}}\sigma\sqrt{t^\ast}}
  =
  z.
  \label{Seq:dwobs_oear_wz_cpue_expect}
\end{equation*}
Thus, unlike naive fitting, CPUE slope nonlinearity does not induce a systematic radial-shrink effect in $(w,z)$ at the expectation level under OEAR.
Its main impact enters through the plug-in distance noise $E_q$, whose contribution is scaled by $1/\beta_{1,\mathrm{cpue}}$ in
\eqref{Seq:dwobs_oear_wz_cpue_def} when expressed in $w$--$z$ units.

\medskip
\noindent
\textbf{Monte Carlo settings.}
I use the same simulation design as in the naive CPUE sensitivity check above (data generated under \eqref{Seq:dwobs_obs_cpue} with the same centering convention and parameter grid).
The only change is the fitting step: for each replicate I apply OEAR fitting (still assuming the baseline observation model \eqref{Seq:dwobs_obs}) and summarize performance using the same metrics as before.

\begin{figure}[H]
\centering
\includegraphics[width=1\linewidth]{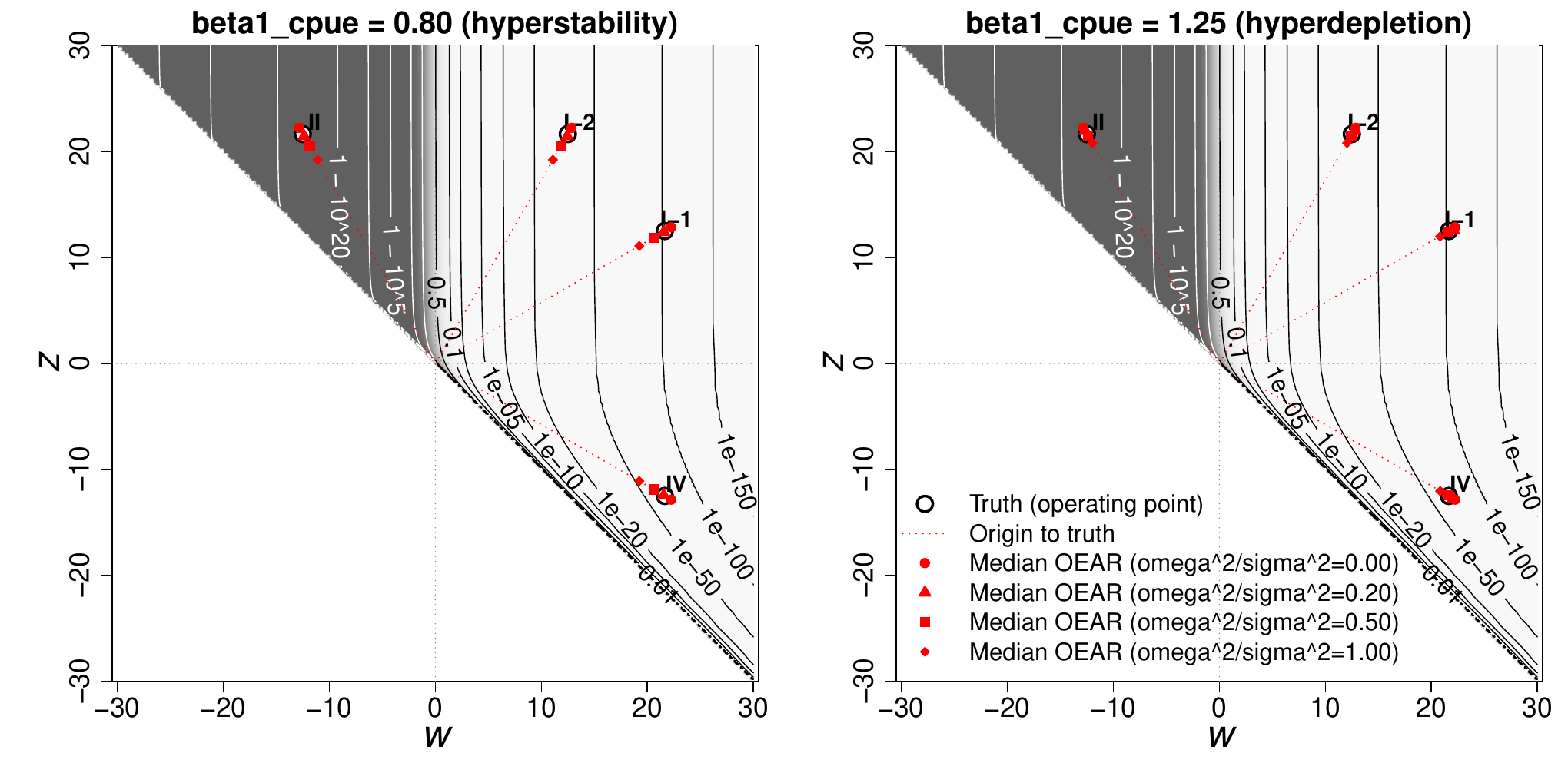}
\caption{CPUE nonlinearity under OEAR fitting ($q=60$): Monte Carlo medians of the OEAR plug-in operating points in the $(w,z)$ plane for $\beta_{1,\mathrm{cpue}}=0.80$ (hyperstability; left) and $\beta_{1,\mathrm{cpue}}=1.25$ (hyperdepletion; right).
Contours show the true extinction probability $G(w,z)$.
For each operating point, the open circles mark the truth, the dotted segments indicate the ray from the origin to the truth, and the red markers show the Monte Carlo median OEAR estimates at each tested observation-error level (indexed by $\omega^2/\sigma^2$).}
\label{fig:wz_oear_cpue_mc_overlay_q60_beta08_125}
\end{figure}

\medskip
\noindent
\textbf{Monte Carlo results (OEAR fitting under CPUE nonlinearity).}
Figure~\ref{fig:wz_oear_cpue_mc_overlay_q60_beta08_125} shows the Monte Carlo medians of the OEAR plug-in operating points in the $(w,z)$ plane for $q=60$ under CPUE nonlinearity, comparing hyperstability ($\beta_{1,\mathrm{cpue}}=0.80$) and hyperdepletion ($\beta_{1,\mathrm{cpue}}=1.25$).
Across operating points and observation-error levels, the OEAR medians remain close to the truth and do not exhibit the pronounced radial contraction seen under naive fitting.
A small but systematic distinction is nevertheless visible between the two CPUE slopes: under hyperstability ($\beta_{1,\mathrm{cpue}}<1$) the medians are shifted slightly toward the origin relative to hyperdepletion ($\beta_{1,\mathrm{cpue}}>1$), consistent with a mildly stronger effective contraction when the CPUE slope is below one.

Tables~\ref{tab:mc_oear_cpue_q30_I12}--\ref{tab:mc_oear_cpue_q60_IIIV} provide the corresponding numerical summary.
As $\omega^2/\sigma^2$ increases, CI widths increase and coverage declines, but the dominant structure is case dependence rather than CPUE slope:
coverage typically ranges from about $0.63$ to $0.84$ in Cases~I-1/I-2/II (with the lower end occurring at $\omega^2/\sigma^2=1$),
whereas it is lower in Case~IV, roughly $0.55$--$0.75$, for both $q=30$ and $q=60$.
Despite these shifts, the \texttt{Degraded} rate remains zero throughout.
Moreover, the fitted risks stay far from the policy threshold on the probability scale (close to $0$ in Cases~I-1/I-2/IV and close to $1$ in Case~II, reported as $1-Q$), and the associated CIs remain sufficiently narrow that they never straddle the decision boundary $G_{\mathrm{target}}=0.1$.

\begin{table}[H]
\centering
\footnotesize
\caption{OEAR fitting: Monte Carlo summary under CPUE nonlinearity ($q=30$, Cases I-1 and I-2).}
\label{tab:mc_oear_cpue_q30_I12}
\setlength{\tabcolsep}{4pt}
\begin{tabular}{lrr rrr rrr}
\hline
Case & $\beta_{1,\mathrm{cpue}}$ & $\omega^2/\sigma^2$ & True $G$ & MC estimate & MC SD & CI width & Coverage & Degraded \\
\hline
\multicolumn{9}{l}{\textit{Case I-1}}\\
\multicolumn{9}{l}{$\beta_{1,\mathrm{cpue}}=0.80$ (hyperstability)}\\
 & 0.80 & 0.00 & $8\times 10^{-104}$ & $1\times 10^{-116}$ & $1\times 10^{-14}$ & $8\times 10^{-56}$ & 0.787 & 0.000 \\
 & 0.80 & 0.20 & $8\times 10^{-104}$ & $3\times 10^{-107}$ & $4\times 10^{-19}$ & $9\times 10^{-51}$ & 0.783 & 0.000 \\
 & 0.80 & 0.50 & $8\times 10^{-104}$ & $8\times 10^{-95}$ & $2\times 10^{-19}$ & $3\times 10^{-44}$ & 0.727 & 0.000 \\
 & 0.80 & 1.00 & $8\times 10^{-104}$ & $2\times 10^{-79}$ & $4\times 10^{-17}$ & $4\times 10^{-36}$ & 0.643 & 0.000 \\
\addlinespace[2pt]
\multicolumn{9}{l}{$\beta_{1,\mathrm{cpue}}=1.00$ (linear)}\\
 & 1.00 & 0.00 & $8\times 10^{-104}$ & $1\times 10^{-116}$ & $1\times 10^{-14}$ & $8\times 10^{-56}$ & 0.787 & 0.000 \\
 & 1.00 & 0.20 & $8\times 10^{-104}$ & $1\times 10^{-110}$ & $3\times 10^{-18}$ & $1\times 10^{-52}$ & 0.788 & 0.000 \\
 & 1.00 & 0.50 & $8\times 10^{-104}$ & $2\times 10^{-101}$ & $1\times 10^{-18}$ & $1\times 10^{-47}$ & 0.762 & 0.000 \\
 & 1.00 & 1.00 & $8\times 10^{-104}$ & $2\times 10^{-89}$ & $1\times 10^{-18}$ & $2\times 10^{-41}$ & 0.708 & 0.000 \\
\addlinespace[2pt]
\multicolumn{9}{l}{$\beta_{1,\mathrm{cpue}}=1.25$ (hyperdepletion)}\\
 & 1.25 & 0.00 & $8\times 10^{-104}$ & $1\times 10^{-116}$ & $1\times 10^{-14}$ & $8\times 10^{-56}$ & 0.787 & 0.000 \\
 & 1.25 & 0.20 & $8\times 10^{-104}$ & $3\times 10^{-113}$ & $4\times 10^{-19}$ & $6\times 10^{-54}$ & 0.792 & 0.000 \\
 & 1.25 & 0.50 & $8\times 10^{-104}$ & $8\times 10^{-107}$ & $4\times 10^{-19}$ & $1\times 10^{-50}$ & 0.782 & 0.000 \\
 & 1.25 & 1.00 & $8\times 10^{-104}$ & $9\times 10^{-98}$ & $4\times 10^{-19}$ & $8\times 10^{-46}$ & 0.744 & 0.000 \\
\addlinespace[2pt]
\hline
\multicolumn{9}{l}{\textit{Case I-2}}\\
\multicolumn{9}{l}{$\beta_{1,\mathrm{cpue}}=0.80$ (hyperstability)}\\
 & 0.80 & 0.00 & $6\times 10^{-36}$ & $2\times 10^{-40}$ & $4\times 10^{-5}$ & $3\times 10^{-16}$ & 0.835 & 0.000 \\
 & 0.80 & 0.20 & $6\times 10^{-36}$ & $5\times 10^{-37}$ & $2\times 10^{-6}$ & $2\times 10^{-14}$ & 0.836 & 0.000 \\
 & 0.80 & 0.50 & $6\times 10^{-36}$ & $8\times 10^{-33}$ & $3\times 10^{-6}$ & $2\times 10^{-12}$ & 0.813 & 0.000 \\
 & 0.80 & 1.00 & $6\times 10^{-36}$ & $9\times 10^{-28}$ & $8\times 10^{-6}$ & $5\times 10^{-10}$ & 0.753 & 0.000 \\
\addlinespace[2pt]
\multicolumn{9}{l}{$\beta_{1,\mathrm{cpue}}=1.00$ (linear)}\\
 & 1.00 & 0.00 & $6\times 10^{-36}$ & $2\times 10^{-40}$ & $4\times 10^{-5}$ & $3\times 10^{-16}$ & 0.835 & 0.000 \\
 & 1.00 & 0.20 & $6\times 10^{-36}$ & $3\times 10^{-38}$ & $3\times 10^{-6}$ & $4\times 10^{-15}$ & 0.839 & 0.000 \\
 & 1.00 & 0.50 & $6\times 10^{-36}$ & $3\times 10^{-35}$ & $3\times 10^{-6}$ & $1\times 10^{-13}$ & 0.830 & 0.000 \\
 & 1.00 & 1.00 & $6\times 10^{-36}$ & $4\times 10^{-31}$ & $4\times 10^{-6}$ & $1\times 10^{-11}$ & 0.795 & 0.000 \\
\addlinespace[2pt]
\multicolumn{9}{l}{$\beta_{1,\mathrm{cpue}}=1.25$ (hyperdepletion)}\\
 & 1.25 & 0.00 & $6\times 10^{-36}$ & $2\times 10^{-40}$ & $4\times 10^{-5}$ & $3\times 10^{-16}$ & 0.835 & 0.000 \\
 & 1.25 & 0.20 & $6\times 10^{-36}$ & $5\times 10^{-39}$ & $2\times 10^{-6}$ & $2\times 10^{-15}$ & 0.841 & 0.000 \\
 & 1.25 & 0.50 & $6\times 10^{-36}$ & $6\times 10^{-37}$ & $2\times 10^{-6}$ & $2\times 10^{-14}$ & 0.836 & 0.000 \\
 & 1.25 & 1.00 & $6\times 10^{-36}$ & $8\times 10^{-34}$ & $3\times 10^{-6}$ & $6\times 10^{-13}$ & 0.823 & 0.000 \\
\addlinespace[2pt]
\hline
\end{tabular}
\vspace{2pt}
\begin{minipage}{0.95\linewidth}
\footnotesize
\emph{Notes:} $\omega^2/\sigma^2$ is the observation-error to process-variance ratio (here $\tau=1$). True $G$ is the true value of the extinction probability $G$. ``MC estimate'' is the Monte Carlo median of the fitted extinction probability, and ``MC SD'' is the replicate SD. CI width and coverage are computed from the $w$--$z$ method. Degraded is defined in \eqref{eq:degraded_def}.
\end{minipage}
\end{table}

\begin{table}[H]
\centering
\footnotesize
\caption{OEAR fitting: Monte Carlo summary under CPUE nonlinearity ($q=30$, Cases II and IV).}
\label{tab:mc_oear_cpue_q30_IIIV}
\setlength{\tabcolsep}{4pt}
\begin{tabular}{lrr rrr rrr}
\hline
Case & $\beta_{1,\mathrm{cpue}}$ & $\omega^2/\sigma^2$ & True $G$ & MC estimate & MC SD & CI width & Coverage & Degraded \\
\hline
\multicolumn{9}{l}{\textit{Case II}}\\
\multicolumn{9}{l}{$\beta_{1,\mathrm{cpue}}=0.80$ (hyperstability)}\\
 & 0.80 & 0.00 & $1-2\times 10^{-36}$ & $1-9\times 10^{-41}$ & $6\times 10^{-9}$ & $2\times 10^{-16}$ & 0.823 & 0.000 \\
 & 0.80 & 0.20 & $1-2\times 10^{-36}$ & $1-2\times 10^{-37}$ & $9\times 10^{-9}$ & $7\times 10^{-15}$ & 0.838 & 0.000 \\
 & 0.80 & 0.50 & $1-2\times 10^{-36}$ & $1-3\times 10^{-33}$ & $5\times 10^{-8}$ & $8\times 10^{-13}$ & 0.809 & 0.000 \\
 & 0.80 & 1.00 & $1-2\times 10^{-36}$ & $1-2\times 10^{-28}$ & $8\times 10^{-7}$ & $2\times 10^{-10}$ & 0.754 & 0.000 \\
\addlinespace[2pt]
\multicolumn{9}{l}{$\beta_{1,\mathrm{cpue}}=1.00$ (linear)}\\
 & 1.00 & 0.00 & $1-2\times 10^{-36}$ & $1-9\times 10^{-41}$ & $6\times 10^{-9}$ & $2\times 10^{-16}$ & 0.823 & 0.000 \\
 & 1.00 & 0.20 & $1-2\times 10^{-36}$ & $1-9\times 10^{-39}$ & $5\times 10^{-9}$ & $2\times 10^{-15}$ & 0.838 & 0.000 \\
 & 1.00 & 0.50 & $1-2\times 10^{-36}$ & $1-7\times 10^{-36}$ & $2\times 10^{-8}$ & $4\times 10^{-14}$ & 0.829 & 0.000 \\
 & 1.00 & 1.00 & $1-2\times 10^{-36}$ & $1-1\times 10^{-31}$ & $1\times 10^{-7}$ & $5\times 10^{-12}$ & 0.795 & 0.000 \\
\addlinespace[2pt]
\multicolumn{9}{l}{$\beta_{1,\mathrm{cpue}}=1.25$ (hyperdepletion)}\\
 & 1.25 & 0.00 & $1-2\times 10^{-36}$ & $1-9\times 10^{-41}$ & $6\times 10^{-9}$ & $2\times 10^{-16}$ & 0.823 & 0.000 \\
 & 1.25 & 0.20 & $1-2\times 10^{-36}$ & $1-1\times 10^{-39}$ & $6\times 10^{-9}$ & $6\times 10^{-16}$ & 0.839 & 0.000 \\
 & 1.25 & 0.50 & $1-2\times 10^{-36}$ & $1-2\times 10^{-37}$ & $1\times 10^{-8}$ & $8\times 10^{-15}$ & 0.838 & 0.000 \\
 & 1.25 & 1.00 & $1-2\times 10^{-36}$ & $1-2\times 10^{-34}$ & $3\times 10^{-8}$ & $2\times 10^{-13}$ & 0.822 & 0.000 \\
\addlinespace[2pt]
\hline
\multicolumn{9}{l}{\textit{Case IV}}\\
\multicolumn{9}{l}{$\beta_{1,\mathrm{cpue}}=0.80$ (hyperstability)}\\
 & 0.80 & 0.00 & $1\times 10^{-68}$ & $3\times 10^{-77}$ & $8\times 10^{-12}$ & $2\times 10^{-40}$ & 0.736 & 0.000 \\
 & 0.80 & 0.20 & $1\times 10^{-68}$ & $1\times 10^{-70}$ & $5\times 10^{-15}$ & $1\times 10^{-36}$ & 0.729 & 0.000 \\
 & 0.80 & 0.50 & $1\times 10^{-68}$ & $2\times 10^{-62}$ & $3\times 10^{-15}$ & $4\times 10^{-32}$ & 0.669 & 0.000 \\
 & 0.80 & 1.00 & $1\times 10^{-68}$ & $2\times 10^{-52}$ & $1\times 10^{-12}$ & $9\times 10^{-27}$ & 0.571 & 0.000 \\
\addlinespace[2pt]
\multicolumn{9}{l}{$\beta_{1,\mathrm{cpue}}=1.00$ (linear)}\\
 & 1.00 & 0.00 & $1\times 10^{-68}$ & $3\times 10^{-77}$ & $8\times 10^{-12}$ & $2\times 10^{-40}$ & 0.736 & 0.000 \\
 & 1.00 & 0.20 & $1\times 10^{-68}$ & $6\times 10^{-73}$ & $2\times 10^{-14}$ & $7\times 10^{-38}$ & 0.743 & 0.000 \\
 & 1.00 & 0.50 & $1\times 10^{-68}$ & $5\times 10^{-67}$ & $1\times 10^{-14}$ & $9\times 10^{-35}$ & 0.704 & 0.000 \\
 & 1.00 & 1.00 & $1\times 10^{-68}$ & $4\times 10^{-59}$ & $1\times 10^{-14}$ & $2\times 10^{-30}$ & 0.640 & 0.000 \\
\addlinespace[2pt]
\multicolumn{9}{l}{$\beta_{1,\mathrm{cpue}}=1.25$ (hyperdepletion)}\\
 & 1.25 & 0.00 & $1\times 10^{-68}$ & $3\times 10^{-77}$ & $8\times 10^{-12}$ & $2\times 10^{-40}$ & 0.736 & 0.000 \\
 & 1.25 & 0.20 & $1\times 10^{-68}$ & $1\times 10^{-74}$ & $5\times 10^{-15}$ & $6\times 10^{-39}$ & 0.747 & 0.000 \\
 & 1.25 & 0.50 & $1\times 10^{-68}$ & $2\times 10^{-70}$ & $4\times 10^{-15}$ & $2\times 10^{-36}$ & 0.728 & 0.000 \\
 & 1.25 & 1.00 & $1\times 10^{-68}$ & $2\times 10^{-64}$ & $4\times 10^{-15}$ & $2\times 10^{-33}$ & 0.685 & 0.000 \\
\addlinespace[2pt]
\hline
\end{tabular}
\vspace{2pt}
\begin{minipage}{0.95\linewidth}
\footnotesize
\emph{Notes:} See Table~\ref{tab:mc_oear_cpue_q30_I12}.
\end{minipage}
\end{table}

\begin{table}[H]
\centering
\footnotesize
\caption{OEAR fitting: Monte Carlo summary under CPUE nonlinearity ($q=60$, Cases I-1 and I-2).}
\label{tab:mc_oear_cpue_q60_I12}
\setlength{\tabcolsep}{4pt}
\begin{tabular}{lrr rrr rrr}
\hline
Case & $\beta_{1,\mathrm{cpue}}$ & $\omega^2/\sigma^2$ & True $G$ & MC estimate & MC SD & CI width & Coverage & Degraded \\
\hline
\multicolumn{9}{l}{\textit{Case I-1}}\\
\multicolumn{9}{l}{$\beta_{1,\mathrm{cpue}}=0.80$ (hyperstability)}\\
 & 0.80 & 0.00 & $8\times 10^{-104}$ & $2\times 10^{-109}$ & $6\times 10^{-40}$ & $3\times 10^{-67}$ & 0.794 & 0.000 \\
 & 0.80 & 0.20 & $8\times 10^{-104}$ & $7\times 10^{-103}$ & $3\times 10^{-36}$ & $7\times 10^{-63}$ & 0.764 & 0.000 \\
 & 0.80 & 0.50 & $8\times 10^{-104}$ & $4\times 10^{-94}$ & $7\times 10^{-29}$ & $4\times 10^{-57}$ & 0.708 & 0.000 \\
 & 0.80 & 1.00 & $8\times 10^{-104}$ & $2\times 10^{-82}$ & $7\times 10^{-27}$ & $2\times 10^{-49}$ & 0.631 & 0.000 \\
\addlinespace[2pt]
\multicolumn{9}{l}{$\beta_{1,\mathrm{cpue}}=1.00$ (linear)}\\
 & 1.00 & 0.00 & $8\times 10^{-104}$ & $2\times 10^{-109}$ & $6\times 10^{-40}$ & $3\times 10^{-67}$ & 0.794 & 0.000 \\
 & 1.00 & 0.20 & $8\times 10^{-104}$ & $1\times 10^{-105}$ & $6\times 10^{-38}$ & $1\times 10^{-64}$ & 0.775 & 0.000 \\
 & 1.00 & 0.50 & $8\times 10^{-104}$ & $9\times 10^{-99}$ & $2\times 10^{-32}$ & $4\times 10^{-60}$ & 0.744 & 0.000 \\
 & 1.00 & 1.00 & $8\times 10^{-104}$ & $3\times 10^{-90}$ & $2\times 10^{-26}$ & $1\times 10^{-54}$ & 0.694 & 0.000 \\
\addlinespace[2pt]
\multicolumn{9}{l}{$\beta_{1,\mathrm{cpue}}=1.25$ (hyperdepletion)}\\
 & 1.25 & 0.00 & $8\times 10^{-104}$ & $2\times 10^{-109}$ & $6\times 10^{-40}$ & $3\times 10^{-67}$ & 0.794 & 0.000 \\
 & 1.25 & 0.20 & $8\times 10^{-104}$ & $1\times 10^{-107}$ & $2\times 10^{-38}$ & $5\times 10^{-66}$ & 0.780 & 0.000 \\
 & 1.25 & 0.50 & $8\times 10^{-104}$ & $1\times 10^{-102}$ & $4\times 10^{-36}$ & $1\times 10^{-62}$ & 0.765 & 0.000 \\
 & 1.25 & 1.00 & $8\times 10^{-104}$ & $3\times 10^{-96}$ & $1\times 10^{-30}$ & $2\times 10^{-58}$ & 0.726 & 0.000 \\
\addlinespace[2pt]
\hline
\multicolumn{9}{l}{\textit{Case I-2}}\\
\multicolumn{9}{l}{$\beta_{1,\mathrm{cpue}}=0.80$ (hyperstability)}\\
 & 0.80 & 0.00 & $6\times 10^{-36}$ & $7\times 10^{-38}$ & $3\times 10^{-13}$ & $1\times 10^{-20}$ & 0.839 & 0.000 \\
 & 0.80 & 0.20 & $6\times 10^{-36}$ & $1\times 10^{-35}$ & $6\times 10^{-12}$ & $2\times 10^{-19}$ & 0.824 & 0.000 \\
 & 0.80 & 0.50 & $6\times 10^{-36}$ & $1\times 10^{-32}$ & $6\times 10^{-10}$ & $2\times 10^{-17}$ & 0.799 & 0.000 \\
 & 0.80 & 1.00 & $6\times 10^{-36}$ & $1\times 10^{-28}$ & $1\times 10^{-9}$ & $6\times 10^{-15}$ & 0.735 & 0.000 \\
\addlinespace[2pt]
\multicolumn{9}{l}{$\beta_{1,\mathrm{cpue}}=1.00$ (linear)}\\
 & 1.00 & 0.00 & $6\times 10^{-36}$ & $7\times 10^{-38}$ & $3\times 10^{-13}$ & $1\times 10^{-20}$ & 0.839 & 0.000 \\
 & 1.00 & 0.20 & $6\times 10^{-36}$ & $1\times 10^{-36}$ & $2\times 10^{-12}$ & $6\times 10^{-20}$ & 0.828 & 0.000 \\
 & 1.00 & 0.50 & $6\times 10^{-36}$ & $2\times 10^{-34}$ & $6\times 10^{-11}$ & $2\times 10^{-18}$ & 0.814 & 0.000 \\
 & 1.00 & 1.00 & $6\times 10^{-36}$ & $2\times 10^{-31}$ & $2\times 10^{-9}$ & $1\times 10^{-16}$ & 0.784 & 0.000 \\
\addlinespace[2pt]
\multicolumn{9}{l}{$\beta_{1,\mathrm{cpue}}=1.25$ (hyperdepletion)}\\
 & 1.25 & 0.00 & $6\times 10^{-36}$ & $7\times 10^{-38}$ & $3\times 10^{-13}$ & $1\times 10^{-20}$ & 0.839 & 0.000 \\
 & 1.25 & 0.20 & $6\times 10^{-36}$ & $2\times 10^{-37}$ & $1\times 10^{-12}$ & $2\times 10^{-20}$ & 0.835 & 0.000 \\
 & 1.25 & 0.50 & $6\times 10^{-36}$ & $1\times 10^{-35}$ & $7\times 10^{-12}$ & $3\times 10^{-19}$ & 0.823 & 0.000 \\
 & 1.25 & 1.00 & $6\times 10^{-36}$ & $1\times 10^{-33}$ & $2\times 10^{-10}$ & $5\times 10^{-18}$ & 0.805 & 0.000 \\
\addlinespace[2pt]
\hline
\end{tabular}
\vspace{2pt}
\begin{minipage}{0.95\linewidth}
\footnotesize
\emph{Notes:} See Table~\ref{tab:mc_oear_cpue_q30_I12}.
\end{minipage}
\end{table}

\begin{table}[H]
\centering
\footnotesize
\caption{OEAR fitting: Monte Carlo summary under CPUE nonlinearity ($q=60$, Cases II and IV).}
\label{tab:mc_oear_cpue_q60_IIIV}
\setlength{\tabcolsep}{4pt}
\begin{tabular}{lrr rrr rrr}
\hline
Case & $\beta_{1,\mathrm{cpue}}$ & $\omega^2/\sigma^2$ & True $G$ & MC estimate & MC SD & CI width & Coverage & Degraded \\
\hline
\multicolumn{9}{l}{\textit{Case II}}\\
\multicolumn{9}{l}{$\beta_{1,\mathrm{cpue}}=0.80$ (hyperstability)}\\
 & 0.80 & 0.00 & $1-2\times 10^{-36}$ & $1-2\times 10^{-38}$ & $2\times 10^{-13}$ & $3\times 10^{-21}$ & 0.838 & 0.000 \\
 & 0.80 & 0.20 & $1-2\times 10^{-36}$ & $1-7\times 10^{-36}$ & $2\times 10^{-12}$ & $1\times 10^{-19}$ & 0.834 & 0.000 \\
 & 0.80 & 0.50 & $1-2\times 10^{-36}$ & $1-6\times 10^{-33}$ & $1\times 10^{-10}$ & $1\times 10^{-17}$ & 0.805 & 0.000 \\
 & 0.80 & 1.00 & $1-2\times 10^{-36}$ & $1-3\times 10^{-29}$ & $7\times 10^{-10}$ & $2\times 10^{-15}$ & 0.737 & 0.000 \\
\addlinespace[2pt]
\multicolumn{9}{l}{$\beta_{1,\mathrm{cpue}}=1.00$ (linear)}\\
 & 1.00 & 0.00 & $1-2\times 10^{-36}$ & $1-2\times 10^{-38}$ & $2\times 10^{-13}$ & $3\times 10^{-21}$ & 0.838 & 0.000 \\
 & 1.00 & 0.20 & $1-2\times 10^{-36}$ & $1-8\times 10^{-37}$ & $6\times 10^{-13}$ & $4\times 10^{-20}$ & 0.842 & 0.000 \\
 & 1.00 & 0.50 & $1-2\times 10^{-36}$ & $1-1\times 10^{-34}$ & $1\times 10^{-11}$ & $9\times 10^{-19}$ & 0.817 & 0.000 \\
 & 1.00 & 1.00 & $1-2\times 10^{-36}$ & $1-6\times 10^{-32}$ & $3\times 10^{-11}$ & $4\times 10^{-17}$ & 0.790 & 0.000 \\
\addlinespace[2pt]
\multicolumn{9}{l}{$\beta_{1,\mathrm{cpue}}=1.25$ (hyperdepletion)}\\
 & 1.25 & 0.00 & $1-2\times 10^{-36}$ & $1-2\times 10^{-38}$ & $2\times 10^{-13}$ & $3\times 10^{-21}$ & 0.838 & 0.000 \\
 & 1.25 & 0.20 & $1-2\times 10^{-36}$ & $1-3\times 10^{-37}$ & $2\times 10^{-13}$ & $2\times 10^{-20}$ & 0.842 & 0.000 \\
 & 1.25 & 0.50 & $1-2\times 10^{-36}$ & $1-8\times 10^{-36}$ & $2\times 10^{-12}$ & $2\times 10^{-19}$ & 0.833 & 0.000 \\
 & 1.25 & 1.00 & $1-2\times 10^{-36}$ & $1-1\times 10^{-33}$ & $4\times 10^{-11}$ & $3\times 10^{-18}$ & 0.815 & 0.000 \\
\addlinespace[2pt]
\hline
\multicolumn{9}{l}{\textit{Case IV}}\\
\multicolumn{9}{l}{$\beta_{1,\mathrm{cpue}}=0.80$ (hyperstability)}\\
 & 0.80 & 0.00 & $1\times 10^{-68}$ & $2\times 10^{-72}$ & $4\times 10^{-27}$ & $4\times 10^{-47}$ & 0.747 & 0.000 \\
 & 0.80 & 0.20 & $1\times 10^{-68}$ & $1\times 10^{-67}$ & $1\times 10^{-26}$ & $5\times 10^{-44}$ & 0.710 & 0.000 \\
 & 0.80 & 0.50 & $1\times 10^{-68}$ & $3\times 10^{-62}$ & $3\times 10^{-21}$ & $2\times 10^{-40}$ & 0.651 & 0.000 \\
 & 0.80 & 1.00 & $1\times 10^{-68}$ & $1\times 10^{-54}$ & $4\times 10^{-19}$ & $4\times 10^{-35}$ & 0.552 & 0.000 \\
\addlinespace[2pt]
\multicolumn{9}{l}{$\beta_{1,\mathrm{cpue}}=1.00$ (linear)}\\
 & 1.00 & 0.00 & $1\times 10^{-68}$ & $2\times 10^{-72}$ & $4\times 10^{-27}$ & $4\times 10^{-47}$ & 0.747 & 0.000 \\
 & 1.00 & 0.20 & $1\times 10^{-68}$ & $1\times 10^{-69}$ & $8\times 10^{-28}$ & $3\times 10^{-45}$ & 0.725 & 0.000 \\
 & 1.00 & 0.50 & $1\times 10^{-68}$ & $3\times 10^{-65}$ & $8\times 10^{-24}$ & $2\times 10^{-42}$ & 0.681 & 0.000 \\
 & 1.00 & 1.00 & $1\times 10^{-68}$ & $1\times 10^{-59}$ & $2\times 10^{-19}$ & $1\times 10^{-38}$ & 0.624 & 0.000 \\
\addlinespace[2pt]
\multicolumn{9}{l}{$\beta_{1,\mathrm{cpue}}=1.25$ (hyperdepletion)}\\
 & 1.25 & 0.00 & $1\times 10^{-68}$ & $2\times 10^{-72}$ & $4\times 10^{-27}$ & $4\times 10^{-47}$ & 0.747 & 0.000 \\
 & 1.25 & 0.20 & $1\times 10^{-68}$ & $6\times 10^{-71}$ & $5\times 10^{-28}$ & $3\times 10^{-46}$ & 0.733 & 0.000 \\
 & 1.25 & 0.50 & $1\times 10^{-68}$ & $2\times 10^{-67}$ & $2\times 10^{-26}$ & $7\times 10^{-44}$ & 0.708 & 0.000 \\
 & 1.25 & 1.00 & $1\times 10^{-68}$ & $9\times 10^{-64}$ & $1\times 10^{-22}$ & $2\times 10^{-41}$ & 0.662 & 0.000 \\
\addlinespace[2pt]
\hline
\end{tabular}
\vspace{2pt}
\begin{minipage}{0.95\linewidth}
\footnotesize
\emph{Notes:} See Table~\ref{tab:mc_oear_cpue_q30_I12}.
\end{minipage}
\end{table}

\subsubsection{Summary}

I conducted Monte Carlo sensitivity checks to assess how observation error affects Criterion~E PVA conclusions under three fitting strategies: naive fitting, OEAR, and a linear-Gaussian state-space model (SSM).
Under additive observation error, naive plug-in operating points shift in the $(w,z)$ plane in a predictable radial-shrink pattern, while OEAR and the SSM largely correct the resulting median bias by accounting for variance inflation (via LRV estimation and Kalman filtering, respectively).
As observation error increases, all methods exhibit wider CIs and somewhat reduced coverage, with the clearest deterioration under naive fitting at high $\omega^2/\sigma^2$.
However, across all tested operating points and error levels, the estimated extinction risks remain far from the policy threshold and the \texttt{Degraded} rate stays essentially zero, so the resulting PVA classification is never altered.
A further CPUE power-law sensitivity check shows that CPUE nonlinearity mainly rescales the same shrinkage geometry (strengthening contraction under hyperstability and weakening it under hyperdepletion) without changing the qualitative conclusions.
Overall, within the scenarios explored here, observation error (and moderate CPUE nonlinearity) does not alter the Criterion~E PVA outcome.

\section{Extinction probability under relaxed process models: sensitivity of the \texorpdfstring{$w$--$z$}{w--z} method}
\label{appendix:env_noise_density}
\setcounter{equation}{0}
\setcounter{figure}{0}
\setcounter{table}{0}

To assess the performance of the $w$--$z$ method under relaxed process models such as colored environmental noise and weak density feedback, I apply both the naive plug-in estimator and the OEAR estimator, using the same drift--Wiener-based operating-point design and the same CI construction as in Appendix~\ref{appendix:obs_error}.
Because finite-sample performance depends on time-series length and the dependence structure of the generating process, neither estimator is uniformly superior.

The analysis requires reference probabilities $G(t^\ast)$ for the data-generating models.
Here $G(t^\ast)=\Pr[T\le t^\ast]$, where $T=\inf\{u\ge 0:X(u)\le x_e\}$ is the first-passage time of the log-population process $X(t)$ to the extinction threshold $x_e$.
Many plausible relaxations of the drift--Wiener baseline can be considered, and obtaining accurate reference values of $G(t^\ast)$ is often challenging.
In the small-risk regime targeted here, direct simulation of extinction events produces no extinctions at feasible replicate sizes and thus reports zero probability; rare-event Monte Carlo would require prohibitively many replicates to observe such tail events with usable precision and is therefore infeasible in practice, even though simulating short time series remains straightforward for Monte Carlo checks of estimators and CIs.

I therefore focus on models for which $G(t^\ast)$ can be computed accurately by analysis combined with stable numerical evaluation, avoiding rare-event simulation in the tail.
Two departures from the drift--Wiener baseline are considered: (i) colored environmental noise (AR(1)-type in discrete time) represented as Ornstein--Uhlenbeck (OU) forcing, and (ii) a weak density-feedback bridge model that adds linear restoring drift to the drift--Wiener baseline.
Both choices are minimal continuous-time extensions that preserve tractability for reference calculations: OU forcing introduces a correlated drift component, while the bridge model introduces mean reversion in $X(t)$.
For the OU-forced model, reference values of $G_{\mathrm{OU}}(t^\ast)$ are obtained by solving the associated two-dimensional backward equation, which remains stable across the correlation regimes considered below.
For the density-feedback bridge model, reference values $G_{\mathrm{Br}}(t^\ast)$ are obtained by solving the corresponding one-dimensional backward equation for the mean-reverting diffusion with an absorbing boundary, using a truncated spatial domain and stability checks analogous to those used in the OU-forcing reference construction.

The drift--Wiener model serves as the baseline throughout this appendix.
As a baseline, the drift--Wiener model assumes
\begin{equation}
  dX(t)=\mu\,dt+\sigma\,dW(t),
  \qquad
  X(0)=x_d,
  \qquad
  T=\inf\{t\ge 0:X(t)\le 0\},
  \label{Seq:DW_base_AR1memo}
\end{equation}
where $X(t)$ denotes the log population size, $W$ is a standard Wiener process, and the convention $x_e=0$ is used for notational simplicity.
In this baseline case, the extinction probability $G(t\mid x_d,\mu,\sigma^2)=\Pr[T\le t]$ has the well-known closed form (see the main text):
\begin{equation}
  G(t\mid x_d,\mu,\sigma^2)
  =
  \Phi\!\left(-\frac{\mu t+x_d}{\sigma\sqrt{t}}\right)
  +
  \exp\!\left(-\frac{2\mu x_d}{\sigma^2}\right)
  \Phi\!\left(-\frac{-\mu t+x_d}{\sigma\sqrt{t}}\right),
  \qquad t>0.
  \label{Seq:DW_closedform_AR1memo}
\end{equation}

\subsection{Colored environmental noise as OU forcing: reference \texorpdfstring{$G_{\mathrm{OU}}(t^\ast)$}{G-OU(tstar)} and estimator performance}
\label{appendix:ar1_env_noise}

Colored environmental variability is represented by adding an Ornstein--Uhlenbeck (OU) forcing term to the drift of the log population size process.
Under the OU-forced generating model, the reference finite-horizon extinction probability is $G_{\mathrm{OU}}(t^\ast):=\Pr[T\le t^\ast]$.
These reference values provide a benchmark for time-series Monte Carlo comparisons of the naive plug-in and OEAR estimators.

Colored environmental forcing is represented by an OU process $\mathcal E(t)$, which captures time-correlated environmental variation in the instantaneous log growth rate (i.e., a random drift component):
\begin{equation}
  d\mathcal E(t)=-\gamma \mathcal E(t)\,dt+\sigma_{\mathcal E}\,dW_1(t),
  \qquad \gamma>0,
  \label{Seq:OU_forcing_AR1memo}
\end{equation}
and the forced log process
\begin{equation}
  dX(t)=\bigl\{\mu+\beta_{\mathcal E}\,\mathcal E(t)\bigr\}\,dt+\sigma\,dW_2(t),
  \label{Seq:forced_X_AR1memo}
\end{equation}
with independent Wiener processes $W_1$ and $W_2$.
Sampling \eqref{Seq:OU_forcing_AR1memo} at equally spaced times with interval $\tau>0$ yields the discrete-time AR(1) representation for $\mathcal E_i=\mathcal E(i\tau)$,
\begin{equation}
  \mathcal E_i=\psi\,\mathcal E_{i-1}+\nu_i,
  \qquad
  \psi=\exp(-\gamma\tau),
  \qquad
  \nu_i\overset{\mathrm{i.i.d.}}{\sim}\mathcal N\!\left(0,\;\frac{\sigma_{\mathcal E}^2}{2\gamma}\bigl(1-\psi^2\bigr)\right).
  \label{Seq:AR1_correspondence_AR1memo}
\end{equation}

The characteristic OU decorrelation timescale is $1/\gamma$.
Note that the AR(1) representation \eqref{Seq:AR1_correspondence_AR1memo} applies to the sampled forcing $\mathcal E_i$, whereas the growth increments $\Delta X_i:=X(i\tau)-X((i-1)\tau)$ combine the integrated forcing over each interval with independent Wiener noise.
Setting $\beta_{\mathcal E}=0$ or $\sigma_{\mathcal E}=0$ removes the forcing term and recovers the drift--Wiener baseline.
Reference values of $G_{\mathrm{OU}}(t^\ast)$ are obtained by stable numerical solution of the associated two-dimensional backward equation across the correlation regimes considered below.

\subsubsection{2D backward-Kolmogorov equation reference for \texorpdfstring{$G_{\mathrm{OU}}(t^\ast)$}{G-OU(tstar)}}
\label{appendix:ou_forcing_2d_pde}

This subsection constructs numerical reference values of $G_{\mathrm{OU}}(t^\ast)$ for the OU-forced model by solving the associated two-dimensional (2D) backward Kolmogorov equation for the Markov state $(X(t),\mathcal E(t))$.
The equation is a partial differential equation (PDE) in the state variables $(x,e)$ and the horizon argument $t$.

\medskip
\noindent\textbf{Model and target probability.}
Consider the OU-forced model \eqref{Seq:OU_forcing_AR1memo} and \eqref{Seq:forced_X_AR1memo} with extinction threshold $x_e=0$.
For $t>0$, define the conditional finite-horizon extinction probability
\begin{equation*}
  G_{\mathrm{OU}}(t,x,e)
  :=
  \Pr\!\left(T\le t \,\middle|\, X(0)=x,\ \mathcal E(0)=e\right),
  \label{Seq:Gou_def_2d}
\end{equation*}
where $T=\inf\{t\ge 0:X(t)\le 0\}$.
The reference probability reported for the OU-forced generating model is the stationary-initial average
\begin{equation*}
  G_{\mathrm{OU}}(t^\ast)
  :=
  \Pr(T\le t^\ast)
  =
  \mathbb E_{\mathcal E(0)}\!\left[\,G_{\mathrm{OU}}\!\left(t^\ast,x_d,\mathcal E(0)\right)\right],
  \label{Seq:Gou_as_average}
\end{equation*}
where $\mathcal E(0)$ is drawn from the stationary distribution of the OU forcing and $X(0)$ is fixed at $x_d$.

\medskip
\noindent\textbf{Scaled variables and SDE used in the implementation.}
The numerical solver uses the scaled variables
\begin{equation*}
  x_{\mathrm{sc}}=\frac{X}{\sigma},\qquad
  \mu_{\mathrm{sc}}=\frac{\mu}{\sigma},\qquad
  e_{\mathrm{sc}}=\frac{\mathcal E}{\sigma},\qquad
  \sigma_e=\frac{\sigma_{\mathcal E}}{\sigma},
  \label{Seq:scaled_vars}
\end{equation*}
so that the diffusion coefficient in the $x_{\mathrm{sc}}$ equation equals $1$.
With these definitions, the OU forcing and the forced log process become
\begin{equation*}
  de_{\mathrm{sc}}(t)
  =
  -\gamma e_{\mathrm{sc}}(t)\,dt
  +\sigma_e\,dW_1(t),
  \label{Seq:OU_forcing_scaled}
\end{equation*}
and
\begin{equation*}
  dx_{\mathrm{sc}}(t)
  =
  \bigl\{\mu_{\mathrm{sc}}+\beta_{\mathcal E}e_{\mathrm{sc}}(t)\bigr\}\,dt
  +dW_2(t),
  \label{Seq:forced_X_scaled}
\end{equation*}
with independent Wiener processes $W_1$ and $W_2$.
The stationary distribution of $e_{\mathrm{sc}}$ is normal with mean $0$ and variance $\sigma_e^2/(2\gamma)$.

\medskip
\noindent\textbf{Backward equation.}
For $x_{\mathrm{sc}}>0$ and $|e_{\mathrm{sc}}|<e_{\max}$, the conditional finite-horizon extinction probability
$G_{\mathrm{OU}}(t,x_{\mathrm{sc}},e_{\mathrm{sc}})$ satisfies
\begin{equation}
  \frac{\partial G_{\mathrm{OU}}}{\partial t}
  =
  \bigl(\mu_{\mathrm{sc}}+\beta_{\mathcal E} e_{\mathrm{sc}}\bigr)\,
  \frac{\partial G_{\mathrm{OU}}}{\partial x_{\mathrm{sc}}}
  -\gamma e_{\mathrm{sc}}\,
  \frac{\partial G_{\mathrm{OU}}}{\partial e_{\mathrm{sc}}}
  +\frac{1}{2}\,\frac{\partial^2 G_{\mathrm{OU}}}{\partial x_{\mathrm{sc}}^2}
  +\frac{\sigma_e^2}{2}\,\frac{\partial^2 G_{\mathrm{OU}}}{\partial e_{\mathrm{sc}}^2},
  \qquad t>0.
  \label{Seq:backward_2d_pde}
\end{equation}
with absorbing boundary at the threshold,
\begin{equation*}
  G_{\mathrm{OU}}(t,0,e_{\mathrm{sc}})=1,
  \qquad t\ge 0,
\end{equation*}
and far-field truncation,
\begin{equation*}
  G_{\mathrm{OU}}(t,x_{\max},e_{\mathrm{sc}})=0.
\end{equation*}
The OU state is truncated by reflecting boundaries,
\begin{equation*}
  \left.\frac{\partial G_{\mathrm{OU}}}{\partial e_{\mathrm{sc}}}(t,x_{\mathrm{sc}},e_{\mathrm{sc}})\right|_{e_{\mathrm{sc}}=\pm e_{\max}} = 0.
\end{equation*}

\medskip
\noindent\textbf{Initialization.}
The exact initial condition at $t=0$ is
$G_{\mathrm{OU}}(0,x_{\mathrm{sc}},e_{\mathrm{sc}})=\mathbf 1\{x_{\mathrm{sc}}\le 0\}$,
which is discontinuous in $x_{\mathrm{sc}}$.
To avoid numerical artifacts, I start the computation at a small $t=\varepsilon$ and set
\begin{equation*}
  G_{\mathrm{OU}}(\varepsilon,x_{\mathrm{sc}},e_{\mathrm{sc}})
  =
  G(\varepsilon \mid x_{\mathrm{sc}},\mu_{\mathrm{sc}},1),
  \label{Seq:init_eps_dw}
\end{equation*}
where the right-hand side is the drift--Wiener closed form \eqref{Seq:DW_closedform_AR1memo}
evaluated in scaled units (diffusion coefficient equal to $1$), i.e., the $\beta_{\mathcal E}=0$
(or $\sigma_{\mathcal E}=0$) plug-in limit used for initialization at the first time slice.
The reference values reported below are insensitive to the particular choice of $\varepsilon$ provided it is taken sufficiently small.

\medskip
\noindent\textbf{Numerical solution and stability checks.}
Equation \eqref{Seq:backward_2d_pde} is solved in \textit{Mathematica} by method-of-lines time stepping with FEM discretization in $(x_{\mathrm{sc}},e_{\mathrm{sc}})$ on the truncated domain $[0,x_{\max}]\times[-e_{\max},e_{\max}]$.
The benchmark value $G_{\mathrm{OU}}(t^\ast)$ is obtained by evaluating the FEM solution at $t^\ast$ and averaging over the stationary distribution of $e_{\mathrm{sc}}(0)$ using a fixed-grid quadrature on $[-e_{\max},e_{\max}]$ (i.e., the stationary normal density restricted to this interval and renormalized), which is stable in the small-risk regime.
Numerical stability is assessed by varying the mesh resolution and the truncation parameters $(x_{\max},e_{\max})$, and by checking that the FEM solution reproduces the 1D drift--Wiener closed form when $\beta_{\mathcal E}=0$.

\subsubsection{Reference calculations under OU forcing}
\label{sec:ref_ou_forcing}

Using the 2D backward-equation construction in Section~\ref{appendix:ou_forcing_2d_pde}, I compute reference values of $G_{\mathrm{OU}}(t^\ast)$ under OU forcing at the Case~I-2 operating point in Table~\ref{tab:obs_sim_points}, which defines the parameter setting used throughout the Monte Carlo comparisons below.
Throughout, I fix $x_d=12$, $t^\ast=100$, and $\tau=1$, and I set
$(\mu,\sigma)=(-0.03215227,\allowbreak 0.07027818)$ with $\beta_{\mathcal E}=0.20$.
To vary persistence, I consider two OU mean-reversion rates, $\gamma\in\{0.8,\,2.0\}$; equivalently, persistence may be indexed by the one-step correlation $\psi=\exp(-\gamma\tau)$ with $\tau=1$.

Forcing strength is indexed by the dimensionless stationary-variance ratio
\begin{equation*}
  \mathrm{Ratio}
  :=
  \frac{\beta_{\mathcal E}^{\,2}\,\mathrm{Var}\{\mathcal E(t)\}}{\sigma^2}
  =
  \frac{\beta_{\mathcal E}^{\,2}}{\sigma^2}\cdot\frac{\sigma_{\mathcal E}^2}{2\gamma},
  \label{Seq:ou_strength_ratio_mid}
\end{equation*}
and I consider four levels $\mathrm{Ratio}\in\{0,0.5,1,2\}$.
Given $(\gamma,\beta_{\mathcal E},\sigma)$, this implies
\begin{equation}
  \sigma_{\mathcal E}^2
  =
  \frac{2\gamma\sigma^2}{\beta_{\mathcal E}^{\,2}}\;\mathrm{Ratio}.
  \label{Seq:sigmaE_from_ratio_mid}
\end{equation}

\begin{table}[H]
\centering
\caption{Reference extinction probabilities $G_{\mathrm{OU}}(t^\ast)$ under OU forcing at the Case~I-2 operating point (2D backward-equation FEM).}
\label{tab:ou_ref_param_sets_I2}
\setlength{\tabcolsep}{6pt}
\renewcommand{\arraystretch}{1.10}
\footnotesize
\begin{tabular}{r r r r r r}
\hline
$\mathrm{Ratio}$ & $\gamma$ & $\beta_{\mathcal E}$ & $\sigma_{\mathcal E}^2$ & $\psi$ & $G_{\mathrm{OU}}(t^\ast)$ \\
\hline
\multicolumn{6}{l}{Higher persistence ($\gamma=0.8$, $\psi\approx 0.449$)}\\
$0$   & $0.8$ & $0.20$ & $0$        & $0.449$ & $5.90\times10^{-36}$ \\
$0.5$ & $0.8$ & $0.20$ & $0.09878$  & $0.449$ & $4.05\times10^{-17}$ \\
$1$   & $0.8$ & $0.20$ & $0.19756$  & $0.449$ & $1.21\times10^{-11}$ \\
$2$   & $0.8$ & $0.20$ & $0.39512$  & $0.449$ & $1.78\times10^{-7}$  \\
\hline
\addlinespace[2pt]
\multicolumn{6}{l}{Lower persistence ($\gamma=2.0$, $\psi\approx 0.135$)}\\
$0$   & $2.0$ & $0.20$ & $0$        & $0.135$ & $5.90\times10^{-36}$ \\
$0.5$ & $2.0$ & $0.20$ & $0.24695$  & $0.135$ & $1.24\times10^{-24}$ \\
$1$   & $2.0$ & $0.20$ & $0.49390$  & $0.135$ & $6.02\times10^{-19}$ \\
$2$   & $2.0$ & $0.20$ & $0.98780$  & $0.135$ & $3.23\times10^{-13}$ \\
\hline
\end{tabular}

\smallskip
\footnotesize
\parbox{0.92\linewidth}{\emph{Note:} $x_d=12$, $t^\ast=100$, and $\tau=1$ are fixed throughout, with $(\mu,\sigma)=(-0.03215227,\allowbreak 0.07027818)$ and $\beta_{\mathcal E}=0.20$.
The values of $\sigma_{\mathcal E}^2$ are computed from \eqref{Seq:sigmaE_from_ratio_mid}, and $\psi=\exp(-\gamma\tau)$ (reported to three decimals).}
\end{table}

\medskip
\noindent\textbf{Colored noise persistence and finite-horizon extinction risk.}
Table~\ref{tab:ou_ref_param_sets_I2} reports the OU-forced reference probability $G_{\mathrm{OU}}(t^\ast)$ at the Case~I-2 operating point.
Within each persistence setting, $G_{\mathrm{OU}}(t^\ast)$ increases with \textit{Ratio}, so \textit{Ratio} serves as a coherent forcing-strength index for ordering finite-horizon extinction risk in the parameter range considered here.

Holding \textit{Ratio} fixed, the less persistent regime ($\gamma=2.0$) yields smaller values of $G_{\mathrm{OU}}(t^\ast)$ than the more persistent regime ($\gamma=0.8$).
This is consistent with the mechanism that persistence in the drift component creates longer runs of unfavorable growth, which increases the chance of hitting the absorbing boundary within the finite horizon even when the stationary variance (as indexed by \textit{Ratio}) is held fixed.

\subsubsection{Monte Carlo evaluation under OU forcing}
\label{sec:mc_ou_forcing}

I evaluate drift--Wiener inference under process misspecification by Monte Carlo experiments in which the data-generating model includes OU environmental forcing as defined above, while inference still fits the drift--Wiener model.
To parallel Appendix~\ref{appendix:obs_error}, I compare two drift--Wiener fitting strategies applied to the same error-free simulated series: (i) the naive plug-in fit and (ii) the OEAR fit, across the forcing-strength levels summarized in Table~\ref{tab:ou_ref_param_sets_I2}.

\medskip
\noindent\textbf{Simulation design.}
I fix the operating point to the Case~I-2 design setting used for the reference calculations in Section~\ref{sec:ref_ou_forcing}.
Specifically, I set $x_d=12$, $t^\ast=100$, and $\tau=1$, and I use $(\mu,\sigma)=(-0.03215227,\,0.07027818)$ with $\beta_{\mathcal E}=0.20$.
I consider the same forcing configurations as in Table~\ref{tab:ou_ref_param_sets_I2}, spanning four forcing-strength levels $\mathrm{Ratio}\in\{0,0.5,1,2\}$ and two persistence settings $\gamma\in\{0.8,\,2\}$ (equivalently $\psi=\exp(-\gamma\tau)$).
For each configuration, I generate $n_{\mathrm{MC}}=50{,}000$ independent replicates of length $q\in\{30,60\}$.

For each replicate, I draw $\mathcal E_0$ from the stationary distribution of the OU forcing and generate the sampled forcing path $\{\mathcal E_i\}_{i=1}^q$ using the exact OU transition \eqref{Seq:AR1_correspondence_AR1memo}.
To match the continuous-time forcing over each sampling interval, I work with the integrated forcing
\begin{equation*}
  I_i
  :=
  \int_{t_{i-1}}^{t_i}\mathcal E(u)\,du,
  \qquad i=1,\ldots,q,
  \label{Seq:mc_ou_I_def}
\end{equation*}
which is generated by an exact Gaussian discretization.
Conditional on $I_i$, I simulate the forced-process increments by
\begin{equation*}
  \Delta X_i
  =
  \mu\tau
  +
  \beta_{\mathcal E} I_i
  +
  \sigma\sqrt{\tau}\,\xi_i,
  \qquad
  \xi_i\overset{\mathrm{i.i.d.}}{\sim}\mathcal N(0,1),
  \qquad i=1,\ldots,q,
  \label{Seq:mc_ou_increment_exactI}
\end{equation*}
and construct the raw path $X^{\mathrm{raw}}_0=0$, $X^{\mathrm{raw}}_i=\sum_{j=1}^i \Delta X_j$.
As in Appendix~\ref{appendix:obs_error}, I shift each path so that the terminal state equals the operating-point distance,
\begin{equation*}
  X_i = X^{\mathrm{raw}}_i - X^{\mathrm{raw}}_q + x_d,
  \qquad i=0,\ldots,q,
  \label{Seq:mc_ou_shift_to_xd}
\end{equation*}
thereby anchoring the projection origin at $X_q=x_d$ in every replicate.

\medskip
\noindent\textbf{Fitting and confidence intervals.}
For each replicate series $\{X_i\}_{i=0}^q$, I fit the drift--Wiener model by the naive plug-in approach and by OEAR exactly as in Appendix~\ref{appendix:obs_error}.
Each method yields $(w_\bullet,z_\bullet)$ at horizon $t^\ast$ and the corresponding plug-in estimate $G(w_\bullet,z_\bullet)$ via the drift--Wiener closed form \eqref{Seq:DW_closedform_AR1memo}.
I construct the nominal 95\% CI for $G$ using the same $w$--$z$ procedure as in Appendix~\ref{appendix:obs_error}.

\medskip
\noindent\textbf{Performance metrics.}
For each configuration, the reference finite-horizon extinction probability under the OU-forced generating model, $G_{\mathrm{OU}}(t^\ast)$, is taken from Table~\ref{tab:ou_ref_param_sets_I2}.
Monte Carlo summaries report the median plug-in estimate $G(w_\bullet,z_\bullet)$, the Monte Carlo standard deviation on the probability scale, the median nominal 95\% CI width, and empirical coverage (the fraction of replicates whose $w$--$z$ CI contains $G_{\mathrm{OU}}(t^\ast)$).

\medskip\noindent\textbf{Monte Carlo patterns under OU forcing (naive and OEAR drift--Wiener fits).}
Table~\ref{tab:mc_ou_forcing_naive_vs_oear_compact} compares naive and OEAR drift--Wiener fits when the generating process includes OU forcing.
To keep the focus on drift--Wiener operating-point summaries under process relaxation, the OU reference ordering in \textit{Ratio} is taken as given from Table~\ref{tab:ou_ref_param_sets_I2}; the Monte Carlo results assess how the two drift--Wiener fits track that reference scale in point and interval summaries.

On the probability scale, both plug-in medians are smaller than the OU reference once \textit{Ratio} is positive.
This is consistent with serial dependence in the drift component modifying finite-horizon tail behavior relative to the drift--Wiener baseline with the same $(\mu,\sigma)$.
Within this shared pattern, the OEAR medians are generally closer to $G_{\mathrm{OU}}(t^\ast)$ than the naive medians, most clearly at larger $q$, indicating that OEAR more effectively adjusts the operating-point summary when forcing contributes non-negligibly.

Empirical coverage is computed for the nominal 95\% $w$--$z$ CI against $G_{\mathrm{OU}}(t^\ast)$.
At $\mathrm{Ratio}=0$, the naive fit attains essentially nominal coverage (e.g., $\mathrm{Cov.}\approx 0.95$ for both $\gamma$ and both $q$), whereas the corresponding OEAR fit yields coverage near $0.83$.
As forcing increases, coverage varies with $(q,\gamma)$, reflecting how the two drift--Wiener fits calibrate the same $w$--$z$ CI under OU relaxation.
For example, at $(q,\gamma)=(60,0.8)$ the naive coverage decreases from $0.642$ to $0.478$ to $0.423$ as $\mathrm{Ratio}$ increases from $0.5$ to $1$ to $2$, whereas the corresponding OEAR coverages are $0.724$, $0.690$, and $0.709$, remaining broadly stable over this range.

Across the intermediate colored-noise regimes examined here, the OU reference probabilities remain far below the decision threshold $G=0.1$ used in Appendix~\ref{appendix:obs_error}, so both naive and OEAR drift--Wiener-based $w$--$z$ summaries lead to the same threshold-based conclusion in this range.

\begin{table}[H]
\caption{Monte Carlo summary under OU forcing (naive vs OEAR).}
\label{tab:mc_ou_forcing_naive_vs_oear_compact}
\centering
\footnotesize
\setlength{\tabcolsep}{3.2pt}
\renewcommand{\arraystretch}{1.05}
\resizebox{\linewidth}{!}{\begin{tabular}{r r r r r r r r r r}
\hline
$q$ & $\gamma$ & Ratio & True $G_{\mathrm{OU}}$ & Naive median & Naive CI width & Naive Cov. & OEAR median & OEAR CI width & OEAR Cov. \\
\hline
30 & 0.8 & 0 & $6\times10^{-36}$ & $8\times10^{-38}$ & $6\times10^{-15}$ & 0.949 & $1\times10^{-40}$ & $3\times10^{-16}$ & 0.828 \\
30 & 0.8 & 0.5 & $4\times10^{-17}$ & $1\times10^{-29}$ & $6\times10^{-11}$ & 0.792 & $5\times10^{-27}$ & $1\times10^{-9}$ & 0.766 \\
30 & 0.8 & 1 & $1\times10^{-11}$ & $2\times10^{-24}$ & $2\times10^{-8}$ & 0.690 & $4\times10^{-20}$ & $2\times10^{-6}$ & 0.750 \\
30 & 0.8 & 2 & $2\times10^{-7}$ & $3\times10^{-18}$ & $1\times10^{-5}$ & 0.632 & $3\times10^{-13}$ & $1\times10^{-3}$ & 0.763 \\
30 & 2 & 0 & $6\times10^{-36}$ & $1\times10^{-37}$ & $7\times10^{-15}$ & 0.950 & $1\times10^{-40}$ & $2\times10^{-16}$ & 0.829 \\
30 & 2 & 0.5 & $1\times10^{-24}$ & $1\times10^{-29}$ & $6\times10^{-11}$ & 0.926 & $5\times10^{-27}$ & $1\times10^{-9}$ & 0.843 \\
30 & 2 & 1 & $6\times10^{-19}$ & $2\times10^{-24}$ & $2\times10^{-8}$ & 0.906 & $5\times10^{-20}$ & $2\times10^{-6}$ & 0.855 \\
30 & 2 & 2 & $3\times10^{-13}$ & $3\times10^{-18}$ & $1\times10^{-5}$ & 0.876 & $2\times10^{-13}$ & $1\times10^{-3}$ & 0.873 \\
\hline
60 & 0.8 & 0 & $6\times10^{-36}$ & $7\times10^{-37}$ & $4\times10^{-20}$ & 0.950 & $4\times10^{-38}$ & $7\times10^{-21}$ & 0.837 \\
60 & 0.8 & 0.5 & $4\times10^{-17}$ & $1\times10^{-28}$ & $6\times10^{-15}$ & 0.642 & $6\times10^{-25}$ & $1\times10^{-12}$ & 0.724 \\
60 & 0.8 & 1 & $1\times10^{-11}$ & $1\times10^{-23}$ & $7\times10^{-12}$ & 0.478 & $1\times10^{-18}$ & $8\times10^{-9}$ & 0.690 \\
60 & 0.8 & 2 & $2\times10^{-7}$ & $1\times10^{-17}$ & $3\times10^{-8}$ & 0.423 & $3\times10^{-12}$ & $3\times10^{-5}$ & 0.709 \\
60 & 2 & 0 & $6\times10^{-36}$ & $8\times10^{-37}$ & $5\times10^{-20}$ & 0.949 & $4\times10^{-38}$ & $7\times10^{-21}$ & 0.838 \\
60 & 2 & 0.5 & $1\times10^{-24}$ & $7\times10^{-29}$ & $4\times10^{-15}$ & 0.915 & $3\times10^{-25}$ & $7\times10^{-13}$ & 0.848 \\
60 & 2 & 1 & $6\times10^{-19}$ & $9\times10^{-24}$ & $6\times10^{-12}$ & 0.879 & $7\times10^{-19}$ & $5\times10^{-9}$ & 0.859 \\
60 & 2 & 2 & $3\times10^{-13}$ & $1\times10^{-17}$ & $3\times10^{-8}$ & 0.841 & $1\times10^{-12}$ & $2\times10^{-5}$ & 0.875 \\
\hline
\end{tabular}}

\smallskip
\footnotesize
\parbox{0.97\linewidth}{\emph{Note:} The design fixes $x_d=12$, $t^\ast=100$, and $\tau=1$, with $(\mu,\sigma)=(-0.03215227, 0.07027818)$ and $\beta_{\mathcal E}=0.20$. Each configuration uses $n_{\mathrm{MC}}=50{,}000$ replicates. ``CI width'' is the median width of the nominal 95\% $w$--$z$ CI, and ``Cov.'' is its empirical coverage (nominal 0.95).}
\end{table}

\subsection{Weak density feedback bridge model: reference \texorpdfstring{$G_{\mathrm{Br}}(t^\ast)$}{G-Br(tstar)} and estimator performance}
\label{appendix:weak_density_model}

To examine the effect of weak density feedback on drift--Wiener inference, I consider a linear mean-reverting extension of the log process,
\begin{equation}
  dX(t)=\Bigl\{\mu+\lambda_{\mathrm{Br}}\bigl(\theta-X(t)\bigr)\Bigr\}\,dt+\sigma\,dW(t),
  \qquad \lambda_{\mathrm{Br}}\ge 0,
  \label{Seq:bridge_SDE_memo2}
\end{equation}
where $\theta$ is a reference log level and $\lambda_{\mathrm{Br}}$ controls the strength of density feedback.
When $\lambda_{\mathrm{Br}}=0$, \eqref{Seq:bridge_SDE_memo2} reduces to the drift--Wiener baseline \eqref{Seq:DW_base_AR1memo}.
For $\lambda_{\mathrm{Br}}>0$, the drift is state dependent; weak density feedback corresponds to small $\lambda_{\mathrm{Br}}$ (equivalently $\beta_1\to 1$ in the discrete-time AR(1) representation below).

\medskip
\noindent\textbf{Discrete time (Gompertz AR(1)) representation.}
For equally spaced observations with interval $\tau>0$, the exact transition of \eqref{Seq:bridge_SDE_memo2} implies the AR(1) form
\begin{equation*}
  X_{i}= \beta_0+\beta_1 X_{i-1}+\xi_i,
  \qquad \xi_i \overset{\mathrm{i.i.d.}}{\sim}\mathcal N(0,\sigma_\xi^2),
  \label{Seq:AR1_bridge_memo2}
\end{equation*}
with
\begin{equation*}
  \beta_1=\exp(-\lambda_{\mathrm{Br}}\tau),
  \qquad
  \beta_0=(1-\beta_1)\,x_{\infty},
  \qquad
  \sigma_\xi^2=\sigma^2\frac{1-\exp(-2\lambda_{\mathrm{Br}}\tau)}{2\lambda_{\mathrm{Br}}},
  \label{Seq:AR1_params_memo2}
\end{equation*}
where the mean-reversion level is
\begin{equation*}
  x_{\infty}=\theta+\frac{\mu}{\lambda_{\mathrm{Br}}}.
  \label{Seq:x_infty_memo2}
\end{equation*}
For unequal sampling intervals, replace $\tau$ by $\tau_i=t_i-t_{i-1}$ to obtain
$\beta_{1,i}=\exp(-\lambda_{\mathrm{Br}}\tau_i)$ and
$\sigma_{\xi,i}^2=\sigma^2\{1-\exp(-2\lambda_{\mathrm{Br}}\tau_i)\}/(2\lambda_{\mathrm{Br}})$.

\subsubsection{Reference calculations under weak density feedback bridge}
\label{sec:ref_bridge}

In this subsection, I compute reference values of $G_{\mathrm{Br}}(t^\ast)$ by a one-dimensional backward equation for $X(t)$, keeping the numerical workflow parallel to the OU-forcing reference construction while avoiding rare-event simulation in the tail regime.

\medskip
\noindent\textbf{Model and target probability.}
Assume the convention $x_e=0$ and fix $X(0)=x_d>0$.
For $t>0$, define the conditional finite-horizon extinction probability
\begin{equation*}
  G_{\mathrm{Br}}(t,x)
  :=
  \Pr\!\left(T\le t \,\middle|\, X(0)=x\right),
  \qquad
  T=\inf\{t\ge 0:X(t)\le 0\}.
  \label{Seq:GBr_def}
\end{equation*}
The reference probability reported for the bridge model is
\begin{equation}
  G_{\mathrm{Br}}(t^\ast):=G_{\mathrm{Br}}(t^\ast,x_d).
  \label{Seq:GBr_ref_def}
\end{equation}

\medskip
\noindent\textbf{One-dimensional backward equation.}
For $x>0$, $G_{\mathrm{Br}}(t,x)$ satisfies the backward Kolmogorov equation associated with \eqref{Seq:bridge_SDE_memo2},
\begin{equation}
  \frac{\partial G_{\mathrm{Br}}}{\partial t}
  =
  \Bigl\{\mu+\lambda_{\mathrm{Br}}(\theta-x)\Bigr\}\,\frac{\partial G_{\mathrm{Br}}}{\partial x}
  +\frac{\sigma^{2}}{2}\,\frac{\partial^{2}G_{\mathrm{Br}}}{\partial x^{2}},
  \qquad t>0,
  \label{Seq:bridge_backward_1d}
\end{equation}
with absorbing boundary and initial condition
\begin{equation*}
  G_{\mathrm{Br}}(t,0)=1,
  \qquad t\ge 0,
  \qquad
  G_{\mathrm{Br}}(0,x)=0\ \ (x>0).
  \label{Seq:bridge_bc_ic}
\end{equation*}

\medskip
\noindent\textbf{Numerical solution and stability checks.}
I solve \eqref{Seq:bridge_backward_1d} on a truncated interval $[0,x_{\max}]$ with far-field truncation
\begin{equation*}
  G_{\mathrm{Br}}(t,x_{\max})=0,
  \qquad t\ge 0,
  \label{Seq:bridge_farfield}
\end{equation*}
and report $G_{\mathrm{Br}}(t^\ast)$ via \eqref{Seq:GBr_ref_def}.
Numerical stability is assessed by varying both the truncation point $x_{\max}$ and the spatial discretization, and by checking that the solution reproduces the drift--Wiener closed form \eqref{Seq:DW_closedform_AR1memo} when $\lambda_{\mathrm{Br}}=0$.

Using the 1D backward-equation construction described above, I compute reference values of $G_{\mathrm{Br}}(t^\ast)$ under the weak density-feedback bridge at the Case~I-2 operating point.
Throughout, I fix $x_d=12$, $t^\ast=100$, and $\tau=1$, and I use the same baseline parameters $(\mu,\sigma)=(-0.03215227,\,0.07027818)$ as in Section~\ref{sec:ref_ou_forcing}.
Density feedback is controlled by $\lambda_{\mathrm{Br}}$, and I consider a small set of weak-feedback levels $\lambda_{\mathrm{Br}}\in\{0.0005,0.001,0.0015,0.002\}$ so that the discrete-time persistence $\beta_1=\exp(-\lambda_{\mathrm{Br}}\tau)$ remains close to $1$.
For concreteness, I set the reference level to $\theta=x_d(=12)$, so that the feedback term switches sign at the operating-point distance.

\begin{table}[H]
\centering
\caption{Reference extinction probabilities $G_{\mathrm{Br}}(t^\ast)$ under weak density feedback at the Case~I-2 operating point (1D backward equation).}
\label{tab:bridge_ref_param_sets_I2}
\setlength{\tabcolsep}{6pt}
\renewcommand{\arraystretch}{1.10}
\footnotesize
\begin{tabular}{r r r r r}
\hline
$\lambda_{\mathrm{Br}}$ & $\beta_1=\exp(-\lambda_{\mathrm{Br}}\tau)$ & $\theta$ & $x_{\infty}$ & $G_{\mathrm{Br}}(t^\ast)$ \\
\hline
$0$       & $1$      & $12$ & -- & $5.90\times10^{-36}$ \\
$0.0005$  & $0.9995$ & $12$ & $-52.3045$ & $2.53\times10^{-38}$ \\
$0.001$  & $0.9990$ & $12$ & $-20.1523$ & $8.58\times10^{-41}$ \\
$0.0015$  & $0.9985$ & $12$ & $-9.4348$ & $2.29\times10^{-43}$ \\
$0.002$  & $0.9980$ & $12$ & $-4.0761$ & $4.80\times10^{-46}$ \\
\hline
\end{tabular}

\smallskip
\footnotesize
\parbox{0.92\linewidth}{\emph{Note:} The design fixes $x_d=12$, $t^\ast=100$, and $\tau=1$ throughout, with $(\mu,\sigma)=(-0.03215227,\,0.07027818)$ (Case~I-2 baseline).
For each $\lambda_{\mathrm{Br}}>0$, $\beta_1=\exp(-\lambda_{\mathrm{Br}}\tau)$ is the corresponding discrete-time persistence and $x_{\infty}=\theta+\mu/\lambda_{\mathrm{Br}}$ is the mean-reversion level  (reported to four decimals).
The baseline row $\lambda_{\mathrm{Br}}=0$ corresponds to the drift--Wiener limit.}
\end{table}

\paragraph{Weak density feedback reduces finite-horizon extinction risk.}
The state-dependent drift $\mu+\lambda_{\mathrm{Br}}(\theta-x)$ makes downward excursions self-limiting: whenever $x<\theta$, the feedback term is positive and partially offsets the negative baseline drift $\mu$, slowing motion toward the absorbing boundary at $0$.
At $x=\theta(=12)$ the drift equals $\mu$, whereas at the boundary it becomes $\mu+\lambda_{\mathrm{Br}}\theta$ (e.g., from $-0.0322$ at $\lambda_{\mathrm{Br}}=0$ to about $-0.0082$ at $\lambda_{\mathrm{Br}}=0.002$).
Because $G_{\mathrm{Br}}(t^\ast)$ is a tail probability, this small drift change propagates on an exponential scale over the finite horizon, which is reflected in the monotone decrease of $G_{\mathrm{Br}}(t^\ast)$ with $\lambda_{\mathrm{Br}}$ in Table~\ref{tab:bridge_ref_param_sets_I2}.

\subsubsection{Monte Carlo evaluation under weak density feedback bridge}
\label{sec:mc_bridge}

I evaluate drift--Wiener inference under weak density feedback by Monte Carlo experiments in which the data-generating model follows \eqref{Seq:bridge_SDE_memo2}, while inference continues to fit the drift--Wiener model.
Except for the data-generation step described below, the fitting, $w$--$z$ operating-point construction, and nominal 95\% CI procedure are identical to Appendix~\ref{appendix:obs_error} and Section~\ref{sec:mc_ou_forcing}.

\medskip
\noindent\textbf{Simulation design and exact discretization.}
I fix $x_d=12$, $t^\ast=100$, and $\tau=1$, use $(\mu,\sigma)=(-0.03215227,\,0.07027818)$ and $\theta=x_d$, and consider $\lambda_{\mathrm{Br}}\in\{0,0.0005,0.001,0.0015,0.002\}$.
For each configuration, I generate $n_{\mathrm{MC}}=50{,}000$ independent replicates of length $q\in\{30,60\}$.

For $\lambda_{\mathrm{Br}}>0$, letting $x_\infty=\theta+\mu/\lambda_{\mathrm{Br}}$, the exact one-step transition implies
\begin{equation}
  X_i^{\mathrm{raw}}
  =
  x_\infty
  +
  \exp(-\lambda_{\mathrm{Br}}\tau)\,\bigl(X_{i-1}^{\mathrm{raw}}-x_\infty\bigr)
  +
  \sigma\sqrt{\frac{1-\exp(-2\lambda_{\mathrm{Br}}\tau)}{2\lambda_{\mathrm{Br}}}}\;\xi_i,
  \qquad
  \xi_i\overset{\mathrm{i.i.d.}}{\sim}\mathcal N(0,1),
  \label{Seq:mc_bridge_exact_transition}
\end{equation}
with $X_0^{\mathrm{raw}}=0$.
For $\lambda_{\mathrm{Br}}=0$, I use the drift--Wiener increment form
\begin{equation*}
  X_i^{\mathrm{raw}} = X_{i-1}^{\mathrm{raw}} + \mu\tau + \sigma\sqrt{\tau}\,\xi_i,
  \qquad i=1,\ldots,q,
  \label{Seq:mc_bridge_lambda0}
\end{equation*}
which matches \eqref{Seq:mc_bridge_exact_transition} in the $\lambda_{\mathrm{Br}}\to 0$ limit.
As in Appendix~\ref{appendix:obs_error}, each replicate is shifted to enforce the operating-point anchoring $X_q=x_d$,
\begin{equation*}
  X_i = X_i^{\mathrm{raw}} - X_q^{\mathrm{raw}} + x_d,
  \qquad i=0,\ldots,q.
  \label{Seq:mc_bridge_shift_to_xd}
\end{equation*}

\medskip
\noindent\textbf{Performance under weak density feedback.}
Table~\ref{tab:mc_bridge_naive_vs_oear_compact} summarizes drift--Wiener inference when the generating process follows the weak density-feedback bridge \eqref{Seq:bridge_SDE_memo2} while inference continues to fit the drift--Wiener model.
Across $\lambda_{\mathrm{Br}}\in\{0,0.0005,0.001$, $0.0015,0.002\}$ and $q\in\{30,60\}$, both procedures produce coherent plug-in summaries as feedback strengthens: the reference $G_{\mathrm{Br}}(t^\ast)$ decreases with $\lambda_{\mathrm{Br}}$, and the drift--Wiener plug-in medians decrease in the same direction.

For $\lambda_{\mathrm{Br}}>0$, both plug-in medians lie below $G_{\mathrm{Br}}(t^\ast)$, and the gap increases with $\lambda_{\mathrm{Br}}$.
Empirical coverage against $G_{\mathrm{Br}}(t^\ast)$ remains close to the nominal $0.95$ level for the naive CI at $q=30$ and declines modestly at $q=60$ as $\lambda_{\mathrm{Br}}$ increases.
Over the same settings, the OEAR CI undercovers more than the naive CI.
In this weak-feedback range, the drift--Wiener approximation is already close to the bridge reference at the operating point, so the OEAR adjustment has little to correct for $G(t^\ast)$ in these experiments.

The bridge reference probabilities in Table~\ref{tab:mc_bridge_naive_vs_oear_compact} are orders of magnitude below the Criterion~E decision threshold $0.1$.
Accordingly, within the weak density-feedback range considered here, both procedures lead to the same threshold-based conclusion, even though systematic differences are visible on the probability scale.

These conclusions are specific to the weak-feedback range and operating-point design considered here; stronger feedback or different horizons may yield larger departures from the drift--Wiener approximation and therefore a different balance between the two plug-in strategies.

\begin{table}[H]
\caption{Monte Carlo summary under weak density feedback (naive vs OEAR).}
\label{tab:mc_bridge_naive_vs_oear_compact}
\centering
\footnotesize
\setlength{\tabcolsep}{3.2pt}
\renewcommand{\arraystretch}{1.05}
\resizebox{\linewidth}{!}{\begin{tabular}{r r r r r r r r r}
\hline
$q$ & $\lambda_{\mathrm{Br}}$ & True $G_{\mathrm{Br}}$ & Naive median & Naive CI width & Naive Cov. & OEAR median & OEAR CI width & OEAR Cov. \\
\hline
30 & 0 & $6\times10^{-36}$ & $7\times10^{-38}$ & $6\times10^{-15}$ & 0.949 & $2\times10^{-40}$ & $3\times10^{-16}$ & 0.829 \\
30 & 0.0005 & $3\times10^{-38}$ & $5\times10^{-43}$ & $2\times10^{-17}$ & 0.951 & $3\times10^{-46}$ & $4\times10^{-19}$ & 0.821 \\
30 & 0.001 & $9\times10^{-41}$ & $2\times10^{-48}$ & $3\times10^{-20}$ & 0.945 & $5\times10^{-52}$ & $5\times10^{-22}$ & 0.805 \\
30 & 0.0015 & $2\times10^{-43}$ & $4\times10^{-54}$ & $4\times10^{-23}$ & 0.932 & $2\times10^{-58}$ & $2\times10^{-25}$ & 0.785 \\
30 & 0.002 & $5\times10^{-46}$ & $2\times10^{-60}$ & $2\times10^{-26}$ & 0.919 & $3\times10^{-65}$ & $8\times10^{-29}$ & 0.766 \\
\hline
60 & 0 & $6\times10^{-36}$ & $7\times10^{-37}$ & $4\times10^{-20}$ & 0.948 & $4\times10^{-38}$ & $7\times10^{-21}$ & 0.838 \\
60 & 0.0005 & $3\times10^{-38}$ & $4\times10^{-42}$ & $2\times10^{-23}$ & 0.947 & $2\times10^{-43}$ & $3\times10^{-24}$ & 0.823 \\
60 & 0.001 & $9\times10^{-41}$ & $1\times10^{-47}$ & $6\times10^{-27}$ & 0.928 & $3\times10^{-49}$ & $5\times10^{-28}$ & 0.796 \\
60 & 0.0015 & $2\times10^{-43}$ & $3\times10^{-53}$ & $1\times10^{-30}$ & 0.903 & $4\times10^{-55}$ & $9\times10^{-32}$ & 0.765 \\
60 & 0.002 & $5\times10^{-46}$ & $6\times10^{-59}$ & $3\times10^{-34}$ & 0.873 & $4\times10^{-61}$ & $1\times10^{-35}$ & 0.733 \\
\hline
\end{tabular}}

\smallskip
\footnotesize
\parbox{0.97\linewidth}{\emph{Note:} $x_d=12$, $t^\ast=100$, $\tau=1$, $(\mu,\sigma)=(-0.03215227, 0.07027818)$, and $\theta=x_d$ are fixed throughout. Each configuration uses $n_{\mathrm{MC}}=50{,}000$ replicates. ``CI width'' is the median width of the nominal 95\% $w$--$z$ CI, and ``Cov.'' is its empirical coverage (nominal 0.95).}
\end{table}

\subsection*{Concluding remarks}

Across the relaxed process models examined in this appendix, neither plug-in strategy is uniformly superior at finite sample sizes.
When departures from the drift--Wiener baseline are mild, the naive and OEAR plug-in summaries can be close, and the resulting $w$--$z$ decision summaries may be practically indistinguishable.

Observation error (including CPUE-motivated nonlinearity) can distort the naive drift--Wiener likelihood fit in systematic ways; these effects are examined separately in Appendix~\ref{appendix:obs_error}.
Here I show that serial dependence in the drift component produces similar distortions.
In dependence-driven settings such as colored environmental noise, the effective-diffusion adjustment underlying OEAR tracks the reference ordering more closely than the naive fit, and the nominal 95\% $w$--$z$ CI remains broadly well behaved across the relaxations considered here.

For practical use, I recommend emphasizing OEAR as the primary plug-in summary in colored-noise settings, while reporting the naive plug-in as a sensitivity check; under the weak density-feedback range examined here, however, OEAR offers no calibration advantage.
 
\section{Eel Harvests by Prefectures in Japan: Seed and Inland Fisheries}
\label{appendix:eel_harvest}
\setcounter{equation}{0}
\setcounter{figure}{0}
\setcounter{table}{0}

This appendix provides the full data background and extended methodological details that underpin the harvest-based analyses in the main text and the subsequent PVA calculations in Appendix~\ref{appendix:eel_pva_sensitivity}.
It consolidates (i) data source descriptions and compilation notes, (ii) the relationship between catch and CPUE for glass eels, (iii) conversion of catch weight to numbers of individuals, (iv) prefectural harvest summaries and figures, (v) recruitment synchrony across the distribution range in East Asia, (vi) construction of the shared national harvest time series for the PVA calculations, and (vii) effort measures used in sensitivity analyses.

\subsection{Fisheries statistics}
\label{sec:S12_1_fisheries_statistics}
\citet{hakoyama2016compilation} digitized and harmonized a comprehensive historical dataset of Japanese fisheries statistics related to the Japanese eel, covering the period from 1894 onward, by compiling multiple official statistical report series.
Key variables include the weight of eels harvested, the weight and number of seeds used for aquaculture, the number of eels stocked, and counts of management units involved in eel fisheries.
These variables are reported at multiple spatial levels (sites such as rivers and lakes, prefectures, inland/coastal domains, and national totals).
For Shizuoka Prefecture, I additionally consulted the \emph{Annual Reports of the Shizuoka Prefectural Fisheries Experiment Station} and the prefectural bulletin \emph{Hamana}, because the prefectural records are recorded in kilograms (allowing finer precision) and include series that are not available from the national official statistics (notably the 2018--2020 coastal seed harvest).
In the analyses below, I focus on the 1957--2020 time series of seed (glass-eel) harvest and inland-eel harvest extracted from this compilation.
The analysis period ends in 2020 because the national official statistics discontinued reporting seed-fishery harvest series (coastal through 2018; inland through 2020) in the \emph{Annual Reports of Catch Statistics on Fishery and Aquaculture in Japan}.

Seed (glass-eel) harvest statistics have been recorded separately for coastal and inland waters.
In coastal waters, reporting was intermittent in several major fishing prefectures.
The pronounced spike in the 1960s coincides with such intermittency, notably in Ibaraki, suggesting that compilation practices may have affected interannual comparability.
In inland waters, recent seed catches consist almost entirely of glass eels, whereas historical inland seed statistics may also have included some elvers; around 1960, substantial elver catches were specifically reported from the Tone River system in Ibaraki and Chiba, a major producing region in Japan.
On the basis of the documented evidence, the inclusion of elvers appears to have been a regional historical feature of inland seed fisheries rather than a nationwide characteristic \citep{HakoyamaEtAl2025_JapaneseEel_R06}.

Such compilation and reporting features can influence long-term trends.
In particular, unusually large elver catches in earlier decades and a coastal series heavily influenced by intermittent reporting from Ibaraki may lead to overestimation of decline rates and hence of extinction risk.
However, I retained the seed statistics without adjustment for possible elver inclusion so as not to understate risk.
This also facilitates direct comparison with the decline-based evaluation under IUCN Criterion~A, highlighting its methodological contrast with Criterion~E.
Details on prefectural patterns and potential aggregation issues are provided below.

Geographic context and the prefecture index map used throughout this appendix are shown in Fig.~\ref{fig:japan_pref_and_inset}.

\subsection{Glass eel indices: harvest and CPUE}
\label{sec:S12_2_glass_eel_indices}
\begin{figure}[H]
\centering
\includegraphics[width=0.72\linewidth]{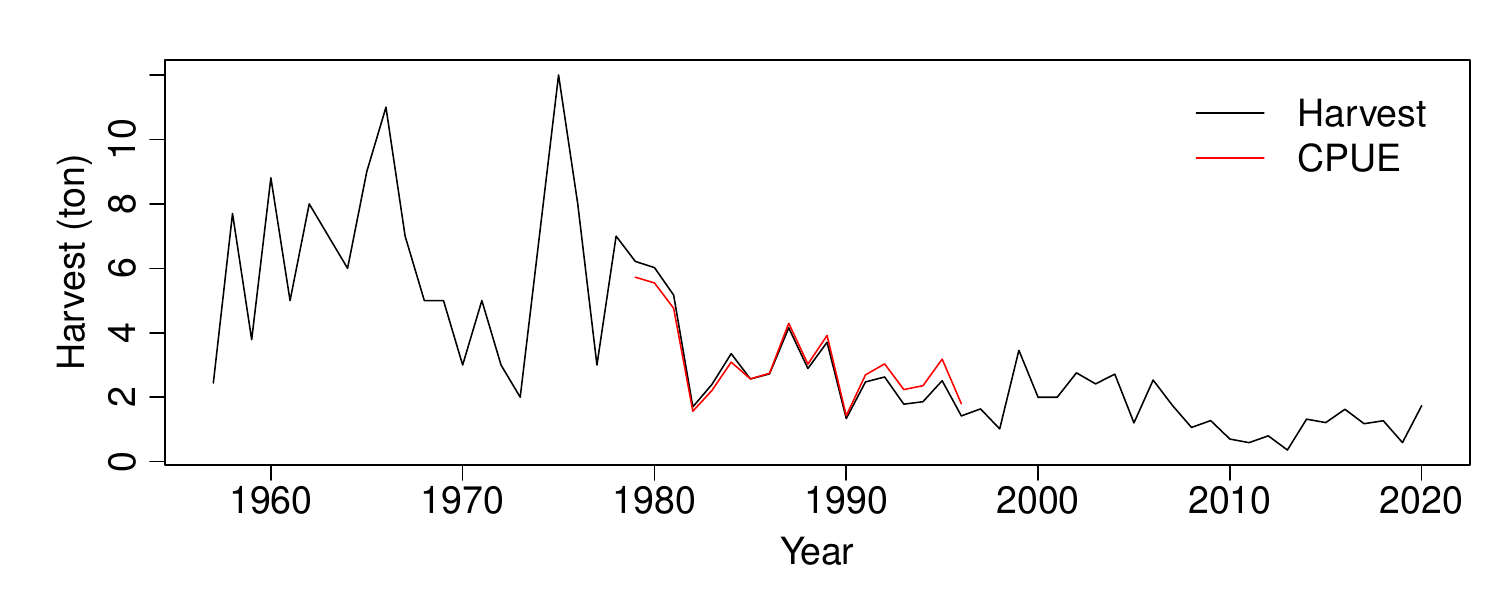}
\caption{Coastal glass eel catch and CPUE in Shizuoka Prefecture.
Catch and CPUE show closely parallel interannual variation in the years of overlap, supporting the use of catch as an abundance index when CPUE is unavailable.}
\label{fig:shizuoka_catch_cpue}
\end{figure}

Glass eel indices in coastal waters are represented by harvest and CPUE (catch per unit effort).
Because CPUE is available only for limited years, mainly from the \emph{Eel Culture Research Council reports} (1977--1997) and with gaps elsewhere, harvest serves as the primary index for long time spans.
Where overlapping years exist, aggregate data from major prefectures show that harvest trends broadly parallel CPUE, suggesting that harvest can serve as a practical proxy when standardized CPUE is unavailable (see also Section~\ref{sec:S12_7_effort_measures}).
Shizuoka Prefecture, which ranks second in total coastal harvest and has the only uninterrupted coastal catch series, provides an illustrative example: in the overlapping years, harvest closely tracks CPUE (Fig.~\ref{fig:shizuoka_catch_cpue}), supporting the use of harvest as a practical index when CPUE is unavailable.

\subsection{Conversion of catch weight to numbers of individuals}
\label{sec:S12_3_conversion_numbers}
Catch weight can be translated to numbers of individuals by dividing by a representative mean body weight for each stage: 0.2~g (glass eels) \citep{HakoyamaEtAl2025_JapaneseEel_R06} and 200~g (yellow and silver eels; representative of the yellow stage, which dominates inland catches) \citep{yokouchi2014demographic}.
Because this conversion is a constant rescaling, it affects only the abundance-scale calibration (and hence $x_d$) and is applied where $x_d$ is constructed in the PVA appendix (Appendix~\ref{appendix:eel_pva_sensitivity}).

\subsection{Prefectural harvest summaries and figures}
\label{sec:S12_4_prefectural_summaries}
\paragraph{Fishery categories.}
Both seed-fishery categories (coastal and inland) correspond mainly to glass eels; potential historical inclusion of elvers in inland seed statistics is discussed in Section~\ref{sec:S12_1_fisheries_statistics}.
Since at least 1978, the seed-fishery season in major aquaculture prefectures has been restricted to winter glass eels \citep{HakoyamaEtAl2025_JapaneseEel_R06}.
Inland eel fisheries primarily target yellow eels and, to a lesser extent, silver eels.

\paragraph{Coding of special symbols.}
The original tables use three non-numeric symbols:
a dash ($-$) denotes \emph{no catch / no response} (the sources do not distinguish these);
``$x$'' denotes \emph{confidential};
and ``$:$'' denotes \emph{prefecture unspecified}.
In the heatmaps these symbols are overlaid as shaded cells so their occurrence can be evaluated alongside numeric values on the log$_{10}$ scale.

\paragraph{Coastal seed fisheries (glass eel).}
Coastal series contain frequent entries marked as no catch or no response, including in prefectures with high average harvest.
This pattern suggests that many such cases reflect non-response rather than confirmed zero catch.
Shizuoka is the only coastal prefecture with a continuous numeric series throughout the period and is among the top producers.
By contrast, several major producers, notably Ibaraki, exhibit intermittent reporting during key periods such as the 1960s, increasing the prevalence of missing values in years when catches were likely positive.

\paragraph{Inland seed fisheries (glass eel and elver).}
Inland series also contain missing-value entries, but their distribution differs from the coastal case.
Prefectures with high average harvest tend to show fewer such gaps, whereas a subset of prefectures report almost exclusively missing values over time, consistent with sustained zero catch or the absence of target fisheries.
Overall, inland seed-fishery data provide more complete numeric coverage than coastal data.

\paragraph{National versus prefectural totals (seed fisheries).}
The sum of prefectural reports broadly tracks the national totals, indicating consistency in most years.
However, if missing entries represent non-response rather than true zero catch, prefectural sums would underestimate actual harvests.
This potential aggregation bias is of greatest concern for coastal data, where gaps are common even in large-producing prefectures; in particular, Ibaraki's reporting pattern raises concern that substantial catches may be missing from the prefectural tables.

\paragraph{Inland eel fisheries (yellow and silver).}
Inland harvests of yellow and silver eels are available for 1957--2020 and show high continuity, with very few missing values or symbol-coded entries.
The national series peaks in the late 1960s at over 3,000 tonnes and declines steadily thereafter to a few hundred tonnes in recent years.
At the prefectural level, early records indicate widespread harvest across many regions, while later decades show contraction toward fewer producing prefectures.
These continuous inland series provide a reliable indicator of long-term decline in exploited inland eel stocks.

\begin{figure}[H]
  \centering

  \begin{subfigure}[t]{0.495\linewidth}
    \centering
    \includegraphics[height=0.33\textheight]{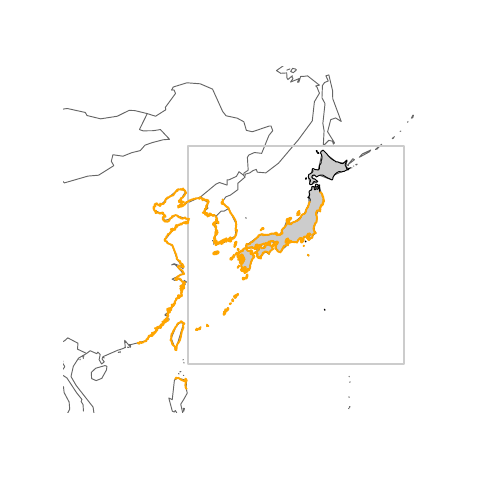}
    \caption{Distribution of the Japanese eel.
The Japanese eel is catadromous, spawning near the West Mariana Ridge, with larvae transported by the Kuroshio Current to East Asian coasts.
Population-genetic studies support broad-scale panmixia across the range \citep{faulks2025panmixia}.}
  \end{subfigure}\hfill
  \begin{subfigure}[t]{0.49\linewidth}
    \centering
    \includegraphics[height=0.33\textheight]{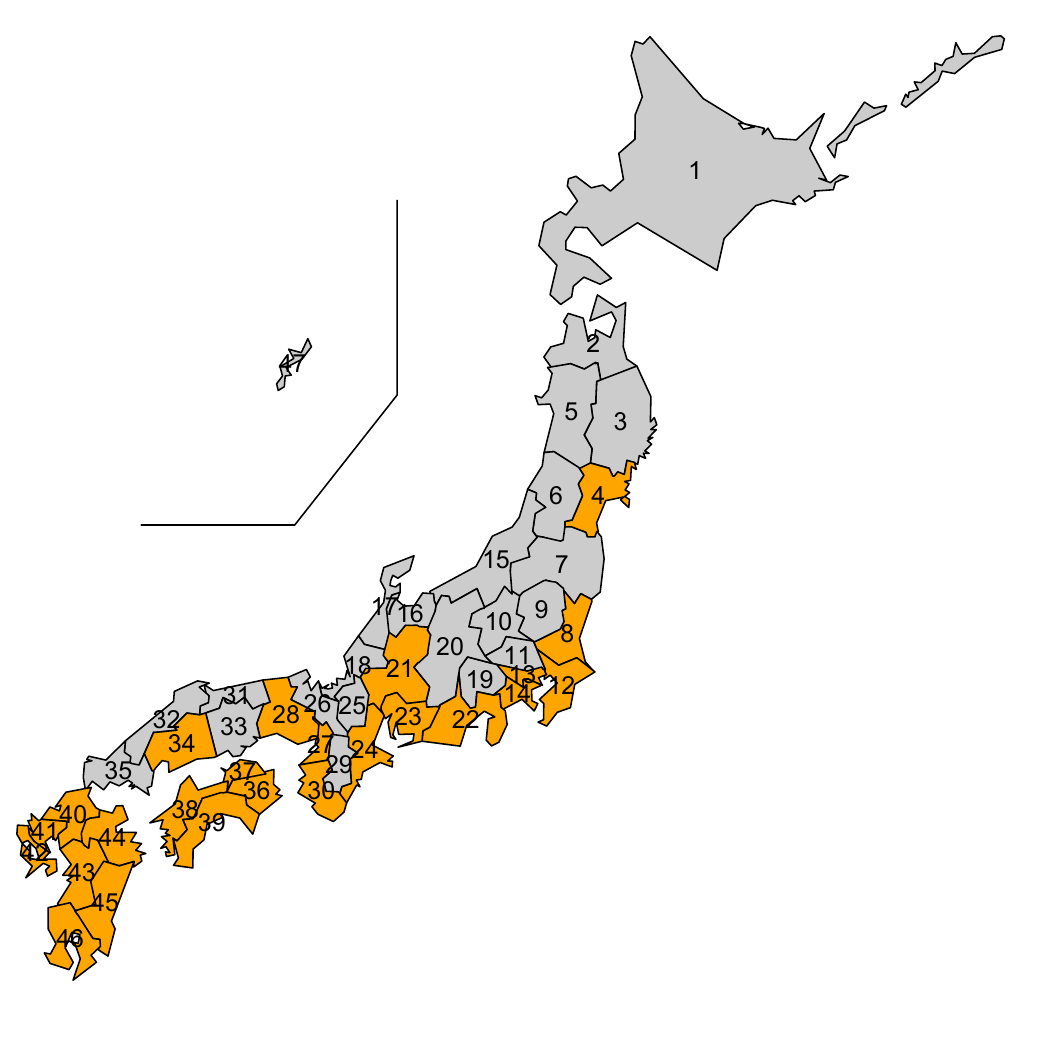}
    \caption{Map of Japan.}
  \end{subfigure}

  \caption{Geographic context and prefecture index map used in this study.
  (a) Geographic distribution of the Japanese eel across its range, redrawn after \citet{han2012larval} and \citet{tsukamoto2003seamounts}; Japan is highlighted.
  (b) Map of Japan. Prefectures shaded in orange indicate those with an active glass eel fishery as of 2020. By contrast, yellow- and silver-stage eel harvest statistics are available for all prefectures except Okinawa. Each prefecture is labeled by its index (1--47) in JIS order:
  1 Hokkaido; 2 Aomori; 3 Iwate; 4 Miyagi; 5 Akita; 6 Yamagata; 7 Fukushima;
  8 Ibaraki; 9 Tochigi; 10 Gunma; 11 Saitama; 12 Chiba; 13 Tokyo; 14 Kanagawa;
  15 Niigata; 16 Toyama; 17 Ishikawa; 18 Fukui; 19 Yamanashi; 20 Nagano;
  21 Gifu; 22 Shizuoka; 23 Aichi; 24 Mie; 25 Shiga; 26 Kyoto; 27 Osaka; 28 Hyogo;
  29 Nara; 30 Wakayama; 31 Tottori; 32 Shimane; 33 Okayama; 34 Hiroshima; 35 Yamaguchi;
  36 Tokushima; 37 Kagawa; 38 Ehime; 39 Kochi; 40 Fukuoka; 41 Saga; 42 Nagasaki;
  43 Kumamoto; 44 Oita; 45 Miyazaki; 46 Kagoshima; 47 Okinawa.
  Both panels were produced in R using the \texttt{NipponMap} and \texttt{maps} packages.}
  \label{fig:japan_pref_and_inset}
\end{figure}

\begin{figure}[H]
\centering
\includegraphics[width=0.72\linewidth]{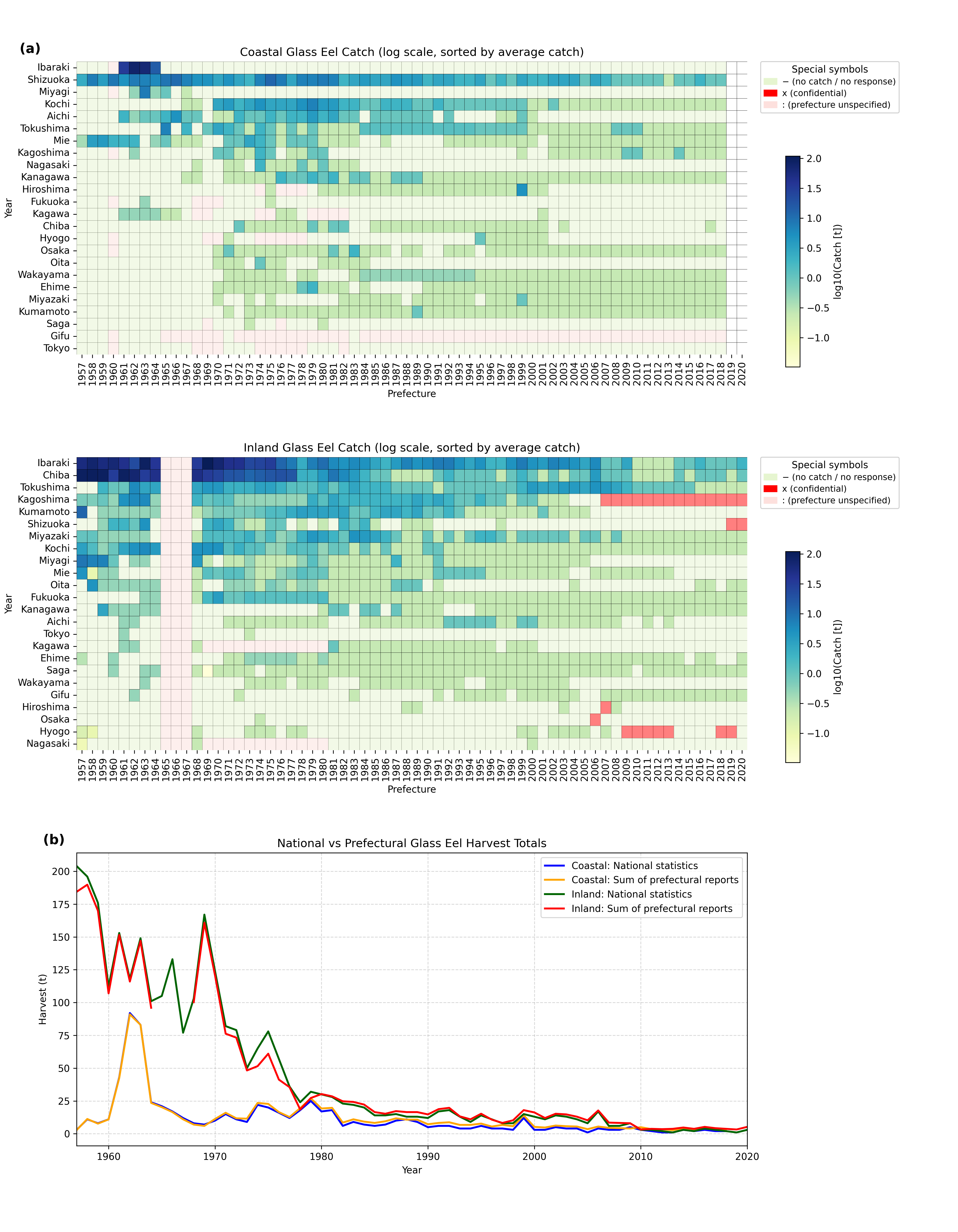}
\caption{Seed-fishery harvests by prefecture in Japan, compiled from \citet{hakoyama2016compilation} and supplemented for recent years using the \emph{Annual Reports of Catch Statistics on Fishery and Aquaculture in Japan}.
Coastal seed fisheries (glass eels), 1957--2018, and inland seed fisheries (glass eels, with possible elver inclusion in early historical statistics, notably around 1960), 1957--2020.
(a) Heatmaps of annual harvest (tonnes) on a log$_{10}$ scale; prefectures (rows) are sorted by mean harvest, and symbol-coded entries in the sources are retained as overlaid cells (see legend).
(b) National totals versus the sum of prefectural reports (tonnes).
The coastal prefectural panel ends in 2018; the Shizuoka coastal series was extended to 2020 for the PVA national series (Section~\ref{sec:S12_6_shared_national_series}).}
\label{fig:glass_eel_catch}
\end{figure}

\begin{figure}[H]
\centering
\includegraphics[width=0.72\linewidth]{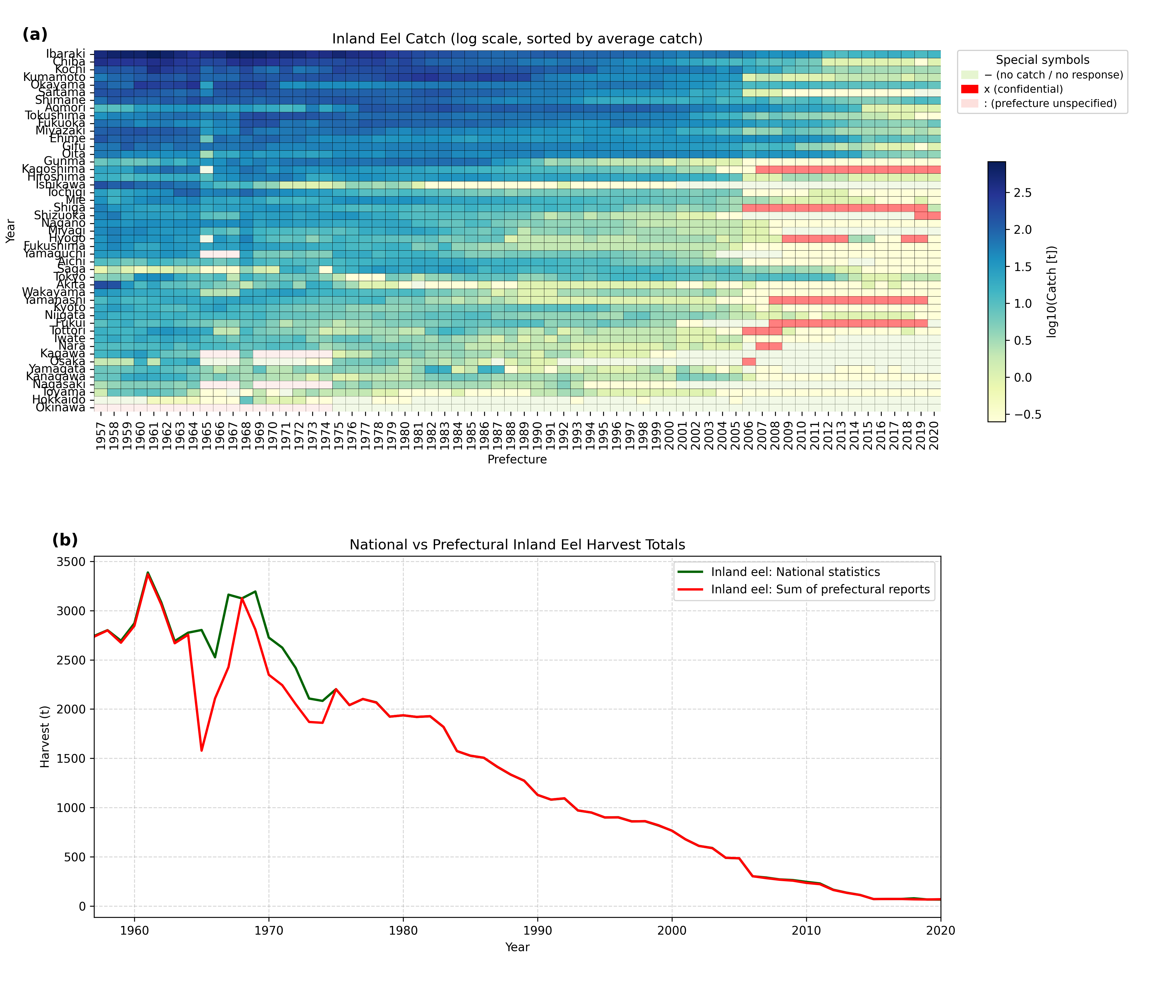}
\caption{Inland eel fisheries (yellow and silver eels), 1957--2020.
(a) Log$_{10}$-scaled heatmap of annual harvest (tonnes) by prefecture; prefectures (rows) are sorted by mean harvest.
(b) National harvest totals through time (tonnes).
Data compiled from \citet{hakoyama2016compilation} and supplemented for recent years using the \emph{Annual Reports of Catch Statistics on Fishery and Aquaculture in Japan}.}
\label{fig:inland_eel}
\end{figure}

\subsection{Recruitment synchrony across the distribution range in East Asia}
\label{sec:S12_5_recruitment_synchrony}

To place Japan-wide harvest indices in a broader context, I summarize interannual covariation in glass-eel catches across the main East Asian range (Fig.~\ref{fig:japan_pref_and_inset}a; China, Japan, Korea, and Chinese Taipei) using catch data reported in the Joint Press Release of the Eighteenth Meeting of the Informal Consultation on International Cooperation for Conservation and Management of Japanese Eel Stock and Other Relevant Eel Species (available at \url{https://www.mofa.go.jp/files/100866466.pdf}).
For Japan, the ``catch of glass eel'' series in the source table is not a direct landing statistic; it is an estimate of domestically caught glass eels computed as the reported input of glass eels into aquaculture ponds minus imports derived from the Trade Statistics.

For the 2011--12 to 2024--25 fishing years, pairwise Pearson correlations are uniformly positive and generally moderate to strong (range: 0.551--0.768; mean off-diagonal $r=0.691$; Table~\ref{tab:informal_consultation_glass_catch_corr}; pairwise sample size: 13--14 years), indicating substantial interannual covariation across the distribution range.

These broad positive correlations indicate that year-to-year fluctuations in glass-eel availability have a substantial component shared across the distribution range.
This supports interpreting Japan-wide harvest time series as reflecting a broadly connected East Asian stock signal rather than a purely local one, while using the Informal Consultation series only as corroborative evidence.

\begin{table}[H]
\centering
\caption{Regional correlations in glass-eel catches reported in the Joint Press Release of the Eighteenth Meeting of the Informal Consultation on International Cooperation for Conservation and Management of Japanese Eel Stock and Other Relevant Eel Species (2011--12 to 2024--25).}
\label{tab:informal_consultation_glass_catch_corr}
\begin{tabularx}{0.9\textwidth}{l>{\centering\arraybackslash}X>{\centering\arraybackslash}X>{\centering\arraybackslash}X>{\centering\arraybackslash}X}
\hline
 & China & Japan & Korea & Chinese Taipei \\
\hline
China & 1.000 & 0.551 & 0.745 & 0.768 \\
Japan & 0.551 & 1.000 & 0.745 & 0.634 \\
Korea & 0.745 & 0.745 & 1.000 & 0.705 \\
Chinese Taipei & 0.768 & 0.634 & 0.705 & 1.000 \\
\hline
\multicolumn{5}{@{}p{0.9\textwidth}@{}}{\footnotesize Years: 2011--12 to 2024--25; mean off-diagonal Pearson $r=0.691$; pairwise sample size: 13--14 years.} \\
\end{tabularx}
\end{table}

For direct use in PVA scale-calibration checks, I also summarize annual glass-eel catches for Japan, China, Chinese Taipei, and Korea from the Joint Press Release of the Eighteenth Meeting of the Informal Consultation (unit: kg; Table~\ref{tab:informal_consultation_4regions_catch}).
In the source table, Japan reports catch in metric tons through 2014--15 and in kilograms thereafter; I converted the ton values to kilograms for consistency.
For China, the source table reports catch only through calendar year 2023 and provides no values for the 2023--24 and 2024--25 fishing years.

\begin{table}[H]
\centering
\caption{Glass-eel catches by region from the Informal Consultation dataset (unit: kg; fishing years 2011--12 to 2024--25).}
\label{tab:informal_consultation_4regions_catch}
\begin{tabular}{lrrrrr}
\hline
Fishing year & Japan & China & Chinese Taipei & Korea & East Asia total \\
\hline
2011--12 & 9,000 & 28,000 & 1,912 & 1,530 & 40,442 \\
2012--13 & 5,200 & 19,500 & 960 & 1,002 & 26,662 \\
2013--14 & 17,400 & 55,000 & 8,250 & 5,489 & 86,139 \\
2014--15 & 15,300 & 20,500 & 1,100 & 4,725 & 41,625 \\
2015--16 & 13,625.2 & 21,000 & 3,060 & 1,830 & 39,515.2 \\
2016--17 & 15,442.4 & 26,500 & 4,500 & 2,717 & 49,159.4 \\
2017--18 & 8,967.5 & 16,000 & 1,100 & 973 & 27,040.5 \\
2018--19 & 3,670.1 & 14,500 & 2,751 & 649 & 21,570.1 \\
2019--20 & 17,112.4 & 50,000 & 5,244 & 4,500 & 76,856.4 \\
2020--21 & 11,333.9 & 38,000 & 6,005 & 3,228 & 58,566.9 \\
2021--22 & 10,344.7 & 29,500 & 1,607 & 2,512 & 43,963.7 \\
2022--23 & 5,660.2 & 40,450 & 1,850 & 2,165 & 50,125.2 \\
2023--24 & 7,110.2 & 13,095$^{\dagger}$ & 1,295 & 1,330 & 22,830.2 \\
2024--25 & 13,400.7 & -- & 6,230 & 7,050 & 26,680.7 \\
\hline
\end{tabular}
\vspace{3pt}
\caption*{\footnotesize\textit{Note.} Values are as reported in the source tables. For Japan, the ``catch of glass eel'' series is not a direct landing statistic; it is an estimate of domestically caught glass eels computed as aquaculture-pond inputs minus imports derived from the Trade Statistics. In the source, Japan reports catch in metric tons through 2014--15 and in kilograms thereafter; ton values were converted to kilograms here for consistency. For China, catch data are available only up to calendar year 2023 (13,095~kg), and the source table provides no values for the 2023--24 and 2024--25 fishing years; accordingly, the China entry in 2023--24 (marked $^{\dagger}$) corresponds to calendar year 2023 and is included for reference. East Asia total is the row-wise sum of available values; in 2024--25 it is based on three regions because China is not reported.}
\end{table}

\subsection{Shared national time series used in PVA}
\label{sec:S12_6_shared_national_series}

The population-viability analyses use national annual harvest time series for
(i) glass-eel seed fisheries (coastal seed and inland seed, combined) and
(ii) inland-eel fisheries (yellow and silver eels, excluding inland seed harvest).
Accordingly, I construct the adopted national seed series by combining an LME-derived coastal-seed component with the official national inland-seed totals, while the adopted inland-eel series is given directly by the official national totals.

To select the adopted source for each component, I used two diagnostics:
(1) whether symbol-coded entries are concentrated in major fishing prefectures, and
(2) the annual discrepancy between the official national total and the sum of the prefectural tables.
Let $\mathcal{H}^{\mathrm{pref}}_{i,t}$ denote harvest (tonnes) for prefecture $i$ in year $t$ as tabulated in the prefectural tables (with symbol-coded entries treated as missing), and let $\mathcal{H}^{\mathrm{official}}_{t}$ denote the corresponding official national total.
I summarize their discrepancy as
\begin{equation*}
D_t=\mathcal{H}^{\mathrm{official}}_{t}-\sum_i \mathcal{H}^{\mathrm{pref}}_{i,t}.
\end{equation*}

For coastal seed, symbol-coded entries occur frequently even in major fishing prefectures (e.g., Ibaraki and Miyagi), and the discrepancy between the official totals and the prefectural sum is typically small over 1957--2018 (mean $D_t=-1.29$~t; median $D_t=-1.50$~t; all years within $\pm 5$~t in this comparison).
Taken together, these patterns suggest that the official coastal series may not consistently recover catches missing from the prefectural tables when prefectural reporting is intermittent, and may inherit under-coverage in years with major-producer gaps.
For this reason, the coastal component is represented by the LME-derived shared series (with level rescaling).

For inland seed and inland eel, symbol-coded entries are less concentrated in major producers, and $D_t$ is strongly positive in periods affected by prefecture-unspecified ($:$) reporting.
For example, inland seed has official totals for 1965--1967 while the prefectural sum is unavailable, and inland eel shows large positive discrepancies in the mid-1960s (e.g., 1965: $D_t=1225.13$~t).
These patterns are consistent with national tabulations including catches that are not allocated to specific prefectures in the published prefectural tables.
Therefore, inland seed and inland eel are represented by official national totals in the adopted PVA series.
Small year-to-year differences around zero are not over-interpreted, because official totals are compiled by summing records at kg precision, whereas prefectural values in historical tables are tabulated in rounded tonnes.

In the LME fits, symbol-coded entries ($-$, $x$, and $:$) are treated as missing and excluded from estimation; in particular, $-$ conflates true zero catch and non-response in the original tables.

For seed, I fit
\begin{equation*}
 \log \mathcal{H}^{\mathrm{pref}}_{i,t}
 =
 \mathfrak{a}_t
 +
 \mathfrak{c}_{\mathrm{Domain}}
 +
 \mathfrak{b}_i
 +
 \zeta_{i,t},
\end{equation*}
where $\mathcal{H}^{\mathrm{pref}}_{i,t}$ is seed harvest (tonnes) in prefecture $i$ and year $t$, $\mathfrak{a}_t$ is a year fixed effect, $\mathfrak{c}_{\mathrm{Domain}}$ is a two-level fixed effect (coastal vs inland), and $\mathfrak{b}_i$ is a prefecture random intercept.
Residual variance of $\zeta_{i,t}$ is allowed to differ between coastal and inland components using \texttt{varIdent} in \texttt{nlme::lme}, i.e., $\mathrm{Var}(\zeta_{i,t}\mid \mathrm{Domain})=\sigma^2_{\mathrm{Domain}}$.
The inland observations contribute to estimation of the shared year effect and the coastal--inland \emph{Domain} offset, but the inland fitted totals are not used in the adopted series.
Model-based back-transformation applies the standard lognormal bias correction using the \emph{Domain}-specific residual standard deviation, and prefectural predictions are summed to obtain annual coastal seed totals.
For coastal seed, the prefectural tables from \citet{hakoyama2016compilation} extend through 2015, and I supplemented the prefectural panel for 2016--2018 using the \emph{Annual Reports of Catch Statistics on Fishery and Aquaculture in Japan}.
The Shizuoka coastal series was further extended to 2020 using the prefectural bulletin \emph{Hamana}.

\begin{figure}[H]
\centering
\includegraphics[width=0.72\linewidth]{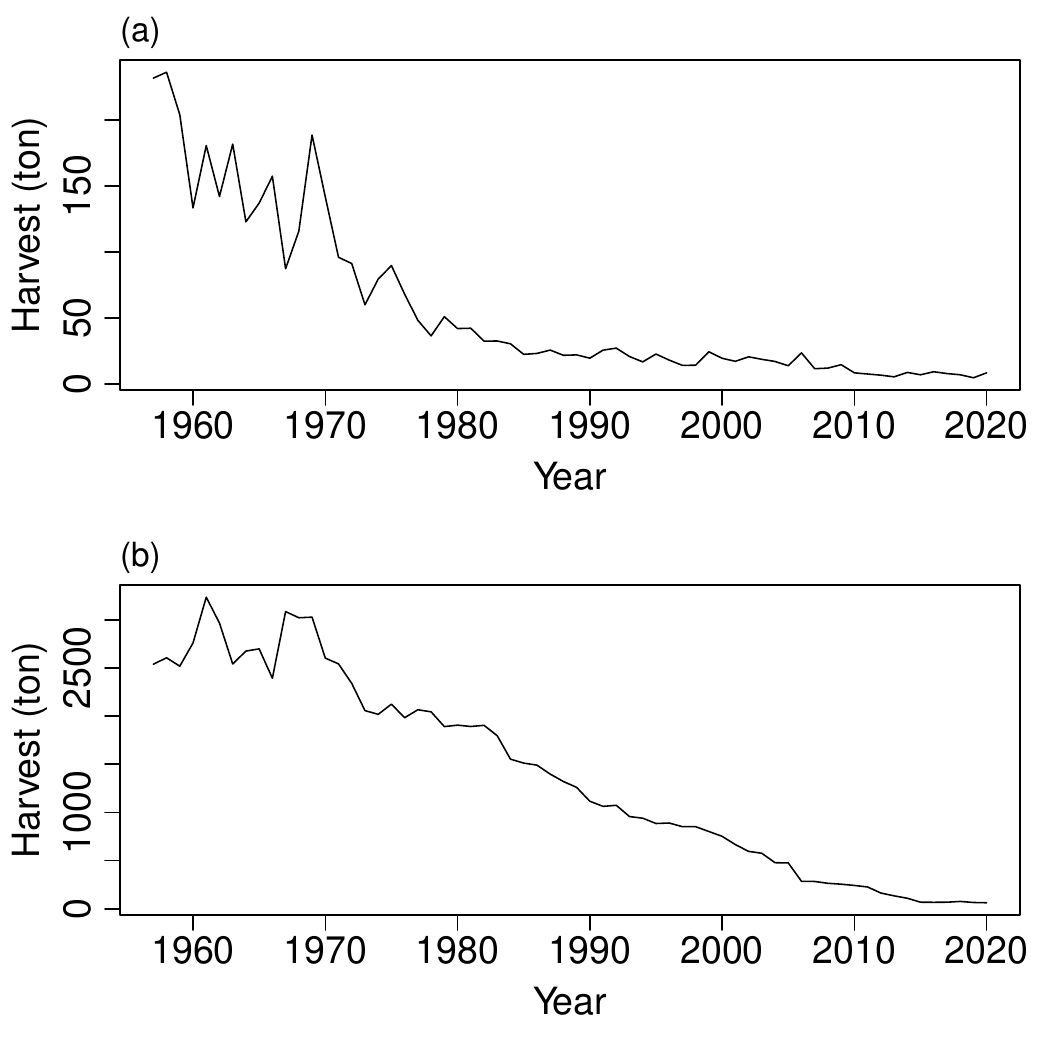}
\caption{Adopted national time series for PVA (tonnes), 1957--2020.
(a) Common glass-eel seed (coastal $+$ inland): the sum of an LME-derived coastal component rescaled to official coastal totals over 1957--2018 and official inland seed totals.
(b) Inland eel harvest (yellow and silver eels): official national totals.}
\label{fig:shared_trends_seed_eel}
\end{figure}

To construct adopted national series for PVA, inland seed and inland-eel harvest use official national totals throughout.
For coastal seed, I adopt the LME-derived series and rescale it by a constant factor $\mathfrak{s}$ estimated by least squares over years where official coastal totals are available (1957--2018):
\begin{equation*}
\check{\mathfrak{s}}
=
\arg\min_{\mathfrak{s}}
\sum_{t=1957}^{2018}
\left(
\mathcal{H}^{\mathrm{official}}_{\mathrm{coastal},t}
-
\mathfrak{s}\,\mathcal{H}^{\mathrm{LME}}_{\mathrm{coastal},t}
\right)^2.
\end{equation*}
This aligns the level to official coastal totals while preserving the LME-derived interannual pattern.
The adopted common-seed series is then
\begin{equation*}
\mathcal{H}^{\mathrm{seed}}_{\mathrm{merged},t}
=
\check{\mathfrak{s}}\,\mathcal{H}^{\mathrm{LME}}_{\mathrm{coastal},t}
+
\mathcal{H}^{\mathrm{official}}_{\mathrm{inland},t}.
\end{equation*}
In this analysis, the estimated scaling factor was $\check{\mathfrak{s}} = 0.5007$ (estimated over 1957--2018, $n=62$).
The adopted national series are shown in Figure~\ref{fig:shared_trends_seed_eel}.

\begin{figure}[H]
\centering
\includegraphics[width=0.78\linewidth]{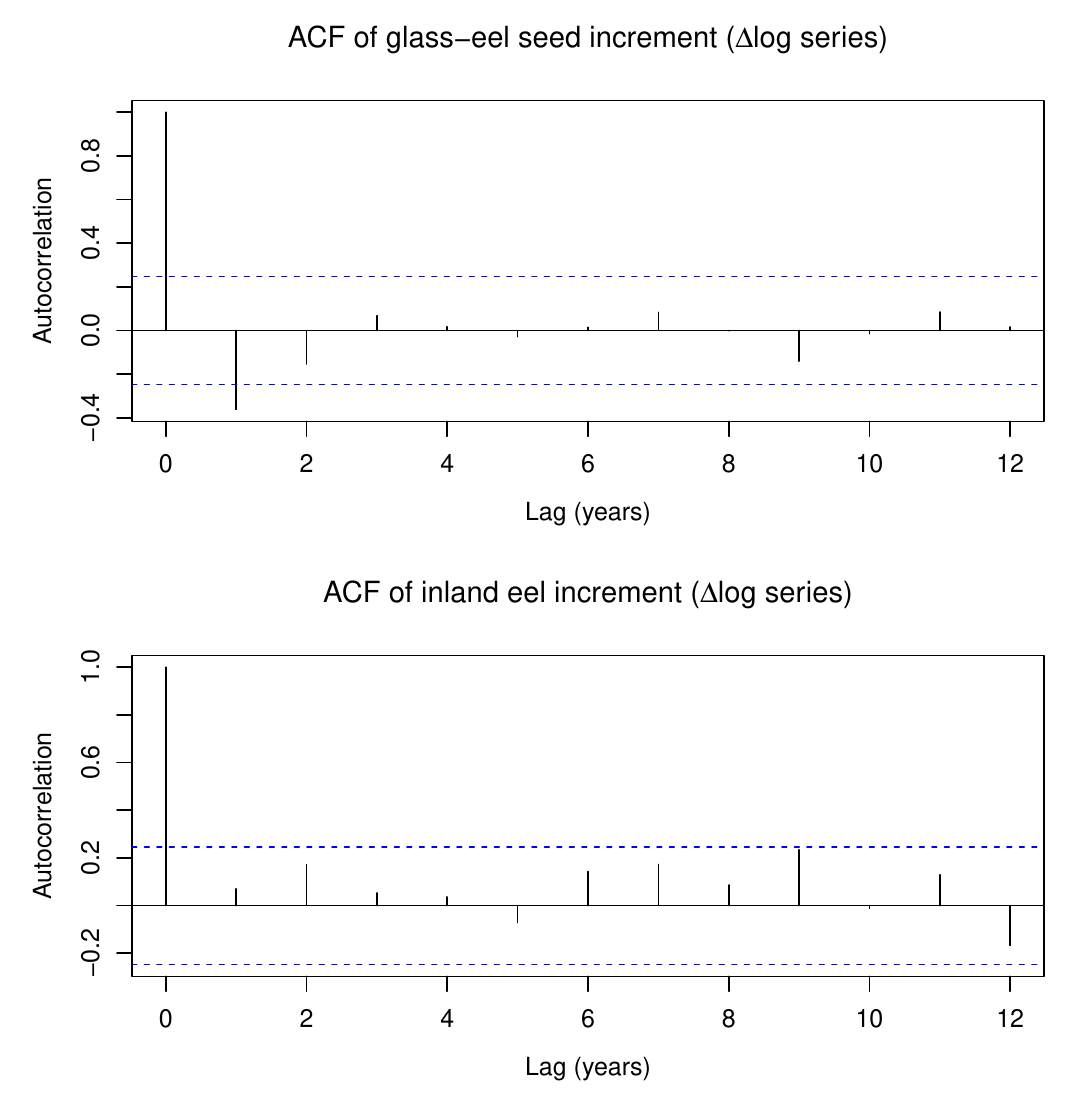}
\caption{Autocorrelation functions (ACFs) of annual log-increments ($\Delta\log \mathcal{H}_t$) for the two adopted national indices (top: glass-eel seed; bottom: inland eel).
Horizontal dashed lines are the default approximate significance bounds returned by \texttt{acf}.}
\label{fig:seed_eel_acf}
\end{figure}

\paragraph{Autocorrelation check within each adopted index.}
To assess serial dependence within each adopted index, I computed the autocorrelation function (ACF) of annual log-increments using \texttt{acf()} in the R \texttt{stats} package (lags 0--12).
For each series $\mathcal{H}_t$, I formed $\Delta \log \mathcal{H}_t=\log \mathcal{H}_t-\log \mathcal{H}_{t-1}$ and computed the ACF over 1958--2020 (63 increments).
As shown in Fig.~\ref{fig:seed_eel_acf}, the seed increment series exhibits a significant negative lag-1 autocorrelation ($\rho_1=-0.362$), whereas autocorrelations at longer lags are small.
The inland-eel increment series shows no clear exceedance of the default approximate significance bounds across lags 1--12.
When such autocorrelation is appreciable, OEAR is expected to offer the clearest advantage over naive drift--Wiener fitting because it targets the long-run variance of the standardized increment series and remains valid under autocovariance.
When autocorrelation is weak, both approaches target essentially the same scale (the long-run variance is close to the variance), but OEAR may be less efficient in finite samples because it estimates additional autocovariance terms; thus the naive approach is sufficient.

\begin{figure}[H]
\centering
\includegraphics[width=0.78\linewidth]{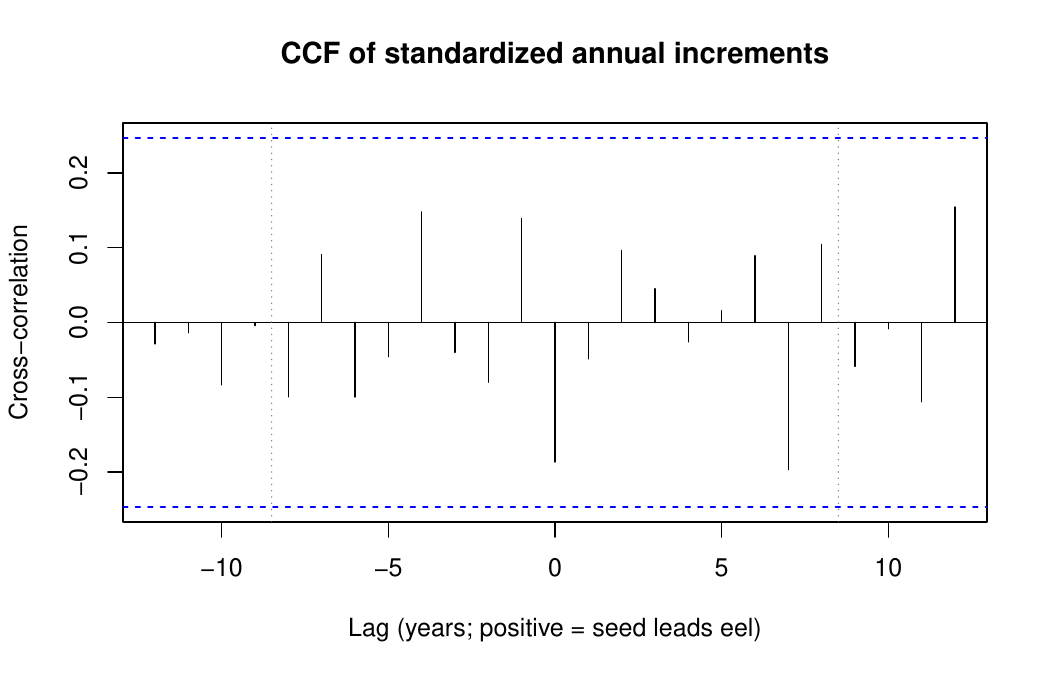}
\caption{Cross-correlation function (CCF) of standardized annual log-increments ($\Delta\log \mathcal{H}_t$) for the two adopted national indices.
The lag axis is defined as seed at year $t$ versus inland eel at year $t+\mathrm{lag}$ ($\mathrm{lag}>0$: seed leads).
Horizontal dashed lines are the default approximate significance bounds returned by \texttt{ccf}.
Vertical dotted lines indicate $\pm 8.5$ years (reference generation-time lag; Appendix~\ref{appendix:generation_time}).}
\label{fig:seed_eel_ccf}
\end{figure}

\paragraph{Lagged cross-correlation check between the two adopted indices.}
Because inland eel harvest aggregates multiple age classes, I evaluated lagged dependence between the two adopted indices using the cross-correlation function (CCF) of standardized annual log-increments.
I computed $\Delta \log \mathcal{H}_t$ and standardized each log-increment series to mean 0 and variance 1 over overlapping years (1958--2020; 63 increments).
I then calculated the CCF using \texttt{ccf()} in the R \texttt{stats} package, defining lag as the association between seed at year $t$ and inland eel at year $t+\mathrm{lag}$ (thus $\mathrm{lag}>0$ means seed leads).
As shown in Fig.~\ref{fig:seed_eel_ccf}, CCF coefficients were small across lags $-12$ to $+12$.
Around the reference generation-time scale (Appendix~\ref{appendix:generation_time}), CCF coefficients at lags $+7$ to $+10$ were $-0.197$, $0.104$, $-0.059$, and $-0.008$, respectively (Fig.~\ref{fig:seed_eel_ccf}).
No lag exceeded the default approximate significance bounds in Fig.~\ref{fig:seed_eel_ccf}.
These lagged associations are therefore interpreted as descriptive diagnostics rather than evidence of strong lead--lag dependence.
Importantly, I do not treat the two indices as independent replicates for inference: each series is analyzed separately, and uncertainty is not reduced by pooling across indices.

\subsection{Effort measures used in sensitivity analyses}
\label{sec:S12_7_effort_measures}

Because long-term catch series may reflect changes in fishing effort as well as abundance, I use available effort records only to calibrate plausible effort-change magnitudes in the sensitivity analyses.
Two sources are used: (i) prefecture-level glass-eel fishery effort tabulations from the Eel Culture Research Council (eel fishing year, December--April), and (ii) fishery-management-entity count series for inland fisheries from the Japanese \emph{Census of Fisheries} (quinquennial; compiled in \citet{hakoyama2016compilation}; described below).

\subsubsection{Fishing effort for glass-eel fisheries in Japan}
\label{appendix:effort_glass_eel}

For the glass-eel fishery, the Council reports two effort measures by prefecture for 1977--1997: scoop-net participants (persons) and set nets (units; \emph{tou}).
Because these measures are in different units, I do not combine them into a single effort index.
For sensitivity calibration, I use scoop-net participants (persons) as the primary effort measure, consistent with the CPUE definition.

\begin{figure}[H]
  \centering
  \includegraphics[width=0.75\linewidth]{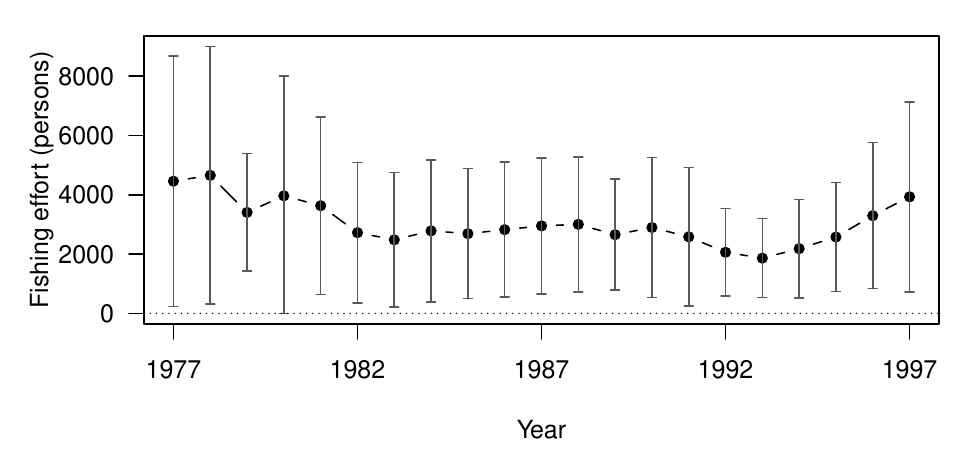}
  \caption{Fishing effort obtainable from the \emph{Eel Culture Research Council} reports for glass-eel fisheries: scoop-net participants (persons), 1977--1997 eel fishing years, shown for the reported prefectures (Chiba, Shizuoka, Aichi, Mie, Tokushima, Kochi, Oita, Miyazaki, Kagoshima).
  Points and error bars show the across-prefecture mean with $\pm 1$ SD.
  The total effort across these prefectures declined from 40{,}126 persons in 1977 to 35{,}386 persons in 1997, corresponding to a geometric mean rate of change of $-0.63\%$ per year (period decline: 11.81\%).}
  \label{fig:scopenet_effort}
\end{figure}

Figure~\ref{fig:scopenet_effort} summarizes scoop-net effort for glass-eel fisheries (persons) reported by prefecture for 1977--1997 eel fishing years.
The prefectures included are Chiba, Shizuoka, Aichi, Mie, Tokushima, Kochi, Oita, Miyazaki, and Kagoshima.
The figure shows the across-prefecture mean with $\pm 1$ SD to calibrate plausible magnitudes of effort change for sensitivity analyses.
Over 1977--1997, the total scoop-net effort (sum across prefectures with available data) declined from 40,126 to 35,386 persons, corresponding to a geometric mean rate of change of
$100\times\left\{\left(\frac{35386}{40126}\right)^{1/(1997-1977)}-1\right\}\approx -0.63\%$
per year.
For the same 9 prefectures, prefecture-specific geometric mean annual rates span roughly $-3.03\%$ to $+2.46\%$ per year (mean $-0.67\%$, SD $2.06\%$), indicating that uncertainty around the aggregate $-0.63\%$ benchmark is on the order of a few percentage points per year.

\subsubsection{Fishing effort for inland eel fisheries in Japan (lake fisheries)}
\label{appendix:effort_lakes}

Inland lake fisheries in Japan have experienced a long-term decline in the number of active fishery management entities (FMEs), consistent with broad structural changes such as aging and limited succession.
The Japanese \emph{Census of Fisheries} (MAFF; quinquennial) tabulates (i) the total number of FMEs engaged in lake fisheries, (ii) the number of those FMEs that reported any catch of yellow/silver eels in lakes, and (iii) the number of eel-catching FMEs that reported yellow/silver eels as a main target species (Table~\ref{tab:fme_trends}).
Because approximately 95\% of lake-fishery FMEs are individual (non-corporate) operations, these counts closely track the number of operating fishers/households.

\begin{figure}[H]
\centering
\includegraphics[width=0.75\linewidth]{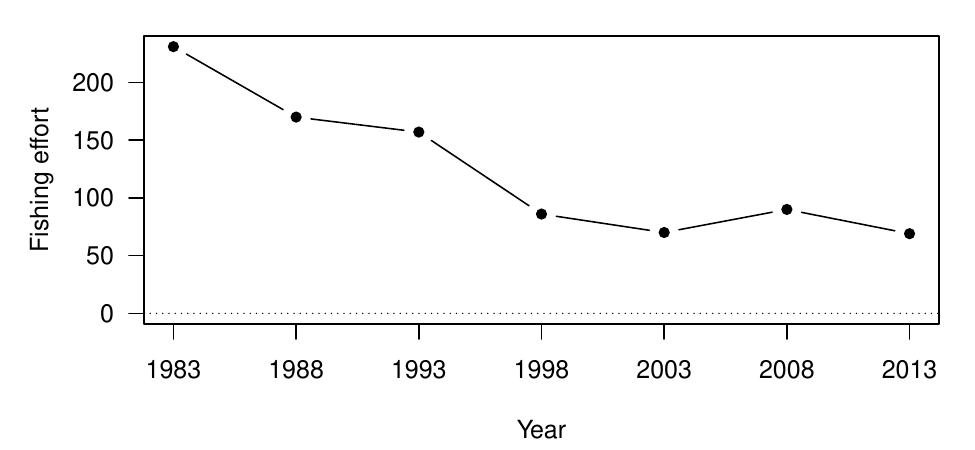}
\caption{Fishing effort obtainable from the Japanese \emph{Census of Fisheries} (MAFF; quinquennial) for lake fisheries: the number of fishery management entities (FMEs) in Japanese lakes that reported yellow/silver eels as a main target species, shown through 2013.}
\label{fig:fme_effort}
\end{figure}

I use the number of eel-targeting FMEs as a fishing-effort measure for yellow/silver eel fisheries in Japanese lakes.
This effort measure declines from 231 in 1983 to 69 in 2013 (Figure~\ref{fig:fme_effort}), corresponding to an average proportional decline of about $3.95\%$ per year over 1983--2013, computed as $100\times\{1-(69/231)^{1/(2013-1983)}\}\approx 3.95\%$.
As a complementary scatter-based summary, a log-linear fit to the 1983--2013 eel-targeting FME series gives an annual change rate of about $-3.99\%$ per year, with an approximate 95\% interval of $[-5.94\%,\,-2.00\%]$.
In the main sensitivity analyses, I use this rate as a plausible magnitude for effort-change scenarios.

\medskip

\begin{table}[H]
\centering
\caption{Fishery management entities (FMEs) engaged in lake fisheries and eel-related activity in Japanese lakes.}
\label{tab:fme_trends}
\begin{threeparttable}
\setlength{\tabcolsep}{4pt}
\begin{tabular}{r r r r r}
\toprule
Year & Lake-fishery FMEs (total) & Eel-catching FMEs & Eel-targeting FMEs & ECPR \\
\midrule
1983 & 6137 & 1640 & 231 & 0.267 \\
1988 & 4961 & 1159 & 170 & 0.234 \\
1993 & 4252 & 1111 & 157 & 0.261 \\
1998 & 3576 & 885  & 86  & 0.247 \\
2003 & 3124 & 646  & 70  & 0.207 \\
2008 & 2850 & 634  & 90  & 0.222 \\
2013 & 2484 & 463  & 69  & 0.186 \\
2018 & 2133 & 444  & --  & 0.208 \\
2023 & 1859 & 360  & --  & 0.194 \\
\bottomrule
\end{tabular}

\begin{tablenotes}[flushleft]
\footnotesize
\item \textit{Notes:} All quantities refer to \emph{lake fisheries in Japan} (inland lake fisheries) as tabulated in the Japanese \emph{Census of Fisheries} (MAFF; quinquennial).
An FME is a fishery management entity.
``Eel-catching FMEs'' are lake-fishery FMEs with any recorded catch of yellow/silver eels in lakes in the census year.
``Eel-targeting FMEs'' are eel-catching lake-fishery FMEs for which yellow/silver eels were recorded as a main target species.
Eel-catching participation ratio (ECPR) is the ratio of eel-catching FMEs to total lake-fishery FMEs.
It is reported as a simple presence/absence-type abundance index for lake eels.
For 2018 and 2023, the eel-targeting count is not available in the published tabulation.
Approximately 95\% of FMEs are individual (non-corporate) operations (household/sole proprietorship).
\end{tablenotes}
\end{threeparttable}
\end{table}

\subsection*{Summary}

This appendix compiles and standardizes harvest-based indices for two life stages of the Japanese eel: glass eels (seed fisheries) and yellow/silver eels (inland eel fisheries), with full prefectural summaries (Figs.~\ref{fig:glass_eel_catch}--\ref{fig:inland_eel}) and the adopted national series for subsequent PVA (Section~\ref{sec:S12_6_shared_national_series}).
Population-genetic studies support broad-scale panmixia across East Asia \citep{faulks2025panmixia}, implying that recruitment variation should be expressed broadly across the distribution range rather than confined to a single locality.
Consistent with this expectation, glass-eel catches reported for China, Japan, Korea, and Chinese Taipei show uniformly positive and generally moderate to strong interannual correlations over 2011--12 to 2024--25 (range: 0.551--0.768; mean off-diagonal $r=0.691$; Table~\ref{tab:informal_consultation_glass_catch_corr}), supporting the interpretation that Japan-wide harvest series track a broadly shared stock signal (Section~\ref{sec:S12_5_recruitment_synchrony}).
Because this multi-region series is substantially shorter than the Japanese historical records (and the Japan values in the source table are estimated rather than direct landings), I use it only as corroborative evidence for broad-scale synchrony and do not adopt it as the primary index for the subsequent analyses.

For glass eels, I also document the relationship between catch and CPUE using Shizuoka, where both series are available (Fig.~\ref{fig:shizuoka_catch_cpue}), providing an empirical check on the interpretability of harvest-based indices.
Across the major glass-eel fishing prefectures with available Council records, the across-prefecture mean scoop-net effort shows only a modest net decline over 1977--1997 (Fig.~\ref{fig:scopenet_effort}), whereas the inland-lake effort proxy for inland-eel fisheries shows a pronounced long-term decline over 1983--2013 (Fig.~\ref{fig:fme_effort}).
This appendix also specifies a simple stage-specific rule for translating catch weight to numbers of individuals (0.2~g for glass eels; 200~g for yellow and silver eels); because this conversion is a constant rescaling, it affects only the abundance-scale calibration (and hence $x_d$) and is applied where $x_d$ is constructed in Appendix~\ref{appendix:eel_pva_sensitivity}.

Several data-handling choices are deliberately conservative in the sense of avoiding optimistic imputation: inland seed statistics can include elvers historically (notably around 1960), which would tend to steepen apparent long-term declines rather than mask them; symbol-coded entries in prefectural seed tables are treated as missing (not as true zero) to avoid conflating non-response with absence of catch; and for components with comparatively complete tabulation, official national totals are used directly (Section~\ref{sec:S12_6_shared_national_series}).

I also compile available effort records for glass-eel and inland lake fisheries only to calibrate plausible magnitudes of effort change for sensitivity analyses, rather than to force a single deterministic effort correction (Section~\ref{sec:S12_7_effort_measures}).
Together, the two-stage harvest indices and their cross-checks provide a coherent and precautionary basis for the trend characterization in the main text and for the PVA sensitivity analyses reported in Appendix~\ref{appendix:eel_pva_sensitivity}.
 
\section{Criterion E thresholds and generation time}\label{appendix:generation_time}
\setcounter{table}{0}

\subsection{Criterion E thresholds and evaluation horizons}

The IUCN Red List includes Criterion E, which relies on a quantitative analysis of extinction probability \citep{IUCN:2012aa}.
Under this criterion, a species is categorized as \textit{Critically Endangered} (CR) if the probability of extinction in the wild is at least 50\% within 10~years or three generations, whichever is longer, up to a maximum of 100~years.
A species qualifies as \textit{Endangered} (EN) if the probability of extinction is at least 20\% within 20~years or five generations, whichever is longer, with the same 100-year maximum.
For the \textit{Vulnerable} (VU) category, the threshold is a probability of at least 10\% within 100~years.

Given the generation time estimated below, three generations equal 25.5~years and five generations equal 42.5~years for the Japanese eel.
Extinction probabilities are therefore evaluated over 25.5, 42.5, and 100~years in accordance with Criterion E.

\subsection{Generation time estimation}

Generation time is a key demographic parameter with several alternative definitions.
In age-structured models, one common measure is the mean age of the parents of the offspring produced by a cohort over its lifetime \citep{caswell2001matrix,coale1972growth}.
This cohort-based measure is equivalent to the average parental age weighted by fertility, and in a stationary population it coincides with the mean parental age at the stable age distribution.
Such measures are conceptually distinct from definitions based on the time required for a population to increase by its net reproductive rate, but they generally provide consistent values when survival is high and growth rates are near unity.

For the Japanese eel \textit{Anguilla japonica}, generation time is complicated by the pronounced difference in maturation age between sexes and by geographic variation in growth.
Males typically mature between 4 and 10~years and females between 3 and 17~years, with mean estimates around 7--10~years depending on locality \citep{Kotake:2005aa,Kotake:2007aa,yokouchi2009biological}.
Comparable studies on the European eel \textit{A. anguilla} have demonstrated geographic variation in both mean age and length at metamorphosis, reflecting fixed developmental trajectories expressed under different environmental conditions rather than adaptive plasticity \citep{VOLLESTAD:1992aa}.
These findings suggest that geographic heterogeneity in age at silvering is to be expected also for the Japanese eel, and that estimates of generation time should integrate across sites to obtain a representative mean.

In this study, I estimated generation time from mean ages of female silver eels collected at multiple sites, weighted by sample size, with an additional $\approx 0.5$~years to account for the estimated migration period from continental waters to the spawning grounds.
The weighted mean age across sites was $\approx 8$~years; adding $\approx 0.5$~years for migration gives $\approx 8.5$~years.
For Criterion E calculations, a generation time of 8.5~years was adopted, yielding evaluation horizons of 25.5, 42.5, and 100~years.
The underlying data are summarized in Table~\ref{tab:silver_eel_age}.

\begin{table}[H]
\centering
\begin{threeparttable}
\caption{Mean age of female silver eels at different locations.}
\label{tab:silver_eel_age}
\begin{tabularx}{\linewidth}{>{\raggedright\arraybackslash}X l c c >{\raggedright\arraybackslash}X}
\toprule
Location & Period & Mean age (yr) & $n$ & Source \\
\midrule
Kaoping River, Taiwan & 1998--2003 & 5.95 & 42 & \citet{Han:2009aa} \\
Amakusa Islands, Japan & -- & 7.80 & 19 & \citet{Kotake:2007aa} \\
Mikawa Bay, Japan & -- & 8.10 & 84 & \citet{Kotake:2007aa} \\
Sanriku Coast, Japan & -- & 8.80 & 5 & \citet{Kotake:2007aa} \\
Mikawa Bay, Japan & Oct--Dec & 7.90 & 132 & \citet{Kotake:2005aa} \\
Hamana Lake, Japan & 2003--2007 & 9.90 & 78 & \citet{yokouchi2009biological} \\
\bottomrule
\end{tabularx}
\begin{tablenotes}[flushleft]\footnotesize
\item $n$ is the sample size (number of individuals).
\end{tablenotes}
\end{threeparttable}
\end{table}

\section{Details of the PVA and sensitivity analyses for Japanese eel}
\label{appendix:eel_pva_sensitivity}
\setcounter{equation}{0}
\setcounter{figure}{0}
\setcounter{table}{0}

This appendix reports Criterion~E PVA estimates for Japanese eel using the two adopted national harvest time series constructed in Appendix~\ref{appendix:eel_harvest}.
For each time series, I evaluate two plug-in strategies for drift--Wiener inference: the naive drift--Wiener likelihood estimator and the observation-error-and-autocovariance-robust (OEAR) effective-diffusion estimator (Appendix~\ref{appendix:obs_error}).
Section~\ref{sec:eel_estimation} in the main text presents only the OEAR-based PVA summaries, whereas the naive-based results are reported here for completeness and comparison.
I then report two sensitivity analyses: (i) deterministic rescaling under sustained proportional effort trends of empirically observed magnitude (distinct from CPUE slope nonlinearity; Appendix~\ref{appendix:obs_error}) and (ii) abundance-scale calibration sensitivity in which $x_d$ is replaced by external scale estimates derived from glass-eel catch totals in the Joint Press Release of the Eighteenth Meeting of the Informal Consultation for the 2019--2020 fishing year (Japan-reported domestic catch, defined as intake minus imports, and the East Asia-wide total).

\subsection{Baseline PVA estimates from the two national series}
\label{appendix:eel_pva_sensitivity_pva}

Appendix~\ref{appendix:eel_harvest} defines two adopted national harvest time series for Japanese eel (1957--\allowbreak2020): glass eel (seed) and inland eel (yellow and silver).
As described in Appendix~\ref{appendix:eel_harvest}, harvest (tonnes) is converted to an individual-scale index using stage-specific mean body mass (0.2~g for the glass-eel (seed) series and 200~g for the inland-eel (yellow and silver) series) to calibrate the threshold-distance term $x_d$ in the drift--Wiener model.

For each series, I apply the drift--Wiener extinction-risk method developed in the main text, evaluating CR, EN, and VU horizons ($t^\ast=25.5,42.5,100$ years; the first two use a generation time of 8.5~years from Appendix~\ref{appendix:generation_time}, and 100 years follows the IUCN upper bound).
Two plug-in strategies are reported: the naive drift--Wiener plug-in applied directly to the annual series and the OEAR effective-diffusion plug-in (Appendix~\ref{appendix:obs_error}), which replaces $\sigma^2$ by a long-run-variance-based diffusion-scale estimate to improve robustness to additive observation error and short-run dependence.
Estimates are summarized in Tables~\ref{tab:estimates_naive} (naive) and \ref{tab:estimates_oear_diag} (OEAR). For both adopted series, the estimated extinction probabilities are far below the Criterion~E thresholds at the CR/EN/VU horizons.

\begin{table}[H]
\centering
\caption{Estimated extinction probabilities (point estimates and 95\% CIs) for the Japanese eel (naive method).}
\label{tab:estimates_naive}

\resizebox{\textwidth}{!}{
\begingroup
\renewcommand{\arraystretch}{3}
\begin{tabular}{ccccccc}
\hline\hline
Time series & $\widehat{\mu}$ & $\widehat{\sigma}^{2}$ & $x_d$ & CR: $\widehat{G}(t^\ast=25.5)$ & EN: $\widehat{G}(t^\ast=42.5)$ & VU: $\widehat{G}(t^\ast=100)$\\
\hline
\shortstack{Glass eel} & \shortstack{-0.052 \\ (-0.124, 0.020)} & \shortstack{0.081 \\ (0.059, 0.121)} & 17.6 & \shortstack{$10^{-29}$ \\ ($9 \times 10^{-42}$, $5 \times 10^{-19}$)} & \shortstack{$10^{-16}$ \\ ($4 \times 10^{-25}$, $10^{-9}$)} & \shortstack{$10^{-5}$ \\ ($7 \times 10^{-12}$, $5 \times 10^{-2}$)}\\
\shortstack{Yellow and\\silver eel} & \shortstack{-0.059 \\ (-0.089, -0.029)} & \shortstack{0.014 \\ (0.010, 0.020)} & 12.6 & \shortstack{$2 \times 10^{-79}$ \\ ($2 \times 10^{-109}$, $3 \times 10^{-52}$)} & \shortstack{$2 \times 10^{-40}$ \\ ($10^{-57}$, $3 \times 10^{-25}$)} & \shortstack{$5 \times 10^{-9}$ \\ ($5 \times 10^{-17}$, $10^{-3}$)}\\
\hline
\end{tabular}
\endgroup
}

\vspace{4pt}
\caption*{\footnotesize\textit{Note.} ML estimates (hats) are shown with 95\% CIs in parentheses. Units: $\widehat{\mu}$ and $\widehat{\sigma}^{2}$ in year$^{-1}$. Let $n_t$ denote the individual-scale index constructed from harvest (Appendix~\ref{appendix:eel_harvest}), and let $y_t:=\log n_t$. Let $n_e$ denote the quasi-extinction threshold and set $x_e:=\log n_e$. Under the naive no-error assumption, the plug-in threshold distance is $\check{x}_d:=y_q-x_e$ with $y_q:=\log n_{t_q}$, equivalently $\check{x}_d=\log(n_{t_q}/n_e)$, and is treated as known because $n_0=n_{t_q}$. Extinction probability is $\widehat{G}(t^\ast)=\Pr[T\le t^\ast]$ with $t^\ast\in\{25.5,\,42.5,\,100\}$ years for CR, EN, and VU. Extinction probabilities are reported in scientific notation. }
\end{table}

\begin{table}[H]
\centering
\caption{Estimated extinction probabilities (point estimates and 95\% CIs) for the Japanese eel (OEAR method).}
\label{tab:estimates_oear_diag}

\resizebox{\textwidth}{!}{
\begingroup
\renewcommand{\arraystretch}{3}
\begin{tabular}{ccccccc}
\hline\hline
Time series & $\widehat{\mu}$ & $\widetilde{\sigma}^{2}$ & $\check{x}_d$ & CR: $\widetilde{G}(t^\ast=25.5)$ & EN: $\widetilde{G}(t^\ast=42.5)$ & VU: $\widetilde{G}(t^\ast=100)$\\
\hline
\shortstack{Glass eel} & \shortstack{-0.052 \\ (-0.124, 0.020)} & \shortstack{0.016 \\ (0.012, 0.024)} & 17.6 & \shortstack{$4 \times 10^{-140}$ \\ ($9 \times 10^{-191}$, $9 \times 10^{-93}$)} & \shortstack{$5 \times 10^{-76}$ \\ ($5 \times 10^{-106}$, $6 \times 10^{-49}$)} & \shortstack{$3 \times 10^{-22}$ \\ ($3 \times 10^{-36}$, $3 \times 10^{-11}$)}\\
\shortstack{Yellow and\\silver eel} & \shortstack{-0.059 \\ (-0.089, -0.029)} & \shortstack{0.016 \\ (0.011, 0.023)} & 12.6 & \shortstack{$4 \times 10^{-70}$ \\ ($10^{-96}$, $6 \times 10^{-46}$)} & \shortstack{$9 \times 10^{-36}$ \\ ($3 \times 10^{-51}$, $4 \times 10^{-22}$)} & \shortstack{$4 \times 10^{-8}$ \\ ($10^{-15}$, $4 \times 10^{-3}$)}\\
\hline
\end{tabular}
\endgroup
}

\vspace{4pt}
\caption*{\footnotesize\textit{Note.} Point estimates are shown with 95\% CIs in parentheses. Here $\widehat{\mu}$ is the ML estimate of the drift; for OEAR (Appendix~\ref{appendix:obs_error}), $\widetilde{\sigma}^{2}$ and $\widetilde{G}$ use tildes, with $\widetilde{\mu}=\widehat{\mu}$. Units: $\widehat{\mu}$ and $\widetilde{\sigma}^{2}$ in year$^{-1}$. Let $y_t:=\log n_t$ denote the log index constructed from harvest (Appendix~\ref{appendix:eel_harvest}). The plug-in threshold distance is $\check{x}_d:=y_q-\log n_e$ with $y_q:=\log n_{t_q}$, equivalently $\check{x}_d=\log(n_{t_q}/n_e)$, and is treated as known because $n_0=n_{t_q}$. Extinction probability is $\widetilde{G}(t^\ast)=\Pr[T\le t^\ast]$ with $t^\ast\in\{25.5,\,42.5,\,100\}$ years for CR, EN, and VU. Extinction probabilities are reported in scientific notation. OEAR diagnostics (AR(1) pre-whitening): Glass eel: $\tilde{\rho}_{\mathrm{pw}}=-0.39$ ($J=4$); Yellow and silver eel: $\tilde{\rho}_{\mathrm{pw}}=0.07$ ($J=1$). The 95\% CI for $\widetilde{\sigma}^{2}$ is a pragmatic approximation obtained by applying the naive drift--Wiener CI formula to $\widetilde{\sigma}^{2}$. }
\end{table}

\subsection{Sensitivity analyses}
\label{appendix:eel_pva_sensitivity_sensitivity}

\subsubsection{Effort-change sensitivity}
\label{appendix:eel_pva_sensitivity_effort}

Because catch can vary with both abundance and fishing effort, I evaluate a deterministic effort-trend sensitivity by rescaling each individual-scale series prior to applying the same PVA pipeline.
I treat each harvest index as proportional to abundance multiplied by a relative effort component and consider log-linear effort trends anchored at 2020.
For a given scenario in Table~\ref{tab:eel_effort_scenarios}, I divide the seed-series index by $\exp\!\left(\frac{\varphi_{\mathrm{seed}}}{100}(t-2020)\right)$ and the inland-eel series by $\exp\!\left(\frac{\varphi_{\mathrm{eel}}}{100}(t-2020)\right)$, where $\varphi_{\mathrm{seed}}$ and $\varphi_{\mathrm{eel}}$ are annual percentage rates.
Because the trends are anchored at 2020, the rescaling factor equals 1 at $t=t_q=2020$; therefore the terminal index value $n_{t_q}$ (and hence the threshold-distance calibration $x_d=\log(n_{t_q}/n_e)$) is unchanged across scenarios, and this sensitivity analysis isolates the effect of plausible sustained effort trends on the inferred drift and diffusion from each rescaled series.
Negative rates represent declining effort through time (historically larger effort in earlier years).

The reference annual rates are taken from Section~\ref{sec:S12_7_effort_measures}.
For glass eel, the effort trend is $-0.63\%$\,yr$^{-1}$ (scoop-net participants; 1977--1997).
For inland eel, the effort trend is $-3.95\%$\,yr$^{-1}$ (eel-targeting FMEs; 1983--2013).
I evaluate four scenarios (Table~\ref{tab:eel_effort_scenarios}): no trend, an observed-rate summary, and two decline variants that bracket empirically observed variation.
For seed, S2 is set to half of the reference decline and S3 is set as a stronger decline within the observed prefecture-level spread reported in Section~\ref{appendix:effort_glass_eel}.
For inland eel, S2 and S3 correspond to the upper and lower bounds of the approximate 95\% log-linear interval reported in Section~\ref{appendix:effort_lakes}.

\begin{table}[H]
\centering
\caption{Effort-trend scenarios used in the PVA sensitivity analysis.}
\label{tab:eel_effort_scenarios}
\begin{tabular}{lcc}
\hline
Scenario & $\varphi_{\mathrm{seed}}$ (\%/year) & $\varphi_{\mathrm{eel}}$ (\%/year) \\
\hline
S0: No effort trend & 0.00 & 0.00 \\
S1: Reference (observed-rate summary) & $-0.63$ & $-3.95$ \\
S2: Mild decline (half of reference) & $-0.315$ & $-2.01$ \\
S3: Strong decline (within observed spread) & $-2.25$ & $-5.95$ \\
\hline
\end{tabular}
\end{table}

For each scenario, both series are analyzed with the same \texttt{naive}/\texttt{oear} settings and CR/EN/VU horizons as in Tables~\ref{tab:estimates_naive}--\ref{tab:estimates_oear_diag}.

Table~\ref{tab:eel_effort_sensitivity_oear_vu} summarizes the resulting VU-horizon extinction probabilities under OEAR.
Among the scenarios considered, the no-trend case (S0) is the most pessimistic; incorporating empirically observed effort declines reduces assessed risk.
In all scenarios, both point estimates and the upper 95\% CI bounds remain below $0.1$.

\begin{table}[H]
\centering
\caption{Effort-trend sensitivity of VU extinction probability estimates under OEAR ($t^\ast=100$ years). Seed denotes the glass-eel (seed) series and inland denotes the inland-eel (yellow and silver) series.}
\label{tab:eel_effort_sensitivity_oear_vu}

\setlength{\tabcolsep}{4pt}
\renewcommand{\arraystretch}{1.15}
\small

\resizebox{\linewidth}{!}{
\begin{tabular}{lcc}
\hline
Scenario & Seed: $\widetilde{G}(100)$ & Inland: $\widetilde{G}(100)$ \\
\hline
S0: No effort trend &
\shortstack{$3 \times 10^{-22}$ \\ ($3 \times 10^{-36}$, $3 \times 10^{-11}$)} &
\shortstack{$4 \times 10^{-8}$ \\ ($10^{-15}$, $4 \times 10^{-3}$)} \\
S1: Reference (observed-rate summary) &
\shortstack{$2 \times 10^{-24}$ \\ ($3 \times 10^{-39}$, $10^{-12}$)} &
\shortstack{$8 \times 10^{-18}$ \\ ($7 \times 10^{-30}$, $10^{-8}$)} \\
S2: Mild decline (half of reference) &
\shortstack{$3 \times 10^{-23}$ \\ ($10^{-37}$, $7 \times 10^{-12}$)} &
\shortstack{$2 \times 10^{-12}$ \\ ($3 \times 10^{-22}$, $2 \times 10^{-5}$)} \\
S3: Strong decline (within observed spread) &
\shortstack{$2 \times 10^{-30}$ \\ ($2 \times 10^{-47}$, $3 \times 10^{-16}$)} &
\shortstack{$2 \times 10^{-24}$ \\ ($4 \times 10^{-39}$, $2 \times 10^{-12}$)} \\
\hline
\end{tabular}
}
\end{table}

\subsubsection{Abundance-scale calibration sensitivity}
\label{appendix:eel_pva_sensitivity_xd}

Because catch represents only a harvested fraction of the underlying abundance, calibrating the threshold distance $x_d$ from catch is precautionary.
Throughout this section, I convert glass-eel catch totals to individual equivalents using the same 0.2~g mean body mass (Appendix~\ref{appendix:eel_harvest}) and set $x_e=\log n_e=0$, so that $x_d=\log(n_{t_q}/n_e)$ is determined by the terminal index scale.

Under this baseline calibration, the adopted national seed series has 8.6~t in 2020, giving $x_d\approx 17.6$.
The Informal Consultation table reports larger glass-eel totals for 2019--2020 (Japan domestic catch: 17.1~t; East Asia total: 76.9~t), which imply $x_d\approx 18.3$ and $x_d\approx 19.8$, respectively.
For reference, in 2024--25 the East Asia total excluding China (Japan, Korea, and Chinese Taipei) is 26.6807~t, corresponding to $x_d\approx 18.7$ (Table~\ref{tab:informal_consultation_4regions_catch}).

I therefore evaluate abundance-scale calibration sensitivity by recomputing risk summaries after replacing only the baseline $x_d$ (17.6) with these externally calibrated values, while holding the drift and diffusion estimates from each adopted national series fixed.

\begin{table}[H]
\centering
\caption{Abundance-scale calibration sensitivity for the glass-eel series: extinction probabilities after replacing the threshold-distance calibration $x_d$ while keeping drift and diffusion estimates fixed.}
\label{tab:eel_xd_sensitivity}
\setlength{\tabcolsep}{4pt}
\renewcommand{\arraystretch}{1.15}
\small
\resizebox{\linewidth}{!}{
\begin{tabular}{llccccc}
\hline
Method & Time series & Calibration & $x_d$ & CR: $G(25.5)$ & EN: $G(42.5)$ & VU: $G(100)$\\
\hline
Naive & \shortstack{Glass eel} & Baseline & 17.6 & \shortstack{$10^{-29}$ \\ ($9 \times 10^{-42}$, $5 \times 10^{-19}$)} & \shortstack{$10^{-16}$ \\ ($4 \times 10^{-25}$, $10^{-9}$)} & \shortstack{$10^{-5}$ \\ ($7 \times 10^{-12}$, $5 \times 10^{-2}$)}\\
OEAR & \shortstack{Glass eel} & Baseline & 17.6 & \shortstack{$4 \times 10^{-140}$ \\ ($9 \times 10^{-191}$, $9 \times 10^{-93}$)} & \shortstack{$5 \times 10^{-76}$ \\ ($5 \times 10^{-106}$, $6 \times 10^{-49}$)} & \shortstack{$3 \times 10^{-22}$ \\ ($3 \times 10^{-36}$, $3 \times 10^{-11}$)}\\
Naive & \shortstack{Glass eel} & Japan (17.1 t) & 18.3 & \shortstack{$4 \times 10^{-32}$ \\ ($5 \times 10^{-45}$, $10^{-20}$)} & \shortstack{$4 \times 10^{-18}$ \\ ($5 \times 10^{-27}$, $10^{-10}$)} & \shortstack{$4 \times 10^{-6}$ \\ ($10^{-12}$, $3 \times 10^{-2}$)}\\
OEAR & \shortstack{Glass eel} & Japan (17.1 t) & 18.3 & \shortstack{$5 \times 10^{-152}$ \\ ($8 \times 10^{-207}$, $10^{-100}$)} & \shortstack{$8 \times 10^{-83}$ \\ ($3 \times 10^{-115}$, $2 \times 10^{-53}$)} & \shortstack{$10^{-24}$ \\ ($2 \times 10^{-39}$, $10^{-12}$)}\\
Naive & \shortstack{Glass eel} & East Asia (76.9 t) & 19.8 & \shortstack{$8 \times 10^{-38}$ \\ ($10^{-52}$, $2 \times 10^{-24}$)} & \shortstack{$2 \times 10^{-21}$ \\ ($2 \times 10^{-31}$, $10^{-12}$)} & \shortstack{$2 \times 10^{-7}$ \\ ($2 \times 10^{-14}$, $9 \times 10^{-3}$)}\\
OEAR & \shortstack{Glass eel} & East Asia (76.9 t) & 19.8 & \shortstack{$9 \times 10^{-180}$ \\ ($2 \times 10^{-244}$, $2 \times 10^{-119}$)} & \shortstack{$10^{-98}$ \\ ($8 \times 10^{-137}$, $5 \times 10^{-64}$)} & \shortstack{$4 \times 10^{-30}$ \\ ($3 \times 10^{-47}$, $4 \times 10^{-16}$)}\\
\hline
\end{tabular}
}
\vspace{4pt}
\caption*{\footnotesize\textit{Note.} Naive rows report $\widehat{G}(t^\ast)$ and OEAR rows report $\widetilde{G}(t^\ast)$. Baseline uses the adopted national glass-eel series terminal-scale calibration, while Japan/East Asia calibrations set $x_d$ from 17.1~t and 76.9~t glass-eel catch equivalents (0.2~g conversion), respectively. The yellow-and-silver-eel series is not recalibrated in this analysis. Probabilities are rounded in scientific notation to 1 significant digit.}
\end{table}

Table~\ref{tab:eel_xd_sensitivity} summarizes abundance-scale calibration sensitivity for the glass-eel (seed) series.
As expected under the drift--Wiener hitting-probability map, increasing the threshold-distance term from the baseline $x_d=17.6$ to $18.3$ (Japan-reported domestic catch) and $19.8$ (East Asia total) decreases estimated extinction probabilities at all horizons, because the quasi-extinction threshold is placed farther from the terminal log-index while the fitted drift and diffusion are held fixed.
Across CR/EN/VU horizons, however, the qualitative conclusion is unchanged: risks remain far below the Criterion~E thresholds under both plug-in strategies.
For example, under the naive estimator, $G(100)$ decreases from $10^{-5}$ at $x_d=17.6$ to $4\times 10^{-6}$ at $x_d=18.3$ and $2\times 10^{-7}$ at $x_d=19.8$, whereas under OEAR it decreases from $3\times 10^{-22}$ to $10^{-24}$ and $4\times 10^{-30}$, respectively.
For reference, the 2024--25 East Asia total excluding China corresponds to $x_d\approx 18.7$ (Table~\ref{tab:informal_consultation_4regions_catch}), which lies between the Japan-based ($x_d=18.3$) and 2019--2020 East Asia-based ($x_d=19.8$) calibrations; its implied risk therefore lies between those two rows and remains far below the Criterion~E thresholds under both plug-in strategies.
Thus, the catch-derived terminal-scale calibration is precautionary in the sense that it yields the largest (most conservative) risk estimates among the calibrations examined, and the inference is robust to plausible alternative abundance-scale calibrations based on the Informal Consultation totals.

\subsection{Precautionary handling and robustness summary}
\label{appendix:eel_pva_sensitivity_precaution}

Some components of the workflow are conservative in the sense that they tend not to understate risk, while other items correspond to common departures from idealized drift--Wiener assumptions.
Table~\ref{tab:precautionary_summary} summarizes these points, separating (A) precautionary elements and (B) robustness checks within the table.

\begin{table}[H]
\centering
\caption{Precautionary elements and robustness checks for the Criterion~E assessment.}
\label{tab:precautionary_summary}
\begin{threeparttable}

\setlength{\tabcolsep}{6pt}
\renewcommand{\arraystretch}{1.15}

\begin{tabular}{>{\raggedright\arraybackslash}p{0.33\textwidth}>{\raggedright\arraybackslash}p{0.55\textwidth}}
\hline
Component & Implication / evidence \\
\hline

\multicolumn{2}{l}{\textit{A. Precautionary elements (tending not to understate risk)}}\\
\hline
Historical elver inclusion in some seed-harvest records (around the 1960s) &
Tends to steepen apparent declines in the seed index rather than mask them, and therefore tends to increase assessed risk (Appendix~\ref{appendix:eel_harvest}). \\
\hline
Index-scale calibration of the threshold distance ($x_d$) &
Because catch represents only a harvested fraction/proxy of the stock, the catch-derived terminal-scale calibration of $x_d=\log(n_{t_q}/n_e)$ is a lower-bound scale calibration and is therefore conservative.
Using larger external glass-eel catch totals increases $x_d$ and decreases assessed risk (Table~\ref{tab:eel_xd_sensitivity}). \\
\hline
No effort correction as the baseline (S0) &
Among the empirically grounded effort-trend scenarios examined, the no-trend baseline (S0) is the most pessimistic; incorporating empirically observed effort declines reduces assessed risk (Table~\ref{tab:eel_effort_sensitivity_oear_vu}). \\
\hline

\multicolumn{2}{l}{\textit{B. Robustness checks (qualitative conclusion unchanged)}}\\
\hline
Estimator choice (naive vs OEAR) and short-run dependence / observation error &
For both adopted national series, extinction probabilities are far below the Criterion~E thresholds under both plug-in strategies (Tables~\ref{tab:estimates_naive}--\ref{tab:estimates_oear_diag}).
OEAR is designed to be robust to additive observation error and short-run dependence by targeting an effective diffusion scale (Appendix~\ref{appendix:obs_error}). \\
\hline
Colored noise and weak density feedback &
Plausible impacts of colored noise and weak density feedback are discussed, and the examined regimes do not reverse threshold-scale decisions (Appendix~\ref{appendix:env_noise_density}). \\
\hline
\end{tabular}

\end{threeparttable}
\end{table}
 
\section{Proof: Criterion A (decline subcriteria) overstates Criterion E for large populations}
\label{appendix:A_over_E_drift_wiener}
\setcounter{equation}{0}

\subsection{Motivation (why compare IUCN Criteria A and E)}

In the original version~1.0 framework of \citet{mace1991assessing}, threat categories were defined by extinction probability over a specified time horizon, and quantities such as population reduction were introduced as practical proxies for that risk.
During subsequent consolidation, however, the version~2.0 drafting group treated the decline rate proxy as an independently sufficient criterion, while explicitly noting a potential inconsistency: because it was not linked to any minimum population size, it could list some ``very large, apparently secure'' populations.
The rationale offered was that the included decline rates should raise concern for almost all populations, and that linking the decline criterion to population size would exclude many taxa with limited census data \citep{mace1992development}.
In addition, \citet{mace2008quantification} appears to overgeneralize theory when stating that, under deterministic exponential decline, population size has little effect on extinction risk, citing \citet{lande1993risks}, even though the \emph{mean time} to reach an extinction threshold increases with initial population size (albeit only logarithmically).
This is a statement about time-to-threshold scaling, not about \emph{extinction probability} within a fixed time horizon; under stochastic dynamics the dependence on initial population size can be stronger.

The current version~3.1 is explicitly framed as an extinction risk classification: extinction is a chance process, and listing in a higher category implies a higher expectation of extinction over the specified time frames \citep{IUCN:2012aa}.
Although the quantitative thresholds in Criteria A to E were set separately, they were calibrated against a common standard and broad consistency among criteria was sought \citep{IUCN:2012aa}.
Consistent with this framing, \citet{mace2008quantification} states that the Red List categories are intended to reflect the likelihood of a species going extinct under prevailing circumstances.
Operationally, the Red List criteria are applied disjunctively: meeting any one of the criteria qualifies a taxon for listing at that level of threat, and the final category is set by the highest criterion met \citep{IUCN:2012aa}.
This matters because, even if Criterion~E yields an accurate estimate of extinction risk under a validated stochastic model, any systematic overestimation by Criterion~A for large initial populations will drive the final category upward under the disjunctive listing rule.

Motivated by this structural and theoretical tension, this appendix provides a formal comparison under a drift Wiener model.
Accordingly, in this appendix the term ``true extinction risk'' refers to the probability implied by the model, $G(t\mid x_d,\mu,\sigma^2)$.
Beyond a sufficiently large initial population size, the decline subcriteria of Criterion~A (a and b) can systematically overestimate the extinction risk quantified by Criterion~E.
To give the result a concrete scale, I also provide a numerical illustration that solves for the initial size at which the Criterion~E equality holds for the CR, EN, and VU thresholds under a simple calibration.

Criterion~A also includes subcriteria based on reductions in habitat extent or quality (e.g., A1c--A2c).
In principle, such habitat-based threats can also be mapped to extinction risk under explicit stochastic population models. For example, \citet{hakoyama2000comparing} compares different risk factors via their effects on the mean time to extinction.
Extending the present comparison to habitat reduction is feasible in the same spirit, but is beyond the scope of this appendix, which restricts attention to decline-based thresholds.

This dependence on initial population size is increasingly salient in recent applications that evaluate extinction risk for extremely large populations, for example in some widely distributed marine taxa, because across such ranges even a logarithmic dependence on initial size can yield nontrivial differences in implied extinction risk.

\subsection{Setup}

Let $X(t)=\log N(t)$ follow the drift Wiener model
\begin{equation*}
  dX(t)=\mu\,dt+\sigma\,dW(t),
  \qquad \mu\in\mathbb R,\ \sigma\in(0,\infty).
\end{equation*}
Under the drift Wiener model, the extinction probability by time $t>0$ is
\begin{equation}
  \Pr[T\le t] = G(t\mid x_d,\mu,\sigma^2)
  =
  \Phi\!\left(-\frac{x_d+\mu t}{\sigma\sqrt t}\right)
  +
  \exp\!\left(-\frac{2\mu x_d}{\sigma^2}\right)
  \Phi\!\left(-\frac{x_d-\mu t}{\sigma\sqrt t}\right),
  \label{Seq:hitting_prob_xd}
\end{equation}
where $T=\inf\{t\ge0\mid X(t)\le x_e\}$, $x_e$ is the fixed log threshold,
$x_0$ is the initial value of $X(t)$, and $x_d=x_0-x_e>0$.

\subsection{Criterion E risk vanishes for large initial populations}

For fixed $(\mu,\sigma^2)$ and fixed $t>0$,
\begin{equation}
  \lim_{x_d\to\infty} G(t\mid x_d,\mu,\sigma^2)=0.
  \label{Seq:G_to_zero_app}
\end{equation}

\paragraph{Proof.}
From \eqref{Seq:hitting_prob_xd}, write
\begin{equation*}
  G(t\mid x_d,\mu,\sigma^2)=G_1(x_d)+G_2(x_d),
\end{equation*}
where
\begin{equation*}
  G_1(x_d)=\Phi\!\left(-\frac{x_d+\mu t}{\sigma\sqrt t}\right),
  \qquad
  G_2(x_d)=\exp\!\left(-\frac{2\mu x_d}{\sigma^2}\right)
           \Phi\!\left(-\frac{x_d-\mu t}{\sigma\sqrt t}\right).
\end{equation*}
As $x_d\to\infty$, the argument of $\Phi$ in $G_1(x_d)$ tends to $-\infty$, hence $G_1(x_d)\to0$.

For $G_2(x_d)$, set $v=(x_d-\mu t)/(\sigma\sqrt t)$.
Then $v\to\infty$ as $x_d\to\infty$, so in particular $v>0$ for all sufficiently large $x_d$.
For such $x_d$, the Mills' inequality $\Phi(-v)\le \phi(v)/v$ applies, so
\begin{equation*}
  0\le G_2(x_d)\le \exp\!\left(-\frac{2\mu x_d}{\sigma^2}\right)\frac{\phi(v)}{v}.
\end{equation*}
Since $\phi(v)=\exp(-v^2/2)/\sqrt{2\pi}$ and
\begin{equation*}
  \frac{v^2}{2}=\frac{(x_d-\mu t)^2}{2\sigma^2 t}
  =\frac{x_d^2}{2\sigma^2 t}-\frac{\mu}{\sigma^2}x_d+\frac{\mu^2 t}{2\sigma^2},
\end{equation*}
the exponent in the bound for $G_2(x_d)$ contains the dominant term $-x_d^2/(2\sigma^2 t)\to-\infty$.
Thus $G_2(x_d)\to0$, proving \eqref{Seq:G_to_zero_app}.

Moreover, the same bound shows that the decay is rapid: for $x_d>|\mu|t$,
Mills' inequality gives
\[
G(t\mid x_d,\mu,\sigma^2)
\le
\frac{\phi\!\left(\frac{x_d+\mu t}{\sigma\sqrt t}\right)}{\frac{x_d+\mu t}{\sigma\sqrt t}}
+\exp\!\left(-\frac{2\mu x_d}{\sigma^2}\right)
\frac{\phi\!\left(\frac{x_d-\mu t}{\sigma\sqrt t}\right)}{\frac{x_d-\mu t}{\sigma\sqrt t}},
\]
so $G(t\mid x_d,\mu,\sigma^2)=O\!\bigl(\exp\{-x_d^2/(2\sigma^2 t)\}\bigr)$ up to polynomial factors in $x_d$.

This does not contradict the ``log scaling'' emphasized by \citet{lande1993risks}, which concerns the scaling of the (mean or typical) time to reach a threshold with initial abundance.
Here the point is different: for a \emph{fixed} finite horizon $t$, the hitting probability is governed by a normal tail, yielding the rapid decay in $x_d$.

\subsection{Decline-based Criterion A ignores \texorpdfstring{$x_d$}{xd}}

Fix a reporting horizon $t^\ast>0$. Under the drift Wiener model,
\begin{equation*}
  X(t^\ast)-X(0)=\mu t^\ast+\sigma W(t^\ast),
\end{equation*}
so the realized proportional change over $[0,t^\ast]$ is
\begin{equation*}
  \frac{N(t^\ast)}{N(0)}=\exp\{X(t^\ast)-X(0)\},
\end{equation*}
a function only of the increment $X(t^\ast)-X(0)$.
Therefore any decline-based rule that depends only on the realized proportional change over $[0,t^\ast]$
is invariant to translating the initial level $x_0$, and hence independent of $x_d=x_0-x_e$.

\subsection{Overestimation beyond a large population threshold}

Let $G_{\mathrm{thr}}\in\{0.5,0.2,0.1\}$ denote the Criterion~E thresholds.
Fix $(\mu,\sigma^2)$ and $t^\ast>0$.
If a decline-based rule over $[0,t^\ast]$ is triggered by an observed decline, that trigger does not change with $x_d$ by the previous subsection.

By \eqref{Seq:G_to_zero_app} with $t=t^\ast$, for each $G_{\mathrm{thr}}$ there exists $x_d^\ast>0$ such that
\begin{equation*}
  x_d\ge x_d^\ast
  \quad\Longrightarrow\quad
  G(t^\ast\mid x_d,\mu,\sigma^2) < G_{\mathrm{thr}}.
\end{equation*}
Thus, increasing $x_d$ does not attenuate the decline signal itself: any decline-based Criterion~A trigger defined in terms of the proportional change over $[0,t^\ast]$ remains possible under the same $(\mu,\sigma^2)$ because it depends only on the increment $X(t^\ast)-X(0)$.
In contrast, \eqref{Seq:G_to_zero_app} shows that the true risk $G(t^\ast\mid x_d,\mu,\sigma^2)$ can be made arbitrarily small by taking $x_d$ sufficiently large.  
Therefore, for sufficiently large initial populations, Criterion~A can be triggered by an observed decline even though the true extinction probability at the same horizon $t^\ast$ lies below the threshold $G_{\mathrm{thr}}$.
Moreover, this discrepancy is not tied to particular numerical choices: for any fixed threshold $G_{\mathrm{thr}}\in(0,1)$ and for any decline threshold used to trigger Criterion~A, there exists a sufficiently large $x_d$ such that $G(t^\ast\mid x_d,\mu,\sigma^2)<G_{\mathrm{thr}}$ while the Criterion~A trigger remains possible under the same $(\mu,\sigma^2)$.
This is the claimed systematic overestimation under the drift Wiener model.

\subsection{Numerical illustration}
\label{appendix:numerical_illustration_A_over_E}

This subsection provides a concrete scale for the log distance $x_d$ at which a decline-based Criterion~A signal and the quantitative Criterion~E thresholds diverge.
Time is measured in years.
Let $t_A>0$ denote the Criterion~A assessment horizon and let $t_E>0$ denote the Criterion~E horizon.
Let $f_A\in(0,1)$ denote the observed proportional threshold that triggers Criterion~A, so that Criterion~A is triggered when $N(t_A)/N(0)\le f_A$.
Let $G_{\mathrm{thr}}\in(0,1)$ denote the Criterion~E extinction probability threshold at horizon $t_E$.

To keep the illustration explicit, the year horizons and thresholds are fixed as follows.
For CR and EN, the chosen Criterion~E horizons correspond to the cases where the generation-length conditions do not extend the time horizon beyond $10$ and $20$~years, respectively.
For CR, set $(t_A,f_A,t_E,G_{\mathrm{thr}})=(10,0.2,10,0.5)$.
For EN, set $(t_A,f_A,t_E,G_{\mathrm{thr}})=(10,0.5,20,0.2)$.
For VU, set $(t_A,f_A,t_E,G_{\mathrm{thr}})=(10,0.7,100,0.1)$.

A representative drift $\mu$ is obtained by matching the median proportional change over the Criterion~A horizon to the corresponding threshold $f_A$.
Since $X(t_A)-X(0)\sim\mathcal N(\mu t_A,\sigma^2 t_A)$, the median of $N(t_A)/N(0)=\exp\{X(t_A)-X(0)\}$ is $\exp(\mu t_A)$.
Thus $\mu$ is set by
\begin{equation*}
  \exp(\mu t_A)=f_A,
  \qquad\text{that is}\qquad
  \mu=\frac{\log f_A}{t_A}.
\end{equation*}

For a given $\sigma$, define $x_d^\ast$ as the (numerically obtained) solution to the Criterion~E equality at the corresponding horizon,
\begin{equation*}
  G(t_E\mid x_d^\ast,\mu,\sigma^2)=G_{\mathrm{thr}},
\end{equation*}
where $G$ is given in \eqref{Seq:hitting_prob_xd}.
Since $G(t_E\mid x_d,\mu,\sigma^2)$ is continuous and strictly decreasing in $x_d$, with $G(t_E\mid 0,\mu,\sigma^2)=1$ and $\lim_{x_d\to\infty}G(t_E\mid x_d,\mu,\sigma^2)=0$, the solution exists and is unique for any $G_{\mathrm{thr}}\in(0,1)$.

Table~\ref{tab:A_over_E_numerical} reports $n_0^\ast$ (and the associated $x_d^\ast$) for representative values $\sigma=0.2$, $0.3$, and $0.5$, with $x_e=0$.
Under the calibration $\exp(\mu t_A)=f_A$, $n_0^\ast$ is the crossover initial size at which the Criterion~E equality holds; for larger initial sizes the corresponding extinction probability at horizon $t_E$ falls below $G_{\mathrm{thr}}$, consistent with the rapid decay in $x_d$ shown above.
Numerically, for $\sigma\in\{0.2,0.3,0.5\}$ this crossover occurs at $n_0^\ast=O(10)$ for CR, $n_0^\ast=O(10\text{ to }10^2)$ for EN, and $n_0^\ast=O(10^3\text{ to }10^5)$ for VU.

Therefore, A and E can agree when abundance is critically low (small $x_d$), but Criterion~A can systematically overstate extinction risk when applied to large populations.

\begin{table}[H]
\centering
\caption{Initial population size $n_0^\ast$ beyond which decline-based Criterion~A can overestimate the true extinction risk under the drift Wiener model.}
\label{tab:A_over_E_numerical}
\begin{threeparttable}
\renewcommand{\arraystretch}{1.3}
\begin{tabular}{lcccc}
\hline\hline
\multirow{2}{*}{Category} & \multirow{2}{*}{$\mu$} &
\multicolumn{3}{c}{$n_0^\ast$ ($x_0^\ast$ in parentheses)} \\
\cline{3-5}
& & $\sigma=0.2$ & $\sigma=0.3$ & $\sigma=0.5$ \\
\hline
CR & -0.16094 & 5.61 (1.7242) & 6.35 (1.8483) & 8.73 (2.1668) \\
EN & -0.06931 & 10.2 (2.3239) & 17.4 (2.8578) & 52.6 (3.9628) \\
VU & -0.03567 & 646 (6.4703) & $3.10\times10^{3}$ (8.0396) & $7.67\times10^{4}$ (11.2473) \\
\hline
\end{tabular}
\caption*{\footnotesize\textit{Note.}
Numerical illustration of $n_0^\ast$ defined by $G(t_E\mid x_d^\ast,\mu,\sigma^2)=G_{\mathrm{thr}}$.
Here the threshold is set to $x_e=0$ (equivalently $n_e=1$), hence $x_d^\ast=x_0^\ast$ and $n_0^\ast=\exp(x_0^\ast)$.
The drift $\mu$ is calibrated by $\mu=\log(f_A)/t_A$.
}
\end{threeparttable}
\end{table}

\section{Notation}
\label{appendix:notation}

\paragraph{Convention.}
Uppercase letters typically denote stochastic processes or random variables, and
lowercase letters denote realized values.
Exceptions (for derived quantities, deterministic transforms, and standard functions) are stated where they appear.

\subsection*{Main text}

\begin{description}

\item[$t$] Time: continuous time in $X(t)$ with $t\in[0,\infty)$, and a generic finite horizon in $G(t)$ and the $(w,z)$ transformation (thus $t>0$ there).

\item[$t_i$] Observation times, $0=t_0<t_1<\cdots<t_q$.

\item[$\tau_i$] Sampling intervals, $\tau_i=t_i-t_{i-1}>0$ for $i=1,\dots,q$.

\item[$q$] Number of increments (the series has length $q+1$).

\item[$t_q$] Observation span (final observation time).

\item[$t^\ast$] A specific finite horizon used for reporting risk (application) or for CI construction, with $t^\ast>0$.

\item[$N(t)$] Population size process on the original scale.

\item[$X(t)$] Log population size process, $X(t)=\log N(t)$.

\item[$n_{t_i}$] Observed population size at time $t_i$.

\item[$x_i$] Observed log population size at time $t_i$, $x_i=\log n_{t_i}$.

\item[$n_0$] Initial population size.

\item[$x_0$] Initial log population size, $x_0=\log n_{0}$.

\item[$\mathbf{n}_{\mathrm{obs}}$] Observed population size time series, $\mathbf{n}_{\mathrm{obs}}=\{n_{t_0}, n_{t_1}, \dots, n_{t_q}\}$.

\item[$\mathbf{N}$]
Random vector of population sizes at observation times,
$\mathbf{N}=\{N_{t_0},\dots,N_{t_q}\}$.

\item[$W(t)$] Wiener process.

\item[$r$] Instantaneous per capita growth rate in the SDE on the original scale (process parameter).

\item[$\sigma^2$] Environmental variance (process parameter).

\item[$\mu$] Drift in the log scale process, $\mu=r-\tfrac12\sigma^2$ (process parameter).

\item[$n_e$] Quasi-extinction threshold.

\item[$x_e$] Log threshold, $x_e=\log n_e$.

\item[$x_d$] Initial log distance to the threshold, $x_d=\log(n_0/n_e)>0$.

\item[$T$] First passage time (extinction time),
$T=\inf\{t\ge0\mid X(t)\le x_e\}$.

\item[$G(t\mid x_d,\mu,\sigma^2)$] Extinction probability by time $t$, $G(t)=\Pr[T\le t]$.

\item[$Q(t\mid x_d,\mu,\sigma^2)$] Survival probability by time $t$, $Q(t)=1-G(t)$.

\item[$w,z$] Dimensionless coordinates (for a generic $t>0$,
$w=(\mu t+x_d)/(\sigma\sqrt{t})$ and
$z=(-\mu t+x_d)/(\sigma\sqrt{t})$).

\item[$G(w,z)$] Extinction probability expressed in $(w,z)$ coordinates.

\item[$Q(w,z)$] Survival probability in $(w,z)$ coordinates, $Q(w,z)=1-G(w,z)$.

\item[$\widehat{\mu}$]
Maximum likelihood estimator of the drift parameter $\mu$.

\item[$\widehat{\sigma}^2$]
Maximum likelihood estimator of the environmental variance $\sigma^2$.

\item[$\widehat{\sigma}$]
Square root of $\widehat{\sigma}^2$.

\item[$\widehat{w},\,\widehat{z}$]
Plug-in estimators of $(w,z)$ evaluated at $t^\ast$,
obtained by replacing $(\mu,\sigma)$ with
$(\widehat{\mu},\widehat{\sigma})$.

\item[$\rho(w,z,k)$]
Correlation between the plug-in estimators $\widehat{w}$ and $\widehat{z}$.

\item[$k$]
Positive design-dependent constant depending on $(q,t_q,t^\ast)$
that determines the correlation structure of $(\widehat{w},\widehat{z})$.
$k=A/D$.

\item[$\delta$]
Noncentrality parameter of the noncentral-$t$ distribution.

\item[$t(\delta,\,q-1)$]
Noncentral $t$ distribution with noncentrality parameter $\delta$
and $q-1$ degrees of freedom.

\item[$S$]
Random variable following a noncentral $t(\delta,q-1)$ distribution.

\item[$s_{\mathrm{obs}}$]
Observed value of the noncentral-$t$ statistic $S$.

\item[$\delta_w,\,\delta_z$]
Noncentrality parameters for the sampling distributions of
$\widehat{w}$ and $\widehat{z}$.

\item[$(\underline{w},\,\overline{w})$]
Lower and upper bounds of an equal-tailed confidence interval for $w$.

\item[$\underline{\delta}_{\mathrm{obs}},\,\overline{\delta}_{\mathrm{obs}}$]
Lower and upper solutions for the noncentrality parameter
used to construct the confidence interval for $w$.

\item[$(\underline{z},\,\overline{z})$]
Lower and upper bounds of an equal-tailed confidence interval for $z$.

\item[$\mathrm{CI}_G$]
Confidence interval for the extinction probability $G(w,z)$
constructed by the $w$--$z$ method,
$\mathrm{CI}_{G}=\bigl(G(\overline{w},\underline{z}),\;
G(\underline{w},\overline{z})\bigr)$.

\item[$G_{\max}$]
Configuration-specific value of the extinction probability $G$ at which the CI width (equivalently, the variance of $\widehat G$) is maximized, outside the exceptional two-peak case.

\item[$\Delta(G)$]
Effect-size index defined as the distance from $G$ to $G_{\max}$,
$\Delta(G)=|G-G_{\max}|$.

\item[$G_{\mathrm{target}}$]
Management threshold for the extinction probability $G(w,z)$.

\item[$\mathcal C$]
Long-run variance (LRV) of $\{U_i\}$,
$\mathcal C:=\sum_{j=-\infty}^{\infty} C_j$.

\item[$C_j$]
Lag-$j$ autocovariance of $\{U_i\}$,
$C_j:=\mathbb E[U_i\,U_{i-j}]$.

\item[$U_i$]
Standardized drift-corrected increment (per unit time),
$U_i:=(\Delta Y_i-\widehat{\mu}_{\mathrm{naive}}\,\tau_i)/\sqrt{\tau_i}$.

\item[$\widetilde{\mathcal C}^{(\varepsilon)}_{\mathrm{NW}}(J)$]
Bartlett (Newey--West) HAC estimator of the residual long-run variance (LRV)
based on the pre-whitened residuals $\{\tilde \varepsilon_i\}$,
with lag truncation $J$.

\item[$\widetilde{\mathcal C}_{\mathrm{NW}}(J)$]
Bartlett (Newey--West) HAC estimator of the LRV $\mathcal C$ for $\{U_i\}$,
with lag truncation $J$, obtained from the pre-whitened estimate via
$\widetilde{\mathcal C}_{\mathrm{NW}}(J):=
\widetilde{\mathcal C}^{(\varepsilon)}_{\mathrm{NW}}(J)/(1-\tilde\rho_{\mathrm{pw}})^2$.

\item[$\widetilde{\sigma}^2$]
OEAR diffusion-scale estimate (per unit time),
$\widetilde{\sigma}^2:=\widetilde{\mathcal C}_{\mathrm{NW}}(J)$.

\item[$J$]
Lag truncation (bandwidth) used in the HAC estimator.

\item[$\Phi(\cdot)$] Standard normal cumulative distribution function.

\item[$\phi(\cdot)$] Standard normal density,
$\phi(x)=\tfrac{1}{\sqrt{2\pi}}\exp(-x^2/2)$.

\item[$\alpha$] Significance level (confidence level $1-\alpha$).

\end{description}

\subsection*{Appendix notation}

\subsubsection*{Numerical Stability of \texorpdfstring{$G(w, z)$}{G(w, z)}}

\begin{description}

\item[$\texttt{DBL\_MAX}$]
Largest finite IEEE 754 double precision number; used in overflow checks.

\item[$\texttt{DBL\_TRUE\_MIN}$]
Smallest positive (subnormal) IEEE 754 double precision number; used in underflow checks.

\item[$R(z)$]
Mills ratio for $z>0$, defined by $R(z)=\Phi(-z)/\phi(z)$.

\item[$R_7(z)$]
Eight term truncation of the asymptotic series for $R(z)$.

\item[$z_{\mathrm{thr}}$]
Switch threshold for using the Mills ratio approximation.

\item[$G_{\mathrm{lin}}(w,z;z_{\mathrm{thr}}),\ Q_{\mathrm{lin}}(w,z;z_{\mathrm{thr}})$]
Double precision evaluations on the linear scale, using the $z_{\mathrm{thr}}$ switch.

\item[$G_{\mathrm{lin,hyb}}(w,z)$]
Hybrid linear scale return value that uses the complementary evaluation via $Q_{\mathrm{lin}}$ when $w<0$.

\item[$\log G_{\mathrm{log}}(w,z;z_{\mathrm{thr}}),\ \log Q_{\mathrm{log}}(w,z;z_{\mathrm{thr}})$]
Double precision evaluations on the log scale (returned as logarithms), using the same $z_{\mathrm{thr}}$ switch.

\item[$a,b,c$]
Log components used in stable transforms:
$a=\log\Phi(-w)$,
$b=(z^2-w^2)/2+\log\Phi(-z)$,
$c=\log\Phi(w)$.

\item[$b_7$]
Large $z$ log component used with $R_7(z)$,
$b_7=\log\phi(w)+\log R_7(z)$.

\item[$G_{\mathrm{ref}},\ Q_{\mathrm{ref}}$]
High precision reference values used for validation.

\item[$\vartheta_0,\vartheta_1$]
Thresholds used for loss of information screening on the probability scale.

\item[$\mathrm{LOI}_0,\ \mathrm{LOI}_1$]
Loss of information flags on the probability scale, based on $\vartheta_0$ and $\vartheta_1$.

\item[$\mathrm{LOI}_{\mathrm{LOG0}},\ \mathrm{LOI}_{\mathrm{LOG\infty}}$]
Loss of information flags on the log scale (values at $0$ or $-\infty$, plus non-finite cases treated as boundary values).

\item[$\varepsilon_{\mathrm{lin}},\ \varepsilon_{\mathrm{log}}$]
Relative error metrics for linear scale and log scale evaluations.

\item[$d_{\mathrm{lin}},\ d_{\mathrm{log}}$]
Digits of accuracy derived from $\varepsilon_{\mathrm{lin}}$ and $\varepsilon_{\mathrm{log}}$.

\item[$G_{\mathrm{thr}}$]
Fixed probability cutoffs for IUCN Criterion E categories,
$G_{\mathrm{thr}}\in\{0.5,0.2,0.1\}$.

\item[$\operatorname{logit}(p)$]
Logit transform, $\operatorname{logit}(p)=\log\{p/(1-p)\}$ for $p\in(0,1)$.

\end{description}

\subsubsection*{Asymptotic Behavior of \texorpdfstring{$G(w,z)$}{G(w,z)} as \texorpdfstring{$z \to \pm\infty$}{z to +/- infinity}}

\begin{description}

\item[$g_0$]
Constant extinction–risk level defining an equal-risk contour,
$G(w,z)=g_0$, with $g_0\in(0,1)$.

\end{description}

\subsubsection*{Slope of Equal-Risk Contours for \texorpdfstring{$w + z > 0$}{w + z > 0}}

\begin{description}

\item[$\ell(w,z)$]
Slope of the equal-risk contour $G(w,z)=g_0$,
defined by implicit differentiation as
$\ell(w,z)=\mathrm{d}z/\mathrm{d}w$.

\item[$h(z)$]
Auxiliary function used in the slope comparison,
$h(z)=\phi(z)-z\,\Phi(-z)$.

\end{description}

\subsubsection*{Details of the Sampling Correlation of \texorpdfstring{$\widehat{w}$}{w-hat} and \texorpdfstring{$\widehat{z}$}{z-hat}}

\begin{description}

\item[$m$] Reparameterized drift statistic, $m=t^\ast\widehat{\mu}$.

\item[$\bar m,\,\sigma_m^2$] Mean and variance of $m$, $\bar m=\mathbb{E}[m]=t^\ast\mu$ and $\sigma_m^2=\operatorname{Var}(m)=t^{\ast2}\sigma^2/t_q$.

\item[$\eta$] Inverse scale statistic, $\eta=\widehat{\sigma}^{-1}$.

\item[$\sigma_\eta^2$] Variance of $\eta$.

\item[$\Xi$] Chi-squared variable, $\Xi=q\,\widehat{\sigma}^2/\sigma^2$.

\item[$A,D$] Design-dependent constants used to express
$\operatorname{Var}(\widehat w)$, $\operatorname{Var}(\widehat z)$, and
$\operatorname{Cov}(\widehat w,\widehat z)$. Ratio $k=A/D$.

\item[$\psi_q$] Ratio of gamma functions used in bounds,
$\psi_q=\Gamma((q-2)/2)/\Gamma((q-1)/2)$.
\end{description}

\subsubsection*{Monte Carlo evaluation of confidence interval methods}

\begin{description}

\item[$n_{\mathrm{MC}}$]
Number of Monte Carlo replicates generated for each parameter setting.

\item[$U,V$]
TMU coordinates,
$U=-\mu\sqrt{t^\ast}/\sigma$ and $V=x_d/(\sigma\sqrt{t^\ast})$.

\item[$\widehat U,\,\widehat V$]
Plug-in estimators of the TMU coordinates $(U,V)$ evaluated at $t^\ast$.

\item[$z_{1-\alpha/2}$]
Standard normal critical value, $z_{1-\alpha/2}=\Phi^{-1}(1-\alpha/2)$.

\item[$H$]
Logit transform of extinction probability, $H=\log\{G/(1-G)\}$.

\item[$\widehat H$]
Plug-in estimate of $H$.

\item[$\mathrm{Var}(\widehat H)$]
Delta-method approximation to the variance of $\widehat H$.

\item[$B$]
Number of parametric bootstrap replicates.

\item[$\{(\mu^{(j)},\sigma^{2\,(j)})\}_{j=1}^B$]
Independent parametric bootstrap draws of $(\mu,\sigma^2)$.

\item[$\{G^{(j)}\}_{j=1}^B$]
Bootstrap replicate extinction probabilities computed from the bootstrap draws.

\item[$\mathrm{CI}^{\mathrm{PPB}}_G$]
Percentile parametric bootstrap confidence interval for $G$, defined by the empirical $\alpha/2$ and $1-\alpha/2$ quantiles of the bootstrap values $\{G^{(j)}\}_{j=1}^B$.

\end{description}

\subsubsection*{Confidence-interval width as a function of data and model parameters}

\begin{description}

\item[$\Sigma$]
Covariance matrix of $(\widehat w,\widehat z)$.

\item[$\Sigma_{ij}$]
$(i,j)$ entry of $\Sigma$, for $i,j\in\{1,2\}$.

\item[$\upsilon(q)$]
Positive scalar factor in $\Sigma$ that depends only on $q$.

\item[$\|v\|_{\Sigma}$]
$\Sigma$-norm of $v\in\mathbb R^2$, defined by $\|v\|_\Sigma=\sqrt{v^\top\Sigma v}$.

\item[$\nabla G(w,z)$]
Gradient of $G$ with respect to $(w,z)$, $\nabla G(w,z)=(G_w,G_z)^\top$.

\item[$h_G(w,z)$]
Wald half width based on the delta method,
$h_G=z_{1-\alpha/2}\,\|\nabla G(w,z)\|_{\Sigma}$.

\item[$s$]
Rescaling factor $s=\sqrt{t^\ast/t_q}$.

\item[$Z_0$]
Standard normal variable, $Z_0\sim\mathcal N(0,1)$.

\item[$\Lambda$]
Positive random scale in the mixture representation,
$\Lambda=\sigma\widehat\sigma^{-1}=\sqrt{q/\chi^2_{q-1}}$, independent of $Z_0$.

\item[$\lambda$]
Realization of the random scale $\Lambda$ in the mixture representation;
a fixed positive scalar used for conditioning, $\lambda>0$.

\item[$\mathcal M_j(\lambda;w,z)$]
Inner conditional moment functional (for $j=1,2$),\\
$\mathcal M_j(\lambda;w,z)=\mathbb E_{Z_0}\!\left[G\bigl(\lambda(w+sZ_0),\,\lambda(z-sZ_0)\bigr)^j\right]$.

\item[$\Phi_2(\cdot,\cdot;\varrho)$]
Bivariate standard normal cdf with correlation $\varrho$.

\item[$\varpi(\lambda)$]
One dimensional reduction parameter,
$\varpi(\lambda)=-\lambda w/\sqrt{1+\lambda^2 s^2}$.

\item[$\varrho(\lambda)$]
Induced correlation parameter,
$\varrho(\lambda)=\lambda^2 s^2/(1+\lambda^2 s^2)$.

\item[$\kappa$]
Distance from the admissible boundary, $\kappa=w+z>0$.

\item[$w_0$]
Unscaled counterpart of $w$, $w_0=(\mu t^\ast+x_d)/\sqrt{t^\ast}$.

\item[$z_0$]
Unscaled counterpart of $z$, $z_0=(-\mu t^\ast+x_d)/\sqrt{t^\ast}$.

\end{description}

\subsubsection*{Determination of the required observation span}
\begin{description}

\item[$P$] True extinction probability at $(w,z)$, $P=G(w,z)$.

\item[\normalfont $G_{\text{upper}}(t_q;w,z,\alpha)$]
Upper confidence bound for $G(w,z)$ at observation span $t_q$, defined by
$G_{\text{upper}}(t_q;w,z,\alpha)=G(w_L,z_U)$.

\item[$\delta_w,\,\delta_z$] Noncentrality parameters for the rescaled estimators.

\item[\normalfont $F_{t(\delta,\,t_q-1)}(\cdot)$]
CDF of the noncentral $t$ distribution with degrees of freedom $t_q-1$ and noncentrality parameter $\delta$.

\item[\normalfont $\delta_{\text{low}},\,\delta_{\text{high}}$]
Lower and upper solutions for the noncentrality parameter determined by
\[
F_{t(\delta_{\text{low}},\,t_q-1)}\!\left(\sqrt{\tfrac{t_q-1}{t^\ast}}\,w\right)=1-\tfrac{\alpha}{2},
\qquad
F_{t(\delta_{\text{high}},\,t_q-1)}\!\left(\sqrt{\tfrac{t_q-1}{t^\ast}}\,z\right)=\tfrac{\alpha}{2}.
\]

\item[$w_L,\,z_U$]
Lower and upper limits for $(w,z)$ induced by $(\delta_{\text{low}},\delta_{\text{high}})$.

\item[$t_q^\ast$]
Required observation span.

\end{description}

\subsubsection*{Observation error and robust inference: the OEAR estimator and implications for \texorpdfstring{$G(w,z)$}{G(w,z)}}

\begin{description}

\item[$Y_i$]
Observed log abundance at time $t_i$, modeled as the true latent log abundance plus additive observation error.

\item[$y_i$]
Realized value of $Y_i$.

\item[$E_i$]
Observation error on the log scale, assumed i.i.d.\ as $\mathcal N(0,\omega^2)$ and independent of the latent process.

\item[$\omega^2$]
Observation error variance on the log scale.

\item[$\check{x}_d$]
Plug-in distance to the extinction threshold, defined by $\check{x}_d:=y_q-x_e$
(here $x_e=0$ so $\check{x}_d=y_q$).

\item[$m_{q\mid q}$]
Filtered (Kalman) mean of $X(t_q)$ given $Y_0,\ldots,Y_q$.

\item[$\check{x}_d^{(m)}$]
Sensitivity-check plug-in distance based on $m_{q\mid q}$:
$\check{x}_d^{(m)} := m_{q\mid q} - x_e$.

\item[$\widehat{\mu}_{\mathrm{naive}}$]
Naive estimator of the drift, obtained by fitting the drift--Wiener model directly to the observed series while treating it as error-free.

\item[$\widehat{\sigma}^2_{\mathrm{naive}}$]
Naive estimator of the environmental variance, obtained under the same error-ignoring diffusion fit.

\item[$\widehat{w}_{\mathrm{naive}}$]
Plug-in estimate of the dimensionless coordinate $w$ from $(\check{x}_d,\widehat{\mu}_{\mathrm{naive}},\widehat{\sigma}^2_{\mathrm{naive}})$.

\item[$\widehat{z}_{\mathrm{naive}}$]
Plug-in estimate of the dimensionless coordinate $z$ from $(\check{x}_d,\widehat{\mu}_{\mathrm{naive}},\widehat{\sigma}^2_{\mathrm{naive}})$.

\item[$\widehat{\mu}_{\mathrm{SSM}}$]
Maximum-likelihood estimator of the drift obtained from a state-space model that explicitly accounts for observation error.

\item[$\widehat{\sigma}^2_{\mathrm{SSM}}$]
Maximum-likelihood estimator of the environmental variance obtained from the same state-space model.

\item[$\widehat{\omega}^2_{\mathrm{SSM}}$]
Maximum-likelihood estimator of the observation-error variance obtained from
the state-space model.

\item[$\widehat{w}_{\mathrm{SSM}}$]
Plug-in estimate of $w$ computed from $(\check{x}_d,\widehat{\mu}_{\mathrm{SSM}},\widehat{\sigma}^2_{\mathrm{SSM}})$.

\item[$\widehat{z}_{\mathrm{SSM}}$]
Plug-in estimate of $z$ computed from $(\check{x}_d,\widehat{\mu}_{\mathrm{SSM}},\widehat{\sigma}^2_{\mathrm{SSM}})$.

\item[$\widetilde{\mu}$]
OEAR drift quantity, equal in the baseline additive-observation-error setting to the naive drift estimate,
$\widetilde{\mu}:=\widehat{\mu}_{\mathrm{naive}}$.

\item[$\widetilde{\sigma}^2$]
OEAR estimator of the long-run variance (per unit time) of the growth increments, constructed by incorporating autocovariances (or, more generally, HAC weighting).

\item[$(\widetilde{w}, \widetilde{z})$]
OEAR plug-in $w$--$z$ coordinates obtained by combining $\widetilde{\mu}$ and $\widetilde{\sigma}^2$ in the $w$--$z$ map.

\item[$\widetilde{G}$]
OEAR plug-in estimate of the extinction probability, obtained by evaluating
$G(w,z)$ at the OEAR operating point $(\widetilde{w},\widetilde{z})$.

\item[$(w_{\mathrm{eff}},\,z_{\mathrm{eff}})$]
Effective $(w,z)$ operating point obtained by evaluating the $(w,z)$ transform at an inflated diffusion scale $\sigma_{\mathrm{eff}}$ (e.g., when observation error is absorbed into the diffusion estimate), while holding $(x_d,\mu,t^\ast)$ fixed:
\[
w_{\mathrm{eff}}=\frac{\mu t^\ast+x_d}{\sigma_{\mathrm{eff}}\sqrt{t^\ast}},
\qquad
z_{\mathrm{eff}}=\frac{-\mu t^\ast+x_d}{\sigma_{\mathrm{eff}}\sqrt{t^\ast}}.
\]

\item[$\varsigma$]
Shrinkage factor, $\varsigma:=\frac{\sigma}{\sigma_{\mathrm{eff}}}\in(0,1)$.

\item[$\mu_c$]
Critical drift value separating deterministic persistence from deterministic threshold crossing over the projection horizon.

\item[$\mathrm{Bias}$]
Bias of an estimator, defined as the difference between its expectation and the true parameter value.

\item[$\mathrm{se}$]
Standard error of an estimator, defined as the square root of its variance.

\item[$\mathrm{SB}_H$]
Standardized-bias index for the logit-transformed extinction probability $H=\log\{G/(1-G)\}$, defined as the bias of $\widehat H$ divided by its standard error.

\item[$\mathrm{SB}_{\sigma^2}$]
Standardized-bias index for the naive variance,
$\mathrm{SB}_{\sigma^2}=\mathrm{Bias}(\widehat\sigma^2_{\mathrm{naive}})/\mathrm{se}(\widehat\sigma^2)$.

\item[$X^{\mathrm{raw}}_i$]
Unshifted latent path generated from increments
$\Delta X_i\sim\mathcal N(\mu\tau,\sigma^2\tau)$, with
$X^{\mathrm{raw}}_0=0$ and
$X^{\mathrm{raw}}_i=\sum_{j=1}^i \Delta X_j$.

\item[$\mathbf V_y$]
Covariance matrix of the observation vector
$y:=(y_1,\ldots,y_q)^\top$
implied by the linear-Gaussian state-space model
\eqref{Seq:dwobs_sde}--\eqref{Seq:dwobs_obs}
under a diffuse treatment of the initial state;
a function of $(\mu,\sigma^2,\omega^2)$.

\item[$m_{i\mid i-1},\, m_{i\mid i}$]
Predictive and filtered means of $X(t_i)$ given
$Y_0,\ldots,Y_{i-1}$ and $Y_0,\ldots,Y_i$, respectively.

\item[$P_{i\mid i-1},\, P_{i\mid i}$]
Corresponding predictive and filtered variances of $X(t_i)$.

\item[$v_i$]
One-step-ahead innovation,
$v_i := y_i - m_{i\mid i-1}$.

\item[$F_i$]
Innovation variance,
$F_i := P_{i\mid i-1} + \omega^2$.

\item[$K_i$]
Kalman gain,
$K_i := P_{i\mid i-1}/F_i$.

\item[$L(\mu,\sigma^2,\omega^2)$]
Likelihood as a function of $(\mu,\sigma^2,\omega^2)$.

\item[$y_i^{\circ}(\mu)$]
Drift-centered observation,
$y_i^{\circ}(\mu):=y_i-\mu t_i$.

\item[$U_i$]
Standardized drift-corrected increment (per unit time),
$U_i:=(\Delta Y_i-\widehat{\mu}_{\mathrm{naive}}\,\tau_i)/\sqrt{\tau_i}$.

\item[$u_i$]
Realized value of $U_i$.

\item[$\mathcal C$]
Long-run variance (LRV) of $\{U_i\}$, $\mathcal C:=\sum_{j=-\infty}^{\infty} C_j$.

\item[$C_j$]
Lag-$j$ autocovariance of $\{U_i\}$, $C_j:=\mathbb E[U_i\,U_{i-j}]$.

\item[$\bar u$]
Sample mean of $\{u_i\}$, $\bar u:=q^{-1}\sum_{i=1}^q u_i$.

\item[$\tilde u_i$]
Centered series used for pre-whitening, $\tilde u_i:=u_i-\bar u$.

\item[$\rho_{\mathrm{pw}}$]
AR(1) pre-whitening coefficient.

\item[$\tilde\rho_{\mathrm{pw}}$]
OLS (conditional Gaussian ML) estimate of $\rho_{\mathrm{pw}}$.

\item[$\varepsilon_i$]
Innovation term in the AR(1) pre-whitening regression.

\item[$\tilde \varepsilon_i$]
Pre-whitened residual sequence,
$\tilde \varepsilon_1:=\tilde u_1$ and
$\tilde \varepsilon_i:=\tilde u_i-\tilde\rho_{\mathrm{pw}}\,\tilde u_{i-1}$ for $i\ge 2$.

\item[$\widetilde C^{(\varepsilon)}_j$]
Lag-$j$ sample autocovariance of the pre-whitened residuals $\{\tilde \varepsilon_i\}$.

\item[$\widetilde{\mathcal C}^{(\varepsilon)}_{\mathrm{NW}}(J)$]
Bartlett (Newey--West) HAC estimator of the residual long-run variance (LRV)
based on the pre-whitened residuals $\{\tilde \varepsilon_i\}$,
with lag truncation $J$.

\item[$\beta_{0,\mathrm{cpue}}$]
Intercept parameter in the CPUE observation model \eqref{Seq:dwobs_obs_cpue} on the log scale.

\item[$\beta_{1,\mathrm{cpue}}$]
Slope (CPUE nonlinearity) parameter in \eqref{Seq:dwobs_obs_cpue}; $\beta_{1,\mathrm{cpue}}<1$ corresponds to hyperstability and $\beta_{1,\mathrm{cpue}}>1$ to hyperdepletion.

\item[$y_e$]
Observation-scale threshold implied by \eqref{Seq:dwobs_obs_cpue},
$y_e:=\beta_{0,\mathrm{cpue}}+\beta_{1,\mathrm{cpue}}\,x_e$.

\item[$\varsigma_{\mathrm{cpue}}$]
CPUE-adjusted shrink factor in the $(w,z)$ plane.

\item[$\sigma_{\mathrm{eff,cpue}}^{2}$]
CPUE-adjusted effective diffusion scale under naive fitting.

\item[$\varsigma_{0}$]
Baseline shrink factor at $\beta_{1,\mathrm{cpue}}=1$,
$\varsigma_{0}=\left\{1+2(\omega^{2}/\sigma^{2})/\tau\right\}^{-1/2}$.

\end{description}

\subsubsection*{Extinction probability under relaxed process models: sensitivity of the \texorpdfstring{$w$--$z$}{w--z} method}

\begin{description}

\item[$\mathcal E(t)$]
Ornstein--Uhlenbeck (OU) environmental forcing process.
It represents colored (time-correlated) variation in the instantaneous log growth rate.

\item[$\gamma$]
OU mean-reversion rate in
$d\mathcal E(t)=-\gamma \mathcal E(t)\,dt+\sigma_{\mathcal E}\,dW_1(t)$;
the forcing correlation time is $1/\gamma$.

\item[$\beta_{\mathcal E}$]
Coupling coefficient on the OU forcing in the log process,
$dX(t)=\{\mu+\beta_{\mathcal E}\mathcal E(t)\}\,dt+\sigma\,dW_2(t)$.

\item[$\sigma_{\mathcal E}$]
Diffusion scale parameter in
$d\mathcal E(t)=-\gamma \mathcal E(t)\,dt+\sigma_{\mathcal E}\,dW_1(t)$.

\item[$W_1,\,W_2$]
Independent Wiener processes driving the OU forcing and the baseline diffusion in $X(t)$, respectively.

\item[$\mathcal E_i$]
Discrete-time forcing at the sampling times, $\mathcal E_i:=\mathcal E(i\tau)$.

\item[$\psi$]
Discrete-time AR(1) coefficient for $\mathcal E_i=\mathcal E(i\tau)$,
$\psi=\exp(-\gamma\tau)$.

\item[$\nu_i$]
Innovation term in the discrete-time AR(1) representation
$\mathcal E_i=\psi\,\mathcal E_{i-1}+\nu_i$.

\item[$\mathcal E_0$]
Initial forcing value, $\mathcal E_0:=\mathcal E(0)$.

\item[$G_{\mathrm{OU}}(t^\ast)$]
Finite-horizon extinction probability under the OU-forced generating model at $t^\ast$, with $\mathcal E(0)$ drawn from the stationary distribution of the OU forcing and $X(0)$ fixed at $x_d$.

\item[$\mathrm{Ratio}$]
Dimensionless forcing-strength index,
$\mathrm{Ratio}:=\beta_{\mathcal E}^{\,2}\mathrm{Var}\{\mathcal E(t)\}/\sigma^2
=\beta_{\mathcal E}^{\,2}\sigma_{\mathcal E}^2/(2\gamma\sigma^2)$.

\item[$I_i$]
Integrated OU forcing over the $i$th sampling interval,
$I_i:=\int_{t_{i-1}}^{t_i}\mathcal E(u)\,du$.

\item[$\xi_i$]
Standard normal innovation used in Monte Carlo generation,
$\xi_i\overset{\mathrm{i.i.d.}}{\sim}\mathcal N(0,1)$.

\item[$x_{\mathrm{sc}}$]
Scaled log-abundance state used in the OU-forcing reference computation,
$x_{\mathrm{sc}}:=X/\sigma$.

\item[$e_{\mathrm{sc}}$]
Scaled OU forcing state used in the OU-forcing reference computation,
$e_{\mathrm{sc}}:=\mathcal E/\sigma$.

\item[$\mu_{\mathrm{sc}}$]
Scaled drift parameter,
$\mu_{\mathrm{sc}}:=\mu/\sigma$.

\item[$\sigma_e$]
Scaled OU forcing diffusion coefficient,
$\sigma_e:=\sigma_{\mathcal E}/\sigma$.

\item[$x_{\max}$]
Far-field truncation point for the backward-equation computations.
The PDE is solved on a truncated domain $x_{\mathrm{sc}}\in[0,x_{\max}]$ for the OU-forced reference, with boundary condition
$G_{\mathrm{OU}}(t,x_{\max},e_{\mathrm{sc}})=0$;
for the 1D bridge reference, the domain is $x\in[0,x_{\max}]$ with $G_{\mathrm{Br}}(t,x_{\max})=0$.

\item[$e_{\max}$]
Truncation half-width for the OU forcing state in the 2D reference computation:
the PDE is solved on $e_{\mathrm{sc}}\in[-e_{\max},e_{\max}]$ with reflecting boundary condition
$\partial G_{\mathrm{OU}}/\partial e_{\mathrm{sc}}=0$ at $e_{\mathrm{sc}}=\pm e_{\max}$.

\item[$\lambda_{\mathrm{Br}}$]
Strength of density feedback (mean-reversion rate) in the bridge model
$dX(t)=\{\mu+\lambda_{\mathrm{Br}}(\theta-X(t))\}\,dt+\sigma\,dW(t)$.
Weak density feedback corresponds to small $\lambda_{\mathrm{Br}}$ (equivalently $\beta_1\to 1$ for fixed $\tau$).

\item[$\theta$]
Reference log level (feedback pivot) in the bridge model; the feedback term $\lambda_{\mathrm{Br}}(\theta-X)$ switches sign at $X=\theta$.

\item[$x_{\infty}$]
Mean-reversion level of the bridge model,
$x_{\infty}:=\theta+\mu/\lambda_{\mathrm{Br}}$.

\item[$\beta_1$]
Discrete-time persistence under equally spaced sampling with interval $\tau$,
$\beta_1:=\exp(-\lambda_{\mathrm{Br}}\tau)$.

\item[$\beta_0$]
Discrete-time intercept under equally spaced sampling with interval $\tau$,
$\beta_0:=(1-\beta_1)\,x_{\infty}$.

\item[$\sigma_\xi^{2}$]
Innovation variance in the exact AR(1) transition of the bridge model,
$\sigma_\xi^{2}:=\sigma^2\{1-\exp(-2\lambda_{\mathrm{Br}}\tau)\}/(2\lambda_{\mathrm{Br}})$ (replace $\tau$ by $\tau_i$ for unequal sampling).

\item[$G_{\mathrm{Br}}(t,x)$]
Finite-horizon extinction probability under the bridge model at horizon $t$, conditional on $X(0)=x$, with $G_{\mathrm{Br}}(t,x):=\Pr(T\le t\mid X(0)=x)$ and $T:=\inf\{u\ge 0:X(u)\le 0\}$.

\item[$G_{\mathrm{Br}}(t^\ast)$]
Reference probability at the operating point,
$G_{\mathrm{Br}}(t^\ast):=G_{\mathrm{Br}}(t^\ast,x_d)$.

\end{description}

\subsubsection*{Eel Harvests by Prefectures in Japan: Seed and Inland Fisheries}

\begin{description}

  \item[$\mathcal{H}^{\mathrm{pref}}_{i,t}$]
  Seed-fishery harvest (tonnes) for prefecture $i$ and year $t$ as tabulated in the prefectural tables, with symbol-coded entries treated as missing (this is the response used in the LME).

  \item[$\mathfrak{a}_t$]
  Year fixed effect in the LME (shared annual signal).

  \item[$\mathfrak{c}_{\mathrm{Domain}}$]
  Coastal/inland fixed effect in the LME.

  \item[$\mathfrak{b}_i$]
  Prefecture random intercept in the LME.

  \item[$\zeta_{i,t}$]
  LME residual term, with Domain-specific variance
  $\mathrm{Var}(\zeta_{i,t}\mid \mathrm{Domain})=\sigma^2_{\mathrm{Domain}}$
  (implemented via \texttt{varIdent} in \texttt{nlme::lme}).

  \item[$\mathcal{H}^{\mathrm{official}}_{t}$]
  Official national total (tonnes) for a given component in year $t$.
  When needed, the component is indicated explicitly, e.g.,
  $\mathcal{H}^{\mathrm{official}}_{\mathrm{coastal},t}$ (coastal seed),
  $\mathcal{H}^{\mathrm{official}}_{\mathrm{inland\ seed},t}$ (inland seed),
  and $\mathcal{H}^{\mathrm{official}}_{\mathrm{inland\ eel},t}$ (inland eel).

  \item[$D_t$]
  Discrepancy between the official national total and the sum of the prefectural tables for the same component in year $t$:
  $D_t=\mathcal{H}^{\mathrm{official}}_{t}-\sum_i \mathcal{H}^{\mathrm{pref}}_{i,t}$.
  This diagnostic is computed separately for each component (coastal seed, inland seed, and inland eel).

  \item[$\mathcal{H}^{\mathrm{LME}}_{\mathrm{coastal},t}$]
  LME-based coastal seed total in year $t$ (sum of back-transformed prefectural predictions over coastal prefectures).

  \item[$\mathfrak{s}$]
  Constant scaling factor applied to $\mathcal{H}^{\mathrm{LME}}_{\mathrm{coastal},t}$ to match the level of $\mathcal{H}^{\mathrm{official}}_{\mathrm{coastal},t}$.

  \item[$\check{\mathfrak{s}}$]
  Least-squares estimate of $\mathfrak{s}$ over 1957--2018.

  \item[$\mathcal{H}^{\mathrm{seed}}_{\mathrm{merged},t}$]
  Adopted national common-seed series used in PVA:
  $\mathcal{H}^{\mathrm{seed}}_{\mathrm{merged},t}
  =\check{\mathfrak{s}}\,\mathcal{H}^{\mathrm{LME}}_{\mathrm{coastal},t}
  +\mathcal{H}^{\mathrm{official}}_{\mathrm{inland\ seed},t}$.

\end{description}

\subsubsection*{Details of the PVA and sensitivity analyses for Japanese eel}

\begin{description}

\item[$\varphi_{\mathrm{seed}}$]
Annual proportional effort-trend rate (\%/year) used to rescale the glass-eel (seed) harvest index in the deterministic effort-trend sensitivity; negative values represent declining effort through time (historically larger effort in earlier years).

\item[$\varphi_{\mathrm{eel}}$]
Annual proportional effort-trend rate (\%/year) used to rescale the inland-eel (yellow and silver) harvest index in the deterministic effort-trend sensitivity.

\item[S0--S3]
Effort-trend scenario labels (Table~\ref{tab:eel_effort_scenarios}) specifying the pair $(\varphi_{\mathrm{seed}},\varphi_{\mathrm{eel}})$ used for rescaling.

\end{description}

\subsubsection*{Proof: Criterion A (decline subcriteria) overstates Criterion E for large populations}

\begin{description}

\item[$t_A$] Criterion~A assessment horizon (years), $t_A>0$.

\item[$t_E$] Criterion~E assessment horizon (years), $t_E>0$.

\item[$f_A$]
Criterion~A proportional threshold.
Criterion~A is triggered when $N(t_A)/N(0)\le f_A$, with $f_A\in(0,1)$.

\item[$x_d^\ast$]
Crossover log distance defined (numerically) by
$G(t_E\mid x_d^\ast,\mu,\sigma^2)=G_{\mathrm{thr}}$.

\item[$n_0^\ast$]
Corresponding crossover initial size on the original scale.
Under the convention $n_e=1$ (equivalently $x_e=0$),
$n_0^\ast=\exp(x_d^\ast)$.

\end{description}


\end{document}